\documentclass[a4paper,11pt]{article}
\pdfoutput=1

\usepackage{jcappub}

\usepackage[T1]{fontenc} 
\usepackage{lineno}
\usepackage{xcolor}
\usepackage{orcidlink}
\usepackage{ulem}
\usepackage{float}
\numberwithin{equation}{section}

\title{\boldmath 
Modeling Gravitational Wave Bias from 3D Power Spectra of Spectroscopic Surveys

}

\author[a,b,e]{Dorsa Sadat Hosseini
\orcidlink{0009-0008-2800-2725},}
\author[a,b,e]{Amir Dehghani \orcidlink{0009-0005-0395-9553},}
\author[c,d]{J. Leo Kim \orcidlink{0000-0001-8699-834X},}  
\author[e,a,*]{Alex Krolewski,}
\author[f]{Suvodip Mukherjee \orcidlink{0000-0002-3373-5236},}
\author[a,b,e]{and Ghazal Geshnizjani \orcidlink{0000-0002-2169-0579}}
\affiliation[a]{Perimeter Institute for Theoretical Physics, Waterloo, ON N2L 2Y5, Canada}
\affiliation[b]{Department of Applied Mathematics, University of Waterloo, Waterloo, ON N2L 3G1, Canada}
\affiliation[c]{Department of Physics, Engineering Physics \& Astronomy, Queen's University, Kingston ON K7L 3N6, Canada}
\affiliation[d]{Arthur B. McDonald Canadian Astroparticle Physics Research Institute, Kingston ON K7L 3N6, Canada}
\affiliation[e]{Waterloo Centre for Astrophysics, University of Waterloo, 200 University Ave W, Waterloo, ON N2L 3G1, Canada}
\affiliation[f]{Department of Astronomy \& Astrophysics, Tata Institute of Fundamental Research, Colaba, Mumbai 400005, India}
\affiliation[*]{CITA National Fellow}

\emailAdd{d2sadath@uwaterloo.ca}
\emailAdd{a8dehgha@uwaterloo.ca}
\emailAdd{leo.kim@queensu.ca}
\emailAdd{akrolews@uwaterloo.ca}
\emailAdd{suvodip@tifr.res.in}
\emailAdd{ggeshnizjani@pitp.ca}

\abstract{
We present a framework for relating gravitational wave (GW) sources to the astrophysical properties of spectroscopic galaxy samples. We show how this can enable using clustering measurements of GW sources to infer the relationship between the GW sources and the astrophysical properties of their host galaxies. We accomplish this by creating mock GW catalogs from the spectroscopic Sloan Digital Sky Survey (SDSS) DR7 galaxy survey. We populate the GWs using a joint host-galaxy probability function defined over stellar mass, star formation rate (SFR), and metallicity. This probability is modeled as the product of three broken power-law distributions, each with a turnover point 
motivated by astrophysical processes governing the relation between current-day galaxy properties and binary black hole (BBH) mergers, such as galaxy quenching and BBH delay time. Given that our analysis is anchored in the specific properties and selection characteristics of the adopted galaxy sample, as well as assumptions regarding the host-galaxy probability functions and BBH merger rate prescriptions, the resulting trends should be regarded as model-dependent.
Within this framework, our results show that GW bias is most sensitive to host-galaxy probability dependence on stellar mass, with increases of up to $\sim \mathcal{O}(10)\%$ relative to galaxy bias as the stellar mass pivot scale rises. We also find a notable relationship between GW bias and SFR: when the host-galaxy probability favors low-SFR galaxies, the GW bias significantly increases. In contrast, we observe no strong correlation between GW bias and metallicity. These findings suggest that the spatial clustering of GW sources is primarily driven by the stellar mass and SFR of their host galaxies and shows how GW bias measurements can inform models of the host-galaxy probability function.}

\begin{document}

\maketitle
\flushbottom

\section{Introduction}
The discovery of gravitational waves (GWs) from a network of GW detectors started a new era of multi-messenger astronomy, which has opened a new avenue to explore the Universe \cite{KAGRA:2013rdx, LIGOScientific:2014pky, VIRGO:2014yos, PhysRevD.88.043007, LIGOScientific:2016aoc, LIGOScientific:2018jsj, LIGOScientific:2020kqk, LIGOScientific:2021aug, KAGRA:2021duu}. The network of GW detectors operated by the LIGO-Virgo-KAGRA (LVK) collaboration targets different astrophysical sources, such as mergers of binary neutron stars (BNS), neutron star-black hole binaries (NSBH), and (stellar) binary black holes (BBHs), with BBHs being the most commonly detected \cite{LIGOScientific:2018mvr, LIGOScientific:2020ibl, LIGOScientific:2021djp}. 
Given that stellar black holes, as remnants of massive stars, are closely related to the history of cosmic star formation, their distribution and evolution could potentially provide valuable clues about the connection between the large-scale structure of dark matter and galaxy formation and evolution. For example, if reliably assessed, observations of BBH mergers could allow for improved estimation of essential cosmological parameters such as the expansion rate \cite{Schutz:1986gp, Oguri:2016dgk, LIGOScientific:2017adf, Chen:2017rfc, LIGOScientific:2018gmd, Mukherjee:2018ebj, DES:2019ccw, Mukherjee:2020hyn, Bera:2020jhx, Mukherjee:2022afz,LIGOScientific:2021aug,Afroz:2024joi,LIGOScientific:2018gmd, Chen:2017rfc,KAGRA:2013rdx} and key astrophysical processes such as the conversion of gas into stars within galaxies \cite{Artale_2019,Di_Carlo_2020,Santoliquido_2021,Vitale_2019}. 
To this end, this paper aims to build upon and extend our earlier research \cite{dehghani2024} by modeling GW sources as tracers of the underlying matter field. The connection between the two is subsequently evaluated using a related `bias' parameter.

This connection is expected as a consequence of galaxies hosting the GW sources and the relation between large-scale distribution of galaxies and dark matter \cite{Mukherjee_2021, Mukherjee:2019oma}. That is, gravitational instability enhances the initial small perturbations, forming bound and virialized structures known as dark matter halos, which preferentially occupy the peaks in the dark matter distribution \cite{1972ApJ...176....1G,2001MNRAS.323....1S}. The degree to which these halos trace the underlying matter distribution is encapsulated by the halo bias parameter
\cite{1986ApJ...304...15B,1996MNRAS.282..347M,Tinker_2010}. Then galaxies form within the densest regions of dark matter halos and serve as robust tracers of dark matter halos, and therefore tracers of the underlying matter density field \cite{Wechsler:2018pic,Benson_2010,1977MNRAS.179..541R,1977MNRAS.179..541R,1993MNRAS.264..201K,1994MNRAS.271..781C}. The relationship that quantifies the galaxy-halo connection is known as galaxy bias and provides information that ranges from basic estimates of host halo mass to more intricate phenomena such as assembly bias and velocity bias \cite{Benson:1999mva, Sheth:1999mn, Cole2000, Peacock:2000qk, ValeOstriker04, Gao:2006qz, Croton:2006ys, Dalal08, Moster_2010, Tinker_2010, Behroozi_2013, Behroozi2019}. Next, within galaxies, neutron stars and black holes arise from the collapse of massive evolved stars, and therefore their populations and merger rates are linked to the ancestral distribution of stars in galaxies. 

The spatial distribution of GW events is additionally expected to be influenced by their formation mechanisms and environment. This includes the physical properties of the host galaxy, such as stellar mass, metallicity, and the SFR, which are also intrinsically linked to the evolution of the underlying dark matter halo \citep{Belczynski_2002, Neijssel2019, O_Shaughnessy_2010, Dominik:2014yma, Mapelli_2017, 2018MNRAS.474.2959G, Cao2018, Fishbach_2018, Santoliquido_2022, Dominik:2012kk, Toffano_2019, Artale_2019, McCarthy_2020,Afroz:2024fzp,2017PhRvD..95l4046G,2019PhRvL.123r1101Y,2020ApJ...893...35D}. Understanding the properties of binary systems such as BBHs from first principles requires detailed modeling of their formation pathways
\cite{Belczynski_2002,Neijssel2019}. 
From the initial stellar population to the eventual coalescence of binary systems \cite{Belczynski_2002}, numerous astrophysical mechanisms, such as stellar winds, mass accretion, supernova kicks, and common envelope evolution, play critical roles \cite{Belczynski_2002,Neijssel2019,Dominik:2012kk}. However, a phenomenological framework for studying the relationship between compact objects and their host environments can offer a complementary way to assess the cosmic evolution of merger rates, the formation channels of binary systems, and their dependence on key astrophysical properties of host galaxies \cite{Mukherjee:2021bmw,Namikawa16, Mukherjee:2018ebj,Mukherjee:2019wcg, Mukherjee:2019oma, Calore:2020bpd, Mukherjee:2020hyn, Mukherjee:2020mha, Scelfo20, Libanore21, Diaz:2021pem, Libanore22, Gagnon23}.

As proposed in our previous paper \cite{dehghani2024}, we study this connection through the GW bias parameter, which measures the relationship between the clustering of GW sources and the underlying density contrast of the matter field. This is an especially promising program given the observational forecast of GW sources in the next generation of GW detectors \cite{Iacovelli_2022, Hild_2011,Abbott_2017,Maggiore_2020}. In \cite{dehghani2024}, we examined the theoretical GW bias parameter based on simulated catalogs of GW sources by populating photometric all-galaxy surveys (2MPZ and WISC). We used a phenomenological model in which the GW host-galaxy probability function only depended on stellar mass.

We then measured the bias of these synthetic GW sources using the angular power spectrum and studied the impact of variations in the host-galaxy probability function. We can thus assess the prospects for using GW bias measurements to better understand the relationship between GWs and their host galaxies.
In this work, we extend that analysis by populating the simulated GW sources into a spectroscopic galaxy survey, the Sloan Digital Sky Survey (SDSS) DR7, which provides detailed information for each galaxy, including stellar mass, SFR, and stellar metallicity. This allows for more complex host-galaxy probability functions that depend simultaneously on all three properties. Moreover, spectroscopic galaxy surveys offer more accurate redshift measurements and a better ability to measure galaxy clustering via the three-dimensional power spectrum. Although this simulated three-dimensional GW catalog is further from what will be available observationally both in terms of number of mergers and sky localization, it allows for a more detailed theoretical analysis of the relationship between the GW sources and the matter field. These assumptions should be interpreted as a theoretical idealization adopted to isolate the intrinsic clustering signal, rather than as a realistic prediction for observed GW source catalog.

The structure of this paper is as follows.
In Section \ref{sec:basic of GW bias}, we review the theoretical GW bias parameter and also summarize the phenomenological framework employed in this study. 
Section \ref{sec:galaxy catalog} presents the spectroscopic galaxy survey used in this analysis, the Sloan Digital Sky Survey (SDSS DR7) Main Galaxy Sample (MGS).
In Section \ref{sec:BBHs}, we describe in detail the host-galaxy probability functions which determine how GW sources are populated within observed galaxies based on stellar mass, SFR and metallicity.
Section \ref{sec:3D power spectra} describes the framework for evaluating the 3D galaxy power spectrum of the mock GW sources. 
In Section~\ref{sec:results}, we measure the best-fit value of the GW bias parameter and show how it is sensitive to variations in the host-galaxy probability function.
Finally, in Section \ref{sec:conclusion}, we summarize our findings and offer concluding remarks.



\section{Modeling GW bias parameter }
\label{sec:basic of GW bias}

\subsection{GW bias parameter}
\label{about GW bias}
One approach to characterizing the way a tracer population tracks another density field is by introducing a bias parameter. For instance, the galaxy bias quantifies the over-density of galaxies, $\delta_g$ with respect to the over-density of the underlying distribution of matter, $\delta$. The over-density for galaxies can be generally expressed as \cite{Desjacques_2018}
\begin{equation}
    \delta_g(x,\tau) = \sum_{W} b_{W}(\tau) W(x,\tau)\label{eq:biasdef1},
\end{equation}
where $W$ is a set of local and non-local operators on the matter over-density, with each having a corresponding bias parameter $b_W$ -- for example,
powers of the density and tidal field, time derivatives of the tidal field and higher-derivative terms. The simplest model for galaxy bias, which is a good approximation on large scales, can be described through a linear and local relationship as

\begin{equation}
    \delta_g=b_g \delta_m, \label{eq:biasdef2}
\end{equation}
where $\delta_g$ and $\delta_m$ are the galaxy and matter over-density respectively and $b_g $ represents a spatially constant galaxy bias. In this study we also work with a spatially constant bias approximation but in general and particularly on small scales, this relation can be non-local as well.\footnote{In our previous study \cite{dehghani2024}, we explored some potential scale dependence of the GW bias in the 2MPZ and WISC surveys and found no significant dependency. In this work, due to small-scale redshift-space distortions, it would be harder to investigate such scale-dependence as our simple empirical RSD model would break down at smaller scales. Given that the RSD effects likely dominate and obscure any potential scale dependencies in the bias parameter, we do not further investigate the scale dependence in this study.} Note if $b_g$ is constant the relation \eqref{eq:biasdef2} remains local both in Fourier space and real space. Therefore, the bias parameter relates the power spectrum of the galaxies $P_g(k,z)$ to the matter power spectrum $P_m(k,z)$ as
\begin{equation}
   P_g(k,z)=  b_g^2 P_m(k,z).
   \label{eq:galbias}
\end{equation}
Similarly, we can define the GW bias parameter to describe how effectively the spatial distribution of the GW sources follows the underlying matter distribution through
\begin{equation}
    \delta_{GW}\equiv b_{\rm GW} \delta_m\,,
\end{equation}
where $\delta_{GW}$ is the over-density field of GW sources, and $b_{\rm GW}$ is the GW bias parameter. Once again this can be written in terms of the power spectra as
\begin{equation}
   P_{GW}(k,z)=  b^2_{GW} P_m(k,z),
\label{eq:gwbias}
\end{equation}
where $P_{GW}(k,z)$ is the corresponding power spectrum of the GW sources. 

In what follows, our approach to determine the GW bias parameter is to fit a phenomenological model to the redshift-space power spectrum of the mock host galaxies catalog for GW sources, consisting of a linear galaxy bias, linear redshift space distortions, and a Finger-of-God velocity dispersion parameter. More details are given in Section \ref{sec:RSD effect}.

\subsection{Phenomenological framework for measuring GW bias parameters using mock GW source catalogs generated from galaxy surveys}
\label{sec:Phenomenological approach}
\begin{figure}
    \centering
    \includegraphics[width=0.95\textwidth]{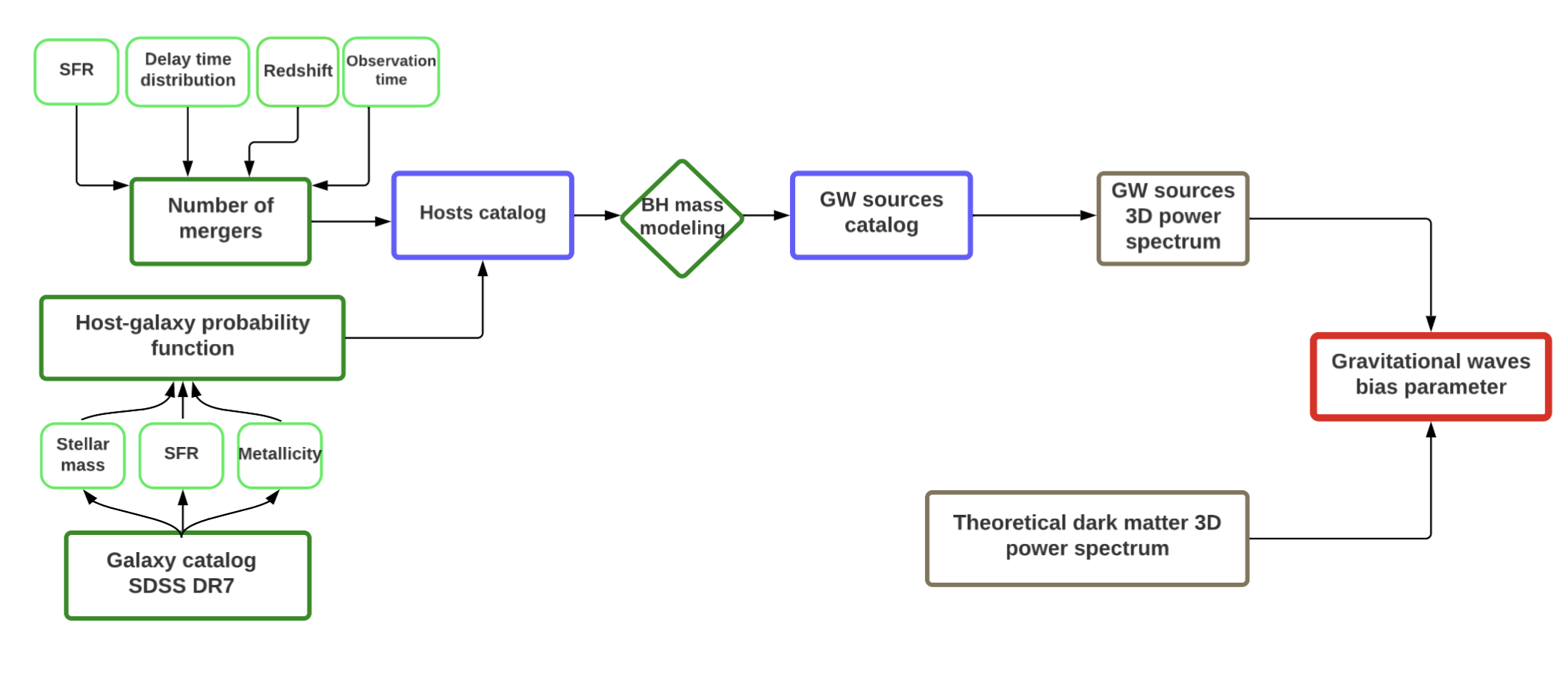}
    \caption{Flowchart summarizing the framework for creating mock GW sources in the SDSS DR7 galaxy sample, and for measuring the GW bias parameter for different parameters of the host-galaxy probability function.
    }
    \label{fig:flowchart}
\end{figure}

In this study, we adopt a \emph{phenomenological} approach to model and compute the GW bias parameter, which is schematically shown in the flow chart in Figure \ref{fig:flowchart}.
Our measurement of the GW bias is based on two essential components: the 3D power spectrum of the distribution of the BBH mergers\footnote{Based on current observations, the majority of GW merger events are expected to originate from BBHs, which form the primary focus of this study. However, it is worth noting that our framework can be readily extended to incorporate binary neutron star (BNS) and neutron star-black hole (NSBH) systems as well.} and the theoretical 3D power spectrum of matter density. We use CAMB (Code for Anisotropies in the Microwave Background) to compute the matter power spectrum with details provided in Section \ref{sec:3D power spectra}.

To calculate the power spectrum of the BBH mergers, we create a mock catalog by selecting their host galaxies from the spectroscopic galaxy survey, SDSS DR7. The initial step involves determining the expected number of mergers occurring at a particular redshift, which depends on the astrophysical processes governing the formation of BBHs, including the properties of the host galaxy. More specifically, the merger rate depends on redshift, the observing time of a detector,the star formation history of the host galaxy and the delay time between the formation of a BH and the subsequent coalescence of two BHs (explained in Section \ref{sec:Num_merger}). As we elaborate later, within our approximations a higher merger rate is essential for reducing the error on the clustering measurement, but it does not directly correlate with the magnitude of the bias parameter.

Next, we used phenomenologically inspired host-galaxy probabilities based on astrophysical properties of the galaxies to select galaxies hosting a BBH merger from the galaxy survey. Taking advantage of the information in the spectroscopic galaxy catalog SDSS DR7, we consider host-galaxy probability functions that can depend on stellar mass, SFR, and metallicity (see Section \ref{sec:assignment}). 
 To compare with our result for photometric redshift surveys \cite{dehghani2024} we first apply the same host-galaxy probability function as that work, which depends only on galaxy stellar mass, and then we let the host-galaxy probability functions vary with SFR and metallicity as well. This three-dimensional selection function enables a more realistic exploration of how GW sources depend on astrophysical properties, extending beyond just the host galaxy's stellar mass into additional, non-degenerate parameter dimensions. Using these probabilities, we then produce mock sirens/host galaxy catalogs. 

After creating the mock siren catalog, we calculate the best-fit GW bias parameter using the galaxy and siren power spectra. We then examine (Section \ref{sec:results}) how the GW bias changes in relation to the parameters governing the dependence of host probability relation to astrophysical properties and the redshift.

A key advantage of this phenomenological approach is its ability to tie black holes directly to observed galaxy characteristics, such as stellar mass, SFR, and metallicity, thereby avoiding uncertainties introduced by the sub-grid astrophysical models used in galaxy simulations. Furthermore, observed galaxy surveys such as SDSS DR7 cover large volumes, allowing robust statistical analyses of galaxy clustering across extensive datasets. In contrast, cosmological hydrodynamical simulations \cite{2022NatAs...6..897S,Nagamine_2021,Naab_2017}, while offering detailed insights into gas properties within galaxies and being more complete (in mass) than observational surveys, are limited to much smaller volumes, which makes large-scale studies of galaxy clustering and its connection to gravitational waves more challenging. Spectroscopic surveys such as SDSS DR7 also enable more precise clustering measurements and allow for the direct determination of galaxy properties like metallicity and SFR. 
Although these surveys are smaller in scale compared to photometric surveys and also biased toward brighter galaxies, the constraints due to selection biases and incompleteness in this work are carefully addressed to minimize the corresponding effects. Having considered these factors, it is worth noting that incorporating simulations instead of observational data in future research could provide complementary insights.

\section{Galaxy catalog}
\label{sec:galaxy catalog}
The Sloan Digital Sky Survey (SDSS DR7) includes the full SDSS-I and SDSS-II  data sets. In these surveys, CCD imaging \cite{1998AJ....116.3040G,2006AJ....131.2332G} was obtained in five bands $(u, g, r, i, z)$ across 11663 deg$^2$, with a uniform photometric calibration described in Ref. \cite{padmanabhan2008improved}. From this data set, a subset of galaxies within an area of $9,380$ deg$^2$ \cite{2009ApJS..182..543A} underwent spectroscopic follow-up as a part of the MGS \cite{2002AJ....124.1810S}. The MGS selection criteria approximately encompass all galaxies with an extinction-corrected r-band Petrosian magnitude $(r_{\textrm{pet}})$ below 17.77 \cite{schlegel1998maps,2002AJ....124.1810S}.
The NYU Value-Added Galaxy Catalog (VAGC) was created to catalog galaxy properties using SDSS DR7 redshifts and spectra \cite{2005AJ....129.2562B}. The NYU-VAGC ``safe0'' galaxy sample specifically includes galaxies with a range of $14.5 < r_{\textrm{pet}} < 17.6$, which cover a total of 7356 deg$^2$. By considering $r_{\textrm{pet}}=14.5$ as the lower limit, it is guaranteed that galaxies with trustworthy SDSS photometry are included, while the upper limit enables us to make a homogeneous selection across the entire area of 7356 deg$^2$ \cite{2005AJ....129.2562B}. For the galaxies with $r_{\textrm{pet}} < 14.5$, saturation in SDSS images can lead to inaccurate galaxy photometry, potentially affecting measurements for the most luminous galaxies.

We use the SDSS DR7 sample because it provides a simple and complete selection of galaxies and robust measurements of their astrophysical properties. Since GW sources can occur in any galaxy, a magnitude selected sample is needed rather than the more complicated color selections used in, e.g.\ BOSS and eBOSS.  The large sky area and volume covered by SDSS-MGS is also desirable, and the similar redshift range to the samples in \cite{dehghani2024} allows us to easily compare results.

\subsection{Sampling from SDSS DR7}
\begin{figure}
\centering 
\includegraphics[width=0.49\textwidth]{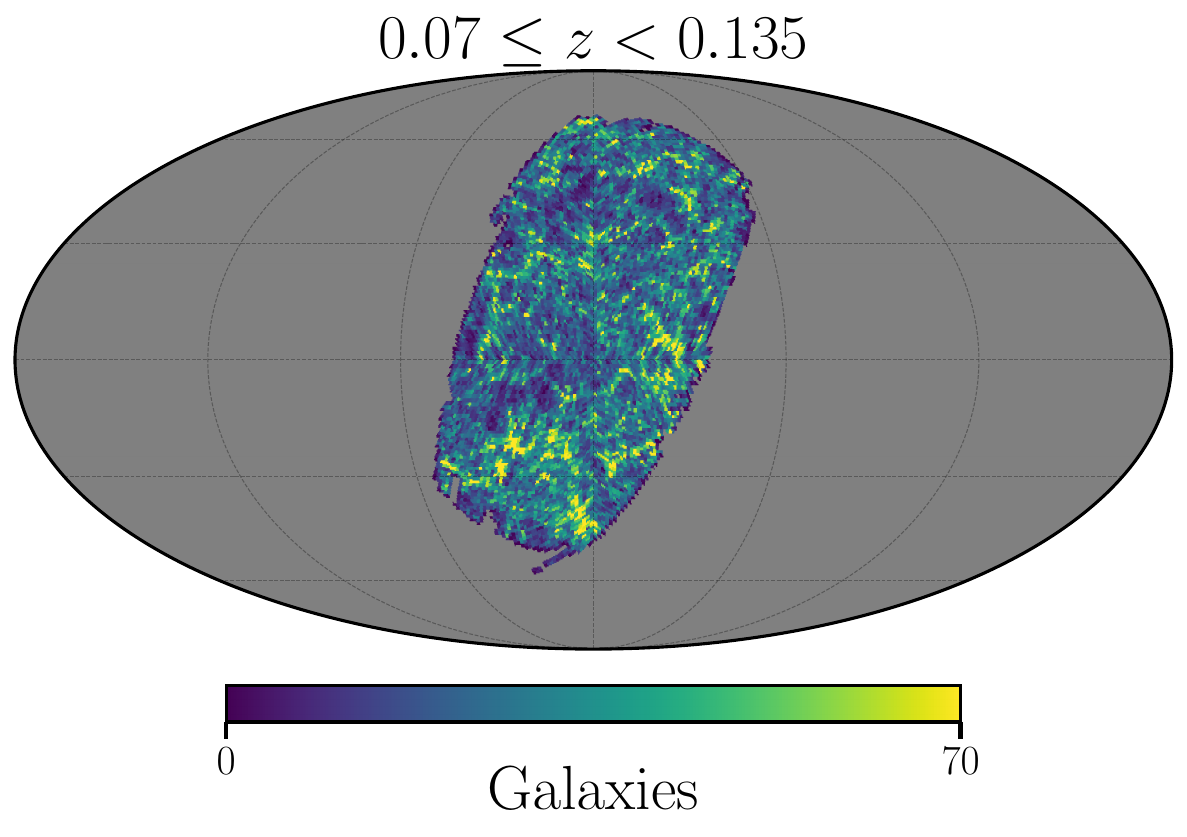}
\includegraphics[width=0.49\textwidth]{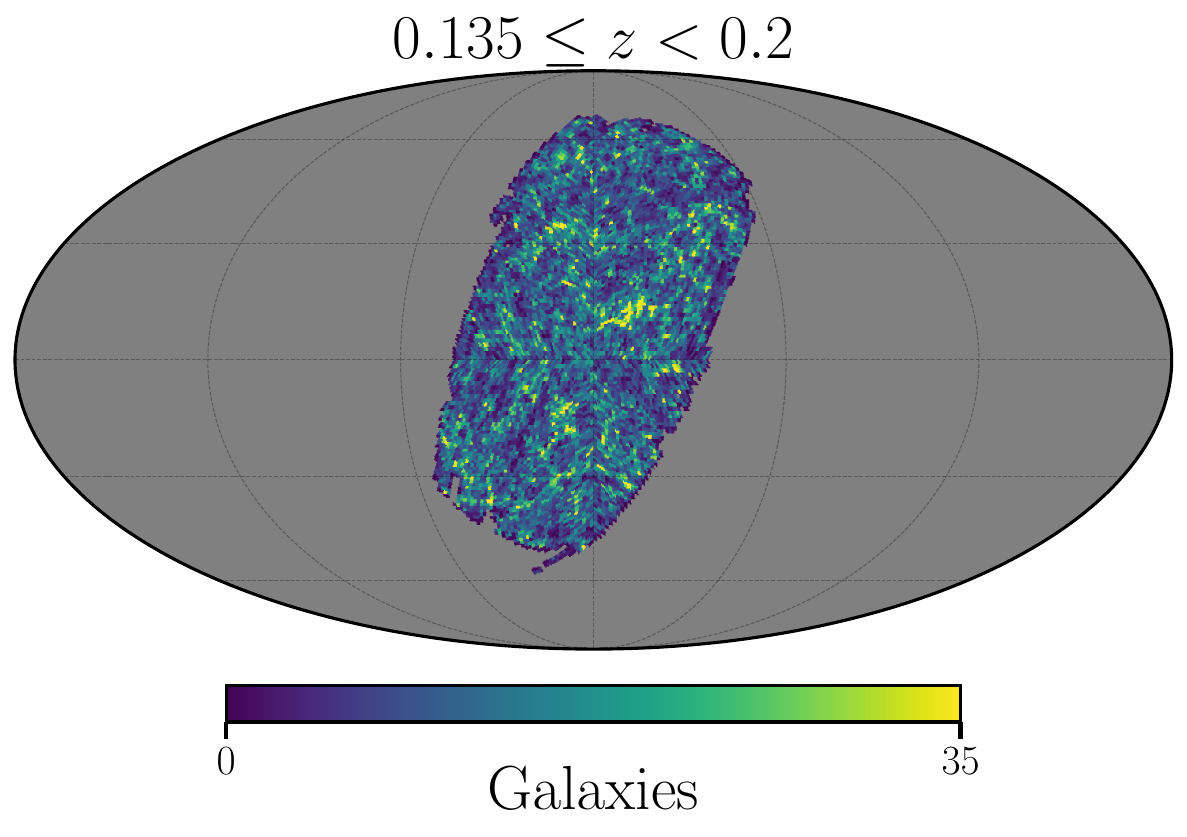}
\includegraphics[width=0.49\textwidth]{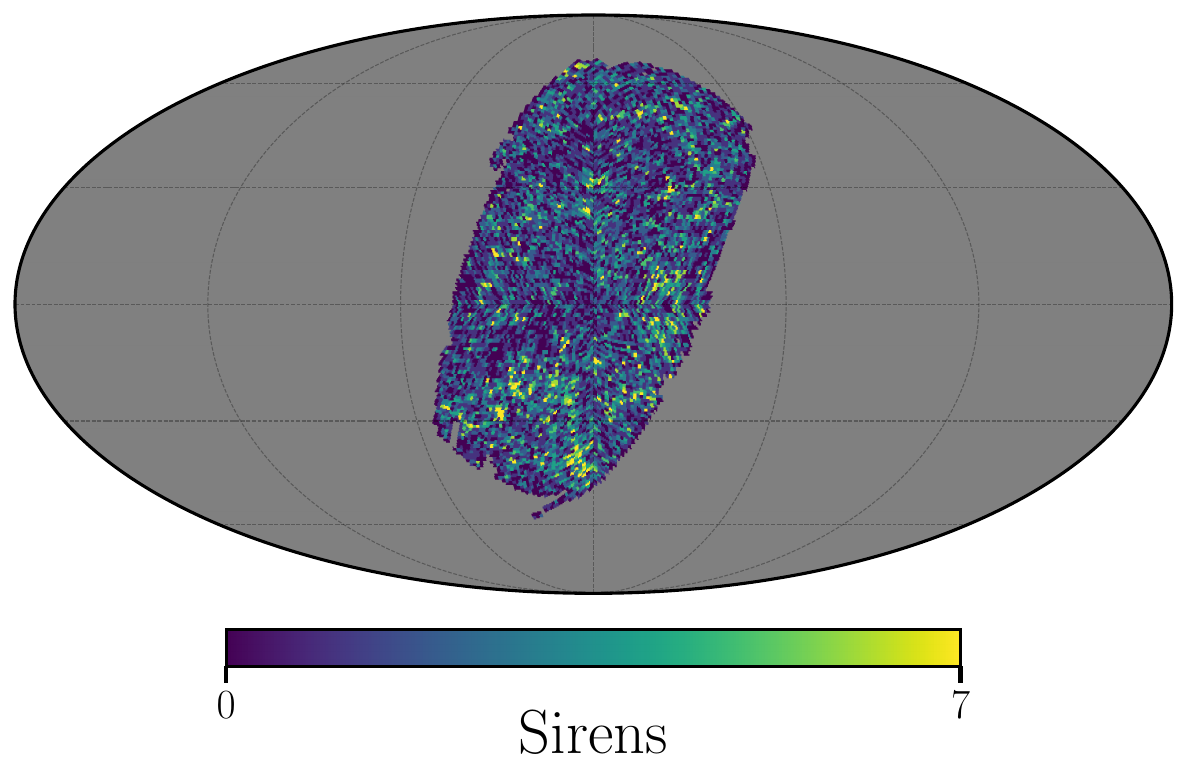}
\includegraphics[width=0.49\textwidth]{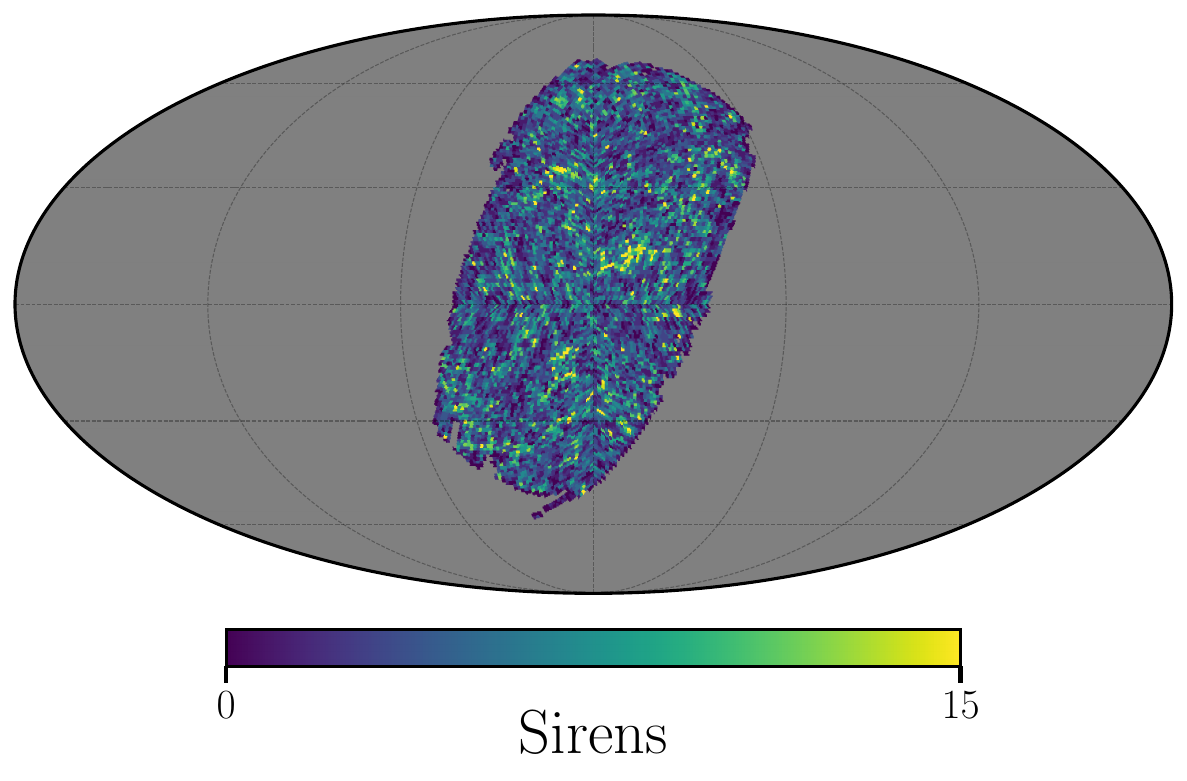}
\caption[scale=0.1]{HEALPix map of the SDSS DR7 galaxy catalogs (top panel) in two different redshift bins $0.07<z \leq 0.135$ (left) and  $0.135 \leq z<0.2$ (right). Bottom panels show maps of mock siren catalogs. All maps are generated with \texttt{NSIDE}$=64$.}
\label{fig:sdss_safe0_gal_count}
\end{figure}

Our galaxy sample has redshift and completeness cuts similar to the SDSS-MGS BAO sample of Ross et al. \cite{Ross_2015} but not identical; they restrict the sample to high-mass, passive galaxies by applying additional color cuts, whereas we want as complete a galaxy sample as possible. 
For our analysis, we focus exclusively on the continuous region within the North Galactic Cap to achieve a more homogeneous density map. Furthermore, we limit our analysis to areas where the completeness exceeds 0.9. Completeness is defined as the fraction of observed galaxies relative to the expected number, accounting for observational effects such as survey geometry and galaxy selection criteria, and is provided as part of the SDSS DR7 galaxy catalogs. In this context, completeness is calculated without considering galaxies missed due to fiber collisions. This approach ensures that the sample remains robust and representative of the true galaxy distribution, consistent with the methodology outlined in the SDSS-MGS BAO sample analysis by Ross et al. \cite{Ross_2015}. 
We also apply a redshift cut of $0.07 \leq z \leq 0.2$.
As a sanity check, we also considered all the cuts applied in Ref. \cite{Ross_2015} and reproduced the corresponding galaxy power spectra to compare with their power spectrum\footnote{Details can be found in Appendix \ref{sec:galpower_comparison}.} (Figure \ref{fig:ross_comparison}).
After applying the aforementioned cuts to the actual data and random data catalogs, the count maps of both galaxies and sirens (the process of generating the mock siren catalog is explained in Section \ref{sec:BBHs}) in the two redshift bins $0.07 \leq z <0.135$ and $ 0.135 \leq z <0.2$ are shown in Figure \ref{fig:sdss_safe0_gal_count}. The number of galaxies in the first and second redshift bins is 237,084 and 98,734, respectively.

\subsection{Physical properties of galaxies in SDSS DR7}
As previously mentioned in Section \ref{sec:Phenomenological approach}, we select host galaxies based on their astrophysical properties, including SFR, stellar mass, and metallicity. In what follows, we briefly describe the procedure for obtaining the value of these parameters for galaxies in the SDSS DR7 catalog.
\begin{itemize}
    \item \textbf{Stellar mass in SDSS DR7:} In SDSS DR7, stellar mass estimation in the NYU VAGC uses both observational data and theoretical models. 
    Specifically, the stellar mass is calculated using the Spectral Energy Distributions (SEDs) of galaxies observed in the SDSS and 2MASS bands, $ugrizJHK$,\footnote{For more information see: \url{http://sdss.physics.nyu.edu/vagc/kcorrect.html}} 
    fitting a stellar population synthesis model for the SED with redshift fixed to the spectroscopic measurement, and finding the best-fit stellar mass \cite{2005AJ....129.2562B,Blanton_2007}.   Note that stellar masses in the catalog are in units of $M_{\odot}/ h^2 $, which we convert to $M_{\odot}$ using $h=0.67$, the value favored by fitting the $\Lambda$CDM model to \textit{Planck} cosmic microwave background (CMB) observations \cite{2020}. 
    \item \textbf{SFR in SDSS DR7:} 
    To calculate the SFR in the SDSS DR7 catalog, we again use the measurements from the NYU VAGC. These parameters correspond to star the formation history obtained from the stellar population synthesis fitting to the SDSS and 2MASS bands:
    \begin{itemize}
        \item{\textbf{B1000}  is the ``birth-rate parameter'' over the past 1 Gyr, as defined by Eq.~16 in \cite{Blanton_2007}. That is, it measures the fraction of total star formation that has occurred over the past 1 Gyr:
        \begin{equation}
            B_{1000} = \frac{\int_0^{\textrm{1 Gyr}} dt \, \textrm{SFR}(t)}{\int_0^{t(z=\infty)} dt \, \textrm{SFR}(t)}
        \end{equation}
        }
        \item{\textbf{INTSFH} is the integral of the star formation rate over cosmic history :
        \begin{equation}
            \textrm{INTSFH} = \int_0^{t(z=\infty)} dt \, \textrm{SFR}(t) \ \text{[in units of } h^{-2} M_\odot \text{]}
        \end{equation}
        }
\end{itemize}
Therefore, the average SFR over the past 1 Gyr for each galaxy is calculated as \cite{2005AJ....129.2562B,Blanton_2007}
    \begin{equation}
        \mathrm{SFR}= \frac{\text{B1000} \times \text{INTSFH}}{h^2 \times 10^9 \textrm{yr}}, 
    \end{equation}
    where the factor of $h^2$ is necessary to convert from INTSFH in units of $h^{-2} M_\odot$ to units of $M_\odot$.
   
\item{\textbf{Metallicity in the SDSS DR7:} metallicity measures the abundance of elements heavier than hydrogen and helium in a star or galaxy. 

    In the SDSS NYU VAGC DR7 catalog, the \textbf{METS} parameter gives the average metallicity from the best-fitting template spectrum, in units of the solar metallicity.}
\end{itemize}

 The k-correct SPS fits have been shown to reproduce SDSS galaxy colours and broadband fluxes with no significant systematic residuals or redshift trends, and to successfully predict independent 2MASS and GALEX fluxes when constrained only by SDSS data, demonstrating that the inferred SFH parameters are consistent with real SDSS observations. Moreover, the physical parameters derived from these fits---such as stellar masses---are in good agreement with independent SDSS spectral-based estimates \cite{2003MNRAS.341...33K}, further validating the underlying SFH modeling. External ultraviolet-based SFR measurements also correlate strongly with SDSS-based SFR indicators \cite{2005AJ....129.2562B}, providing an additional observational cross-check. Finally, the SDSS galaxies in our sample reproduce the expected SFR-stellar mass main-sequence structure (Fig \ref{fig:SFR_vs_stellarmass}, top panels), offering an internal validation that the adopted SFR estimates behave consistently with well-established empirical trends.

\begin{figure}
\centering
\includegraphics[width=0.7\textwidth]{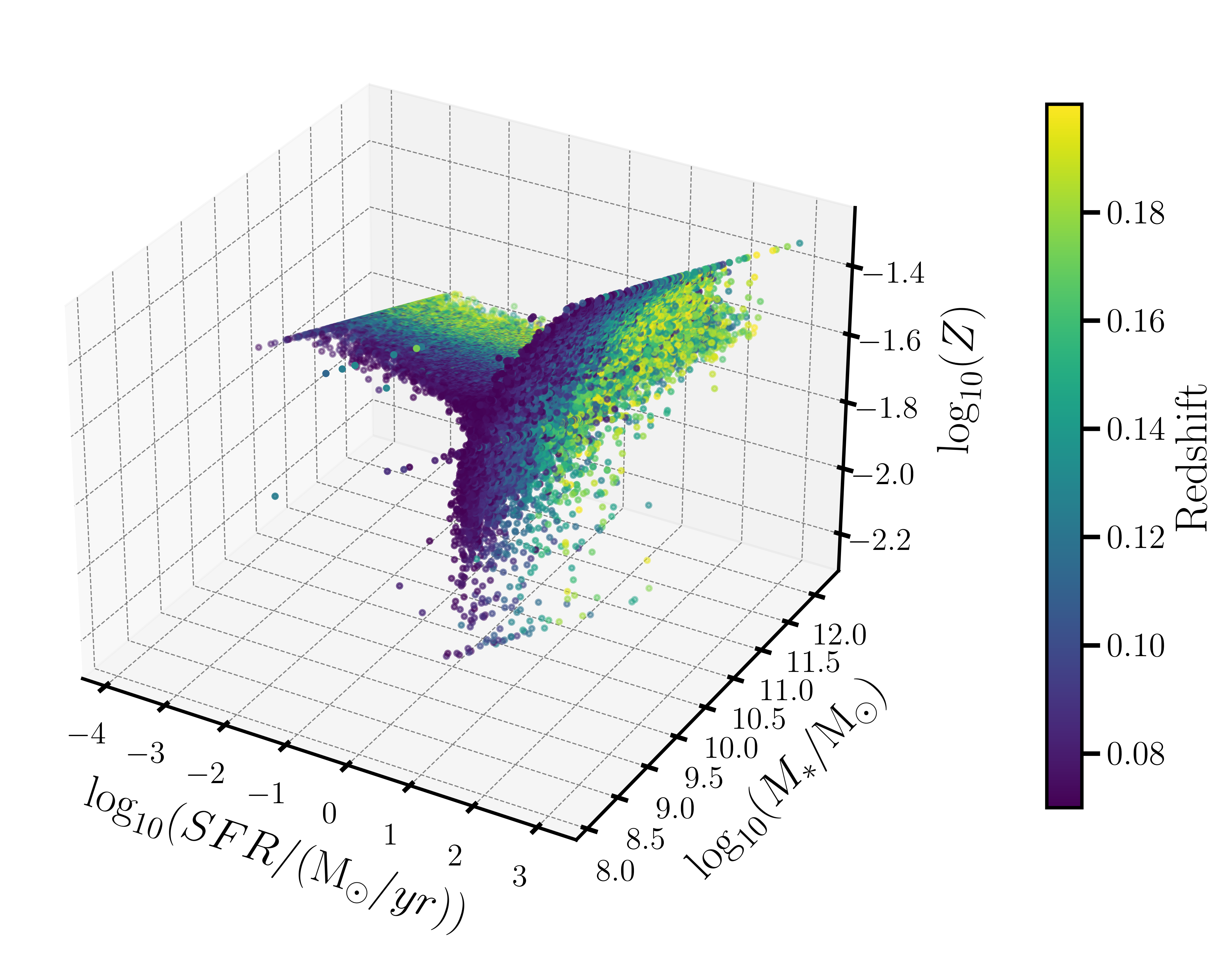}
\caption{The distribution of stellar mass ($M_\star$), star formation rate (SFR), and metallicity ($Z$) of the galaxy sample in the redshift range of $0.07 \leq z <0.2$.}
\label{fig:SFR_M_Z_hists}
\end{figure}

\begin{figure}
\includegraphics[width=.49\textwidth]{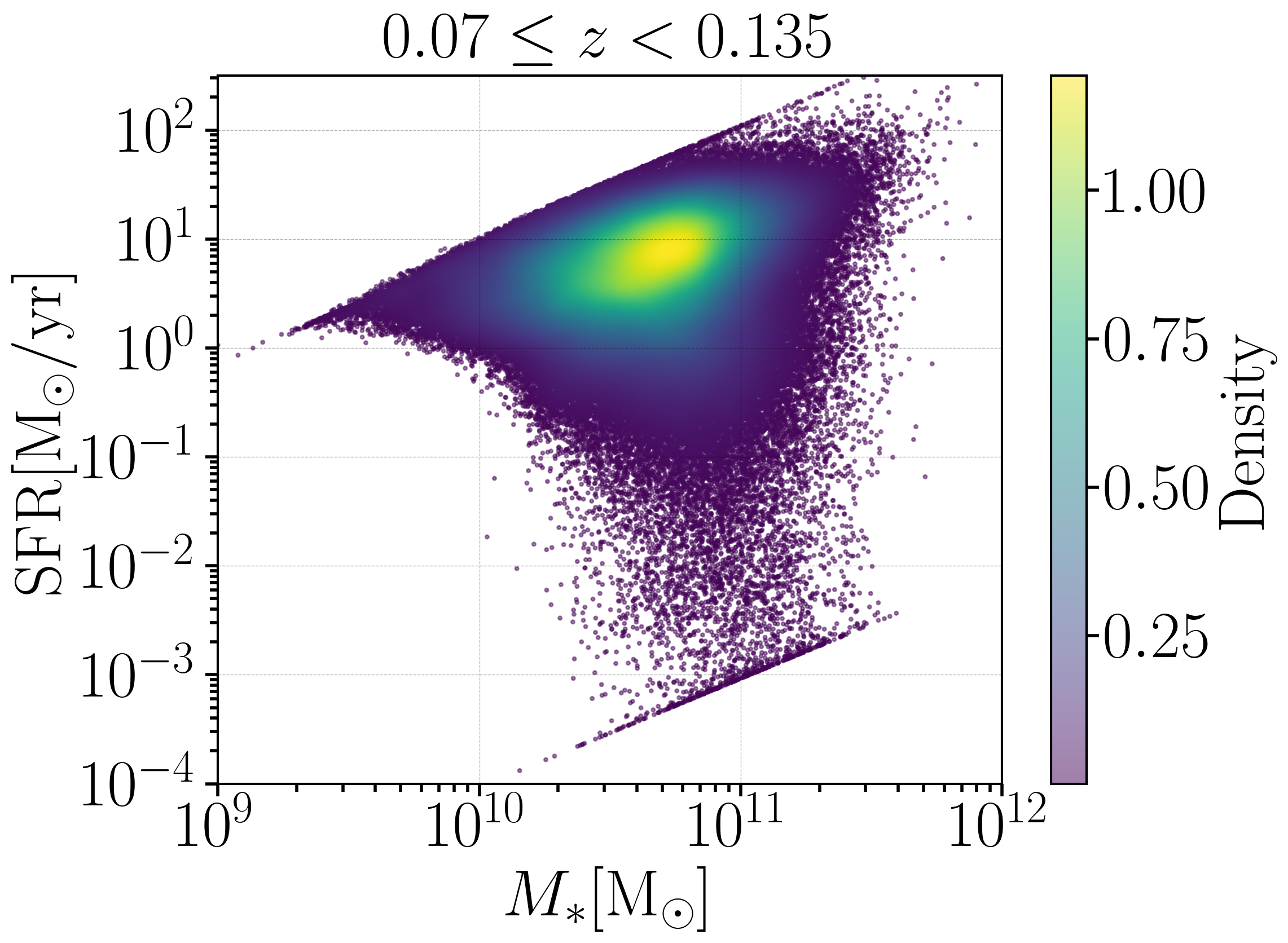}
\includegraphics[width=.49\textwidth]{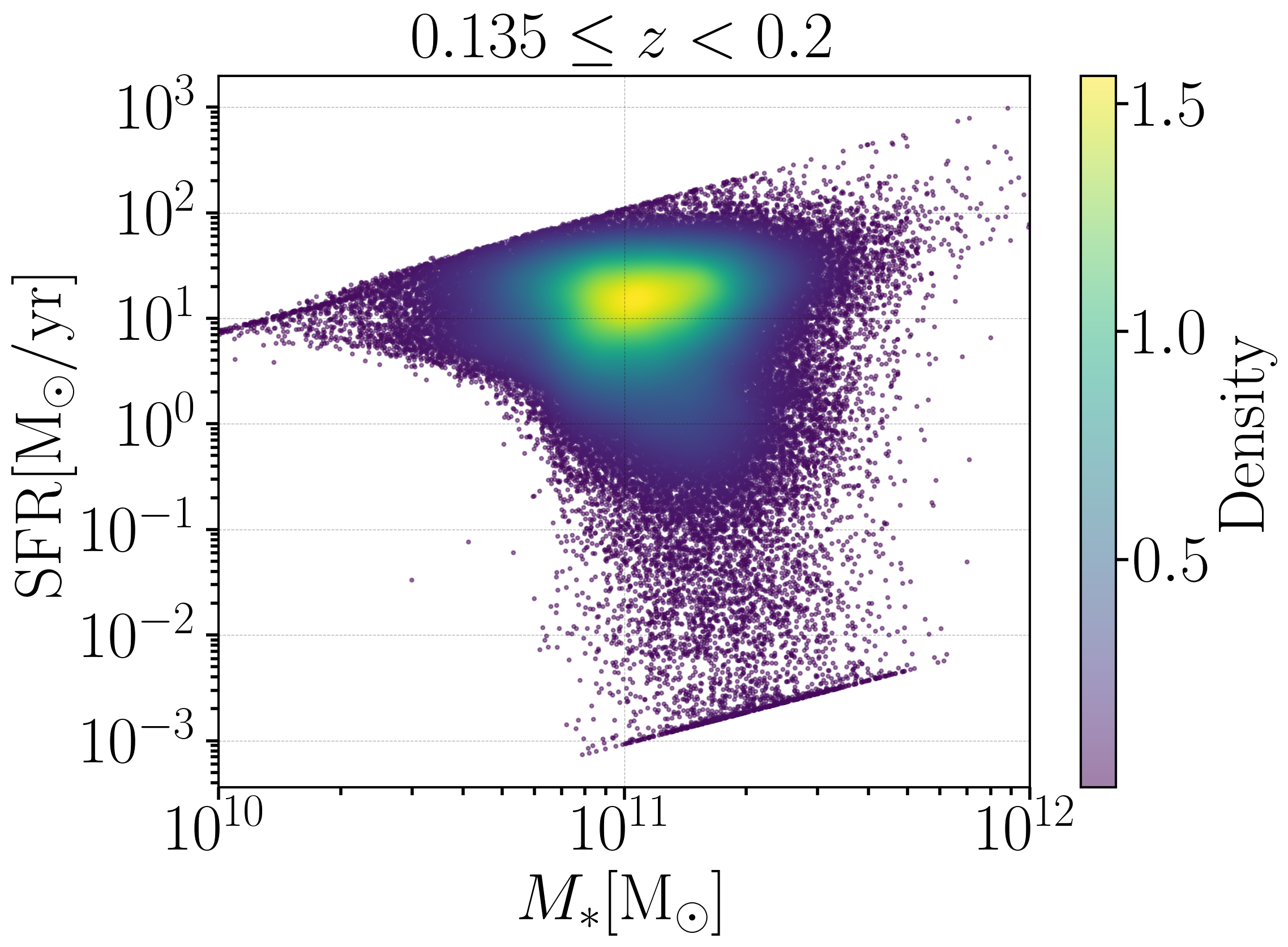}

\includegraphics[width=.49\textwidth]{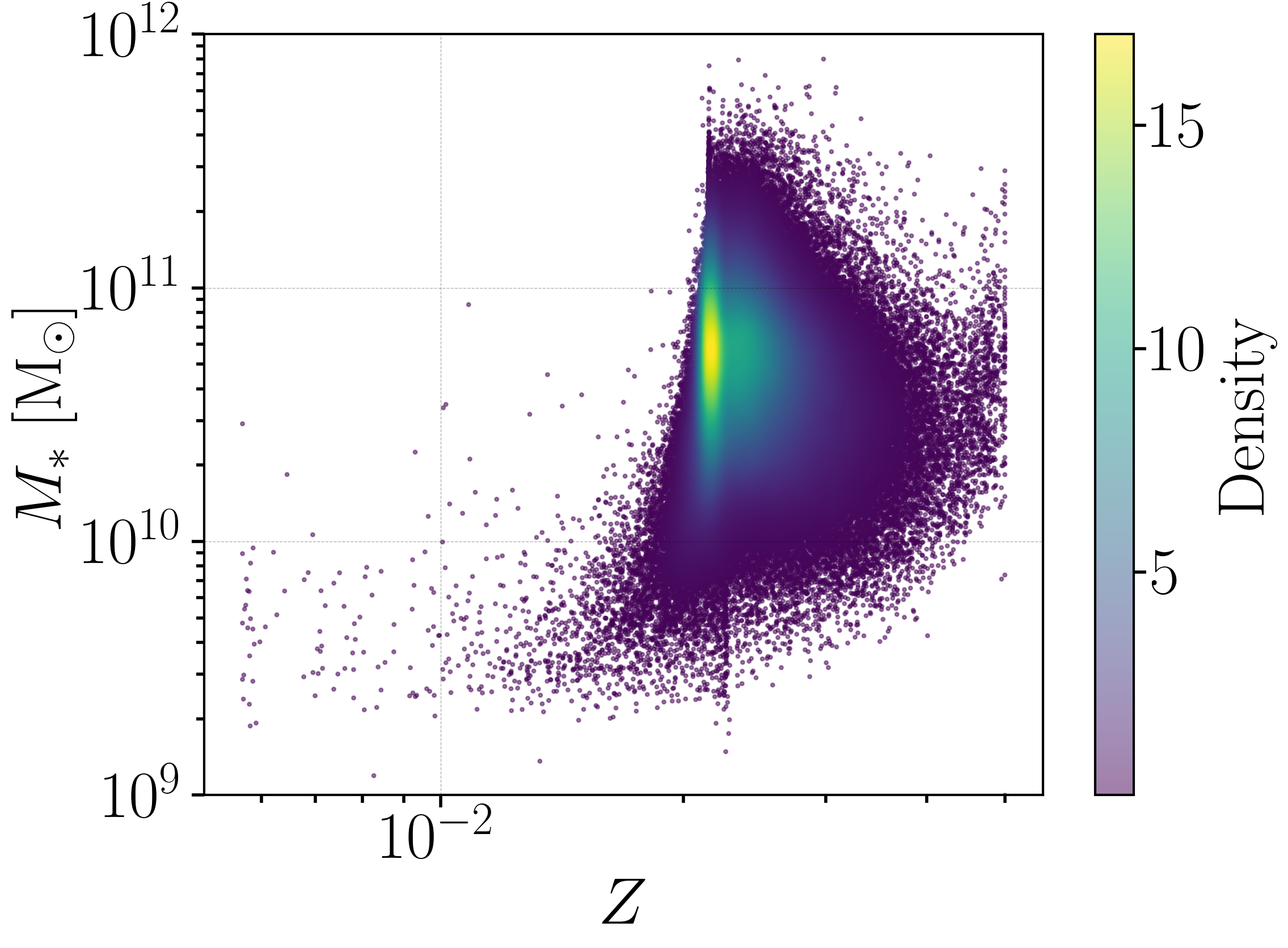 }
\includegraphics[width=.49\textwidth]{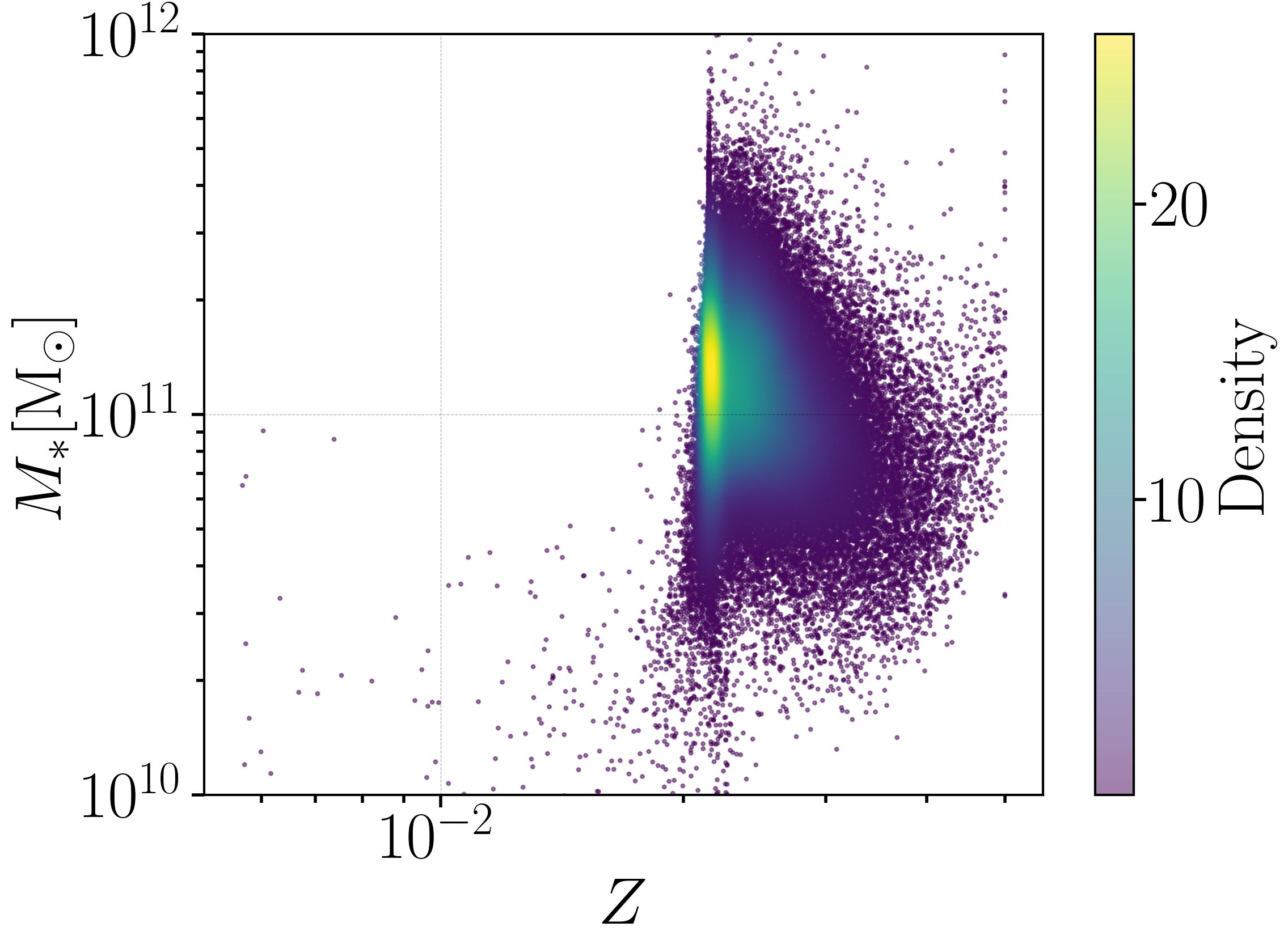}
\includegraphics[width=.49\textwidth]{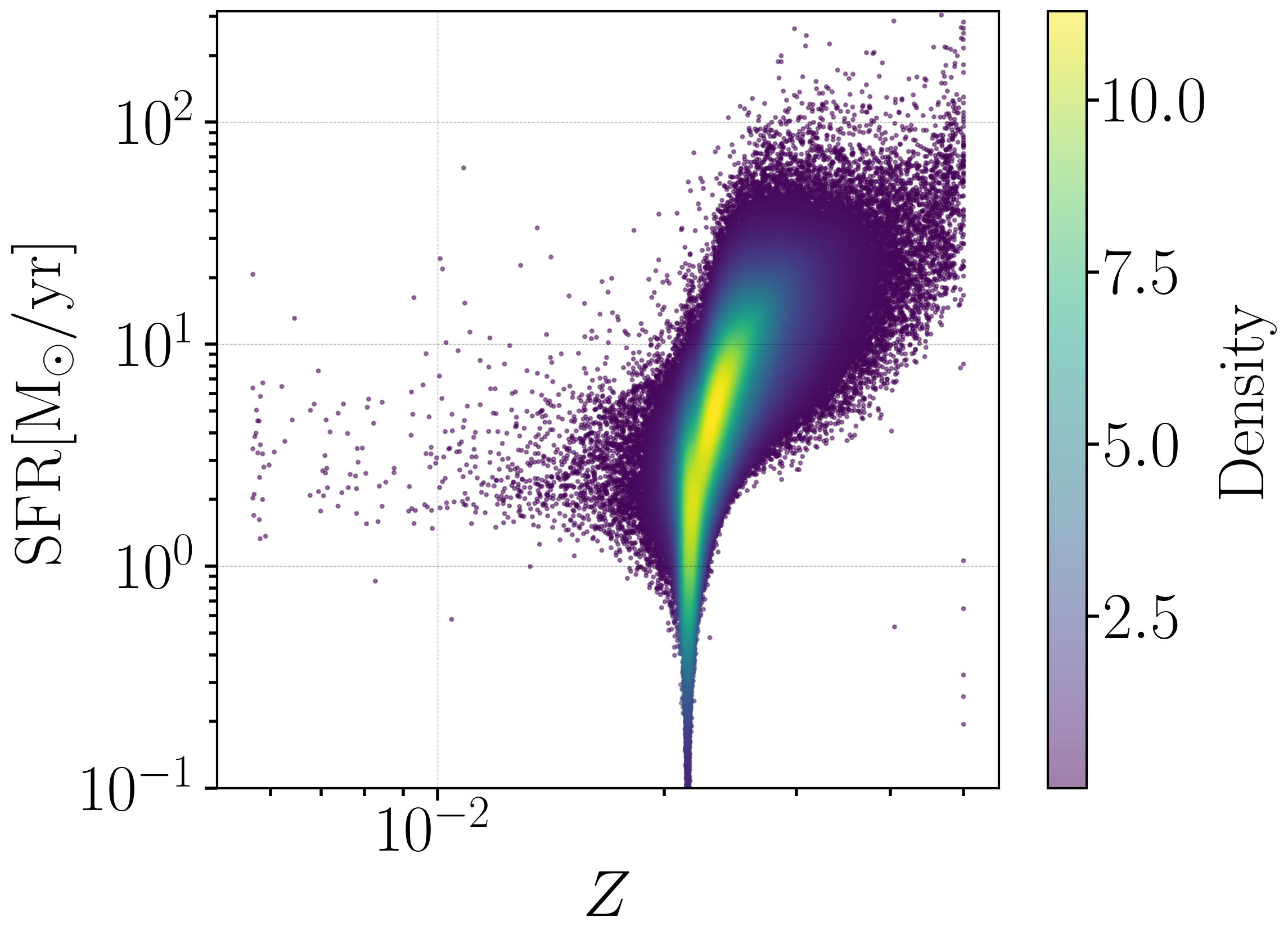}
\includegraphics[width=.49\textwidth]{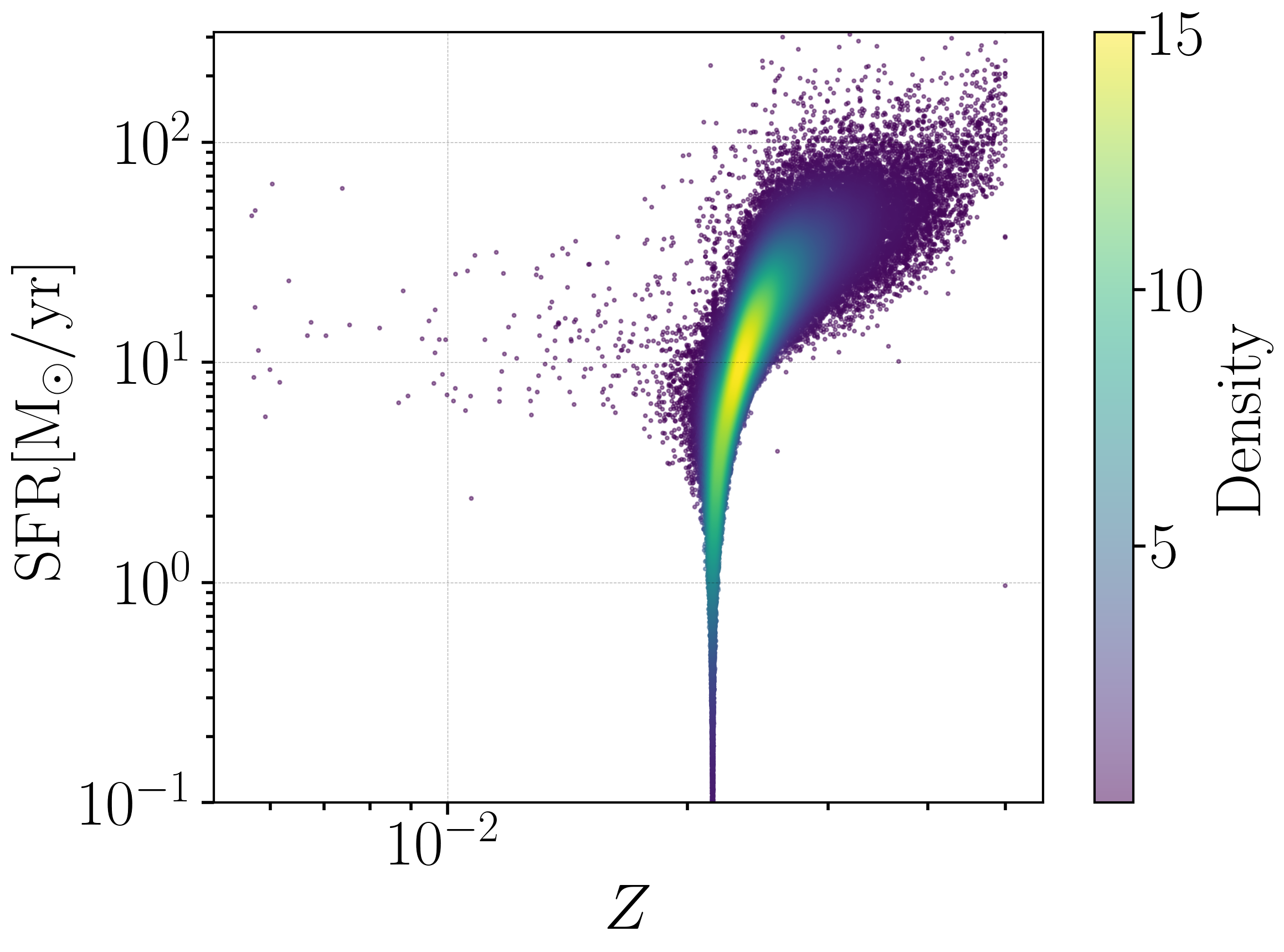}
\caption{Projected density distributions of galaxies in the SDSS DR7 catalog across two redshift bins: $0.07 \leq z < 0.135$ (left column) and $0.135 \leq z < 0.2$ (right column). Each row shows a different pairwise relationship among stellar mass $M_\star$, star formation rate (SFR), and metallicity $Z$. Color bar represents the galaxy density (from an unnormalized kernel density estimate), with brighter regions indicating higher number densities.}
\label{fig:SFR_vs_stellarmass}
\end{figure}

To obtain a better visual sense of the distribution of these properties for SDSS DR7 galaxies, Figure \ref{fig:SFR_M_Z_hists} shows the 3D distribution of stellar mass, SFR, and metallicity within the redshift range $0.07 \leq z <0.2$ while  Figure \ref{fig:SFR_vs_stellarmass} shows the 2D projections. 

Note that as we are using observed galaxy catalogs to construct the siren catalogs, there are inevitably limitations arising from the depth and completeness of the observational data in our result. Most importantly, our measured GW bias will be influenced by incompleteness in stellar mass, which we will further discuss in Sections \ref{sec:mzr_result} and \ref{sec:mzsfrr_result}.
Just a visual inspection of the plots in Figure \ref{fig:SFR_vs_stellarmass}, especially those in the upper panel seem to indicate the galaxy sample is incomplete for stellar masses below $10^{9}-10^{10} M_\odot$ across both redshift ranges. This can be better quantified as follows. According to \cite{2016MNRAS.459.2150W}, the minimum stellar mass required for a galaxy to be reliably detected and included in the sample for redshifts $z < 0.06$ is $ 1.3 \times 10^{10} M_\odot$. We can use this threshold as a calibration value to estimate the corresponding stellar mass completeness limits at higher redshifts by applying an empirical scaling relation and based on the fact that the detection is limited by the apparent brightness of the galaxies. More specifically, given that the apparent flux of a galaxy decreases with the inverse of the luminosity distance squared, and that stellar mass and luminosity are broadly correlated, we assume that the stellar mass completeness threshold also falls approximately as inverse of the luminosity distance squared. This enables us to estimate the stellar mass completeness limits at $z = 0.07$, $0.135$, and $0.2$ to be approximately $1.7 \times 10^{10}$, $6.3 \times 10^{10}$, and $1.3 \times 10^{11}\ M_\odot$, respectively.

\begin{figure}[H]
\centering
\includegraphics[width=.3\textwidth]{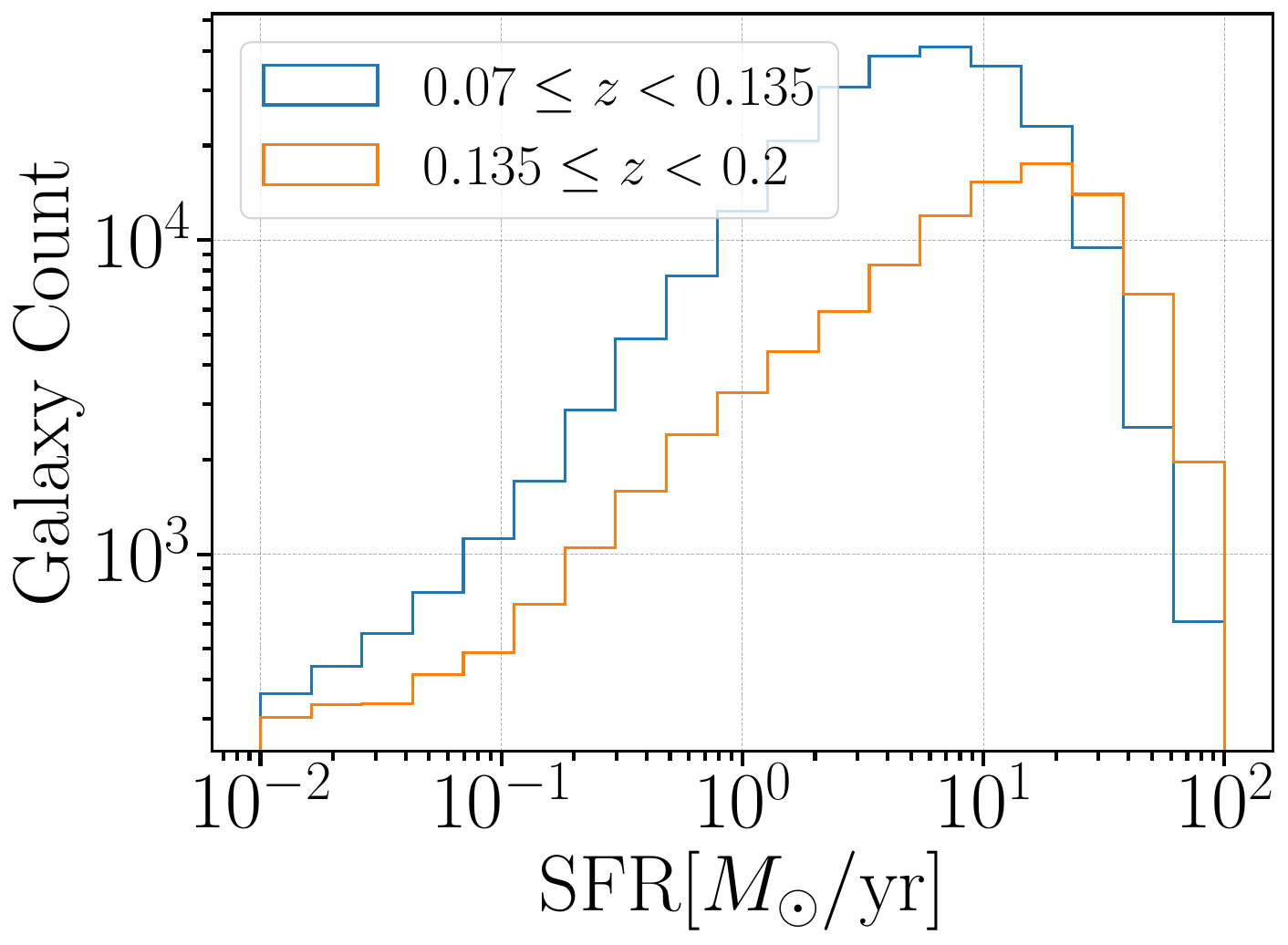}
\includegraphics[width=.3\textwidth]{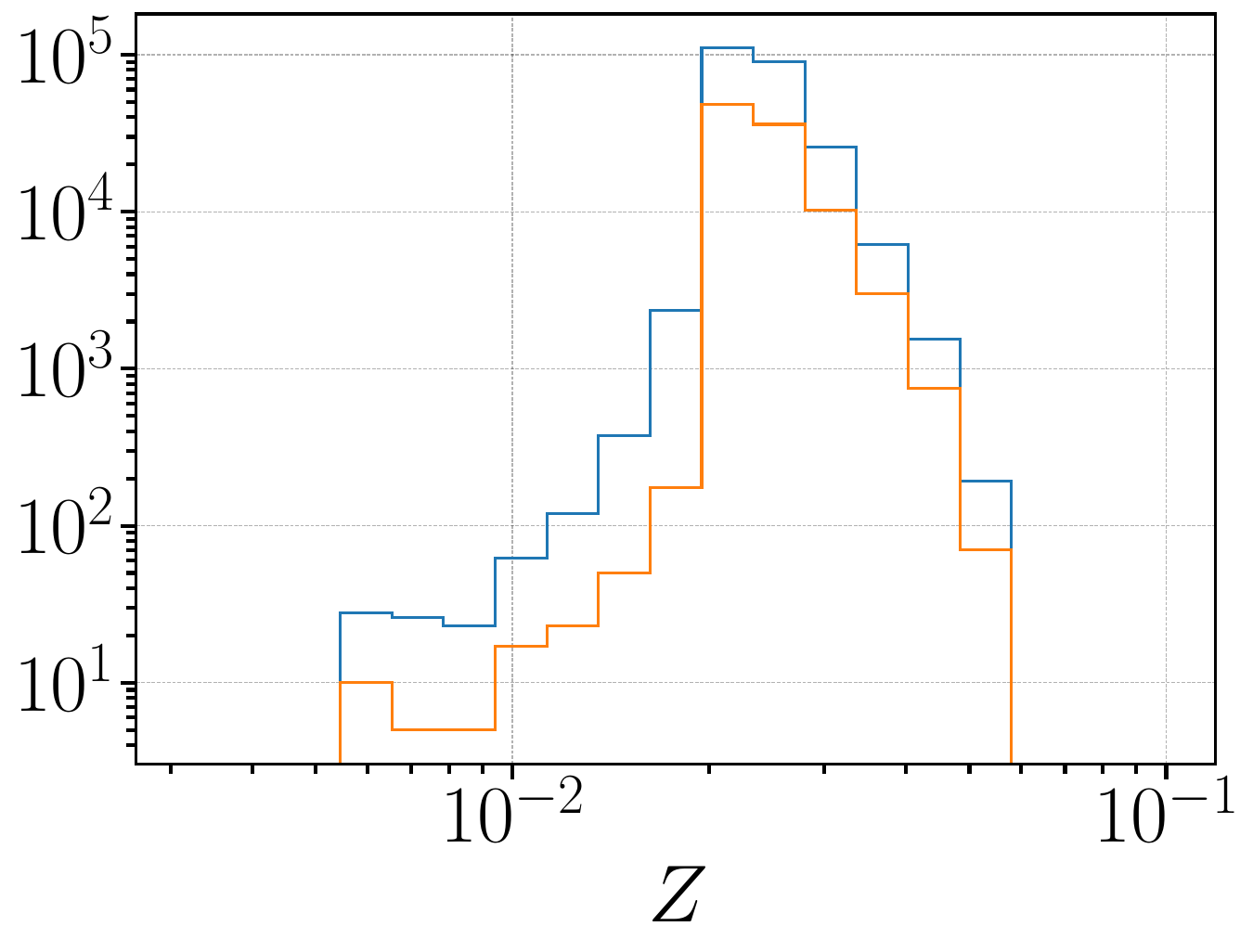}
\includegraphics[width=.3\textwidth]{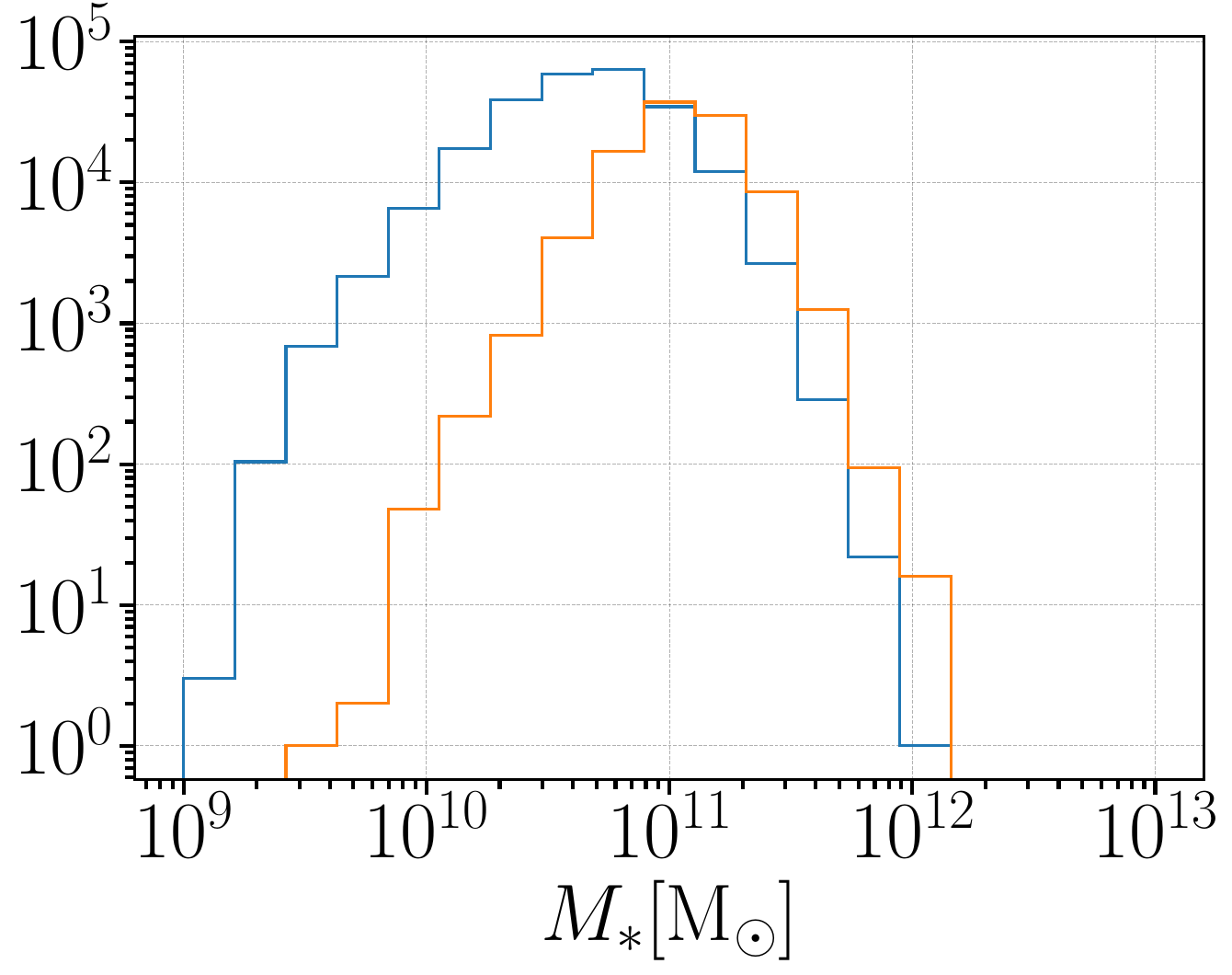}
\caption{Histogram of galaxy properties: stellar mass, SFR and metallicity in two redshift bins $0.07 \leq z <0.135$ and $0.135 \leq z <0.2$.}
\label{fig:SFR_M_Z_dist}
\end{figure}

\begin{figure}[H]
    \centering
    \includegraphics[width=0.8\textwidth]{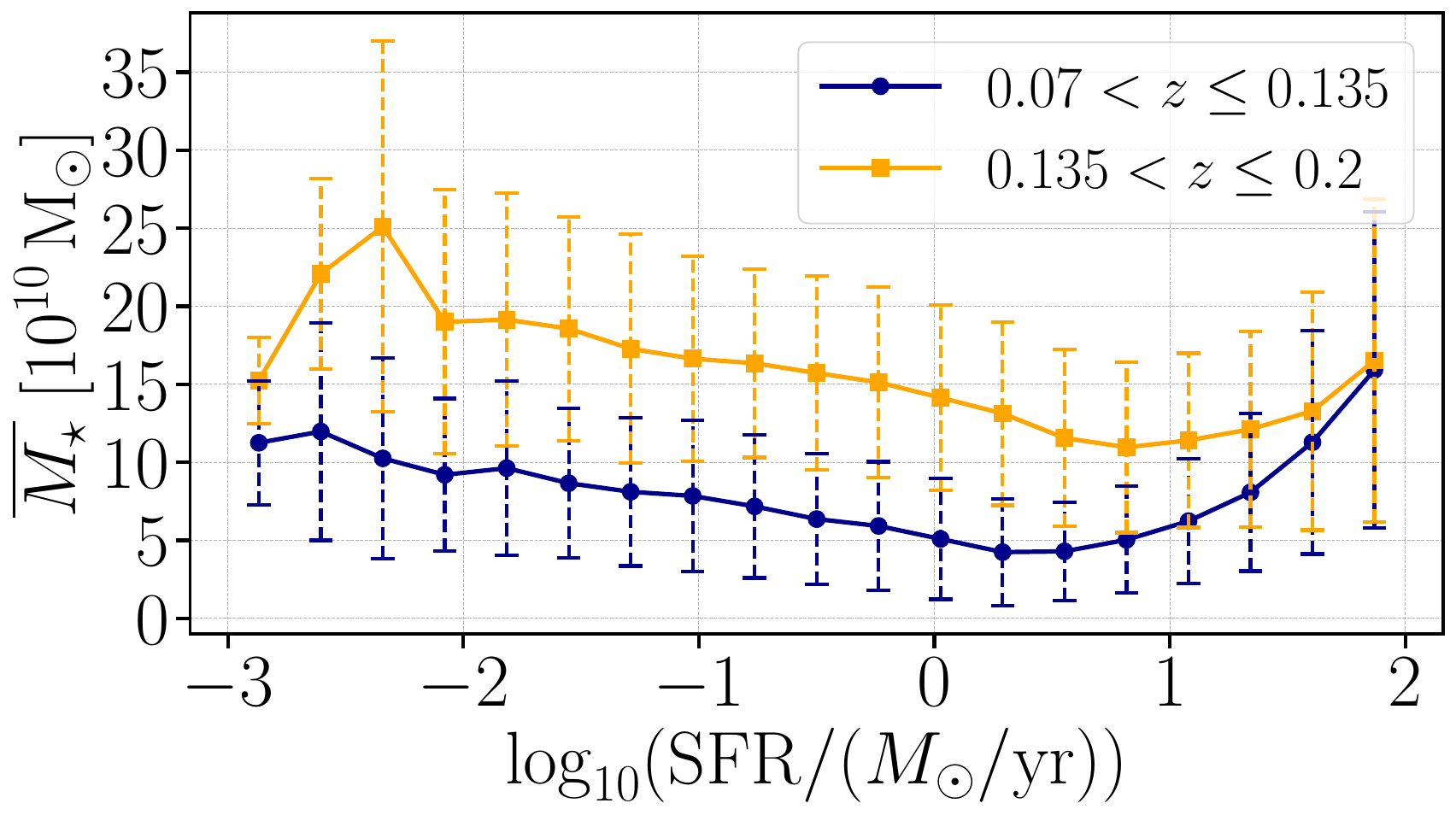}
    \caption{Mean stellar mass as a function of star formation rate (SFR) in two redshift bins $0.07 < z \leq 0.135$ and $0.135 < z \leq 0.2$. Galaxies are binned in $\log_{10}(\mathrm{SFR}/M_\odot\,\mathrm{yr}^{-1})$ from $10^{-3}$ to $10^{2}$, and the average stellar mass $\overline{M_\star}$ is computed within each bin. Error bars represent the standard deviation of the stellar mass distribution within each SFR bin, illustrating the intrinsic scatter of galaxy stellar masses at fixed SFR.
    }
    \label{fig:mean_mass_vs_sfr}
\end{figure}

 As another peculiar feature of the catalog, in the middle and bottom plots, we can observe a sharp spike at a metallicity value of around $Z=0.021$. We traced back this feature to the underlying template fitting methodology of the \texttt{ kcorrect} software \cite{Blanton_2007}. The software fits a linear combination of five fixed galaxy templates to each galaxy's photometry. For galaxies that fall on the metallicity spike, we found that the fit is dominated almost entirely by the fourth template — with the corresponding coefficient being significantly nonzero while the others are nearly zero. According to Figure 4 in \cite{Blanton_2007}, the metallicity associated with this fourth template at the present-day lookback time is approximately $Z=0.021$. Thus, we believe the spike in the histogram does not indicate a failure of the fit, but rather reflects the discrete nature of the template set: in these cases, the fit selects a single dominant template whose metallicity is directly imprinted in the result. 
 Finally, to better understand the overall distribution of individual properties, we also plotted the 1D histograms in  Figure \ref{fig:SFR_M_Z_dist} 
and mean stellar mass for different values of star formation rates
in Figure \ref{fig:mean_mass_vs_sfr}. 

Figure \ref{fig:mean_mass_vs_sfr} shows the average stellar mass of galaxies across different SFR bins, separated into two redshift bins \(0.07 < z \leq 0.135\) and \(0.135 < z \leq 0.2\). It illustrates the regime of transition between quiescent galaxies with SFR $\lesssim \mathcal{O}(1)$ $M_\odot$ yr$^{-1}$, where galaxies with a lower star formation rate tend to have higher stellar mass, and the star-forming main sequence at SFR $\gtrsim \mathcal{O}(1)$ $M_\odot$ yr$^{-1}$, where SFR and stellar mass are correlated, approximately consistent with previous measurements \cite{2015ApJS..219....8C}. This figure plays a crucial role in interpreting the results in the GW bias measurements, as discussed in section \ref{sec:GW bias parameter vs star formation rate}.

\section{Generating mock GW source catalogs from the galaxy catalog}
\label{sec:BBHs}

\subsection{Merger rates and counts for binary black holes (BBHs)}
\label{sec:Num_merger}
In this section, we describe our approach to determining the merger rate and calculating the merger counts in our GW catalogs. We summarize the key points of the method described in Section 4 of \cite{dehghani2024}, which also shows plots of the merger rates and the number of mergers as a function of redshift.

The number of BBH mergers at a specific redshift dictates the sample size of host galaxies, which affects the error on our estimation of the GW bias, but theoretically it is not correlated with the value of the bias itself.
Therefore, in producing our siren catalog, discussed later, we do not change the parameters controlling the merger rate. Instead, we focus on optimizing the ``observing times'' to effectively measure the GW bias with a small error bar.

We assume that the BBH merger rate density $R_\text{GW}$ \cite{O_Shaughnessy_2010,Dominik_2012,Belczynski_2002,Dominik_2015,Mapelli_2017,2018MNRAS.474.2959G,Cao2018,Fishbach_2018,Santoliquido_2022,Mukherjee_2021} at redshift $z_m$, is controlled by the cosmic SFR density $R_\mathrm{SFR}$ at the formation redshift $z_f$, and the delay time probability distribution $P_D(t_d)$, where $t_d$ characterizes the delay time between the formation of stars, $t_f$, and the merger of the compact objects, $t_m$. Hence, the merger rate density of GW sources at a given merger redshift of $z_m$, can be written as
\begin{align}
    R_\text{GW}(z_m) = \mathcal{A}_0 \int_{z_m}^\infty P_D(t_d) \frac{d t_f}{dz_f} R_\text{SFR}(z_f) ~ dz_f. \label{eq:merger_rate}
\end{align}
The redshift dependence of $R_\text{SFR}(z_f)$ can be estimated analytically using the best-fit formula to observation \cite{Madau:1996hu} as
\begin{align}
\label{eq:sfr}
    R_\text{SFR}(z) = \psi(z) = 0.015 \frac{(1+z)^{2.7}}{1 + [(1+z)/2.9]^{5.6}} M_\odot \text{ yr}^{-1} \text{ Mpc}^{-3}.
\end{align}
$\mathcal{A}_0$ in Eq. \eqref{eq:merger_rate} denotes the normalization factor so that the local merger rate of GW sources at the present day redshift $ R_\text{GW}(z_m)|_{z_m=0} $ matches the observed value of $\mathcal{R}_\text{BBH} = 23.9^{+14.3}_{-8.6}$ Gpc$^{-3}$~yr$^{-1}$~ \cite{abbott2020gw190425}. This implies that
\begin{equation*}
    \mathcal{A}_0= {\mathcal{R}_\text{BBH}\over \int_{0}^\infty P_D(t_d) \frac{d t_f}{dz_f} R_\text{SFR}(z_f) ~ dz_f}.
\end{equation*}
The other parameter needed to evaluate the merger rate in Eq. \eqref{eq:merger_rate} is the delay time probability $P_D(t_d)$. While there are some theoretical studies on this, the exact behavior of the delay time distribution is not well constrained \cite{O_Shaughnessy_2010,Dominik_2012,Toffano_2019,Artale_2019,McCarthy_2020,Banerjee2010,Mukherjee_2021}. In this work we model the delay time distribution as an inverse power-law function in $t_d$ with a minimum threshold for delay time, $t_{d,\mathrm{min}}$, 
\begin{align}
    P_D(t_d) \propto \Theta(t_d - t_{d,\text{min}}) t_d^{-\kappa}.
    \label{eq:ptd}
\end{align}
Here $\kappa$ is a non-negative free parameter and $\Theta(t)$ is the Heaviside function such that for $t_d<t_{d,\mathrm{min}}$, the probability distribution is zero. Lastly, the factor $dt_{f} /dz$ in Eq.\eqref{eq:merger_rate} is the Jacobian for change of variable from time coordinate time to redshift which can be calculated as 
\begin{align}
    \frac{dt_f}{dz_f} = \frac{1}{H_0} \frac{1}{(1+z_f) E(z_f)}, 
\end{align}
where 
\begin{align}
    E(z) = \sqrt{ \Omega_{m}(1+z)^3 + \Omega_{r}(1+z)^4 + \Omega_\Lambda + \Omega_K (1+z)^2 },
\end{align}
and we consider a cosmology with vanishing curvature and radiation density parameters (i.e. $\Omega_K,~ \Omega_{r} \simeq 0$). In addition, $\Omega_{m},\Omega_{\Lambda}$ and $H_0$ are taken from the data release of the Planck Collaboration for the Flat $\Lambda$CDM (Lambda Cold Dark Matter)  model \cite{2020A&A...641A...6P}. 

Once the merger rate density is obtained, one can integrate it over the volume to compute the total number of mergers in a given redshift bin of width $\Delta z$ and midpoint $z_a$, 
\begin{align}
    N_{GW} (z_a) = T_\mathrm{obs} \int_{z_a - \Delta z/2}^{z_{a}+ \Delta z/2} \frac{dV}{dz} \frac{R_{GW}(z)}{1+z}  ~ dz. 
    \label{eq:N_GWcalculate}
\end{align}
Here, the indices of the midpoint redshifts are labeled as $a\in\{1,\cdots, N_\mathrm{bins}\}$ where $N_\mathrm{bins}$ is the number of bins in the redshift range of the galaxy survey. $T_\mathrm{obs}$ denotes the observation time and depends on the running time of the GW detectors. 
As mentioned above, the total number of mergers should not alter the theoretical value of GW bias. Additionally, within our framework for modeling and calculating GW bias, the explicit formulation of delay time is only involved in the derivation of the total number of BBH mergers. Consequently, the variation of the parameters $\kappa$ and $t_{\textrm{d,min}}$ defined in Eq.~\eqref{eq:ptd} does not change our result and, for simplicity, we fix them as $\kappa=1$ and $t_{d,\textrm{min}}=500$ Myr for all the GW samples. However, as we will elaborate in the following subsection, the distribution of delay times is expected to have an indirect effect on the shape of the host-galaxies probability and hence GW bias. Therefore, while our sample sizes are based on the merger rate calculation assuming shorter delay times that peaked around $500$ Myr, we will explore variations in host-galaxy probability that account for
longer delay times by modifying the shape of the GW host-galaxy probability function.

\subsection{Selecting host galaxies for BBH mergers using astrophysical host-galaxy probability functions}
\label{sec:assignment}

In Section \ref{sec:Num_merger}, we computed the expected number of BBH mergers at each specific redshift. In what follows, we develop a general statistical framework to estimate the probability for galaxies to host BBH mergers based on their astrophysical properties. We refer to these probability functions as \textit{GW host-galaxy probability function} or for short, the \textit{host-galaxy probability function} (in agreement with the terminology of \cite{Artale_2020} and \cite{dehghani2024}).

In general, we do not expect the BBH formation to be influenced by non-host galaxies and therefore the full joint host-galaxy probability function is a product of individual host-galaxy probabilities.
There is an ongoing area of research on the relation between BBH merger rates and the astrophysical properties of their host galaxies
\cite{2023arXiv231203316V, Artale_2019,Santoliquido_2022,2023MNRAS.523.5719R,2024MNRAS.530.1129P,Artale_2019b,Toffano_2019,Cao2018}.
  
In this study, we consider three important astrophysical properties governing the host-galaxy probabilities: the stellar mass $M_*$, star formation rate SFR, and metallicity $Z$. For simplicity we also assume the joint probability distribution $P_\mathrm{host}(\mathrm{GW}|M, \mathrm{SFR}, Z)$ which we refer to as $M_*$-$Z$-$\rm SFR$  host-galaxy probability function, is a product of probabilities corresponding to each of these properties individually,
\begin{align}
     P_\mathrm{host}(\mathrm{GW}|M_*,\mathrm{SFR},Z)= P_\mathrm{host}(\mathrm{GW}|M_*)P_\mathrm{host}(\mathrm{GW}|\mathrm{SFR})P_\mathrm{host}(\mathrm{GW}|Z). \label{eq:joint_prob_hosts}
\end{align}
We then model $P_\mathrm{host}(\mathrm{GW}|M_*)$, $P_\mathrm{host}(\mathrm{GW}|\mathrm{SFR})$, and $P_\mathrm{host}(\mathrm{GW}|Z)$ as individual host-galaxy probabilities associated with stellar mass, SFR, and metallicity, respectively. 
Of course, as these properties are not uncorrelated, it is expected that our result will have some degeneracy directions or planes in the parameter space governing their shapes. Note that, although stellar mass and SFR show a visible correlation in Figure 4, the relation displays substantial intrinsic scatter and includes a mixture of star-forming and quiescent galaxy populations. This motivates our choice of a factorized host-galaxy probability as a flexible phenomenological description, rather than attempting to model a single smooth joint probability surface. 
What follows describes our formulation of these probabilities to populate the BBHs through galaxies guided by the physical principles governing BBH formation and merger events.

\subsubsection{\texorpdfstring{$M_*$-$Z$-$\rm SFR$}{M*-Z-SFR} host-galaxy probability function} 

\paragraph{$M_*$ host-galaxy probability function, $P_\mathrm{host}(\mathrm{GW}|M_*)$:} 
The stellar mass of a galaxy at any given time is determined by its prior stellar mass and star formation history. At the beginning of a galaxy's life, due to hierarchical clustering and more frequent merging with other galaxies, the SFR is often higher, leading to galaxies accumulating more stellar mass by forming new stars. 
Therefore, since the BBH merger rate is also expected to be correlated with the host galaxy's SFR during formation, it should also be positively correlated with stellar mass. In fact, a strong positive correlation has been shown in population synthesis simulations such as \cite{Artale_2019}. 
However, this relationship can be affected by other factors, such as the delay times between BH formation and BBH mergers. 
To understand this, note that the BBH mergers occurring at $z=[0-0.2]$ could have formed at early times (higher redshift for longer delay times) or late times (lower redshift for shorter delay times). On the other hand, we know that for massive galaxies, after experiencing rapid star formation at high redshifts, are quenched at late times \cite[e.g.,][]{Salim_2007, Peng_2010}. This leads to a bimodal stellar mass--SFR relation characterized by low-mass, high-SFR star-forming galaxies and high-mass, low-SFR quiescent galaxies \cite{Salim_2007}.
This suggests that in the case of BBH mergers, if delay times on average are shorter than quenching timescales, BBH numbers at high stellar mass should be suppressed, since BBHs formed early would have already merged. This suppression is expected around $10^{11} M_\odot$ based on the empirical SFR-mass relation.\footnote{Some studies have also found that there can be an opposite increasing trend in the relation between the merger rate per galaxy and the stellar mass at late times for $M_*>10^{10.5} M_\odot$ which might be due to stellar and AGN feedback \cite{Artale_2019}. In this work we do not consider such trends.}
However, if the average delay times are long, then there should not be a suppression of merger rates in massive galaxies, as most are merging after quenching.

\begin{figure}
\includegraphics[width=.45\textwidth]{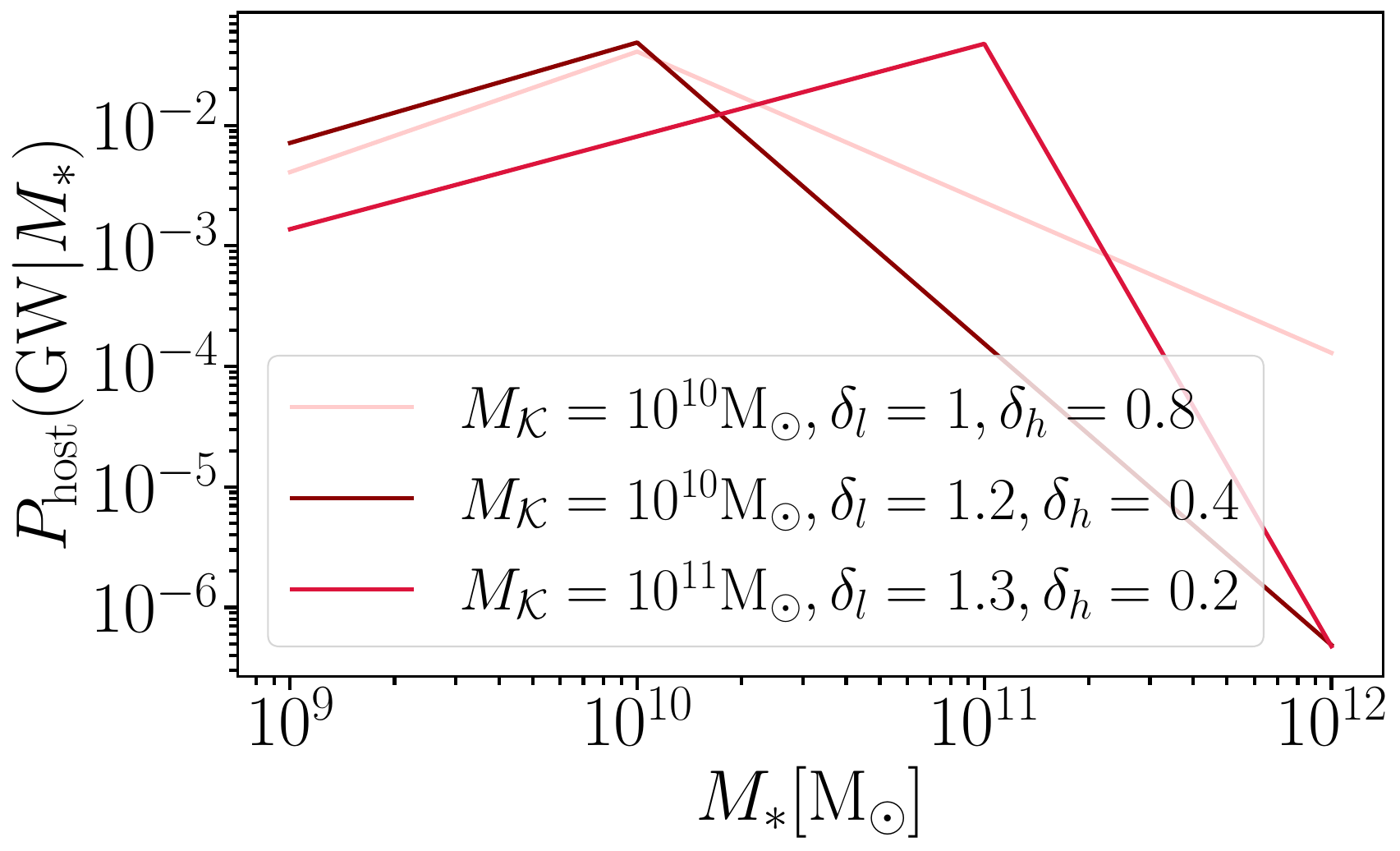}
\includegraphics[width=.45\textwidth]{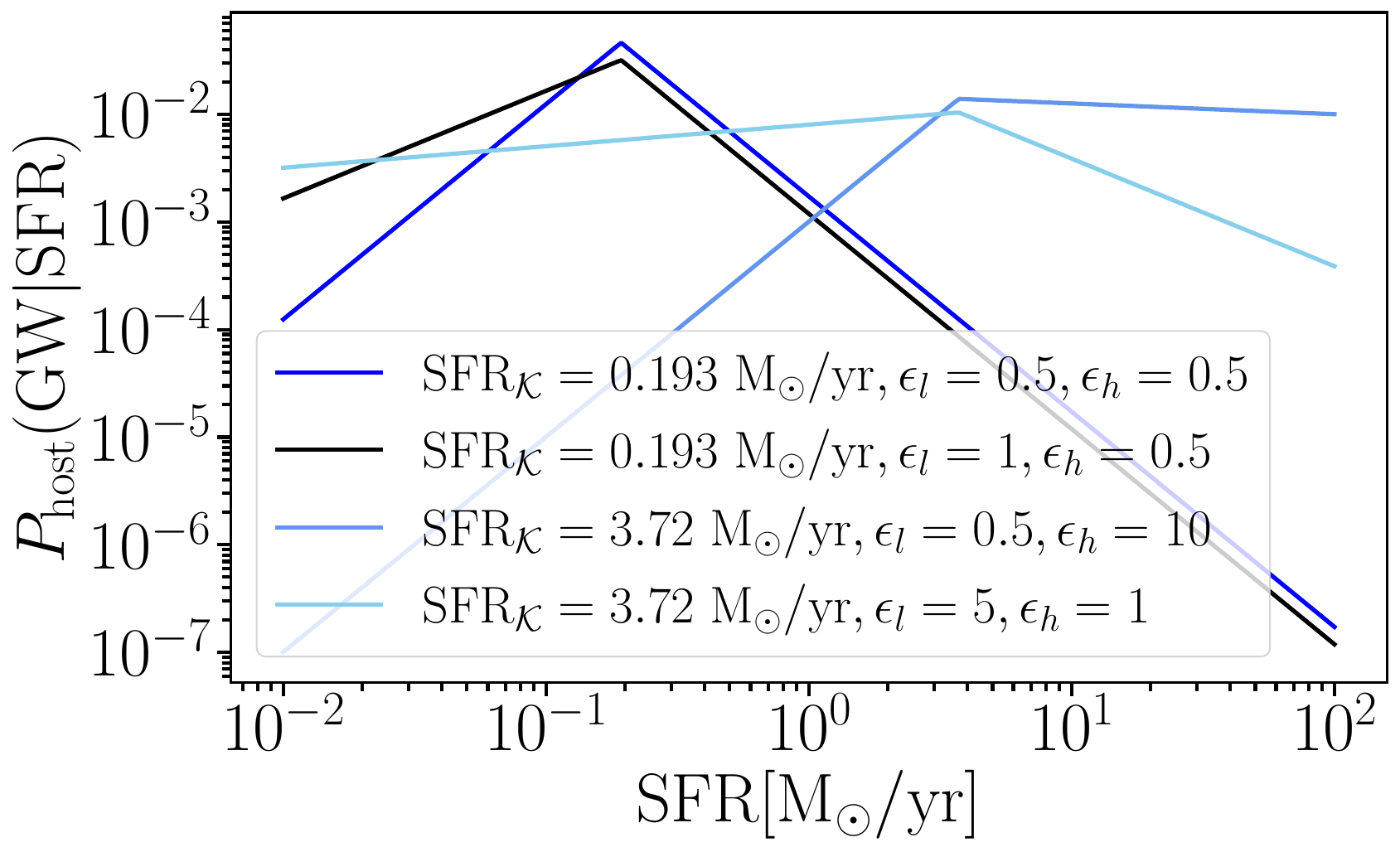}
\includegraphics[width=.45\textwidth]{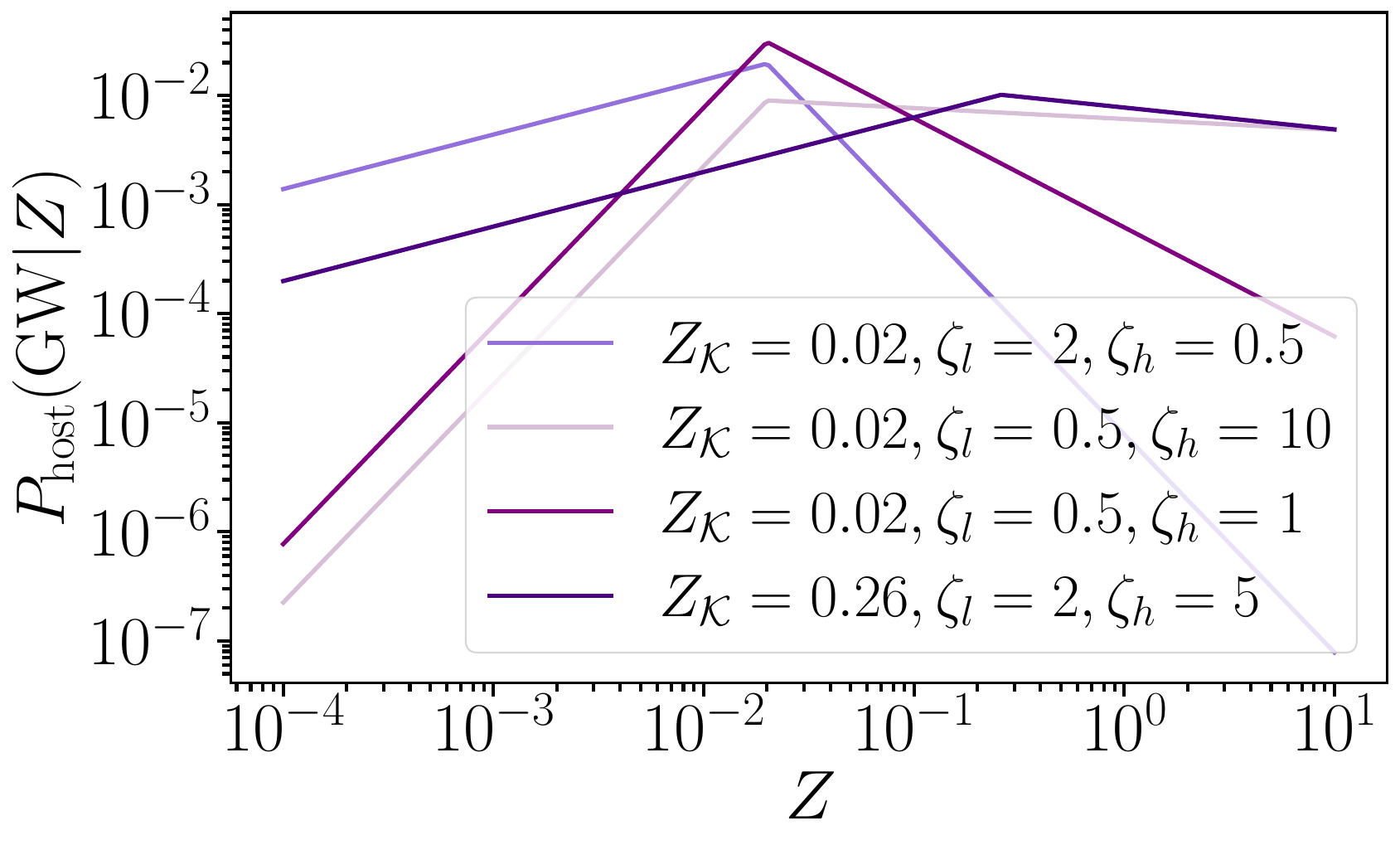}
\centering
\caption{$M_*$, SFR and $Z$ host-galaxy probability functions for a variety of parameters controlling the break point and upper and lower slopes.
}
\label{fig:MZsfR_selec_func}
\end{figure}

Considering these factors, we use a broken power-law for the $M_*$ host-galaxy probability function, reflecting the potential impact of suppressed SFR history in high-mass galaxies. Note that, in the case of considering longer delay times, the absence of a suppression is automatically implemented by variables controlling the peak and slopes of the power-law to make the host-galaxy probability function a monotonically increasing function of stellar mass
\cite{Toffano_2019,Cao2018,Artale_2019,2023MNRAS.523.5719R}. The $M_*$ host-galaxy probability function is parameterized as follows
\begin{align}\label{eq:MZR_function}
    P_{\rm host}(\mathrm{GW}|M_*)=
    \begin{cases}
    A_{M}10^{(\log(M_*)-\log(M_\mathcal{K}))/\delta_l}, \quad & \text{for}\,\, 7 \leq \log(M_*) \leq \log(M_\mathcal{K}),\\
    A_{M}10^{(\log(M_\mathcal{K})-\log(M_*))/\delta_h}, \quad & \text{for} \,\, \log(M_*) \geq \log(M_\mathcal{K}), 
     \end{cases}
\end{align}
where $M_*$ is the stellar mass of the host galaxy obtained from the catalog and $M_\mathcal{K}$ sets the turn-around point. The slopes of the low and high $M_*$ parts for the broken power-law are adjusted by $\delta_l$ and $\delta_h$ and $A_M$ is the normalization factor. In this study, we consider many different parameters choices of the host-galaxy probability function governing the stellar mass dependence in the following ranges:  $M_\mathcal{K}\in[10^9,10^{12}]~M_\odot$,  $\delta_l \in [0.5, 10]$ and
$\delta_h \in [0.5,4.5]$ (see the top left panel of Figure \ref{fig:MZsfR_selec_func} for a few examples).

\paragraph{SFR host-galaxy probability function, $P_\mathrm{host}(\mathrm{GW}|\mathrm{SFR})$:}

As mentioned previously, there is good evidence that indicates a significant correlation between the probability of a galaxy undergoing a merger and its stellar mass since it is a strong indicator of the galaxy's star formation history. However, as the relationship can be more complex, further insight can be obtained by introducing an additional probabilistic dependence to the SFR in the merger event. The merger rate of BBHs may be tied to the SFR of their host galaxies, which could have different behavior at low and high SFRs. For example, in a sample of low redshift galaxies for early-type, low-SFR galaxies, the merger rate is influenced by the cumulative star formation history rather than ongoing star formation. The compact binaries formed during those early, intense star-forming periods (regardless of the value of ongoing SFR) may or may not be merging, depending on the delay times. As a result, the merger rate in low-SFR galaxies correlates more strongly with stellar mass than SFR, since their merger history reflects past stellar populations. However, in the case of late-type, high-SFR galaxies, active star formation leads to the formation of massive stars, which evolve into compact binaries. These galaxies are efficient at forming BBHs, but due to the delay times, many of BBHs formed in these galaxies remain unmerged. At intermediate SFRs, the merger rate is shaped by both ongoing star formation and the legacy of past star formation, underscoring the complex relationship between SFR, stellar mass, and merger rates. Therefore, a transition from low- to high-SFR can reflect a shift in merger rate dominance, from early-type galaxies to late-type galaxies. This transition can be seen as a consequence of the interplay between cosmic star formation, galaxy assembly, and the delay time distribution of BBHs mergers \cite{Artale_2019b}. We model this interplay  in the SFR host-galaxy probability function as 

\begin{align}
    P_{\rm host}(\rm GW | \rm SFR) = 
    \begin{cases} 
        A_{\rm SFR} 10^{(\log({\rm SFR}) - \log({\rm SFR_\mathcal{K}})) / \epsilon_l}, & \text{for } \log(\rm SFR) \leq \log(\rm SFR_\mathcal{K}), \\[10pt]
        A_{\rm SFR} 10^{(\log({\rm SFR_\mathcal{K}}) - \log({\rm SFR})) / \epsilon_h}, & \text{for } \log(\rm SFR) > \log(\rm SFR_\mathcal{K}),
    \end{cases}
    \label{eq:FMR_function}
\end{align}
where SFR$_\mathcal{K}$, and SFR are respectively the pivot SFR scale, and the measured SFR, which for each galaxy is extracted from the catalog. $\epsilon_l$ and $\epsilon_h$ control the growing and decaying regime of the host probability and $A_{\rm SFR}$ is also the normalization factor. In this study, we consider $\ \mathrm{SFR}_\mathcal{K}\in[10^{-2},100]~M_\odot/\rm yr$, $\epsilon_l\in [0.5,10]$ and $\epsilon_h\in\{1,\infty\}$, where $\epsilon_h=\infty$ means the decaying slope is zero. 

\paragraph{$Z$ host-galaxy probability function, $P_\mathrm{host}(\mathrm{GW}|Z)$:}
Another property of the host galaxies which could correlate with BBH merger events is their metallicity. Metallicity can influence the formation
of binary black holes (BBHs) by affecting the evolutionary pathways of their massive stellar progenitors. In low-metallicity environments—typically prevalent at earlier cosmic times—stellar winds are weaker, causing massive stars to retain more of their mass. This increases the likelihood of forming more massive black holes, thereby improving the efficiency of BBH formation~\cite{Santoliquido_2022}. In contrast, in high-metallicity environments—more typical of the later universe—strong stellar winds lead to greater mass loss, resulting in lower-mass remnants and a suppression of BBH formation and merger rates.
Therefore, this effect should result in host-galaxy probability decreasing with metallicity at formation time. However, if the delay time between BBH formation and merger is long, these systems may merge at much later epochs, so low-metallicity host galaxies have evolved to be more massive and metal rich. Interestingly, low-metallicity progenitors also appear to be more associated with short delay times, leading to BBH mergers that occur shortly after formation ~\cite{Fishbach_2021,Chruslinska_2018}. This intricate interplay between metallicity, star formation history, and delay time, though theoretically complex, is a key factor underlying the observed redshift evolution of BBH merger rates~\cite{Rauf_2023,Mannucci2010}.\footnote{For additional details, see~\cite{Artale_2019,2023MNRAS.523.5719R,Salim_2007}.} Note that, for our purposes, in galaxy surveys at low redshift, metallicity-driven differences in BBH merger environments are less apparent. Most galaxies in this regime have already undergone substantial chemical enrichment, resulting in a narrower metallicity range. Combined with limited sensitivity to low-metallicity or faint host galaxies, this observational bias complicates efforts to isolate the impact of metallicity from other factors such as star formation rate or delay time. These trends are more evident at higher redshifts or in simulations that track chemical evolution over cosmic time.
Still, to test the potential impact of the host-galaxy correlation with metallicity, analogous to stellar mass and SFR, we represent the influence by a broken power law for the $Z$ host-galaxy probability function as follows,
\begin{align}
    P_{\rm host}(\rm GW|Z)=   
    \begin{cases}    
      A_{Z}10^{(\log(Z)-\log(Z_\mathcal{K}))/\zeta_l}, \quad &\text{for}\,\, -4 \leq \log(Z) \leq \log(Z_\mathcal{K}),\\
    A_{Z}10^{(\log(Z_\mathcal{K})-\log(Z))/\zeta_h}, \quad &\text{for}\,\, \log(Z) > \log(Z_\mathcal{K}),
    \end{cases}
     \label{eq:M_Z_SFR function}
\end{align}
where $Z_\mathcal{K}$ is the pivot point and Z is the metallicity of galaxies read from the catalog. $A_{Z}$ is the normalization factor. In this study, we consider $Z_\mathcal{K}\in[0.01,0.3]$,  $\zeta_l \in [0.5,10]$, and $\zeta_h \in [0.5,10]$.

Finally, by taking the product of all three host-galaxy probability functions defined above for SFR, $Z$, and $M_*$ and directly extracting their values from the catalog (without relying on observational scaling relationships), we can calculate the overall probability of host galaxies based on their combined astrophysical properties using Eq.~\eqref{eq:joint_prob_hosts}. This provides a very direct and comprehensive strategy for the selection of host galaxies such that the complex underlying physics of the BH-galaxy relation is encoded in the shape of these three host-galaxy probability functions. By varying the parameters governing their shape, we are able to cover different BH formation scenarios, and investigate how much GW bias changes for different parameters. Furthermore, this formulation also allows us to consider scenarios in which only one of these three astrophysical properties are important. In particular, we also consider a scenario in which we marginalize over \( \mathrm{SFR} \) and \( Z \), populating the BBHs in the catalog solely based on  $P_{\rm host}(\mathrm{GW}|M_*)$, to compare the result with our previous paper \cite{dehghani2024}.



\section{3D power spectrum estimation}
\label{sec:3D power spectra}

\subsection{Measuring the anisotropic power spectrum and covariance}
\label{sec:3dpower in catalog}
In this section, we present our derivation of the 3D power spectrum for galaxy/host-galaxy samples selected from the SDSS DR7 catalog described in Section \ref{sec:galaxy catalog}. The power spectra of the galaxy and host-galaxy catalogs are computed using \verb|pypower| \footnote{\url{https://github.com/cosmodesi/pypower}} (from the \verb|Cosmodesi| package), which is designed for the calculation of galaxy auto- and cross-power spectra, along with the estimation of associated window functions. \verb|pypower| can calculate the power spectra of both periodic boxes and of realistic survey geometries, 
 operating in either flat-sky or plane-parallel RSD configurations, and incorporating first-order odd wide-angle corrections when computing the window-convolved power spectrum from a galaxy survey \cite{BeutlerMcDonald20}.
More specifically, we used the \verb|CatalogFFTPower| algorithm \cite{Hand2017}, which takes the galaxy catalog (with position coordinates provided in terms of right ascension, declination, and redshift) as an input and provides the associated power spectrum multipoles using the Yamamoto estimator \cite{Yamamoto06}. This algorithm requires the redshift-dependent number density for the data catalog, $n(z)$\footnote{Note that, the SDSS DR7 galaxy sample has data and random files. Data files contain the actual observed data of galaxies. Random files are used to correct for observational biases and to account for the survey's selection functions.}. We first  calculate the redshift histogram of the randoms catalog (using the \verb|RedshiftHistogram| algorithm) and then adjust the redshift distribution to match that of the data catalog. 
Next we compute their corresponding FKP weight using the following formula\footnote{The term ``FKP weight'' refers to the weight assigned to each galaxy or object in the catalog based on the FKP (Feldman, Kaiser, and Peacock) method to optimally measure the power spectrum for a sample varying in redshift \cite{Feldman_1994}.}
\begin{equation}
    w_{\textrm{FKP}}=\frac{1}{1+n(z) \times P_0},
\end{equation}
where $P_0=10^4 h^{-3}$ Mpc$^{3}$. We have examined the impact of different FKP weighting schemes on the inferred GW bias by comparing cases in which the same FKP weights are applied to both galaxies and sirens with cases in which separate FKP weights are used for each tracer. As discussed in Appendix \ref{sec:random_sel}, these choices can lead to small but systematic offsets at the level of a few percent in the inferred bias, with the effect being more noticeable in the higher redshift bin $0.135 \leq z < 0.2$. While these differences do not alter the qualitative trends of our results, they are relevant for the interpretation of bias measurements. To provide the cleanest and most stable comparison throughout the analysis, we therefore adopt the galaxy FKP weights for both galaxies and sirens in all subsequent GW bias estimations, to ensure that galaxies and sirens have an identical effective redshift.

With these ingredients in place, one can proceed to calculate the 3D power spectrum using the \verb|CatalogFFTPower| algorithm. 
We use interlacing and paint galaxies to the grid using the Triangular Shaped Cloud (TSC) interpolation to allow accurate recovery of the power spectrum up to the Nyquist frequency \cite{Jing05,Sefusatti15}.
Specifically, because of redshift-space distortions, isotropy is broken in the observed power spectrum and it is therefore a function of both wavenumber $k$, cosine of the angle between the wavevector and the line-of-sight direction $\mu$, or can be equivalently expressed in the basis of Legendre multipoles of the line-of-sight. More details are provided in the \verb|Pypower|\footnote{\url{https://pypower.readthedocs.io/en/latest/}} and \verb|Nbodykit|\footnote{\url{https://nbodykit.readthedocs.io/en/latest/cookbook/convpower.html}} \cite{Hand2018} documentations.

The window-convolved power spectrum covariance is calculated analytically using \verb|thecov| software \cite{Wadekar19,Alves24}.\footnote{\url{https://github.com/cosmodesi/thecov}} We compute the window-convolved Gaussian contribution to the covariance, using the appropriate randoms for the SDSS DR7 geometry. We find good agreement between the analytic covariance and the mock-based covariances from the SDSS DR7 clustering analysis of \cite{Ross_2015}.

As the last remark of this section, to calculate the power spectra of sirens (host galaxies), as previously mentioned, we first need to generate the siren catalog from the galaxy catalog. To do so, we first calculate the number of mergers in each redshift bin using Eq. \eqref{eq:N_GWcalculate}. We then populate the galaxy catalog with potential hosts of binary mergers using  $M_*$-$Z$-$\rm SFR$ host-galaxy probably model mentioned in Section \eqref{sec:assignment}. To form the host catalogs, we select galaxies without replacement.
When using the merger rate normalization $\mathcal{A}_0$ for observing times of order one year to a decade, the number of predicted hosts in each stellar mass-SFR-metalicity bin extracted from our galaxy sample would be very small. This would produce a siren catalog that is too sparse and therefore the theoretical prediction for GW bias would be strongly shot-noise dominated. On the other hand, if we increase $T_\mathrm{obs}$ too much to improve the theoretical estimate, then all the galaxies in certain ranges of stellar mass and SFR of the catalog are selected as hosts. Especially because galaxies are selected without replacement, this would bias the mean host galaxy properties (stellar mass, star formation rate, and metallicity) away from the true value due to observational incompleteness of the catalogs.
To address this we increase the observing time in our calculations (typically around two orders of magnitude) for each choice of underlying host-galaxy model parameters up to a point that the effective number of mergers is large enough to avoid shot-noise domination yet not so large that it selects a significant fraction of galaxies in any stellar-mass, metallicity, or SFR bin, thereby reflecting impact of the host-galaxy probability selection.

\subsection{Measuring the GW bias from the power spectrum multipoles}
\label{sec:RSD effect}
Galaxies are observed in redshift space, and their observed positions are influenced by peculiar velocities and gravitational infall in addition to cosmological expansion. This causes two types of anisotropic clustering along the line of sight: apparent elongation due to random motion on small scales (Fingers of God (FoG) \cite{Jackson1972} and large-scale squashing along the line-of-sight (Kaiser effect) due to galaxy motions toward over-densities (see Ref.\ \cite{1987MNRAS.227....1K}). 
Taking into account the redshift space distortion due to the Kaiser effect, the power spectrum in redshift space can be shown as \cite{percival2013largescalestructureobservations} 
\begin{equation}
    P^s_{XX}(k, \mu) = P^r_{XX} - 2 \mu^2 P^r_{X\theta} + \mu^4 P^r_{\theta \theta} ,
\end{equation}
where if we set the subscript to $X=g$ it refers to galaxies and if it is set to $X=GW$ it indicates GW host galaxies. The $s,\,r$ superscripts indicate the redshift-space and the real-space field, respectively, $\theta$ denotes the divergence of the peculiar velocity field, and finally $\mu$ is the cosine of the angle between the wave vector $k$ and the line of sight.
Using $\delta_X = b_X \delta_m$ and $\theta = -f \delta_m$ in the linear regime, where $b_X$ is the galaxy (or GW sources) bias and $f$ is the linear growth rate $d\ln D/d\ln a$, we find
\begin{equation}
    P_{XX}(k, \mu) = b_X^2 P_m + 2 \mu^2 b_X f P_m + \mu^4 f^2 P_m\,,
\end{equation}
with $P_m$ denoting the matter power spectrum.
Small-scale redshift space distortions (the FoG effect) can be phenomenologically modeled as
\cite{Ross_2016}
\begin{equation}
    P_{XX}(k, \mu) = \left(b_X^2 P_m + 2 \mu^2 b_X f P_m + \mu^4 f^2 P_m\right) \left(1+\frac{k^2 \mu^2 \sigma_p^2}{2}\right)^{-2},
    \label{eq:p_gg with RSD,FOG}
\end{equation}
where $\sigma_p$, the velocity dispersion parameter is estimated to be $\sigma_p = 400$ km s$^{-1}$ based on constraints from SDSS MGS \cite{Howlett_2015, Ross_2016}. 
 The FoG term affects the damping of the power spectrum, particularly the quadrupole-to-monopole (defined below) ratio. Higher $\sigma_p$ leads to stronger suppression of anisotropies, while lower values cause overestimation. 
 In our fits to the GW bias, we allow $\sigma_P$ to be a free parameter.

  It is convenient to decompose $P_{XX}(k, \mu)$ into the Legendre basis, including the monopole ($P_{XX,0}$), quadrupole ($P_{XX,2}$), and hexadecapole ($P_{XX,4}$). 
 The signal-to-noise ratio on the hexadecapole is poor, so we will only consider the monopole and quadrupole hereafter. The monopole and quadrupole are then calculated by integrating $P(k, \mu)$ against Legendre polynomials in $\mu$:
\begin{equation}
    P_{XX,0}(k) = \int  \left(b_X^2 P_m + 2 \mu^2 b_X f P_m + \mu^4 f^2 P_m\right) \left(1  +\frac{k^2 \mu^2 \sigma_p^2}{2}\right)^{-2}d\mu.
    \label{eq:monopole}
\end{equation}

\begin{equation}
    P_{XX,2}(k) = \int \frac{1}{2} \, \left(3 \mu^2 - 1)  (b_X^2 P_m + 2 \mu^2 b_X f P_m + \mu^4 f^2 P_m\right) \left(1  +\frac{k^2 \mu^2 \sigma_p^2}{2}\right)^{-2} d\mu.
    \label{eq:quadropole}
\end{equation}
in which $X$ can again be either galaxies or GW sources.

To measure the GW bias, we start by computing the theoretical monopole and quadrupole, using the Code for Anisotropies in the Microwave Background (\verb|CAMB|) to compute the nonlinear matter power spectrum. We convolve the monopole and the quadrupole using the relevant window function which accounts for survey geometry effects and first-order wide angle effects \cite{BeutlerMcDonald20}. We then vary the free bias parameter and find the best-fit GW bias that minimizes $\chi^2$ given the data, the covariance, and the theory model. 
The uncertainties in the bias and $\sigma_p$ are estimated by evaluating the likelihood over a two-dimensional grid of parameter values around their best-fit estimates. For each combination of \(b_X\) and \(\sigma_P\), we compute the model power spectrum and its corresponding \(\chi^2\) with respect to the observed data. These \(\chi^2\) values are converted into a likelihood surface, which is then marginalized to obtain one-dimensional posterior distributions. The \(1\sigma\) uncertainties are defined as the intervals between the 16th and 84th percentiles of the marginalized distributions.

The 3D power spectrum of both GW sources and galaxies, is computed over a range of wave numbers up to a maximum value of $k_{\mathrm{max}} = 0.5\,h\,\mathrm{Mpc}^{-1}$, with a bin width of $\Delta k = 0.02\,h\,\mathrm{Mpc}^{-1}$. This results in $k$-bins defined by $k_{\mathrm{edges}} = \{k_0, k_1, ..., k_N\}$, where the total number of bins is given by $N = \mathrm{int}[(k_{\mathrm{max}} - k_{\mathrm{min}})/\Delta k]$.



\section{GW bias results}
\label{sec:results}
In this section, we present our findings on GW bias using the procedures outlined in the previous sections. Our goal is by varying the parameters that govern the shape of host-galaxy probabilities to explore how the correlation between the distribution of GW sources and the astrophysical environments of their host galaxies affects the GW bias. As a sanity check, we first test our framework by
randomly selecting sirens from galaxy samples rather than using any host-galaxy probability. The details are shown in Appendix \ref{sec:random_sel}. We would also like to point out that observational selection effects can alter the magnitude of the galaxy bias, which in turn affects the absolute value of the GW bias. Therefore, we always present our results in terms of the difference between the GW and galaxy bias. 
Moreover, all the error bars represented in the figures in this section correspond to the \(1\sigma\) marginalized uncertainties on GW bias (between the 16th and 84th percentiles of the posterior), obtained by fitting the theoretical monopole and quadrupole to the measured power spectrum using the window–convolved Gaussian covariance described in Sec.~\ref{sec:3D power spectra}. These error bars therefore reflect the statistical uncertainty (cosmic variance and shot noise) for a single realization of the mock siren catalog.

\subsection{GW bias exclusively derived from $M_*$ host-galaxy probability ($M_*$ model)} 
\label{sec:mzr_result}

\begin{figure}
\includegraphics[width=.47\textwidth]{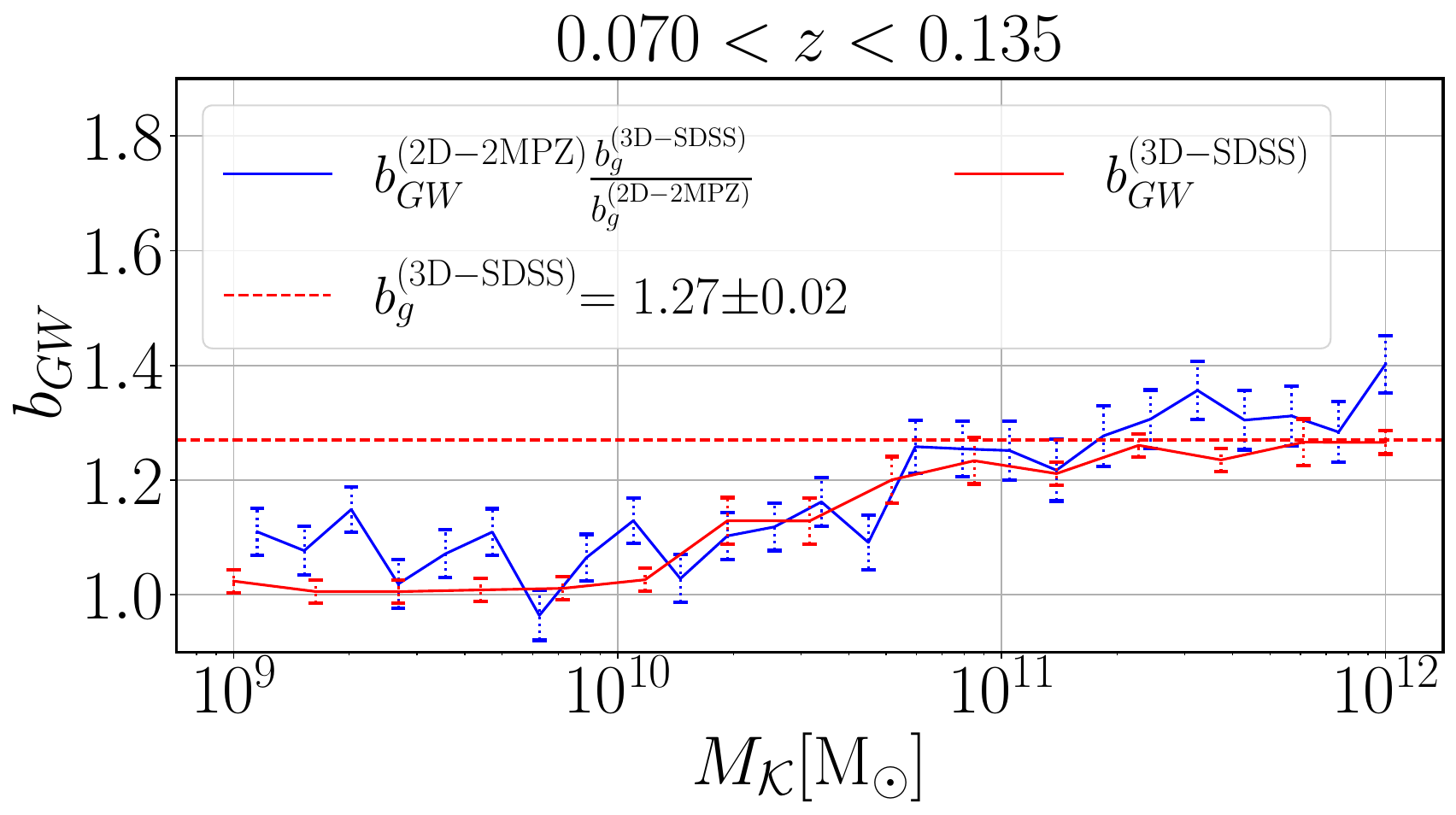}
\includegraphics[width=.47\textwidth]{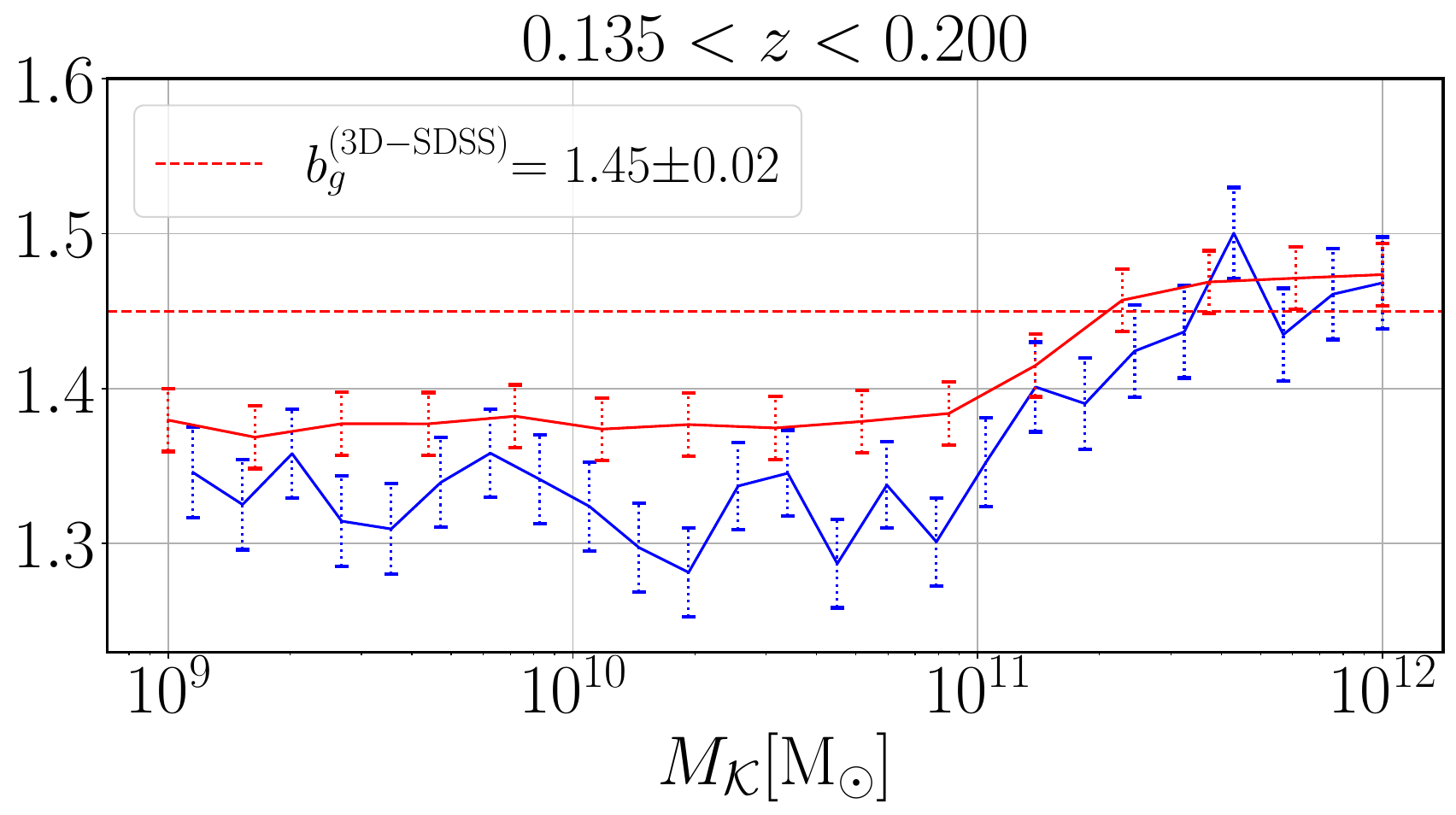}
\caption{Comparison between our results and those in \cite{dehghani2024}. Here we compare the GW bias computed using angular galaxy clustering in 2MPZ to the bias computed using 3D galaxy power in SDSS. We use two different redshift bins, $0.07 \leq z <0.135$ and $0.135 \leq z <0.2$. The 2MPZ bias (blue line) is rescaled to match the overall galaxy bias of SDSS (red line) as explained in the text. The agreement between GW hosts populated in 2MPZ and SDSS is good, up to small differences in the second redshift bin at $M_{\mathcal{K}} < 10^{11}$ $M_{\odot}$ due to differences in the galaxy survey completeness. Here we use a GW host-galaxy probability function which depends on $M_*$ only with $\delta_l=4$, and $\delta_h=0.5$.
} 
\label{fig:sdssvs2mpz_bias}
\end{figure}

In order to compare our results with those of our previous paper \cite{dehghani2024}, we start with a simpler model where the GW host-galaxy probability function only depends on the stellar mass, similar to the model used there. In other words, we only consider $P(\mathrm{GW}|M*)$ in Eq~\eqref{eq:joint_prob_hosts} and marginalize over SFR and $Z$.
The result of populating the sirens according to this prescription is presented in Figure \ref{fig:sdssvs2mpz_bias}. As the figure shows, there is good agreement between the calculated GW bias of the 3D power spectrum using populated SDSS DR7 sirens and the 2D power spectrum sirens populated in 2MPZ (following the methods of \cite{dehghani2024}). Note that the average galaxy biases are different between SDSS and 2MPZ. In the first bin, they are 1.27 and 1.34 for SDSS and 2MPZ, and in the second bin they are 1.45 and 1.78, respectively.
This matches the trend found in stellar mass: the mean stellar mass in 2MPZ is $7.6 \times 10^{10}$ $M_\odot$ and $16.1 \times 10^{10}$ $M_\odot$ in the two bins, while SDSS has smaller stellar masses of $5.7 \times 10^{10}$ $M_\odot$ and
$12.7 \times 10^{10}$ $M_\odot$.
Due to this offset in the galaxy biases, we scale the GW biases from 2MPZ by the ratio of the galaxy biases from the two catalogs. This rescaling removes the amplitude offset caused by the different galaxy biases to allow us to more easily test consistency of the trend between GW bias and pivot mass using either SDSS or 2MPZ. In general, the GW biases obtained from the two methods are in good agreement within 1-2$\sigma$ (68-95 \% C. I.) range.
The increase in GW bias parameter with $M_\mathcal{K}$ for the particular host-galaxy probability functions that we display ($\delta_l=4$ and $\delta_h=0.5$) is due to the fact that for lower values of $M_{\mathcal{K}}$ the distribution favors less massive galaxies, which have smaller halo masses and thus lower biases \citep{Behroozi_2013,Moster_2010}. Similarly, the convergence to $b_g$ at higher values of $M_{\mathcal{K}}$ is due to the fact that the host-galaxy probability distribution in this limit has a very shallow slope and is close to a uniform distribution (see \cite{dehghani2024} for more details).   

\subsection{GW bias from the $M_*$-$Z$-$\rm SFR$ host-galaxy probability ($M_*$-$Z$-$\rm SFR$ model)}
\label{sec:mzsfrr_result}

Next, we consider the full host-galaxy probability function depending on all three properties: $P(\mathrm{GW}|M_*)$, $P(\mathrm{GW}|\mathrm{SFR})$, and $P(\mathrm{GW}|Z)$. We aim to isolate the impact of each on GW bias by fixing two of these distributions at a time and varying the parameters of the third.

\subsubsection{Dependency on parameters of $P(\mathrm{GW}|M_*)$}
\label{sec:GW bias parameter vs stellar mass}
\begin{figure}

\includegraphics[width=.49\textwidth]{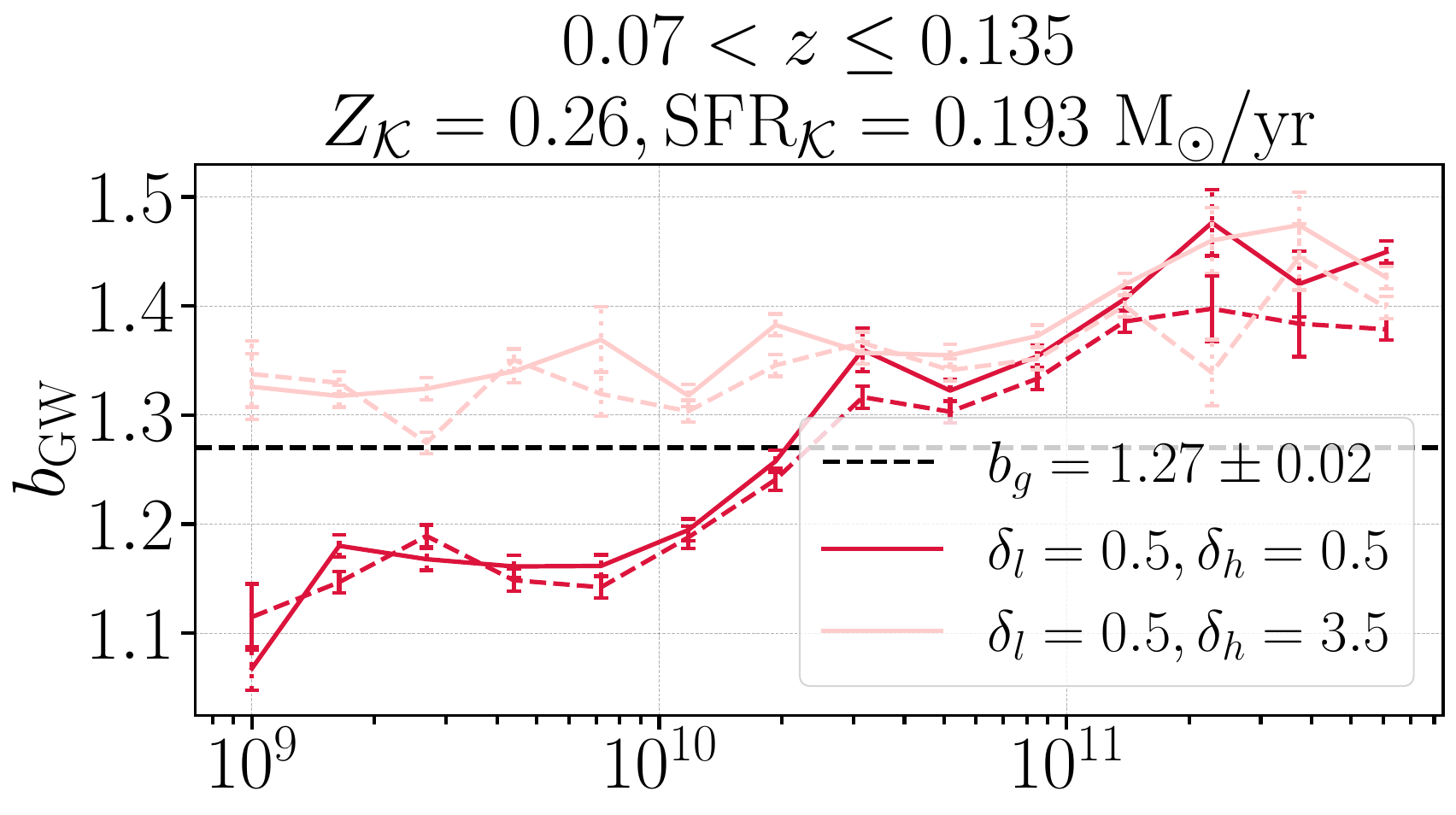}
\includegraphics[width=.49\textwidth]{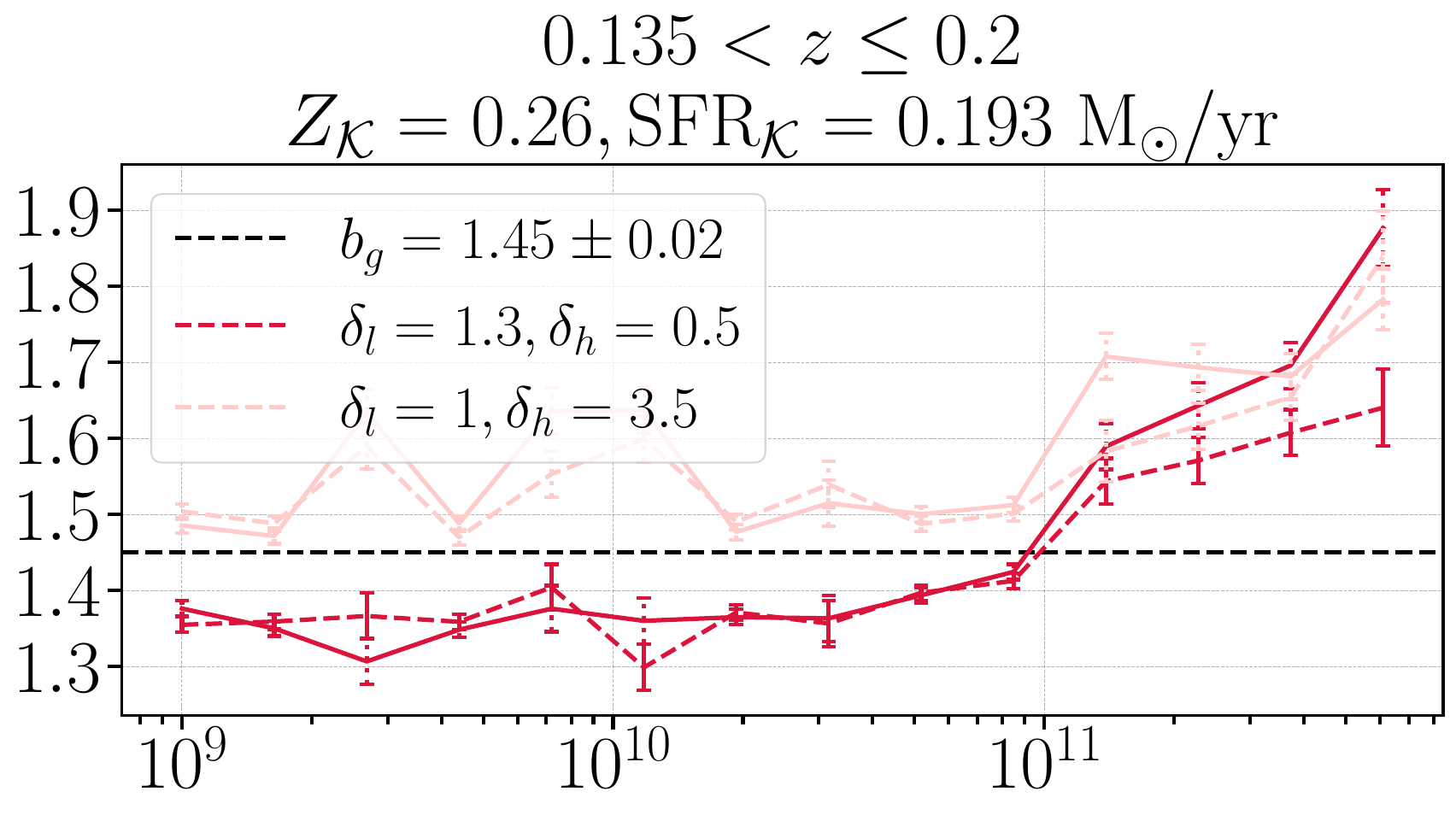}

\includegraphics[width=.49\textwidth]{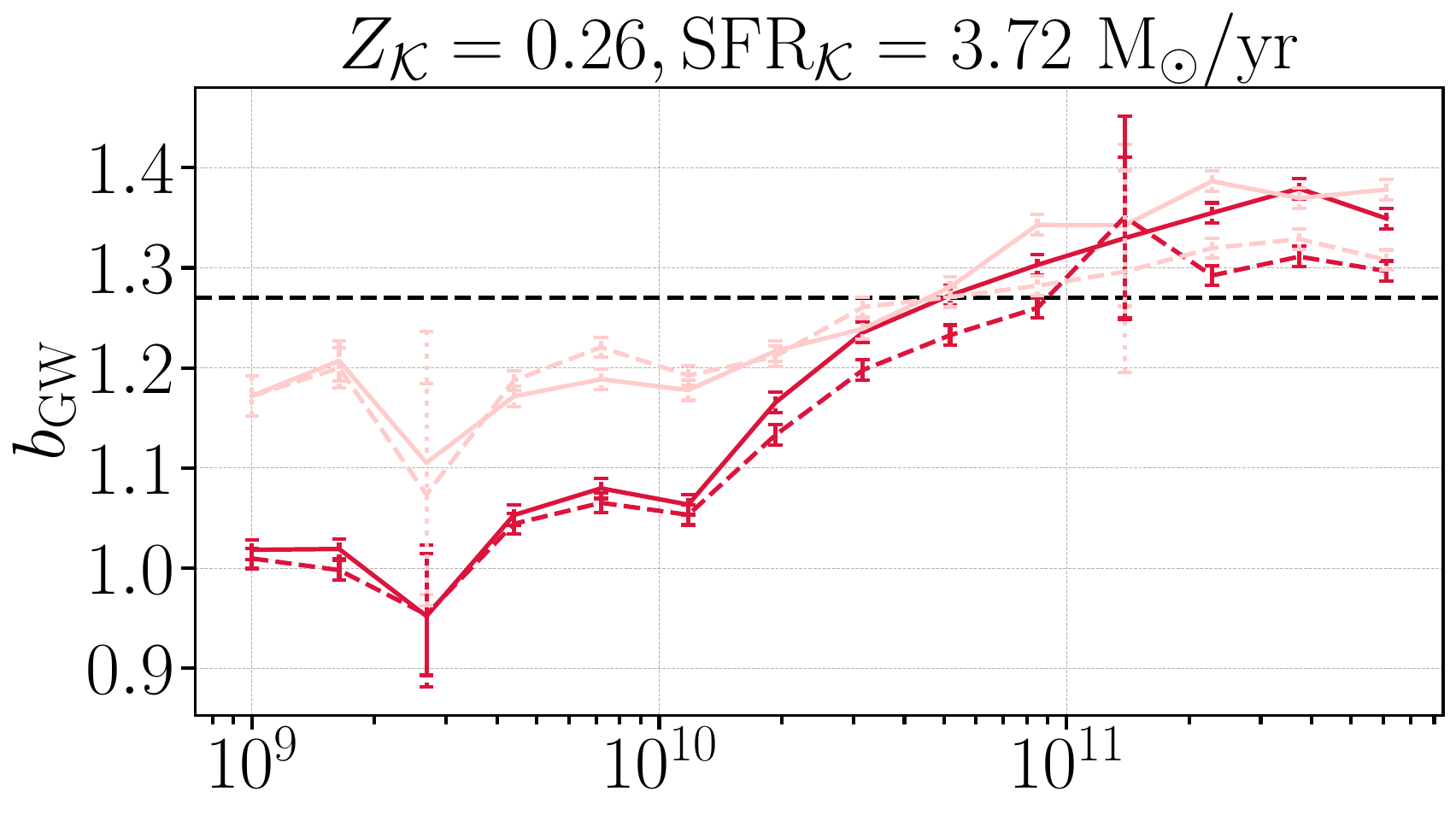}
\includegraphics[width=.49\textwidth]{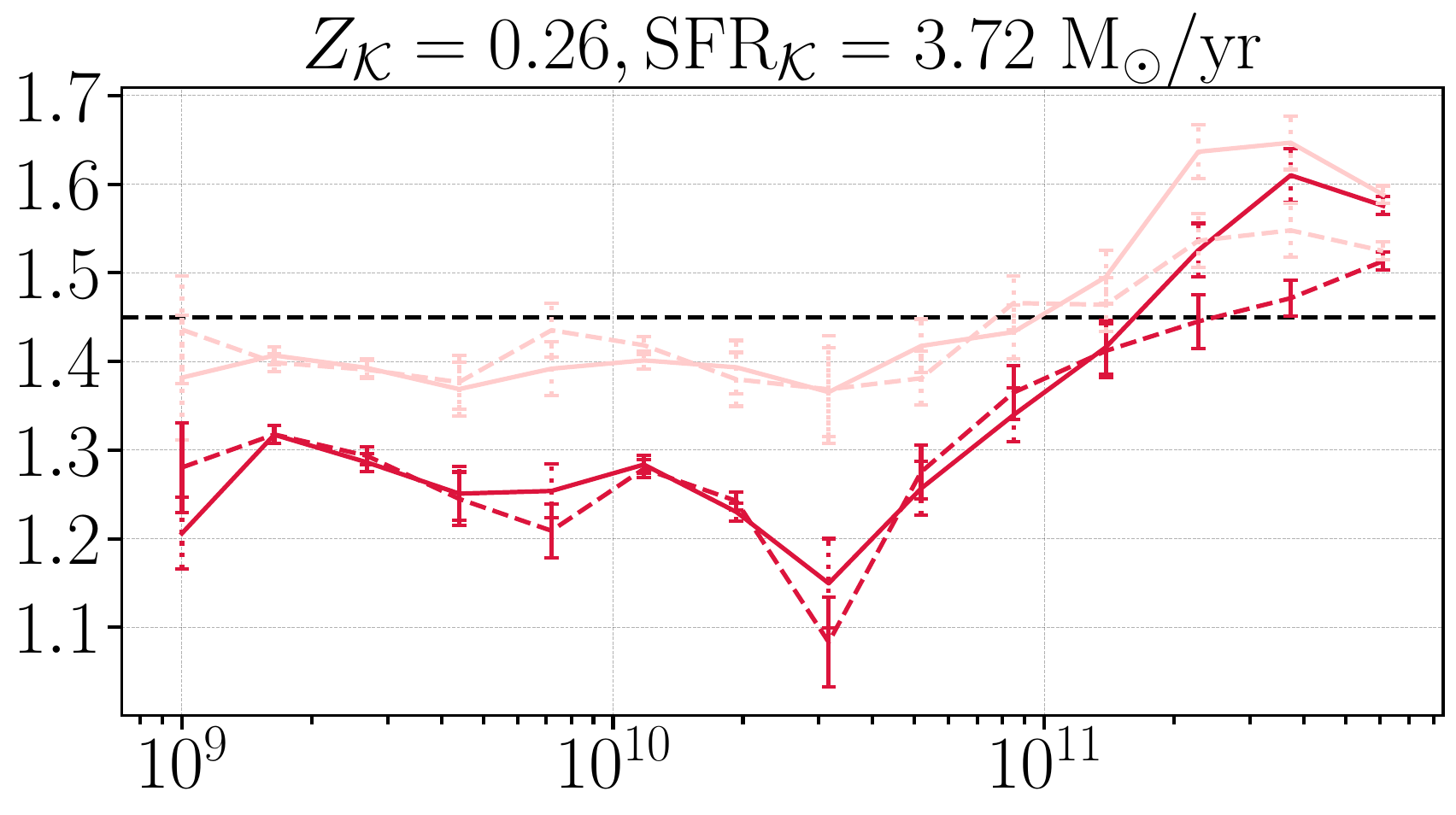}

\includegraphics[width=.49\textwidth]{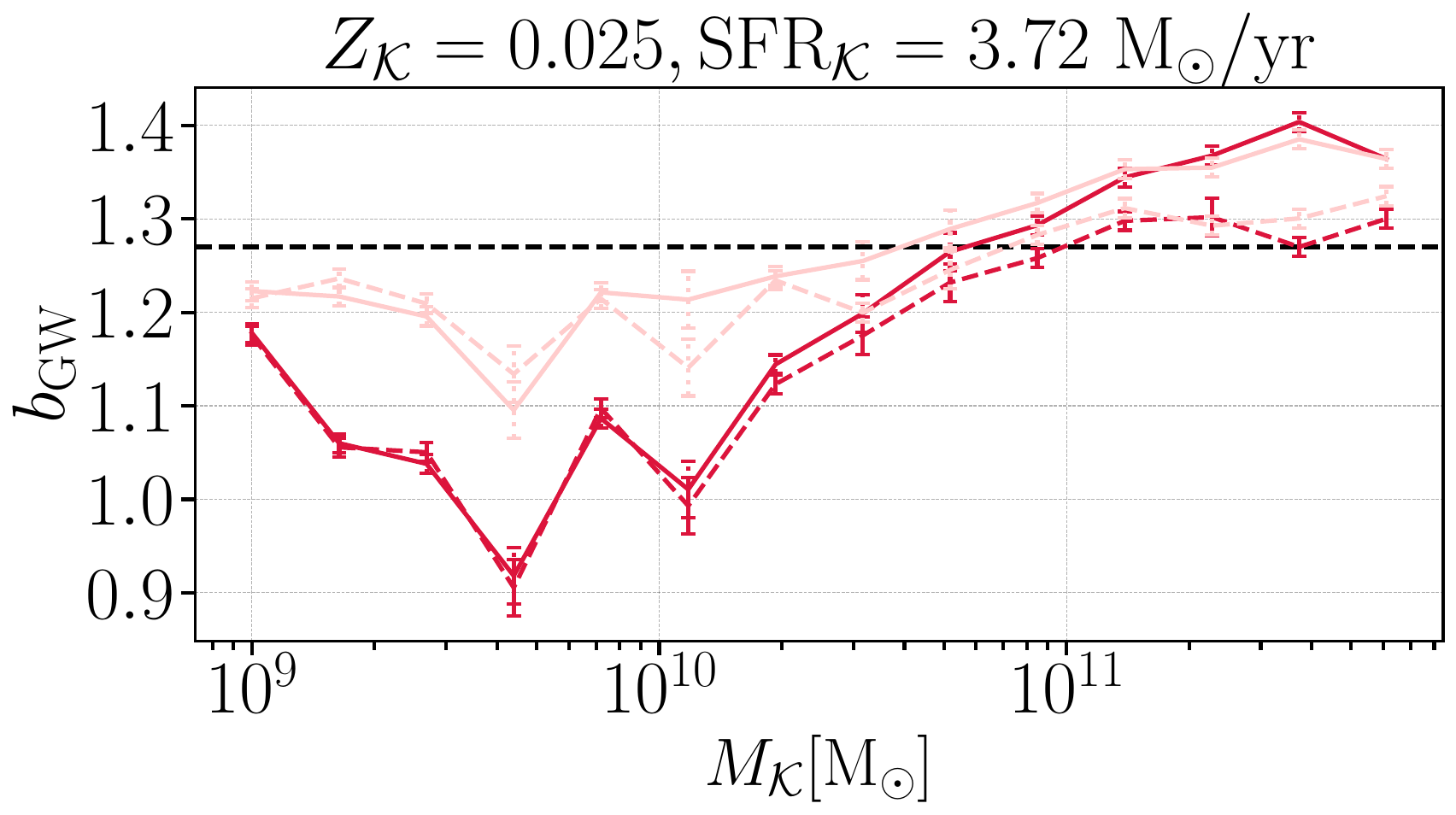}
\includegraphics[width=.49\textwidth]{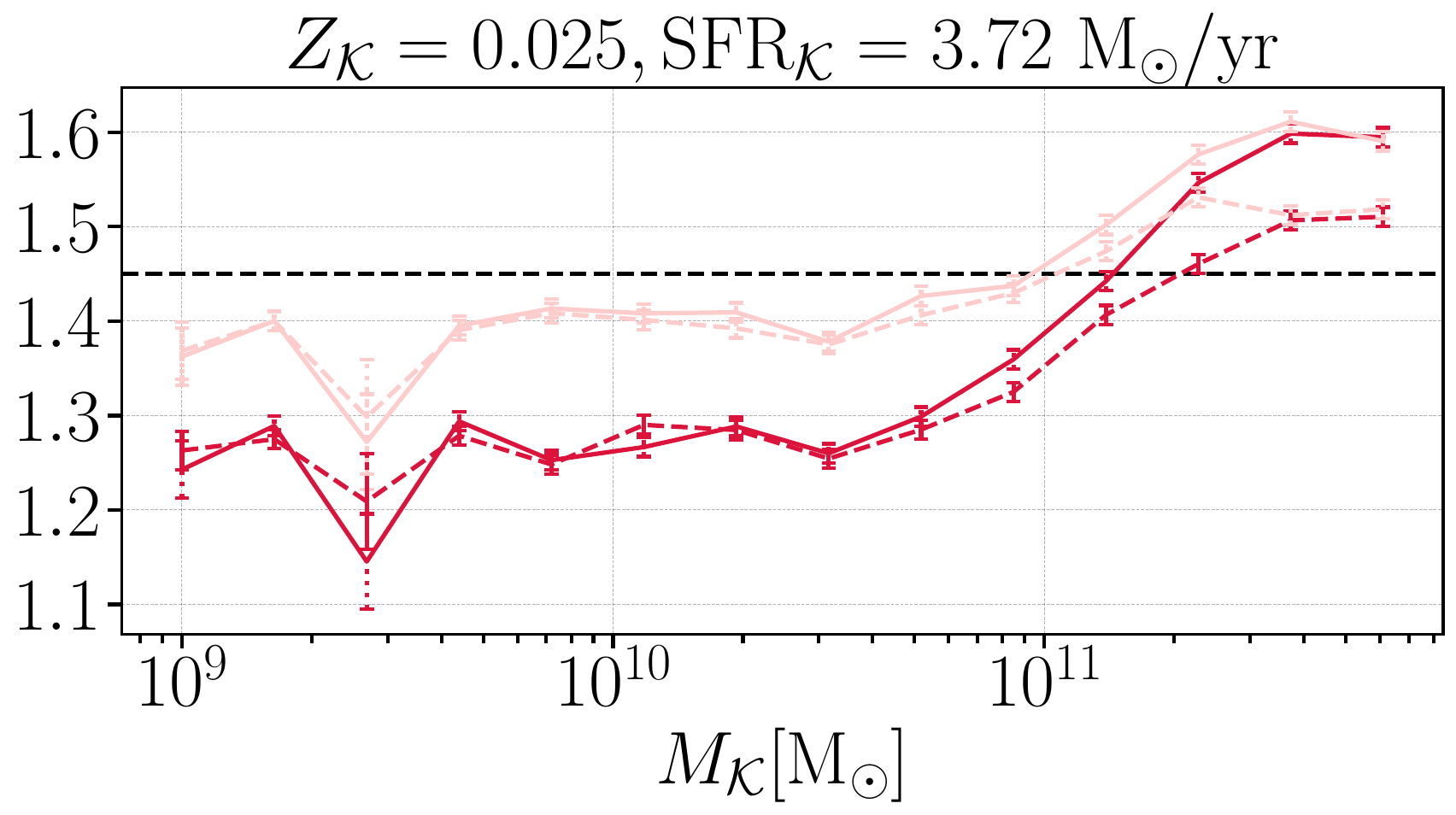}

\includegraphics[width=.49\textwidth]{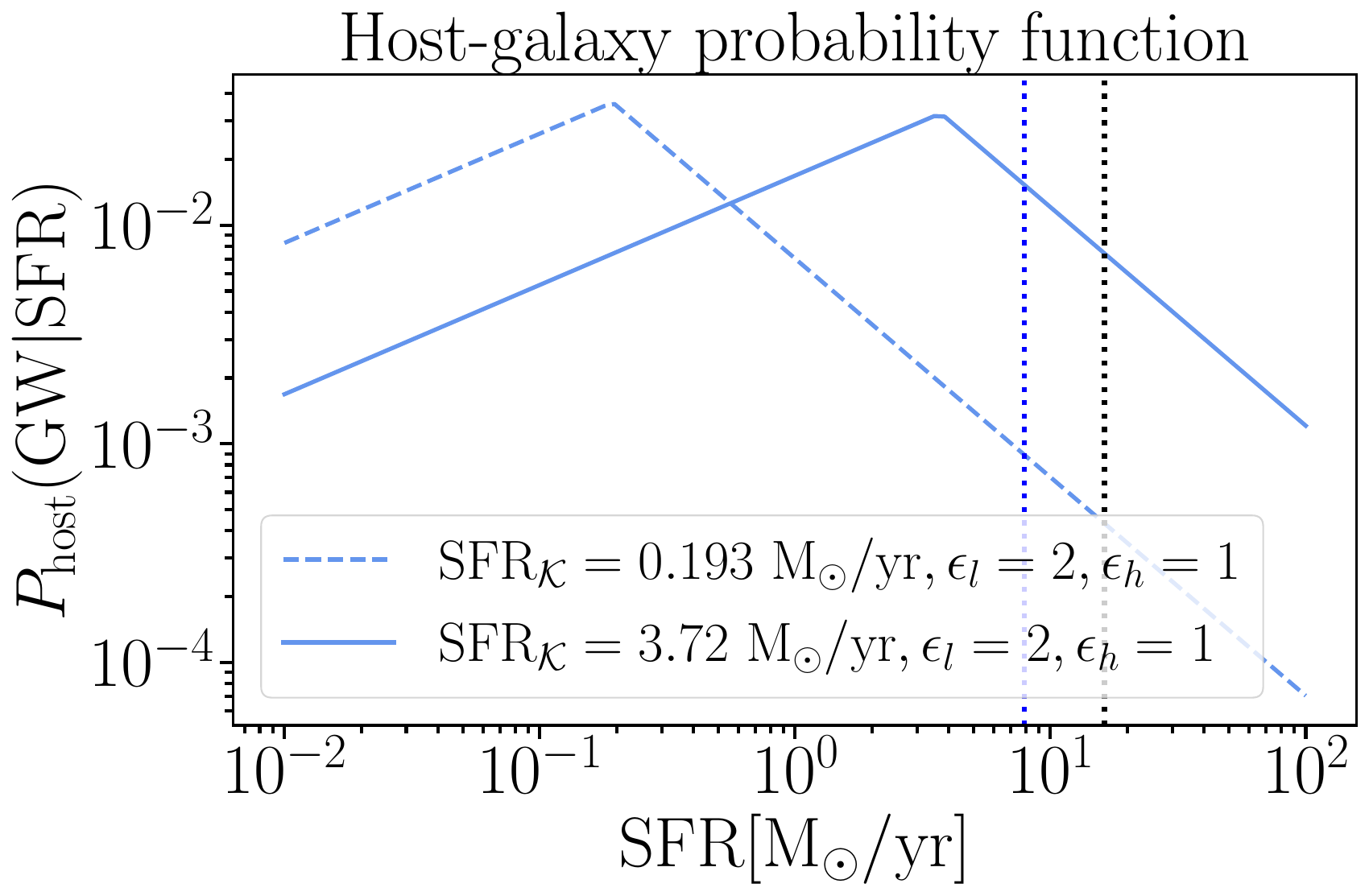}
\includegraphics[width=.49\textwidth]{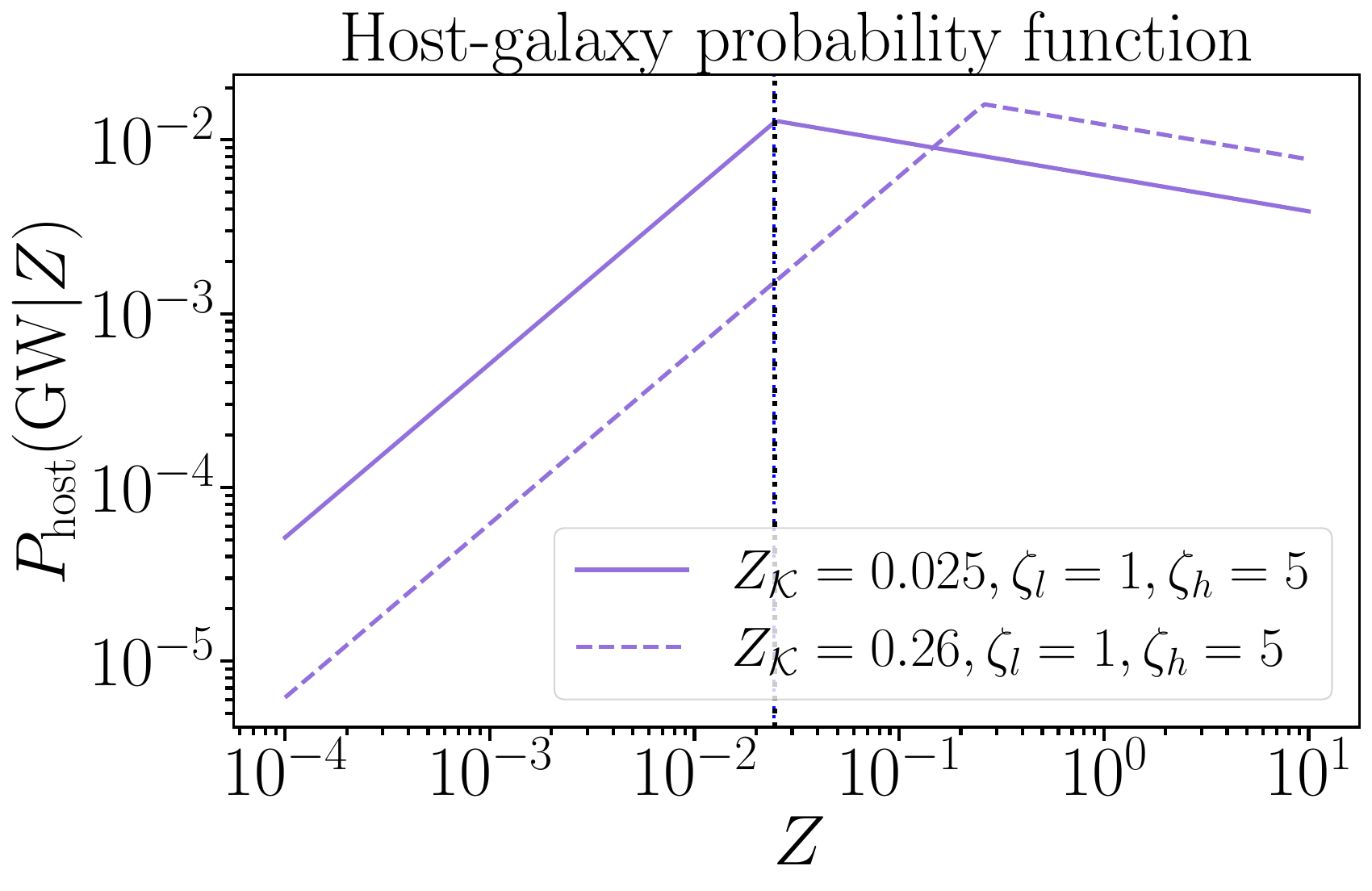}

\caption{GW bias parameter as a function of stellar mass break of the host-galaxy probability, $M_\mathcal{K}$ at redshift $0.07 \leq z <0.135$ (left) and $0.135 \leq z <0.2$ (right). Within each plot, we show the impact of changing the growing slope ($\delta_l$) and the falling slope ($\delta_h$) of $P(\mathrm{GW}|M_*)$. Different rows show the impact of first changing the power-law break in star formation rate selection, $\rm SFR_\mathcal{K}$, and next the metallicity, $Z_\mathcal{K}$, set to the values indicated in the plot titles. The slopes of the SFR and metallicity portions of the GW host-galaxy probability are fixed at  $\epsilon_l=2$, $\epsilon_h=1,~ \zeta_l=1,~ \zeta_h=5$. The last row shows the shape of the $P(\mathrm{GW}|\mathrm{SFR})$, and $P(\mathrm{GW}|Z)$ functions used in upper plots. In the bottom-left plot, the dotted vertical blue and black lines show the mean SFR of the galaxies in the first and second redshift bins, which are $7.9~[M_\odot\mathrm{\ yr}^{-1}]$ and $ 16.28~[M_\odot\mathrm{\ yr}^{-1}]$, respectively. In the bottom-right plot, the blue and black vertical dotted lines (coinciding) show the mean metallicity of the galaxies in the first and second redshift bins which are almost the same ($0.0245$ and $0.0246$).
}
\label{fig:DMbias_mzsfrvsmk}
\end{figure}

\begin{figure}
\includegraphics[width=.49\textwidth]{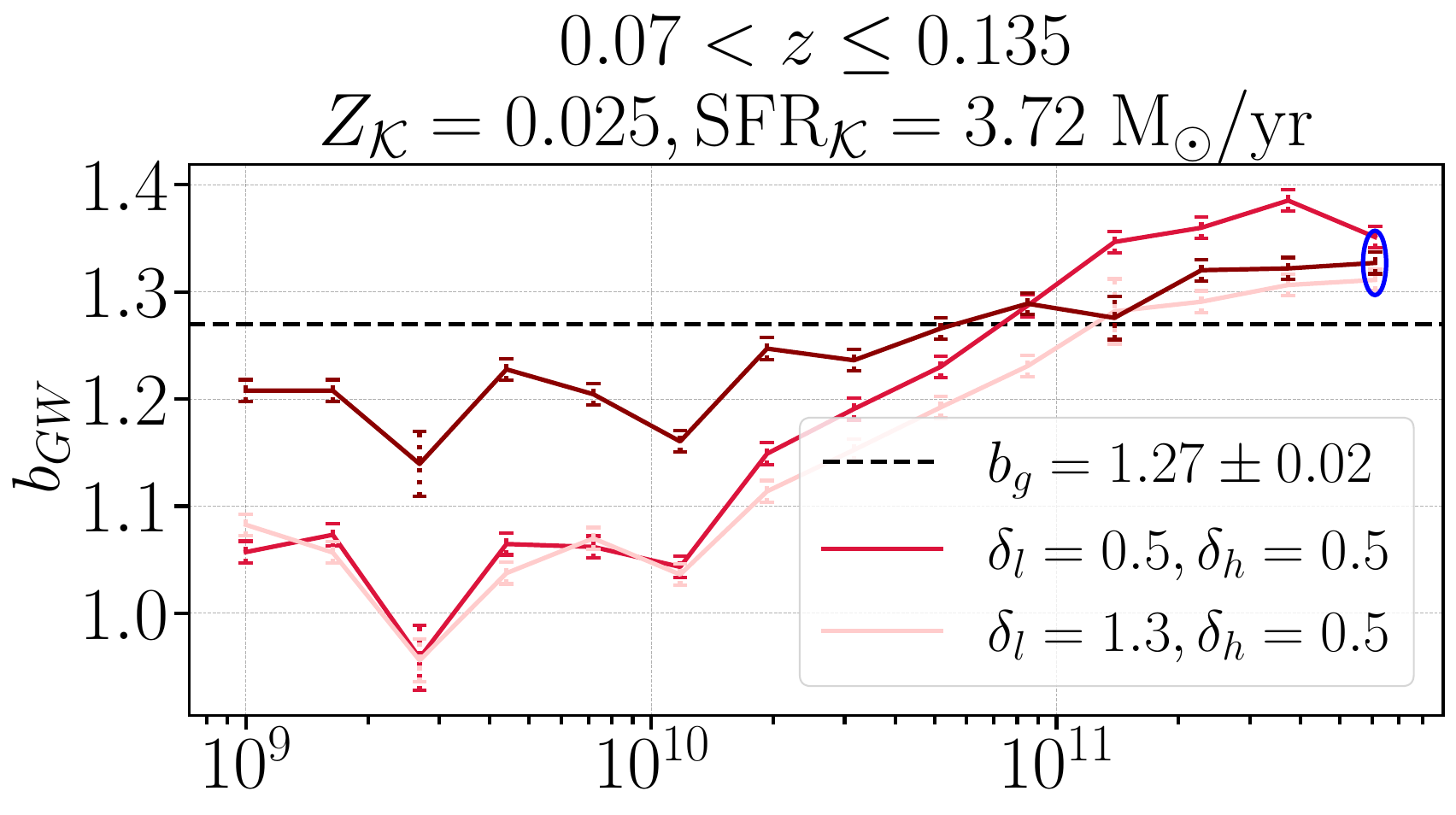}
\includegraphics[width=.49\textwidth]{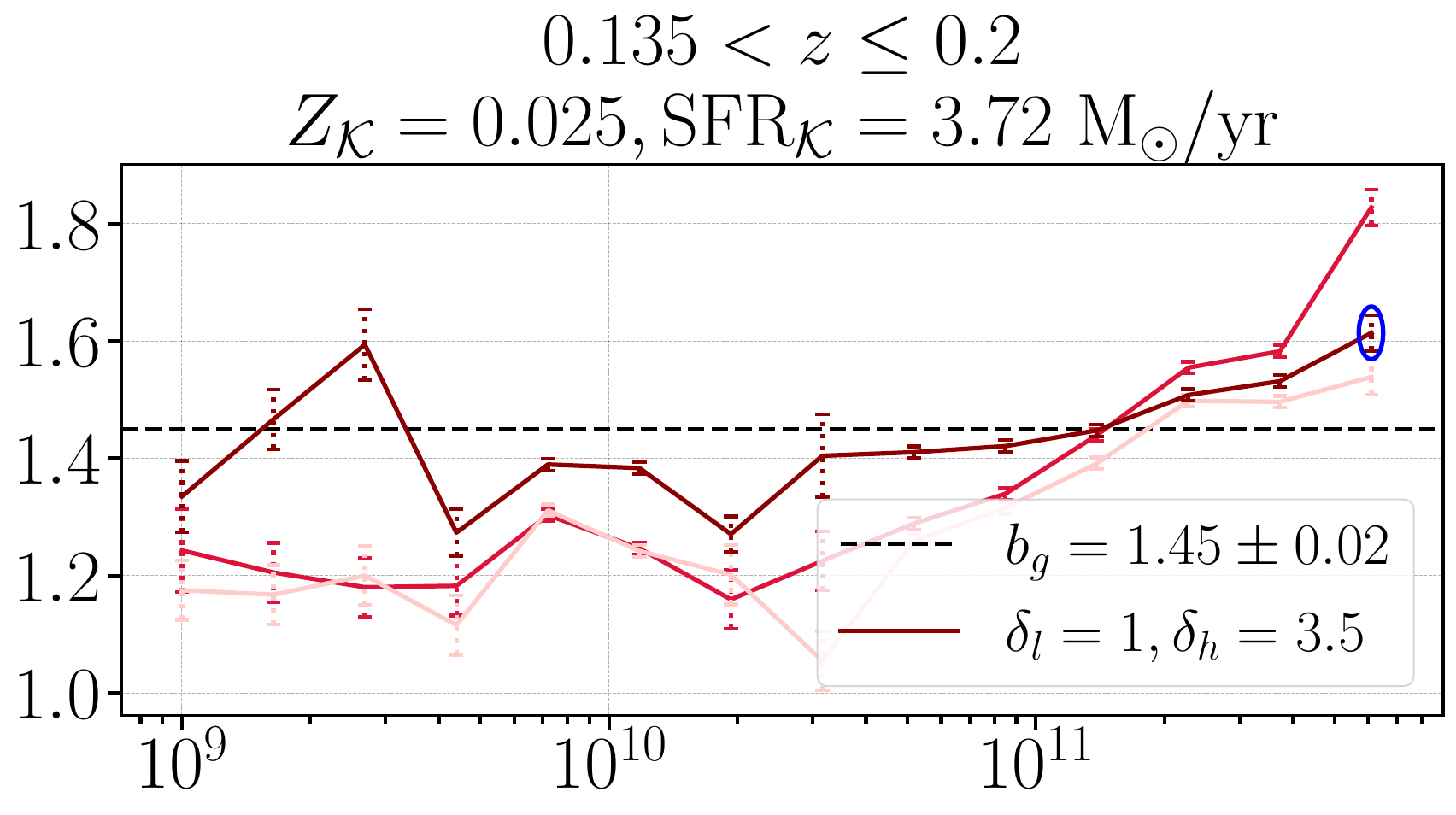}
\includegraphics[width=.49\textwidth]{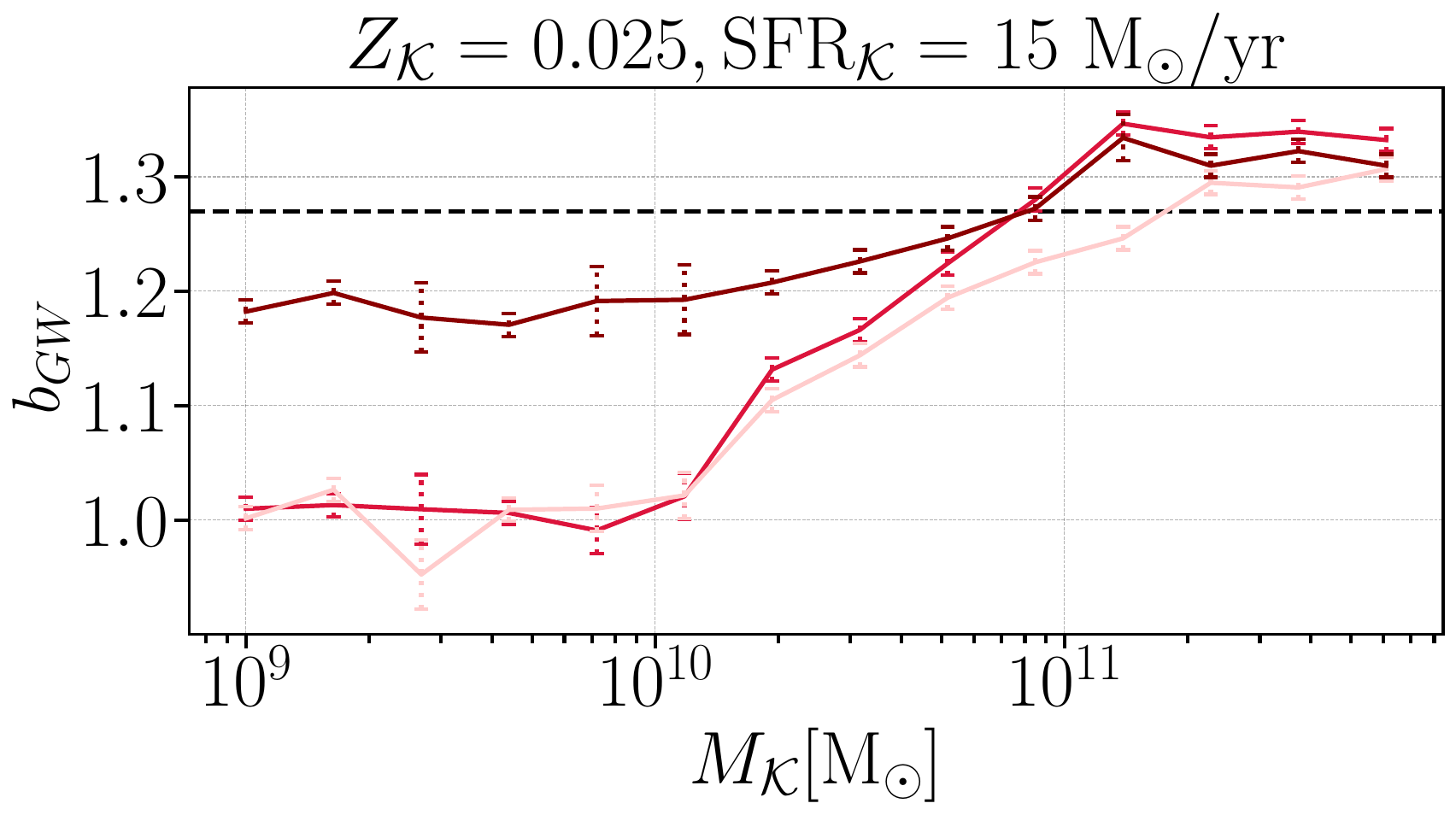}
\includegraphics[width=.49\textwidth]{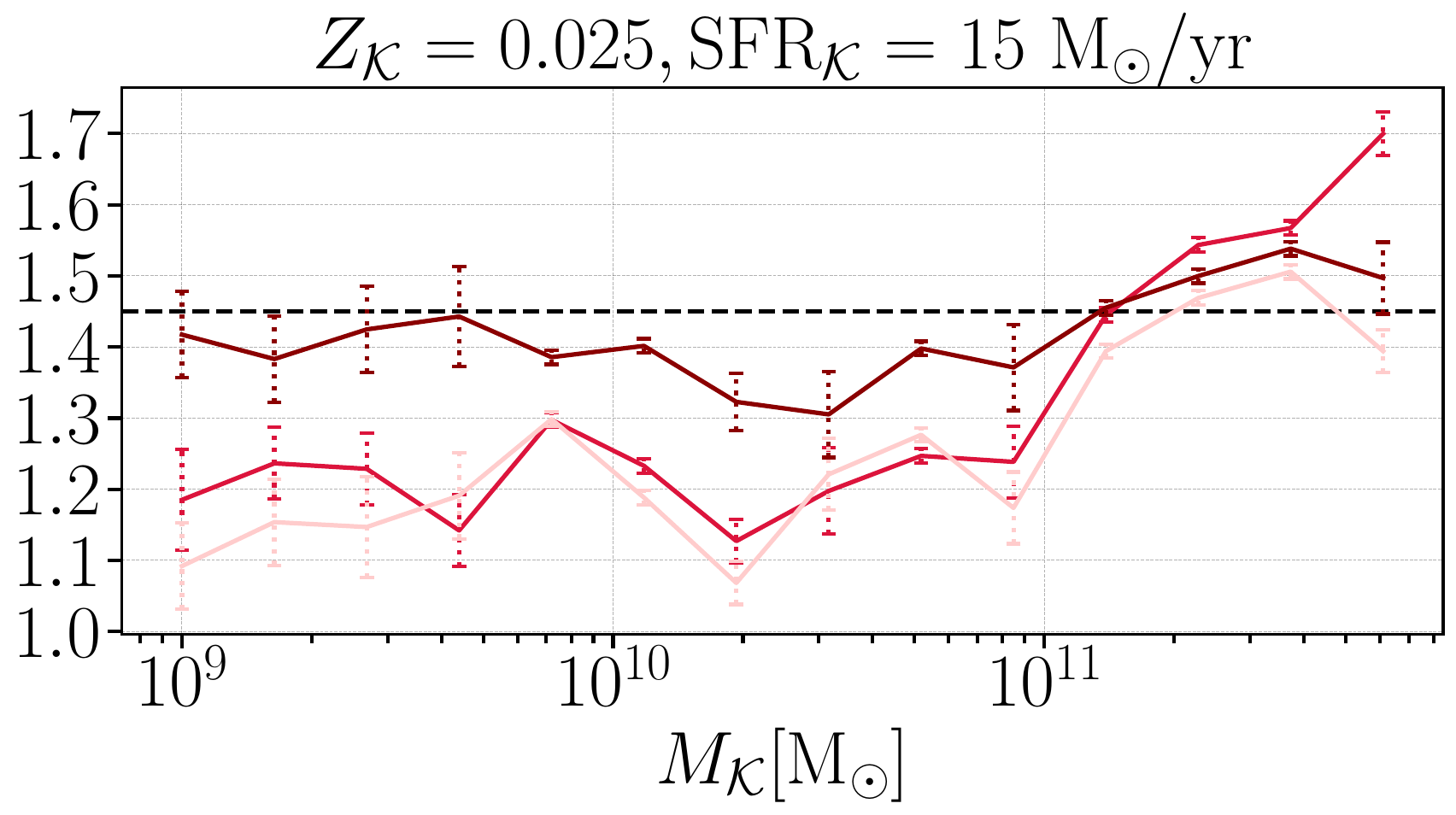}

\includegraphics[width=.49\textwidth]{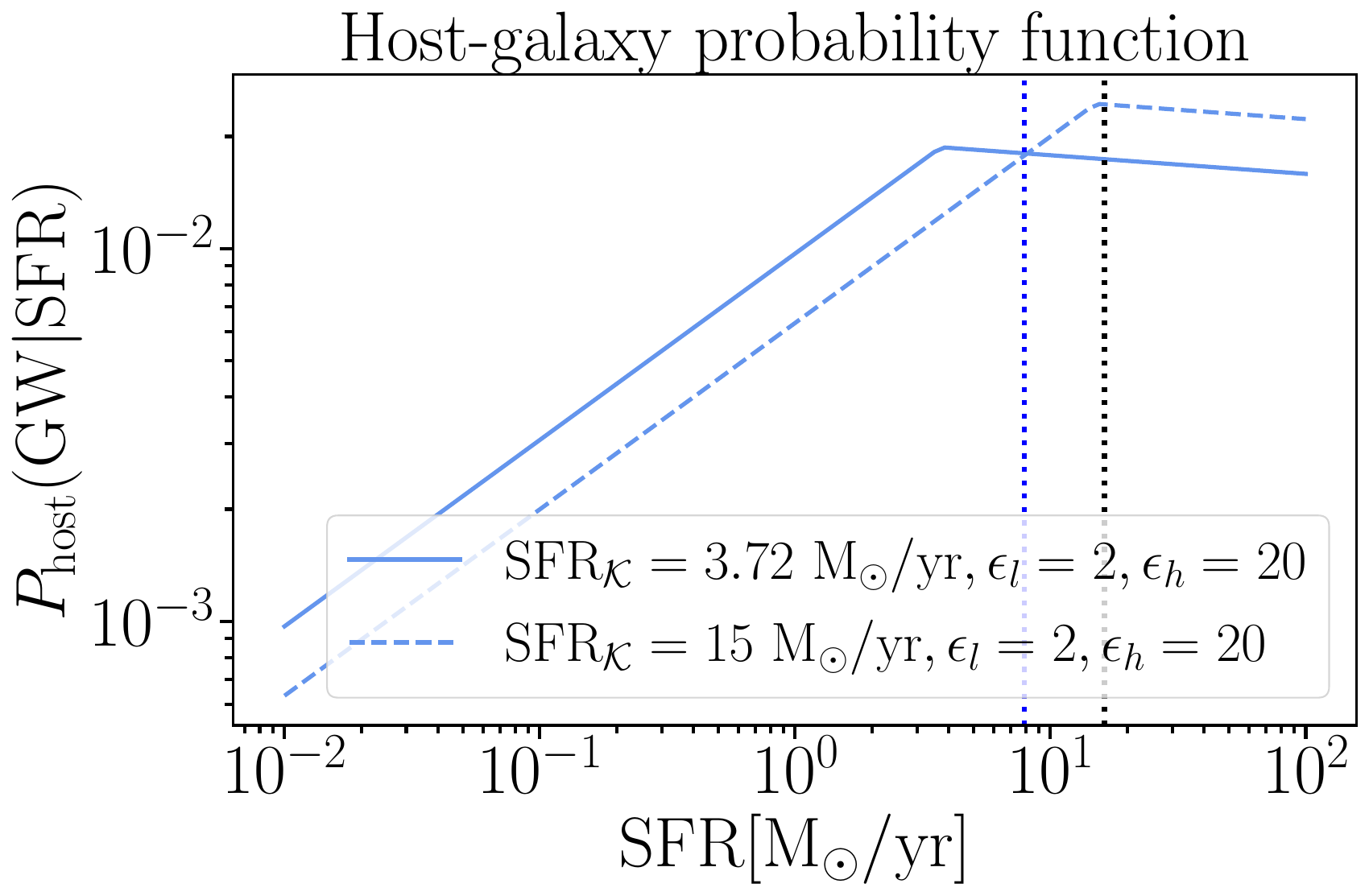}
\includegraphics[width=.49\textwidth]{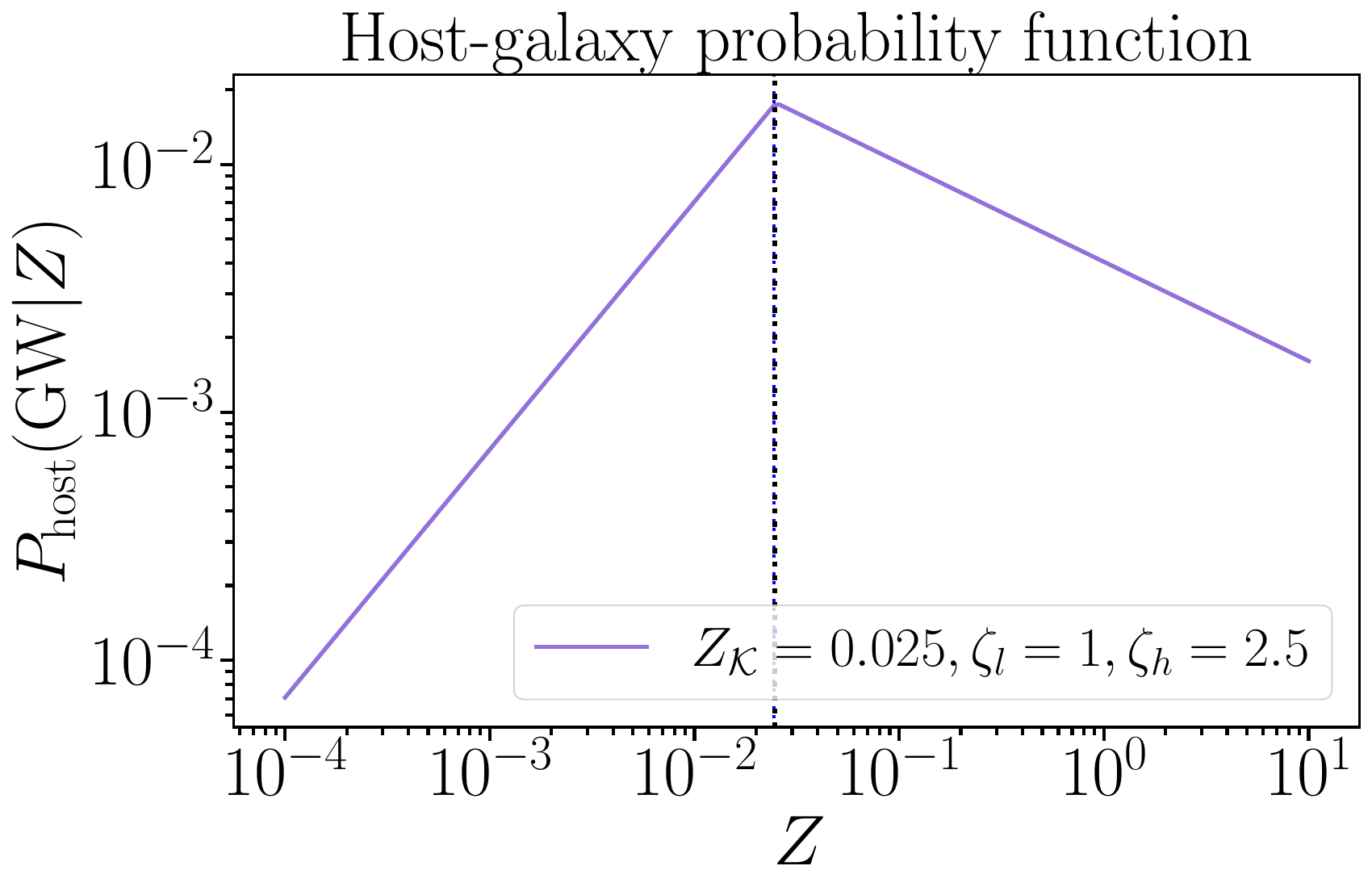}
\caption{Similar to \ref{fig:DMbias_mzsfrvsmk}, GW bias parameter as a function of stellar mass break in the host-galaxy probability $M_\mathcal{K}$ within redshift $0.07 \leq z <0.135$ (left) and $0.135 \leq z <0.2$ (right), considering additional shapes for SFR and metallicity selection. Specifically, here we are considering SFR selections with shallower high-mass slopes and the metallicity dependence of $P(\mathrm{GW}|Z)$ is chosen to be similar to what was obtained in \cite{Artale_2019}. 
The slopes are fixed by setting  $\epsilon_l=2$, $\epsilon_h=20,~ \zeta_l=1,~ \zeta_h=2.5$. The last row shows the shape of the $P(\mathrm{GW}|\mathrm{SFR})$, and $P(\mathrm{GW}|Z)$ functions used in upper plots. In the bottom-left plot, the dotted vertical blue and black lines show the mean SFR of the galaxies in the first and second redshift bins. In the bottom-right plot, the blue vertical dotted line shows the mean metallicity of the galaxies in the first and second redshift bins.
In the top panels of the figure, the rightmost points on the maroon curves, highlighted by blue ovals, represent the GW bias values corresponding to galaxy host probabilities similar to those reported in \cite{Artale_2019}.
}
\label{fig:DMbias_mzsfrvsmk_eph20}
\end{figure}

We start by examining the GW bias dependence on the parameter $M_\mathcal{K}$ of $P(\mathrm{GW}|M_*)$, for different combinations of $\delta_l$, $\delta_h$ choices while also fixing the parameters corresponding to $P(\mathrm{GW}|\mathrm{SFR})$ and $P(\mathrm{GW}|Z)$ to several potential scenarios. 
Figures \ref{fig:DMbias_mzsfrvsmk} and \ref{fig:DMbias_mzsfrvsmk_eph20} illustrate how $b_{\textrm{GW}}$ varies as a function of $M_{\mathcal{K}}$ for these cases. The bottom panels show the shapes and values of the parameters for $P(\mathrm{GW}|\mathrm{SFR})$ and $P(\mathrm{GW}|Z)$).  

The parameters in Figure \ref{fig:DMbias_mzsfrvsmk_eph20} were specifically set so we could also observe the GW bias results when the host-galaxy probability is in the vicinity of the regimes obtained in \cite{Artale_2019}, which measured the host-galaxy probability function from cosmological simulations integrated with a particular population synthesis model. In that study, a log-linear fit was used to relate the BBH merger rate per galaxy to the properties of the hosts and showed that at low redshift, stellar mass is the primary factor influencing merger rates. Their results also indicated that the slopes of the host probability distribution can vary depending on the binary type and generally yielded positive slope for the stellar mass dependence, negative slope for metallicity and almost close to flat for SFR.\footnote{The host-galaxy parameters can roughly be translated from their coefficients into our notation for BBHs at $z=0.1$ as follows: for 1D stellar mass fitting, $\delta_l = 1/\alpha_1\sim 1.23$ when $M_\mathcal{K}\sim 10^{12}M_\odot$. In the case of 3D $M_*$-$Z$-$\rm SFR$ fitting, we have $\delta_l = 1/\gamma_1\sim 1$ for $M_\mathcal{K}\sim 10^{12}M_\odot$, $\zeta_h = 1/\gamma_3\sim 2.5$ for $Z_\mathcal{K}\sim 3\times 10^{-3}$, along with an approximately constant SFR or $\epsilon_{h} = 1/\gamma_2 \sim 20 \gg 1, \,\rm SFR_\mathcal{K}\sim 3.72$ $M_\odot\mathrm{\ yr}^{-1}$.}

The trends in Figures \ref{fig:DMbias_mzsfrvsmk} and \ref{fig:DMbias_mzsfrvsmk_eph20} can be summarized as follows.  

\begin{itemize}

    \item {\textbf{General behavior:} Even though $P(\mathrm{GW}|\mathrm{SFR})$ and $P(\mathrm{GW}|Z)$ are no longer uniform, the overall trends are still similar to the results of \cite{dehghani2024} and $M_*$ model considered in Figure \ref{fig:sdssvs2mpz_bias}. The distinctive feature of all is that increasing $M_{\mathcal{K}}$ on average increases the GW bias. This is again as expected the consequence of the relationship between stellar mass and halo mass \citep{Behroozi_2013,Moster_2010}, i.e., galaxies with more stellar mass live in more massive and biased dark matter halos. Another feature of $b_\mathrm{GW}$-$M_\mathcal{K}$ is that (also similar to Figure \ref{fig:sdssvs2mpz_bias}) at lower values of $M_\mathcal{K}$, the GW bias is almost constant up to a point from which the GW bias starts growing and then at high stellar mass values, the GW bias is almost constant again.
    
    The constant $b_{\rm GW}$ at low $M_\mathcal{K}$ is a natural consequence of incompleteness at the low-mass end of the stellar mass distribution in observations. At this end, once $M_\mathcal{K}$ drops far below the peak of stellar mass distribution, further lowering $M_\mathcal{K}$ has insignificant effect on the selection, i.e.\ galaxies are chosen with the same power-law slope above $M_\mathcal{K}$, since there are very few galaxies below $M_\mathcal{K}$. 
    This regime roughly corresponds to $M_\mathcal{K} < 10^{10}$ $M_\odot$ in the first redshift bin and $2 \times 10^{10}$ $M_\odot$ in the second bin. These are within the estimated stellar mass completeness limit extrapolated from the results of \cite{2016MNRAS.459.2150W}, $1.7 \times 10^{10}$ $M_\odot$ at $z = 0.07$ and $6.2 \times 10^{10}$ $M_\odot$ at $z = 0.135$. We therefore emphasize that this feature at the lower $M_\mathcal{K}$ is purely an observational effect based on our selection of GW sources from a stellar mass-limited sample, and does not correspond to an accurate physical scale representing a transition in the relationship between GW bias and $M_\mathcal{K}$. 
    
    In most cases (especially those where selection is not too sharp), we can observe $b_{\rm GW}$ approaching a constant at the high-mass end as well. We believe that this feature corresponds to a more physical scale (we refer to it as $\hat M_{\star}$) characterizing the transition in the Schechter function between the power-law low mass regime and the exponentially dropping high-mass end of the mass function.
    Once $M_\mathcal{K}$ exceeds $\hat M_{\star}$, the number of galaxies above the pivot mass is exponentially suppressed and the bias asymptotes to a constant value. This transition occurs at about $2 \times 10^{11}$ $M_\odot$ in both redshift bins; this is close to $\hat M_{\star}=1.3 \times 10^{11}$ $M_\odot$ from fits to the stellar mass function \cite{Thanjavur_2016}.}
    
    \item \textbf{Effect of slopes for $M_*$ selection on GW bias:}
    Figures \ref{fig:DMbias_mzsfrvsmk} and \ref{fig:DMbias_mzsfrvsmk_eph20} also show how the different sets of slopes chosen as $\delta_l=0.5,1,1.3$ and $\delta_h=0.5 ,3.5$ can influence the GW bias. 
    An increase in $\delta_h$ results in a shallower decaying slope in the host-galaxy probability function for galaxies with $ M_\mathcal{K}< M_*$. This implies a less preferential populating of the lower-mass galaxies over the higher-mass galaxies, and subsequently a higher GW bias. Changing $\delta_h$ is most significant in the low-$M_\mathcal{K}$ regime. This is because if $M_\mathcal{K}$ is above the mean of stellar mass distribution, changing $\delta_h$ changes the host-galaxy probability for fewer galaxies with masses higher than $M_\mathcal{K}$, and therefore has less impact on the GW bias. In contrast, $\delta_l$ plays the opposite role, with slightly higher bias for lower $\delta_l$ values in the high-$M_\mathcal{K}$ part (for example, compare solid and dashed lines of the same color in Figure \ref{fig:DMbias_mzsfrvsmk}).

     \item \textbf{Effect of different $\rm SFR_\mathcal{K}$ and $Z_\mathcal{K}$ while varying $M_\mathcal{K}$:} While varying $M_\mathcal{K}$, several sets of parameters for the SFR and metallicity were considered. More specifically $(\rm SFR_\mathcal{K},\, \epsilon_l,\, \epsilon_h,\, Z_\mathcal{K},\, \zeta_l, \,\zeta_h)$ were taken as $(0.193,\,2,\, 1,\, 0.26,\, 1, \,5)$, $ (3.72, \,2, \,1,\,0.26,\, 1, \,5)$ and $(3.72, \,2, \,1,\,0.025,\, 1, \,5)$ in the first, second, and third panels of Figure \ref{fig:DMbias_mzsfrvsmk}, respectively, while they were $(3.72, \,2, \,20,\,0.025,\, 1, \,2.5)$ and $(15, \,2, \,20,\,0.025,\, 1, \,2.5)$ in the first and second panels of Figure \ref{fig:DMbias_mzsfrvsmk_eph20}.
     In these scenarios, the values of $SFR_\mathcal{K}$ and $Z_\mathcal{K}$ were picked to account for different cases where the selection of the host galaxy is primarily influenced by one side of the probability functions or the combination of both. Note that the average SFR of galaxies in the first and second redshift bins is $7.9  M_\odot\mathrm{\ yr}^{-1}$ and $16.28  M_\odot\mathrm{\ yr}^{-1}$ , respectively, while the average metallicity in the first and second redshift bins is $0.0245$ and $0.0246$, respectively.
   
    General trends indicate that the change in $\rm SFR_\mathcal{K}$ has a greater impact on the $b_\mathrm{GW}$-$M_\mathcal{K}$ relationship than $Z_\mathcal{K}$. For example, in Figure \ref{fig:DMbias_mzsfrvsmk} for the same slopes, the lower value of $\rm SFR_\mathcal{K}$ leads to a higher overall bias (compare the changes between the upper and middle panels with the difference between the second and third row). This could be due to the fact that for small values of $\rm SFR_\mathcal{K}\lesssim 1$ given our choice of $\epsilon_l=1$, sirens preferentially populate galaxies with very low SFR (quenched galaxies), which in both redshift bins have higher mean stellar masses (see Figure \ref{fig:mean_mass_vs_sfr}) and thus higher galaxy bias (see also the top panel of Figure \ref{fig:SFR_vs_stellarmass}). In contrast to that changing the metallicity parameter $Z_{\mathcal{K}}$ has almost no impact (compare the middle and bottom panels in Figure \ref{fig:DMbias_mzsfrvsmk}) since metallicity is weakly correlated with the mean of stellar mass (see the middle panel of Figure \ref{fig:SFR_vs_stellarmass}). As we pointed out earlier, this is not unexpected given that the galaxies in these bins also have a narrow range of metallicity. 
    However, it is important to note that this behavior may also reflect a limitation of our phenomenological model, which may not fully capture the complex dependence of GW host selection on stellar metallicity. Future observational efforts with spectroscopic follow-up and improved metallicity measurements could provide deeper insights into this dependence and help refine the modeling of host-galaxy properties.
    
    We also measure the GW bias for host-galaxy probability functions close to those obtained in \cite{Artale_2019}. The metallicity dependence of $P(\mathrm{GW}|Z)$ in Figure \ref{fig:DMbias_mzsfrvsmk_eph20} ($\zeta_h\sim 2.5, ~Z_K=0.025$) is chosen to be close to what was obtained in \cite{Artale_2019}. Matching $P(\mathrm{GW}|SFR)$ to their results requires a shallow decreasing slope, $\epsilon_h=20$. Finally, we need to consider a low value for SFR$_\mathcal{K}$ $M_\odot\mathrm{\ yr}^{-1}$ to consider the regime where only the decreasing slope $\epsilon_h$  is relevant.
   We highlight these points with blue ovals; they are the rightmost points of the series colored in maroon in the top panel of Figure~\ref{fig:DMbias_mzsfrvsmk_eph20}.
   In these cases, we see that $b_{\textrm{GW}}$ can be about a few percent higher than $b_g$.
    
    \item {\textbf{Redshift dependency:}  In every scenario illustrated in Figures \ref{fig:DMbias_mzsfrvsmk} and \ref{fig:DMbias_mzsfrvsmk_eph20}, we observe that the GW bias has higher values in the higher redshift range ($0.135 \leq z <0.2$) compared to the lower redshift range ($0.07 \leq z <0.135$). This is similar to the findings of \cite{dehghani2024} showing that the dependence of GW bias on redshift mostly reflects the trend observed in the redshift dependence of galaxy bias, which rises as redshift increases. This is a purely observational effect due to increasing incompleteness at higher redshifts, and so the survey preferentially picks out higher-mass galaxies and thus higher galaxy bias. This observation seems to hold roughly for all selections of host-galaxy probabilities, but there may be some mild dependence in the redshift evolution of $b_{\rm GW}/b_g$ depending on the parameters of the host-galaxy probability. We leave this for future studies.
    }
\end{itemize}
Lastly, for the interested reader, we also present an example of a more comprehensive 3D visualization of the dependence of the GW bias ($b_\mathrm{GW}$) on key parameters of the host-galaxy probability function $M_*$, including $\delta_l$, $\delta_h$ and $M_\mathcal{K}$ given a fixed shape for the SFR and metallicity host-galaxy probabilities in Figure \ref{fig:3ddmbias_MZSFRv2_sfr3.72_mc1}. However, as we can see it is much harder to decipher the trends in the 3D plots.  

\subsubsection{Dependency on parameters of  $P(\rm GW|\mathrm{SFR})$}
\label{sec:GW bias parameter vs star formation rate}
\begin{figure}

\includegraphics[width=.49\textwidth]{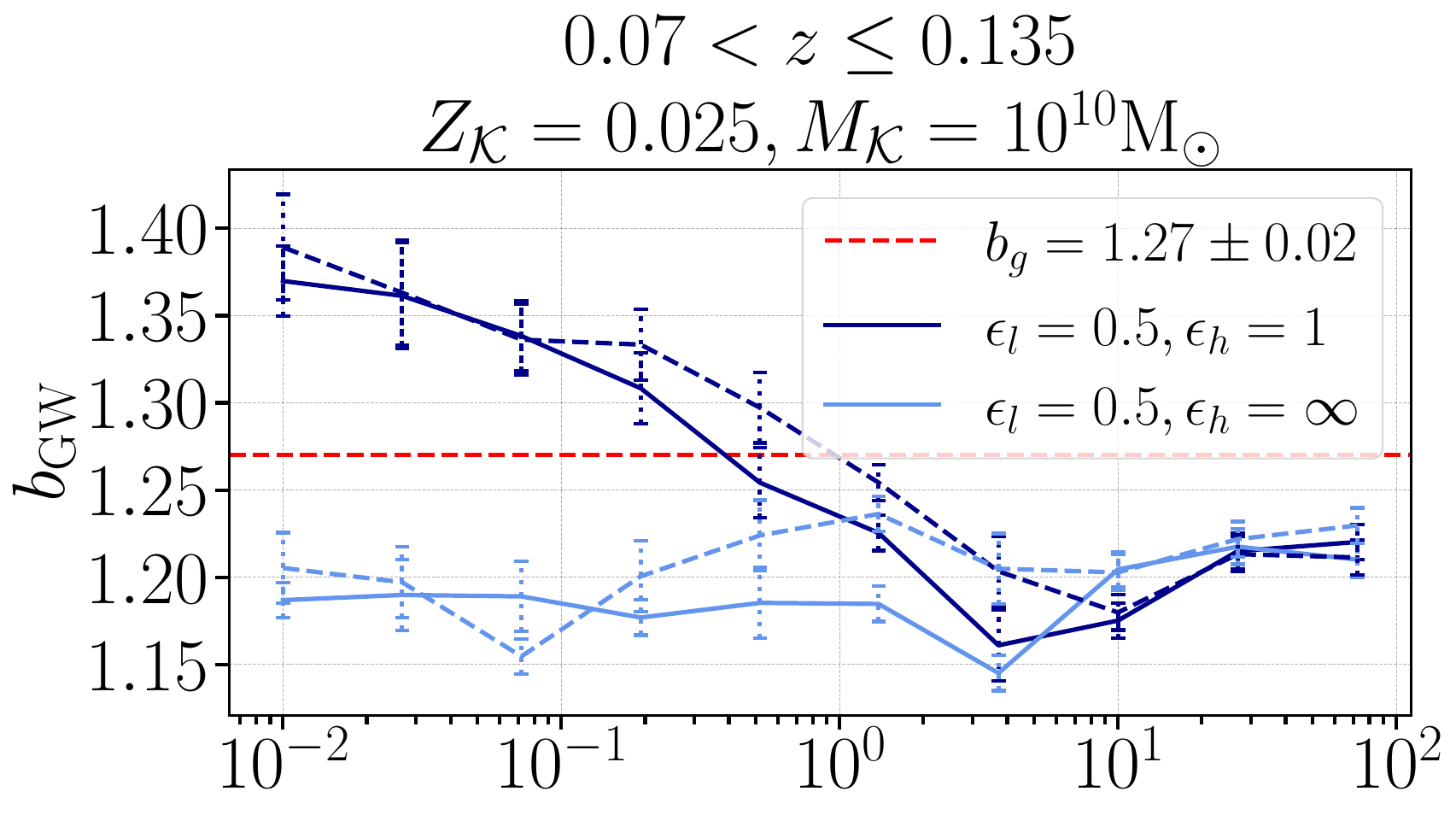}
\includegraphics[width=.49\textwidth]{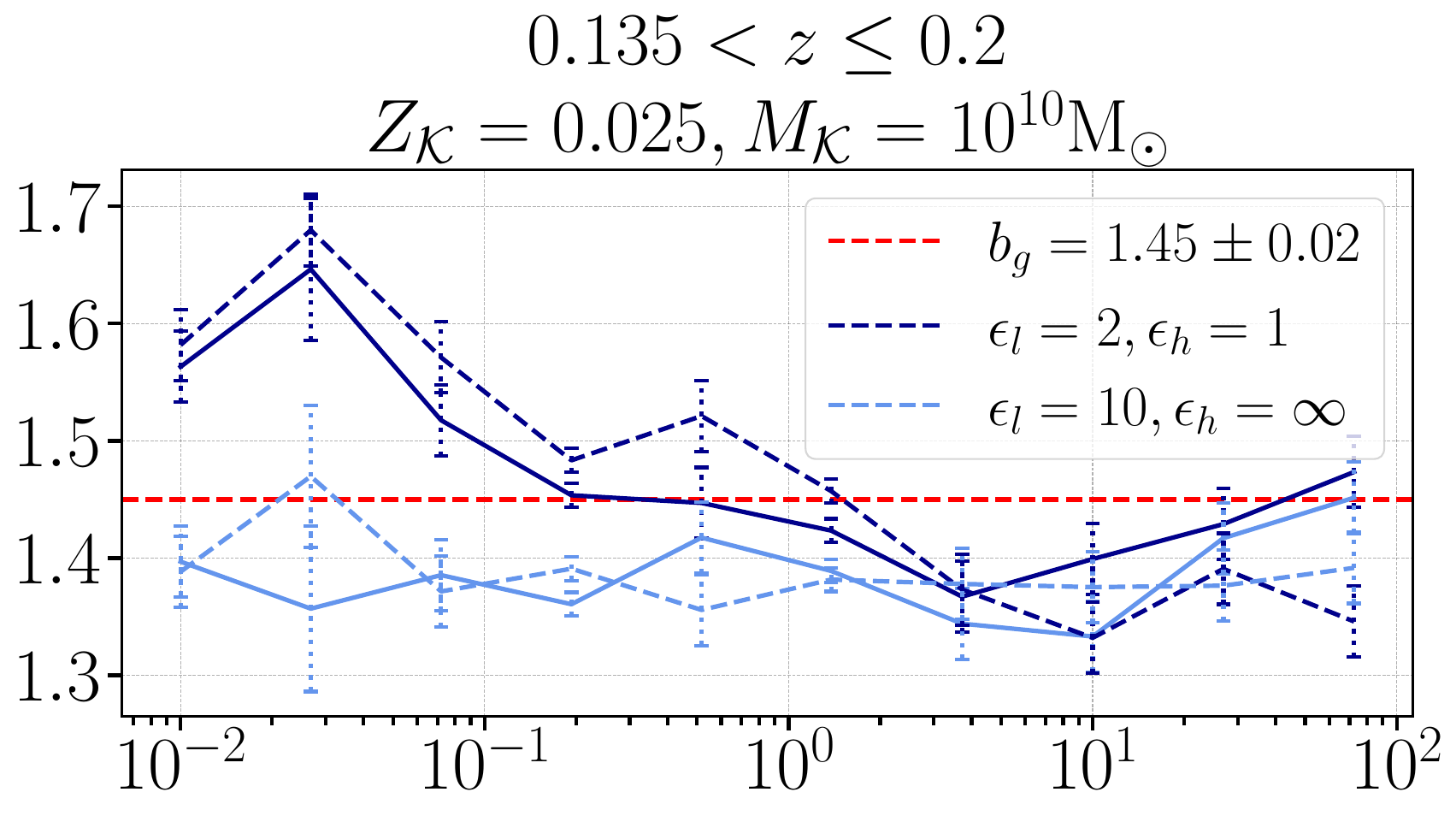}

\includegraphics[width=.49\textwidth]{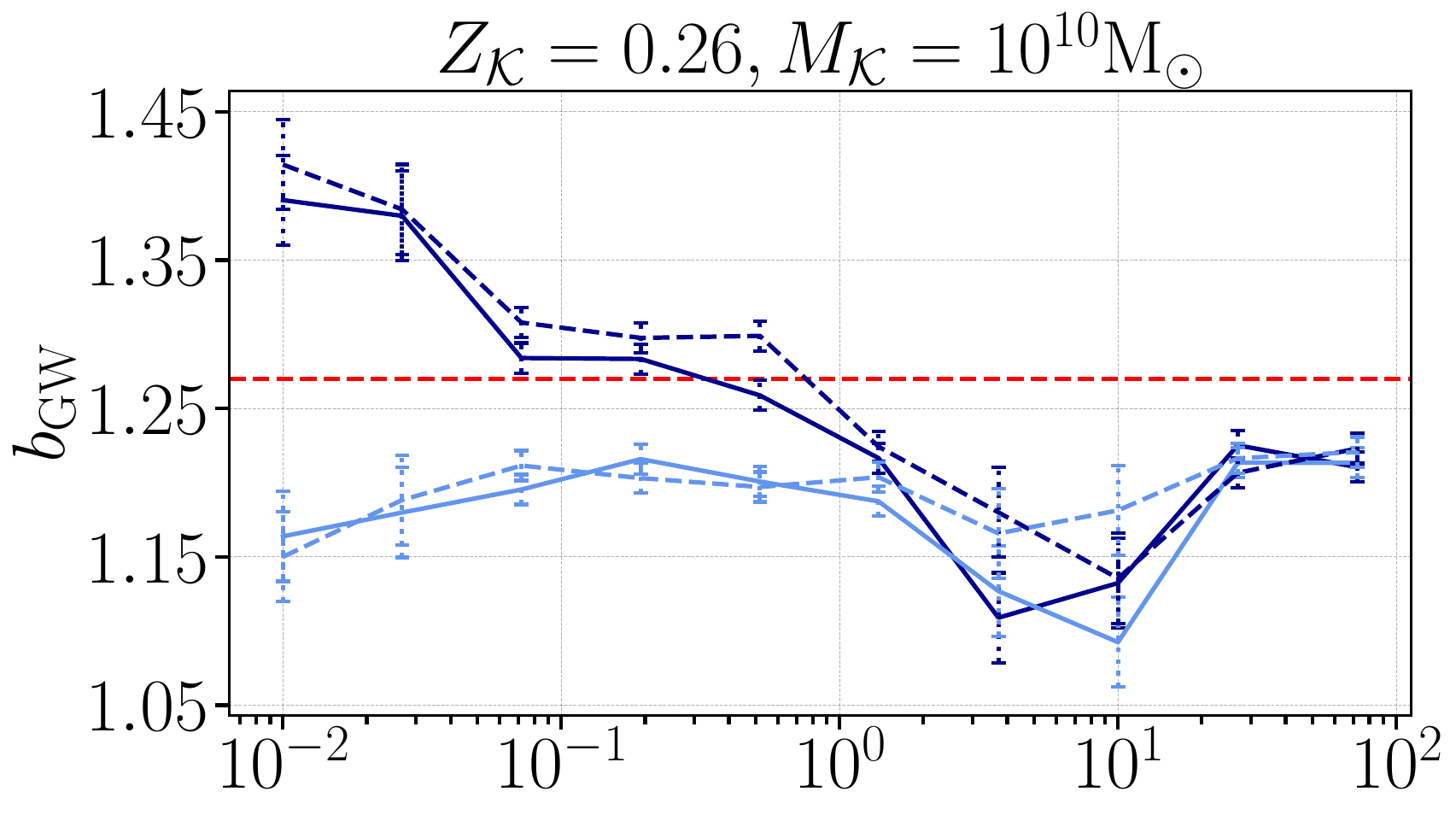}
\includegraphics[width=.49\textwidth]{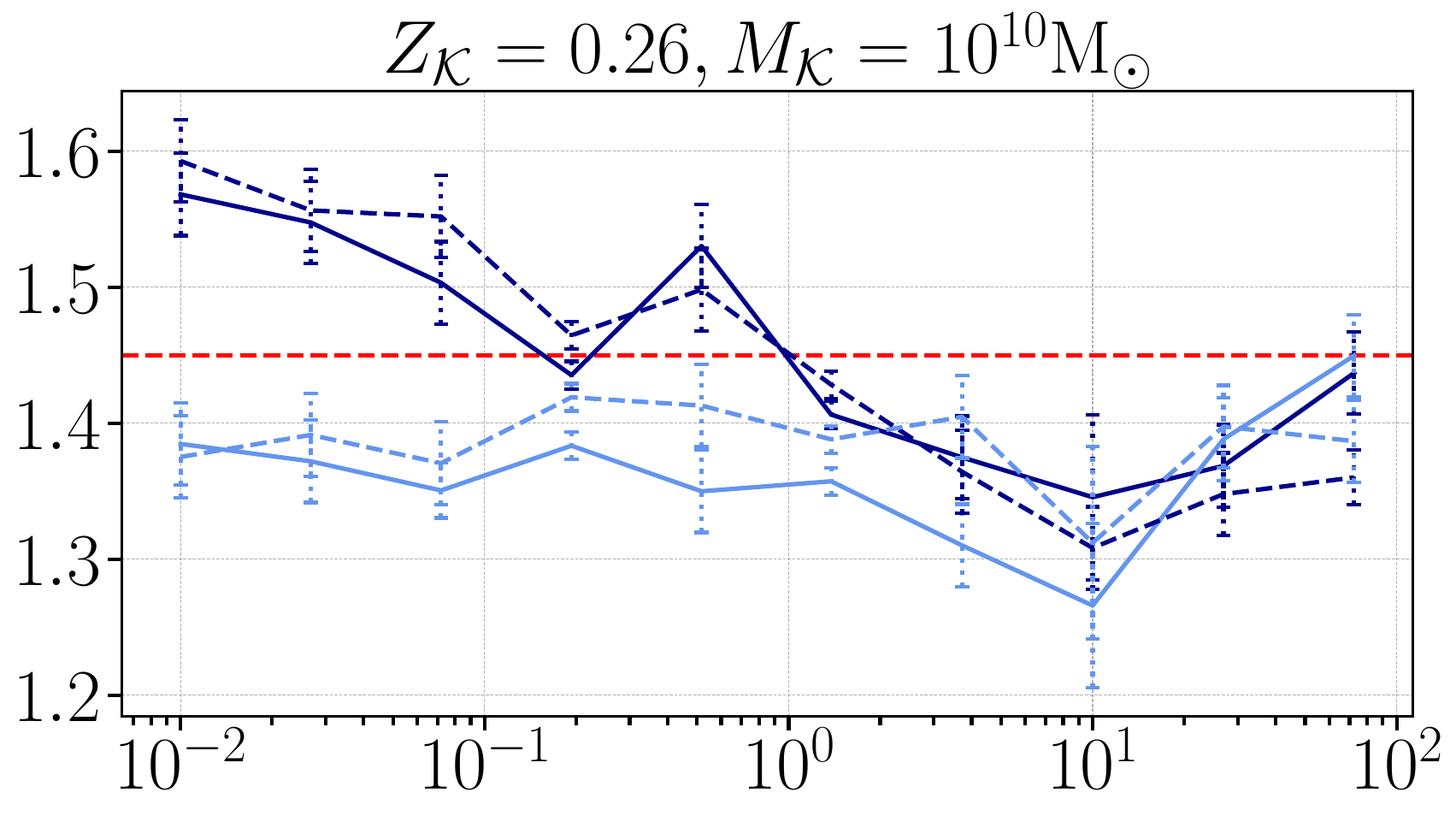}

\includegraphics[width=.49\textwidth]{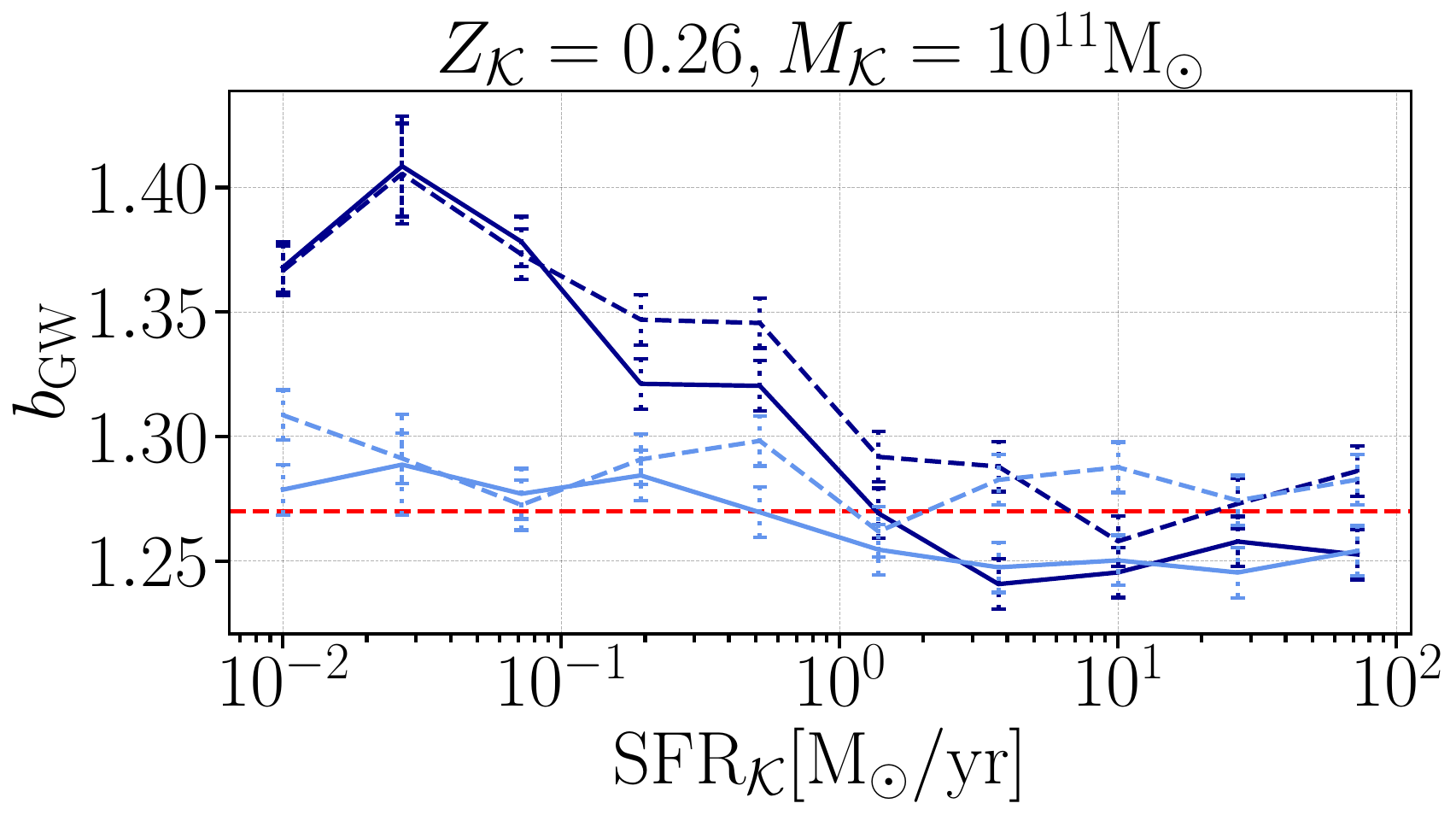}
\includegraphics[width=.49\textwidth]{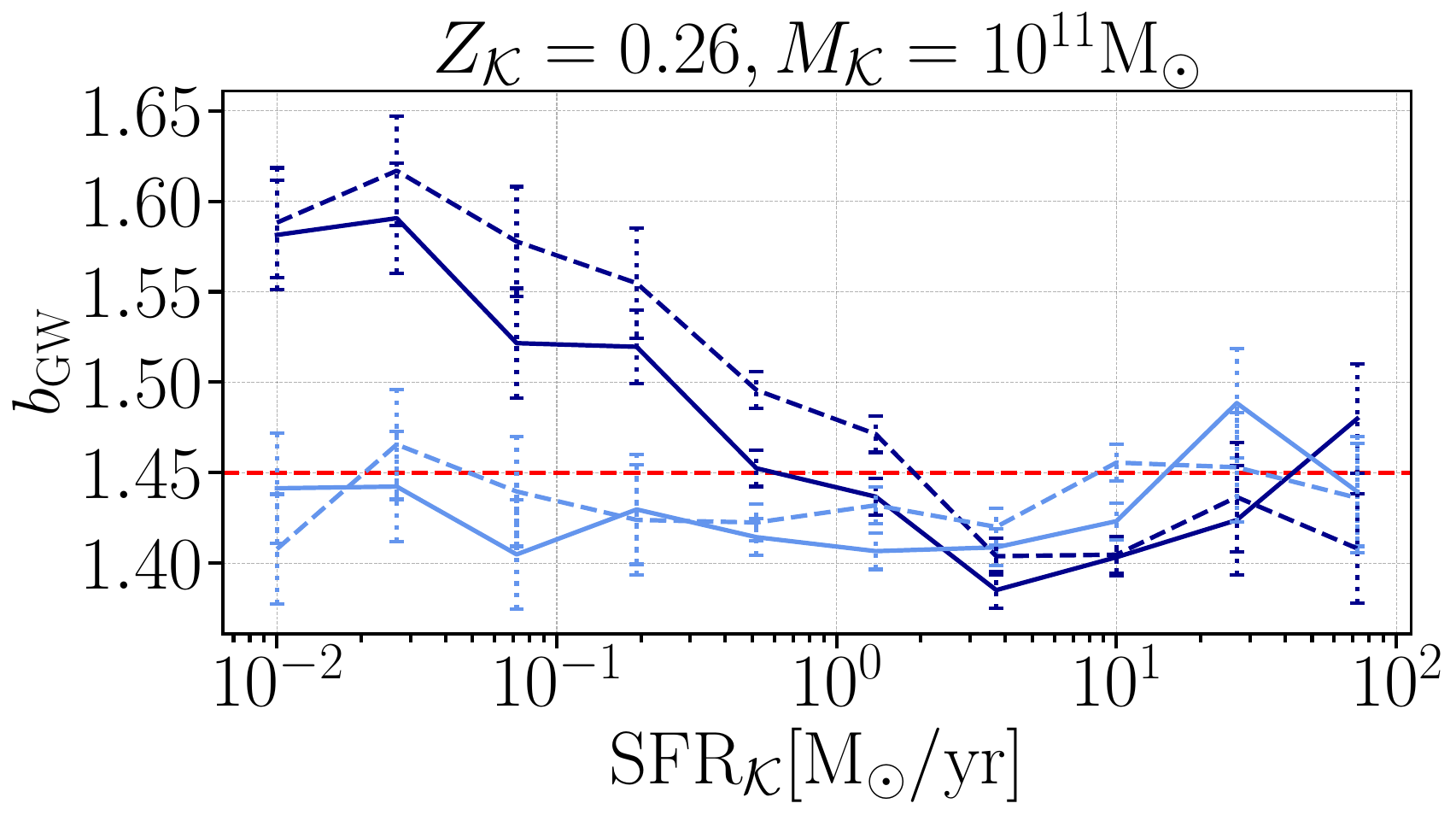}

\includegraphics[width=.49\textwidth]{Paper-II-fig/mzsfrvsZv2_function.pdf}
\includegraphics[width=.49\textwidth]{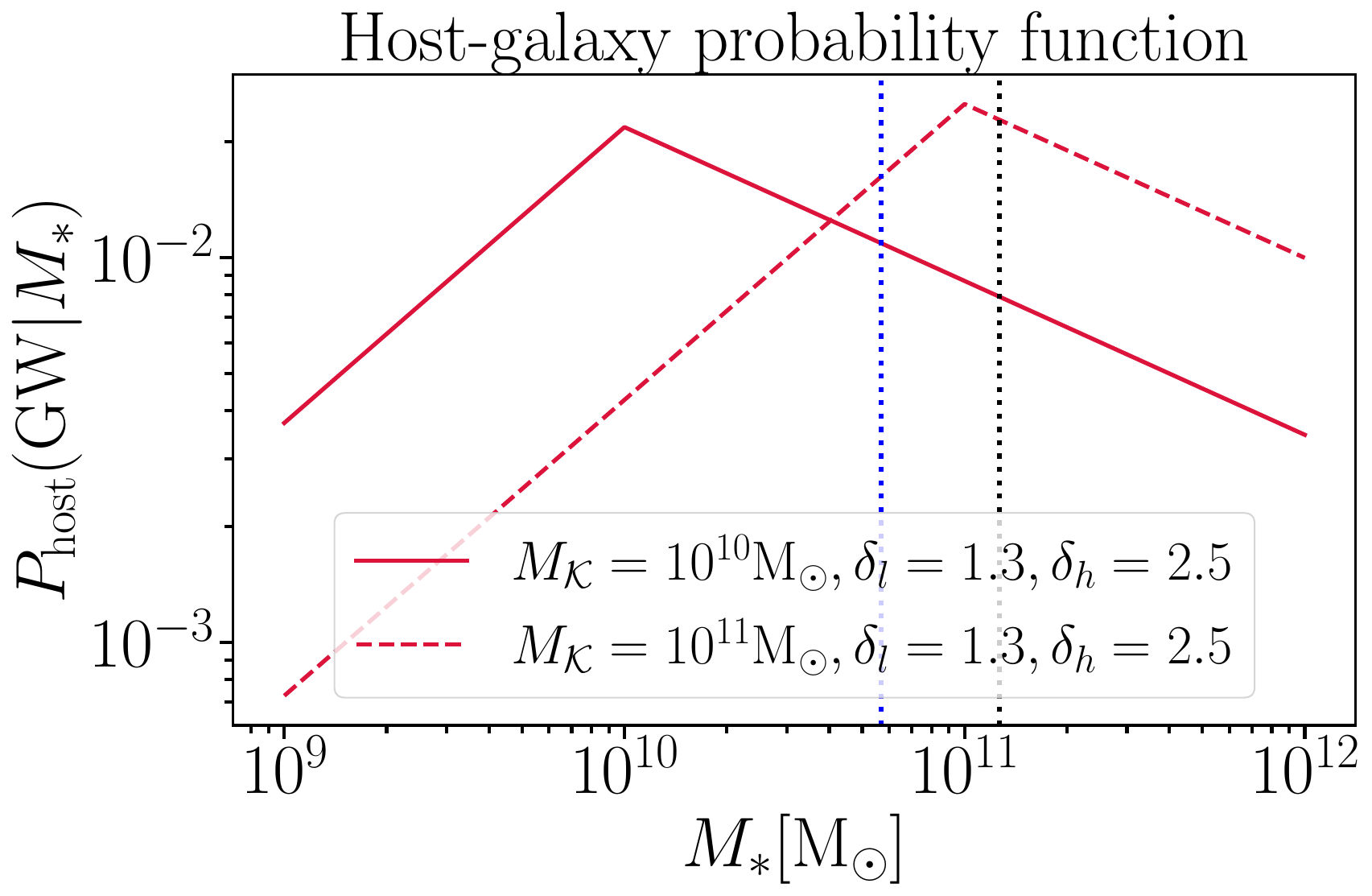}

\caption{GW bias parameter as a function of star-formation rate  turnover, $\mathrm{SFR}_\mathcal{K}$, within redshift $0.07 \leq z <0.135$ (left) and $0.135 \leq z <0.2$ (right). Within each panel, the different line styles and colors indicate different values of the SFR slopes through the parameters $\epsilon_l$ and $\epsilon_h$. The solid blue lines correspond to $\epsilon_l = 0.5$ and $\epsilon_h = 1$, and the solid light blue line indicates a model where $\epsilon_l = 0.5$ but the inverse of the falling slope, $\epsilon_h$, goes to $\infty$ (i.e.\ there is no SFR-dependence in the host-galaxy probability function above the turnover). The dashed lines display the impact of changing $\epsilon_l$,  compared to solid lines of the same color.
Different rows of panels show the impact of varying the transition points for stellar mass  $M_\mathcal{K}$ and the metallicity  $Z_\mathcal{K}$.
We have fixed the other parameters to $\delta_l=1.3$, $\delta_h=2.5$, $\zeta_l=1$, and $\zeta_h=5$. The last row shows the shape of the fixed $P(\mathrm{GW}|M_*)$, and $P(\mathrm{GW}|Z)$  functions used here. In the bottom-left plot, the dotted vertical line shows the mean metallicity of the galaxies in the first and second redshift bins with $\overline{Z}\sim 0.024$. In the bottom-right plot, the blue and black vertical dotted lines show the mean stellar mass of galaxies in the first and second redshift bins, which are $5.67 \times 10^{10} \,\rm{M}\odot$ and $1.27 \times 10^{11} \,\rm{M}\odot$, respectively.}
\label{fig:DMbias_mzsfrvssfrk}
\end{figure}

 In this section, we focus into the impact of varying the shape of $P(\mathrm{GW}|\mathrm{SFR})$ i.e., $\epsilon_l$, $\epsilon_h$ and $\mathrm{SFR}_\mathcal{K}$ for several distinct but fixed configurations of $P(\mathrm{GW}|M_*)$, and $P(\mathrm{GW}|Z)$ through specific choices of $\delta_l$, $\delta_h$, $M_\mathcal{K}$, $\zeta_l$, $\zeta_h$, and $Z_\mathcal{K}$, see bottom panel of Figure \ref{fig:DMbias_mzsfrvssfrk}. The results are shown in Figure \ref{fig:DMbias_mzsfrvssfrk} and the following points are some highlights:
 \begin{itemize}
     \item \textbf{General behavior:} Based on our discussion in section \eqref{sec:assignment} we expect that if the delay times are long, the merger events in low redshifts are more probable in early type galaxies with low SFR. The population synthesis simulations of \cite{Artale_2019} found that for a 3D $M_*$-$Z$-$SFR$ power-law fit of the host-galaxy probability, for SFR the fit has a negative power of around $-0.051$, which in our parameterizations translates into $\epsilon_h\sim 20 \gg 1$ and $\rm SFR_\mathcal{K}\ll \overline{\rm SFR}$ so that the host-galaxy probability is dominated by the decreasing slope.
     Therefore, it would be more interesting to explore the behavior of $b_{\rm GW}$ in the lower end neighborhood of $\textrm{SFR}_\mathcal{K}$ ($< \overline{\rm SFR}$) which allows us to primarily assess the impact of the decreasing side of $P(\mathrm{GW}|\mathrm{SFR})$ on $b_{\rm GW}$. 
    Meanwhile, the limit of $\epsilon_h=\infty$ corresponds to  uniform $\rm SFR$ host selection for galaxies with $\rm SFR>SFR_\mathcal{K}$ \footnote{Note in our parametrization the slope is $-1/\epsilon_h$.}. As we see in this limit (dashed and solid light blue lines in Figure \ref{fig:DMbias_mzsfrvssfrk}), the general trend is that the GW bias does not vary much with $\rm SFR_\mathcal{K}$, which is expected since there are few galaxies with $\rm SFR< SFR_\mathcal{K}$ and the host-galaxy probability dependence on SFR is very weak. On the other hand, when we set $\epsilon_h=1$ (i.e. linearly decaying) as $\textrm{SFR}_\mathcal{K}$ decreases, $b_{\textrm{GW}}$ increases (dashed and solid dark blue lines in Figure \ref{fig:DMbias_mzsfrvssfrk}). This is once again because for low $\textrm{SFR}_\mathcal{K}\lesssim \mathcal{O}(1)$ and a steeper negative slope, the selection of host galaxies gets dominated by those with very low SFR that have a higher mean stellar mass (see the top panel of Figure \ref{fig:SFR_vs_stellarmass} and also Figure \ref{fig:mean_mass_vs_sfr}). Similarly for $\textrm{SFR}_\mathcal{K} \gtrsim \mathcal{O}(10)$, and $\epsilon_l\lesssim 1$ (solid lines) since the host-galaxy selection preferentially populates high SFR values, we observe in most cases $b_{\rm GW}$ increasing with $\textrm{SFR}_\mathcal{K}$, which is also consistent with the the rise of mean stellar values seen in Figure \ref{fig:mean_mass_vs_sfr}. This is more pronounced in the second redshift bin, since the mean SFR of the galaxies is $16.28~[M_\odot\mathrm{\ yr}^{-1}]$ higher than $10~[M_\odot\mathrm{\ yr}^{-1}]$, so there are more high-mass galaxies in this end that are picked up as $\textrm{SFR}_\mathcal{K}$ increases above 10 $M_\odot\textrm{\ yr}^{-1}$.
    
     \item \textbf{Effect of different $M_{\mathcal{K}}$ and $Z_{\mathcal{K}}$ on the GW bias:}   
     As we have seen so far, overall the GW bias seems to have the strongest correlation with stellar mass selection. We can further see this trend in Figure \ref{fig:DMbias_mzsfrvssfrk} in $b_{\rm GW}$ moving up or down when changing $M_\mathcal{K}$ (from second panel to third), while it does not show a significant change when varying $Z_\mathcal{K}$ (from top panel to second). This is despite the fact that the values of $Z_\mathcal{K}$ we have considered are each in very different regions of the metallicity distribution. 
     The impact of changing $M_\mathcal{K}$ is also consistent with what we have discussed before. We can see this more distinctively when considering $\epsilon_h =\infty$ and $\rm SFR_{\mathcal{K}}\lesssim 1$, corresponding to uniform SFR host selection for most hosts. For the lower value of $M_\mathcal{K}= 10^{10}$ $M_\odot \ll \overline{M_*}$,  $b_{\textrm{GW}}$ is less than the average galaxy bias (solid and dashed light blue lines in the top two panels of Figure \ref{fig:DMbias_mzsfrvssfrk}).    
    However, when $M_\mathcal{K}$ increases to $10^{11}$ $M_\odot$, which is close to the mean stellar mass of galaxies (light blue lines in the third panel of Figure \ref{fig:DMbias_mzsfrvssfrk}),
    $b_{\textrm{GW}}$ approaches the average value of the galaxy bias. Although less distinctive, there is a similar upward shift around a few percent in the case of $\epsilon_h =1$.  
 \end{itemize}

\subsubsection{Dependency on parameters of $P(\mathrm{GW}|Z)$}
\begin{figure}
\centering
\includegraphics[width=.49\textwidth]{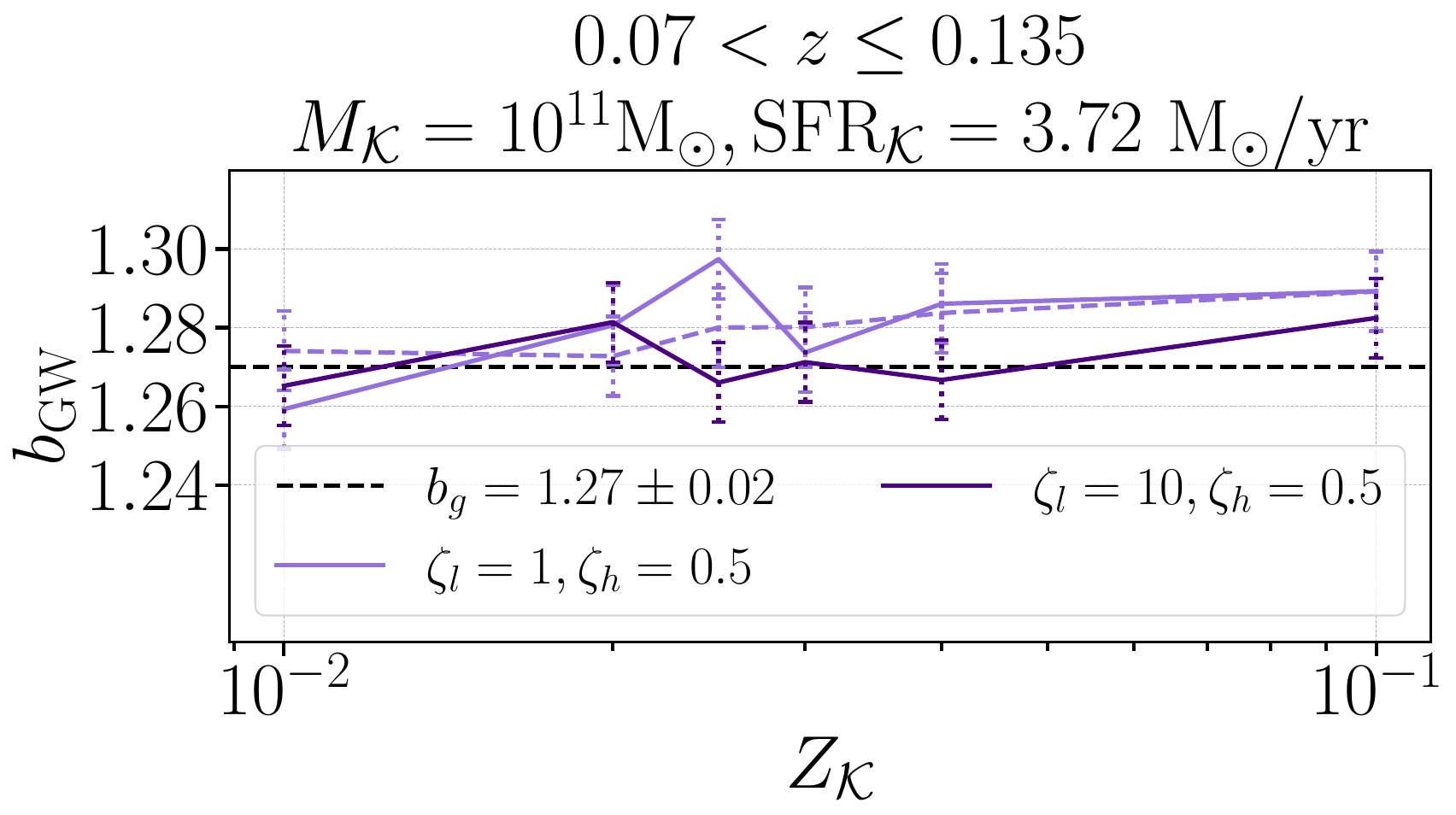}
\includegraphics[width=.49\textwidth]{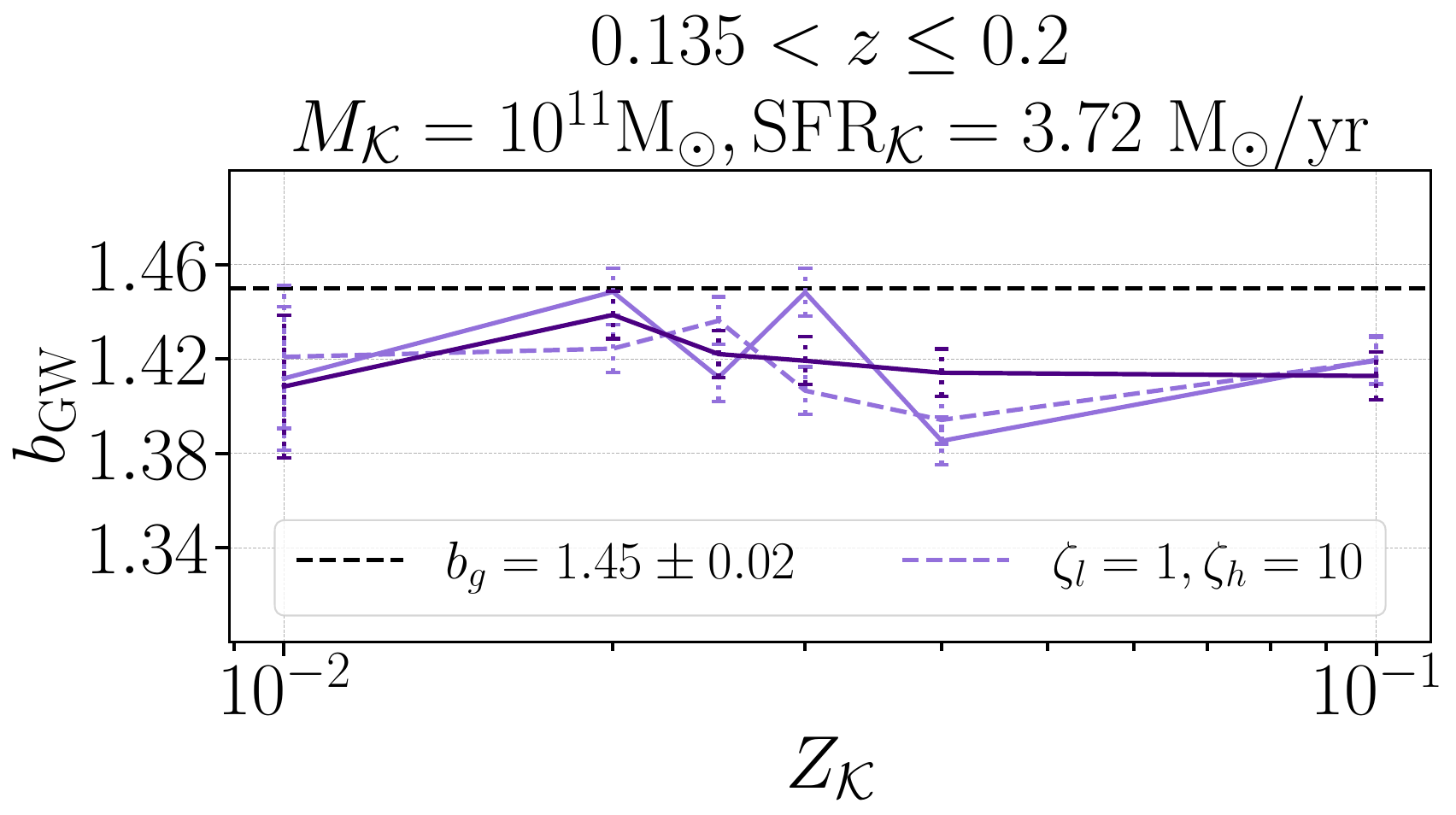}

\includegraphics[width=.49\textwidth]{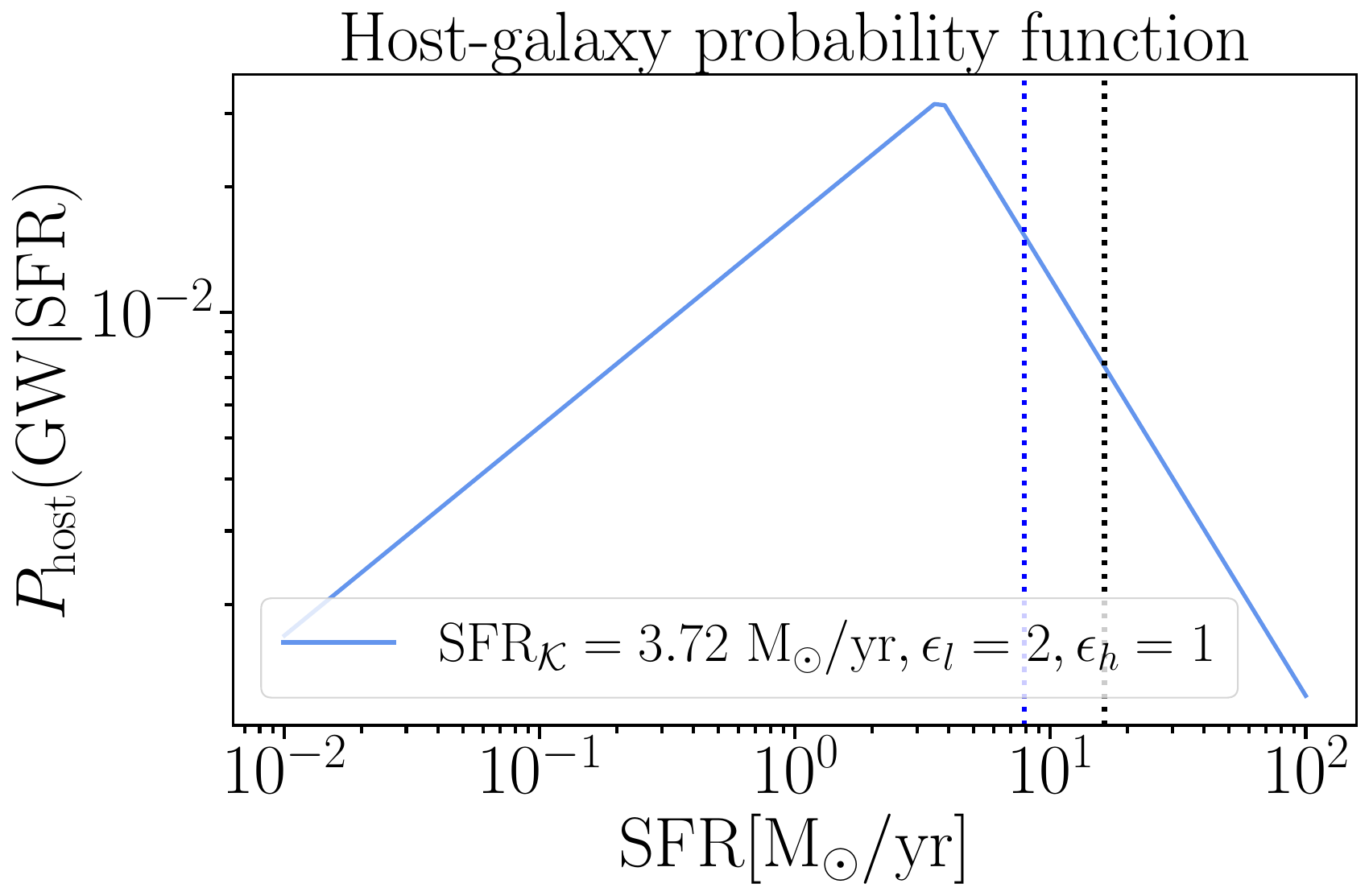}
\includegraphics[width=.49\textwidth]{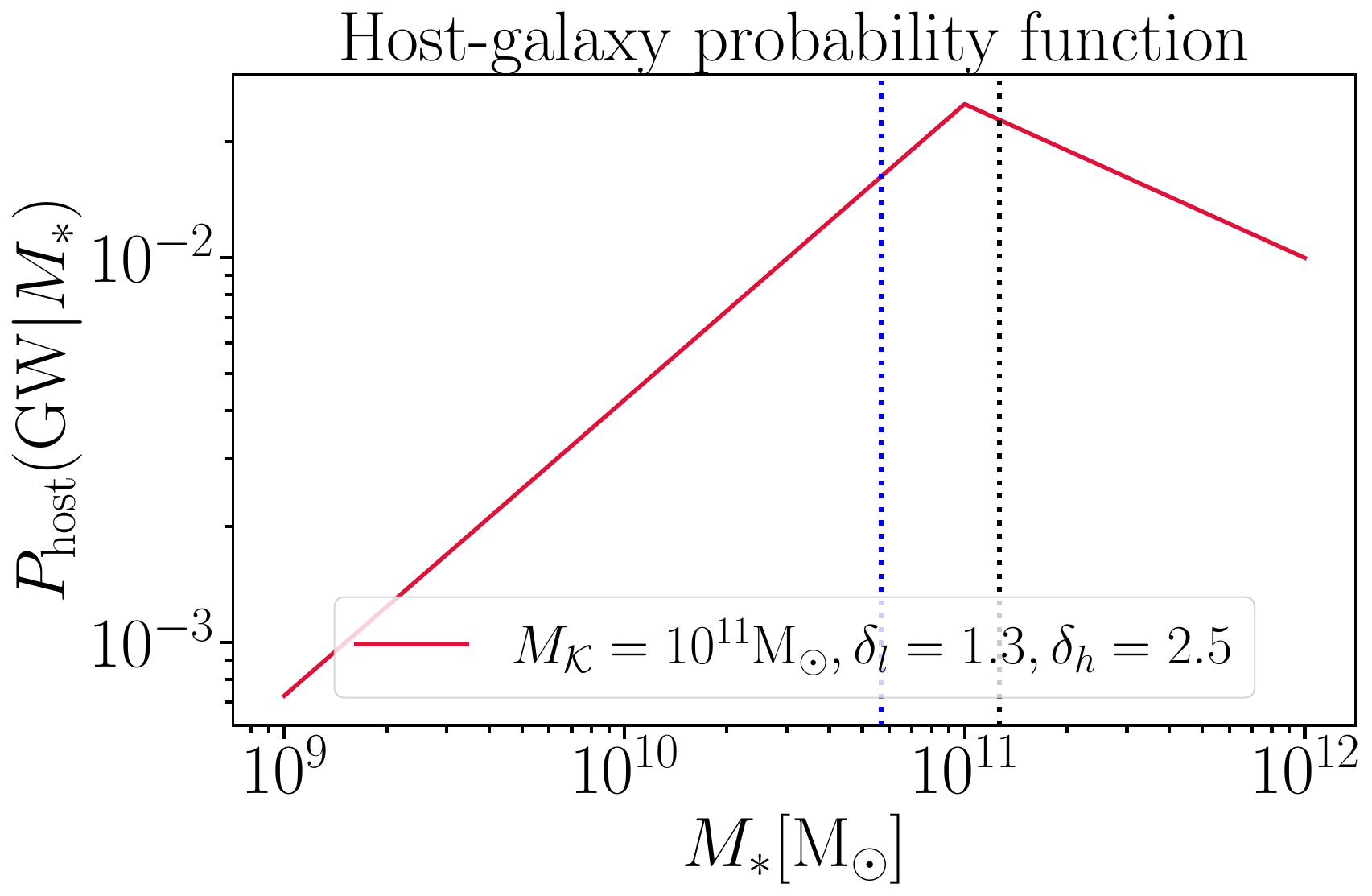}

\caption{GW bias parameter as a function of $Z_\mathcal{K}$ for the $0.07 \leq z <0.135$ (left) and $0.135 \leq z <0.2$ (right) bins. Here we only consider a single value of $M_\mathcal{K}$ and $\textrm{SFR}_{\mathcal{K}}$ as the plots are similar when these change.
The different colors and line styles show the impact of changing the metallicity slopes $\zeta_l$ and $\zeta_h$. Note that, $\delta_l=1.3 , \delta_h=2.5 , \epsilon_l=2 , \epsilon_h=1$. The second row shows fixed $P(\mathrm{GW}|M_*)$, and $P(\mathrm{GW}|\mathrm{SFR})$  functions used here. In the bottom-left plot, the dotted vertical blue and black lines show the mean SFR of the galaxies in the first and second redshift bins $7.9 M_\odot\mathrm{\ yr}^{-1}$ and $16.28 M_\odot\mathrm{\ yr}^{-1}$ ,respectively. In the bottom-right plot, the blue and black vertical dotted lines show the mean stellar mass of galaxies in the first and second redshift bins ($5.67 \times 10^{10}\, M_\odot$ , $1.27 \times 10^{11} \, M_\odot $), respectively.
}
\label{fig:dmbias_mzsfrvsz}
\end{figure}

Finally, we fix the parameters of $P(\mathrm{GW}|M_*)$, and $P(\mathrm{GW}|\mathrm{SFR})$ (i.e., $\delta_h$, $\delta_l$, $M_\mathcal{K}$ and $\epsilon_l$,$\epsilon_h$, $SFR_\mathcal{K}$) and vary the parameters of $P(\mathrm{GW}|Z)$ (i.e., $\zeta_l$, $\zeta_h$, $Z_\mathcal{K}$ ). Figure \ref{fig:dmbias_mzsfrvsz} shows the GW bias parameter across different values of $Z_\mathcal{K}$.  We set the SFR and stellar shapes host-galaxy as $\mathrm{SFR}_\mathcal{K}=3.72~[M_\odot\mathrm{\ yr}^{-1}]$, and $M_\mathcal{K}=10^{11}[M_\odot]$ with slopes of $\mathcal{O}(1)$ (see the bottom panel of Figure \ref{fig:dmbias_mzsfrvsz}). The metallicity slopes are taken from  
$(\zeta_l,\zeta_h)\in \{(1,0.5),\, (10, 0.5),\, (1, 10)\}$.
As can be seen, the bias exhibits some fluctuations within both redshift ranges; however, the results show an insignificant correlation between the GW bias and $Z_\mathcal{K}$. 
The various styles of lines and colors represent different slopes considered in the $M_*$-$Z$-SFR model ($\zeta_l$ and $\zeta_h$),
and show that the GW bias is not very sensitive to these slopes either. We can also note that in the second redshift bin $b_{\rm GW}$ is slightly lower than the galaxy bias, which is expected since in this case $M_\mathcal{K}$ is also slightly lower than the mean stellar mass of the galaxies. 

To summarize our exploration of the parameter space of the host-galaxy probability shapes, within the $M_*$--$Z$--SFR model, the GW bias is most sensitive to stellar mass, followed by star formation rate (SFR), and is least sensitive to metallicity. 
As mentioned previously, the observed insensitivity may stem from simplifications in our phenomenological model, which may not fully account for the intricate role of stellar metallicity in GW host selection. Enhanced metallicity measurements from future spectroscopic surveys could offer critical insights and improve host-galaxy modeling. 
 This reduced sensitivity to metallicity probably stems from the relatively small variation in metallicity across the low-redshift universe considered in this study and also from the selection effects of the survey. 
Before ending our result section, let us also add that as we discuss in Appendix \ref{sec:BH mass}, we also conducted some preliminary investigation of the potential correlation between clustering of BBH merger events and their masses. To be more specific, in our model the dependence of BBH masses on the galaxy properties enters through a logarithmic metallicity-dependent PISN upper mass cutoff, so it was expected that there would be no significant correlation between BBH chirp mass and $b_{GW}$.  In other words, this result was primarily intended as a consistency check for our pipeline and we leave a more in depth investigation of  this correlation to future work.


\section{Conclusion}
\label{sec:conclusion}
In this paper, we presented a comprehensive framework to connect gravitational wave (GW) sources to the underlying matter distribution by modeling the GW bias parameter. We achieve this by incorporating galaxy properties from the Sloan Digital Sky Survey (SDSS DR7) spectroscopic survey to  GW bias modeling to study the dependence of the GW bias on the astrophysical properties of BBH host galaxies such as stellar mass, star formation rate (SFR) and metallicity. 

For this to be done, we first populated mock GW sources within galaxies in the SDSS DR7 catalog (see Section \ref{sec:galaxy catalog} for more details), using a host-galaxy probability function denoted as the $M_*$-$Z$-SFR model. 
We then measured the anisotropic power spectrum of the mock GW sources and
calculated the GW bias parameter by comparing the power spectrum with a simple linear bias model with RSD described by the Kaiser model and a fingers-of-god damping term. We then assessed the impact and sensitivity of the GW bias parameter to the astrophysical properties of the galaxies hosting the mergers under different assumptions for parameters of the host-galaxy probability function.

Our detailed analysis indicates that stellar mass dependence of host-galaxy probability has the strongest and consistent correlation with GW bias. In general, when increasing the pivot point in the stellar mass probability function, $M_\mathcal{K}$ (from $ 10^{9}$ $M_\odot$ to $10^{12} M_\odot$) the GW bias parameter also increases. In some scenarios, we observe that the GW bias parameter can increase even up to $\sim30\%$ above the corresponding galaxy bias (see Figure \ref{fig:DMbias_mzsfrvsmk}). This trend aligns well with our previous results \cite{dehghani2024}, obtained based on a stellar mass–only host-galaxy probability function and using the photometric galaxy surveys 2MPZ and WISCland.

Our analysis also suggests a notable relationship between the GW bias and the SFR. We observed that when the host-galaxy probability has preference in selecting low-SFR galaxies ($\epsilon_h\lesssim 1$), it induces a substantial increase in GW bias at the lowest values of the pivot point in the SFR probability function, $SFR_\mathcal{K}\lesssim \mathcal{O}(10^{-1})\, M_\odot\mathrm{\ yr}^{-1}$ compared to the higher values of $SFR_\mathcal{K}$ $( \gtrsim \mathcal{O}(1)\, M_\odot\mathrm{\ yr}^{-1})$ where the host-galaxy probability has no preference in selecting low-SFR galaxies (see Figure \ref{fig:DMbias_mzsfrvssfrk}). We still see that the overall magnitude of $b_{GW}$ can change significantly upward or downward depending on the value of $M_\mathcal{K}$. The behavior of $b_{\rm GW}$ at low $SFR_\mathcal{K}$ and $\epsilon_h\lesssim 1$ can be due to the fact that lower star-forming galaxies are quenched high stellar mass galaxies, which tend to be older and hence reside in halos with inherently higher biases.

Finally, we did not observe any significant and meaningful correlation between GW bias and host-galaxy probability dependence on the metallicity of the host galaxy (see Figure \ref{fig:dmbias_mzsfrvsz}). The observed insensitivity to metallicity may be partly attributed to the redshift limitations of the galaxy surveys employed, as well as to simplifications in our host-galaxy probability function, which may fail to fully capture the dependence of GW host selection on stellar metallicity. Next-generation spectroscopic surveys, equipped with high-precision metallicity diagnostics, will be crucial for accurately characterizing the role of chemical enrichment in the environments that host gravitational wave sources.
Let us also highlight that stellar mass, SFR, and metallicity are inherently interdependent. Consequently, future explorations of the complete nine-dimensional parameter space defining our host-galaxy probability model might uncover additional complexities and potential degeneracies. Furthermore, we would like to emphasize again that our goal in this paper was to construct computable phenomenological models of GW bias by mapping GW sources to galaxies and exploring their dependence on host-galaxy properties, while a realistic observational forecast will require independent frameworks that incorporate GW selection effects, sky localization error and instrumental noise. It is also worth noting for the low-redshift range considered here, future GW observations are expected to have a selection function close to unity, limiting the impact of detector-related incompleteness.

As part of concluding remarks, it is worth pointing out that our result at this stage is among the pioneering efforts to connect the vast scope of astrophysics involved in modeling the distribution of GW events as tracers of matter density and large-scale structure. In particular, to the best of our knowledge, this is the first study, in using spectroscopic surveys to connect galaxy properties to GW bias modeling. Although this approach is still in its infancy and has revealed certain limitations due to observational constraints and the numerous unknowns in astrophysics that impact parameterization, it already offers significant insights into the behavior of GW bias and how it connects to underlying astrophysics. Notably, it highlights the critical role of understanding the environmental effects on the formation of stellar black holes and delay-time distribution on shape of the host-galaxy probability and consequently its impact on GW bias. Conversely, if we ever experimentally measure $b_{\rm GW}$, our findings suggest how it could be interpreted in terms of host galaxy probabilities, thus providing clues on BBH formation and delay times. From a cosmological perspective, this is even more crucial, since any statistical inference of cosmological parameters, including estimating \( H_0 \) through cross-correlations with galaxies or other astrophysical tracers, relies on GW bias \cite{Mukherjee:2019wcg, Mukherjee:2020hyn, Diaz:2021pem, Afroz:2024joi, Zazzera:2025ord}. Our results thus offer a framework to easily accommodate and account for the astrophysical effects involved. There remain many avenues for future research to extend our current framework to model the GW bias parameter for binary neutron stars and neutron star-black hole binary mergers which will give deeper insight into the host-galaxy connection.

\section*{Acknowledgments}
We express our gratitude to the Sloan Digital Sky Survey (SDSS) Data Release 7 (DR7) for providing the galaxy catalog. The computational aspects of this research were executed on the computing clusters at the Perimeter Institute (PI) for Theoretical Physics with support from the Canada Foundation for Innovation. In this work, several Python packages have been utilized, namely: \verb|Pypower|, \verb|Astropy| \footnote{\url{http://www.astropy.org}}\cite{astropy:2013, astropy:2018, astropy:2022}, \verb|NumPy| \footnote{\url{http://www.numpy.org}}, \verb|SciPy|\footnote{\url{http://www.SciPy.org}}, \verb|Healpy|\footnote{\url{https://healpy.readthedocs.io/}} and \verb|Matplotlib|\footnote{\url{http://www.Matplotlib.org}}.
The research carried out by DSH, AD, and GG is supported by the Discovery Grant from the Natural Science and Engineering Research Council of Canada (NSERC). DSH, AD, AK, and GG are also supported by the Perimeter Institute for Theoretical Physics (PI). Research at PI is supported by the Government of Canada through the Department of Innovation, Science and Economic Development Canada and led by the Province of Ontario through the Ministry of Research, Innovation and Science.
JLK is supported by the NSERC CGS-D and the Arthur B. McDonald Canadian Astroparticle Physics Research Institute. AK was supported as a CITA National Fellow by the Natural Sciences and Engineering Research Council of Canada (NSERC), funding reference \#DIS-2022-568580.
The work of SM is a part of the $\langle \texttt{data|theory}\rangle$ \texttt{Universe-Lab} which is supported by the TIFR and the Department of Atomic Energy, Government of India.

 \section*{Data Availability}
The data underlying this article will be shared at request to the authors.

\appendix

\section{3D power spectrum of the SDSS DR7 galaxy catalog--validation}
\label{sec:galpower_comparison}
\begin{figure}
\centering 
\includegraphics[width=.49\textwidth]{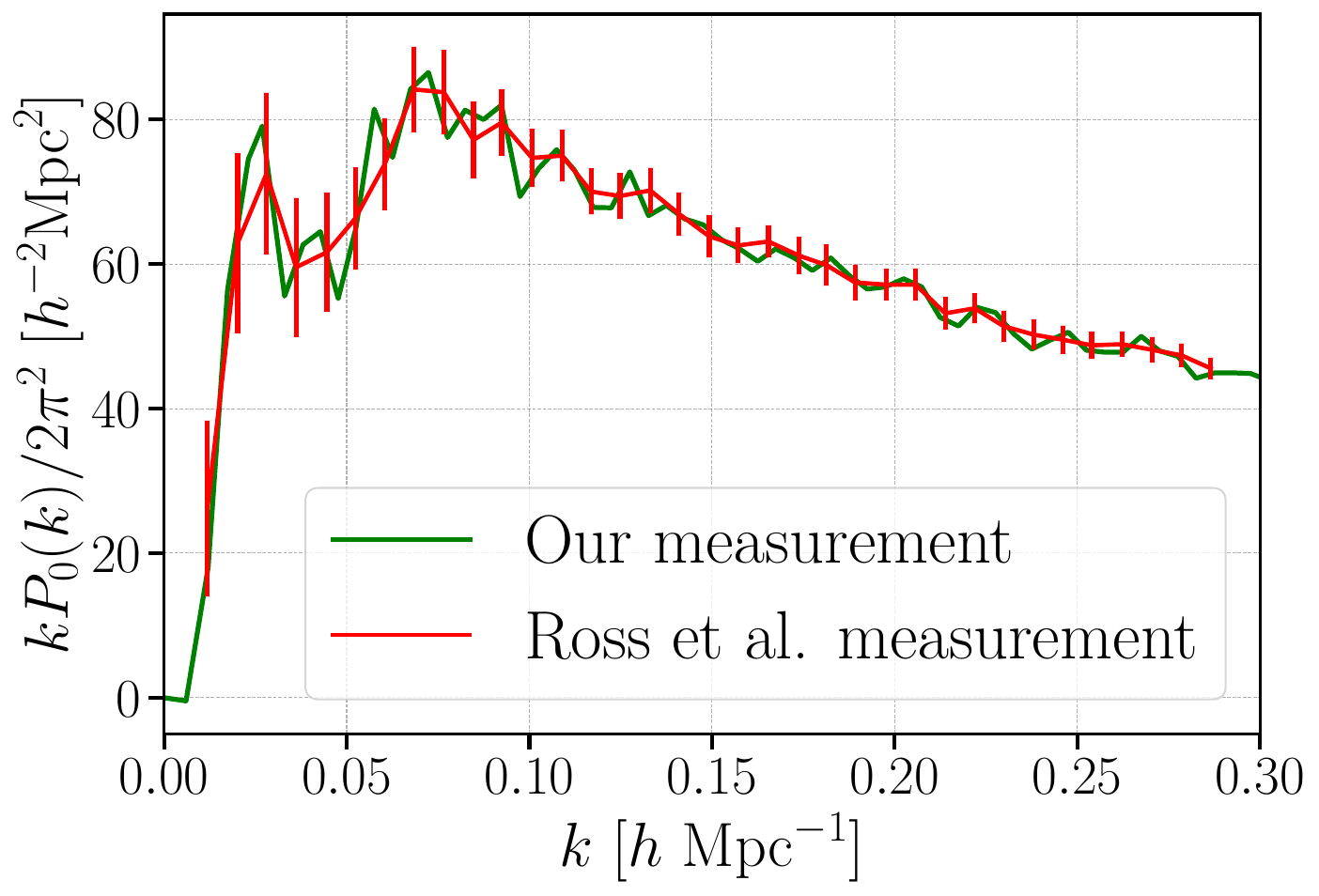}
\includegraphics[width=.49\textwidth]{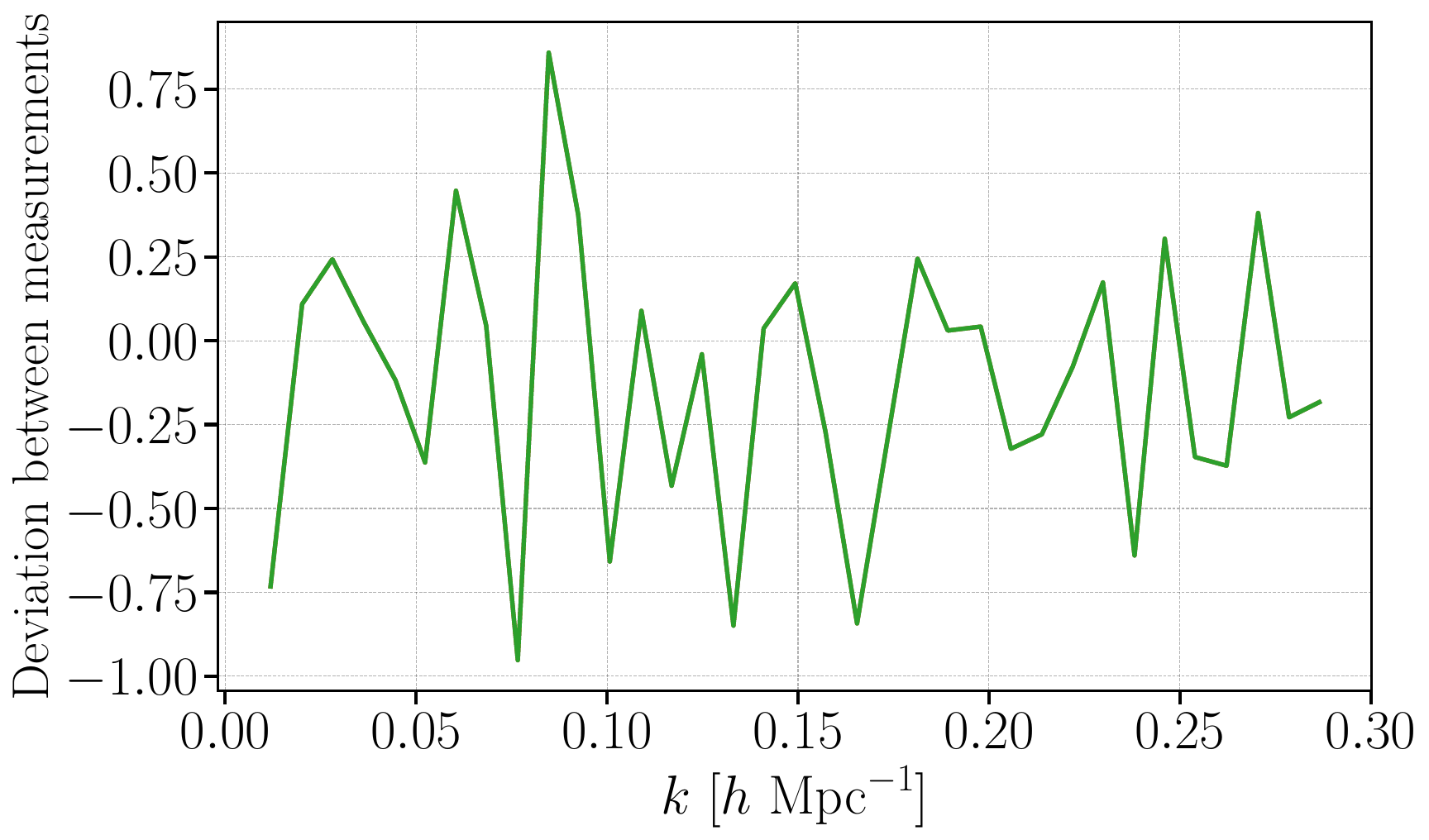}

\caption{Comparison of our galaxy power spectrum measurements with Ross et al.\ \cite{Ross_2015} (in red). The figure shows our estimate of the power spectrum monopole overlaid with Ross et al.’s measurement (red line and points with error bars). The right panel shows the deviation in units of the error bar, which is generally $\lesssim 0.25\sigma$. The  agreement indicates that our power spectrum and covariance estimates are consistent with those from Ross et al.}

\label{fig:ross_comparison}
\end{figure}

As a sanity check, we first apply all the cuts considered in \cite{Ross_2015} to reproduce a similar galaxy power spectrum. That is, all the cuts we have also considered in our base analysis except for the North cap cuts, but still including the color cuts ($M_r<-21.2 ,\, g-r>0.8$) which make the sample more homogeneous (as a function of redshift) and enhance the clustering amplitude. This increases the mass of the halos that host the galaxy. For more details, see \cite{Ross_2015}.
Applying all the mentioned cuts to the galaxy data and random catalogs (which are dispersed inside the window with equal surface density and outside the mask), Figure \ref{fig:ross_comparison} shows our reproduced power spectrum compared to the one calculated by Ross et al. and the difference between the two in units of the error bar, validating our work. 

\section{Null test for GW bias based on uniform $M_*$-$Z$-$SFR$ host-galaxy probability}
\label{sec:random_sel}
The random selection was performed as a sanity check for our selection (populating) pipeline. For randomly selected host galaxies (sirens sample) out of the galaxy catalog, the sirens' distribution should follow the distribution of galaxies and hence the ratio of GW and galaxy bias should be fluctuating around 1. Naturally, we expect that the higher amplitude for selection, which is translated into higher $T_{\rm obs}$ in our pipeline, would reduce the scatter. At the same time, since observed galaxies are not uniformly distributed in stellar mass or other properties and especially since there are incompleteness limitations in the survey, as we explained also in the main text, if $T_{\rm obs}$ too high it can skew the sampling. 
Therefore, we performed the null test for the cases of $T_{\rm obs}=900\, \rm{yr}$ and $T_{\rm obs}=140 \, \rm{yr}$ for the first and second redshift bins, respectively, translating to selecting six percent of the observed galaxies in each bin. Table \ref{table:sirnum} illustrates the number of galaxies in the catalog and our random sirens in two redshift bins.

\begin{table}
\centering
\begin{tabular}{| c | c | c |} 
 \hline 
 Redshift & Number of galaxies &  Number of sirens\\[0.5ex]
 \hline
 $0.07< z \leq 0.135$ & 237084 & 15375 \\
 $0.135< z \leq 0.2$ & 98734 & 5955 \\
 \hline
\end{tabular}
\caption{Number of galaxies and sirens in each redshift bin. Number of sirens is calculated from Eq~\eqref{eq:N_GWcalculate} by setting $T_{\rm obs}=900$ years for the first redshift bin and $T_{\rm obs}=140$ years for the second redshift bin. In both redshift bins we chose 6\% of galaxies to be sirens.}
\label{table:sirnum}
\end{table}

\begin{figure}
\includegraphics[width=.49\textwidth]{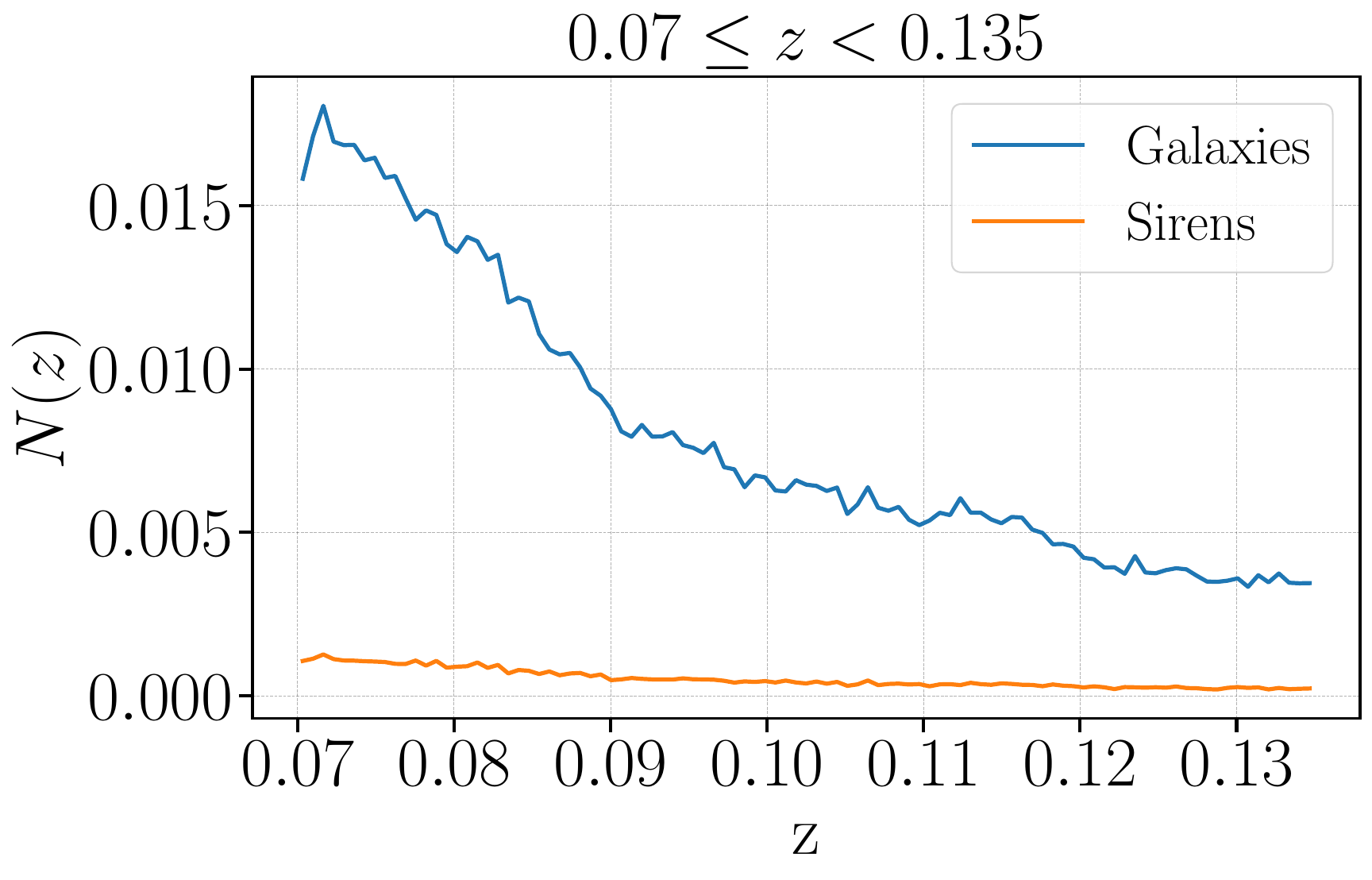}
\includegraphics[width=.49\textwidth]{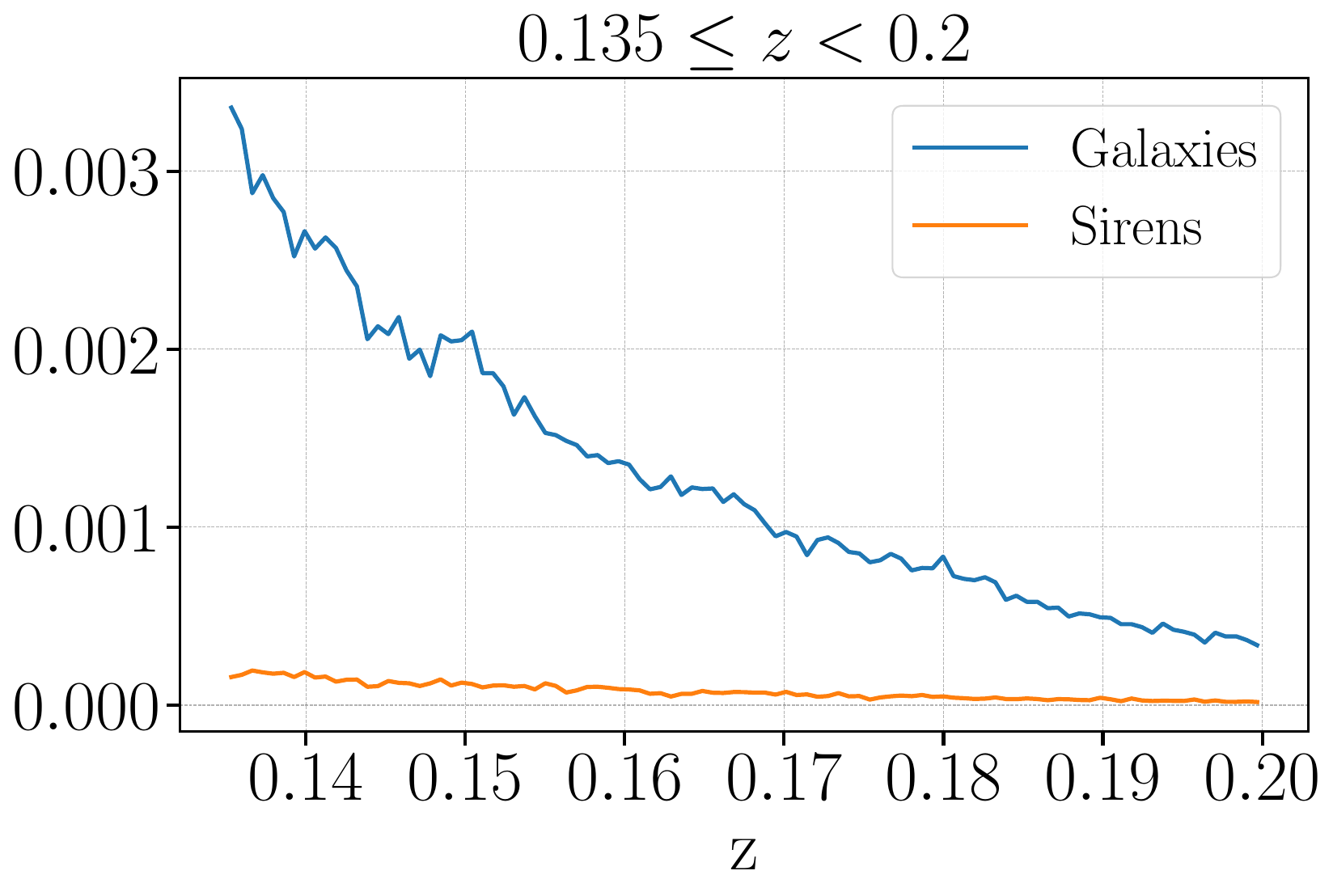}
\caption{Comoving number density in units of $(\mathrm{Mpc}/h)^{-3}$ of galaxies and random sirens as a function of redshift in first redshift bin $0.07 \leq z<0.135$ and second redshift bin $0.135 \leq z<0.2$. $T_{\rm obs}$  is considered to be 900 and 140 for the first and second redshift bins, respectively. }
\label{fig:n(z)}
\end{figure}

In addition to the effect of the amplitude, we realized that there can be an additional offset between the two biases due to the FKP weights. As we can see in Figure \ref{fig:n(z)}, in both redshift bins galaxies and the random sirens have different number densities, so their corresponding FKP weights are also different, leading to different effective redshifts, which can result in
different effective biases, leading to
a few percent offset between galaxy and siren power spectra in the most extreme cases.

\begin{figure}
\includegraphics[width=.49\textwidth]{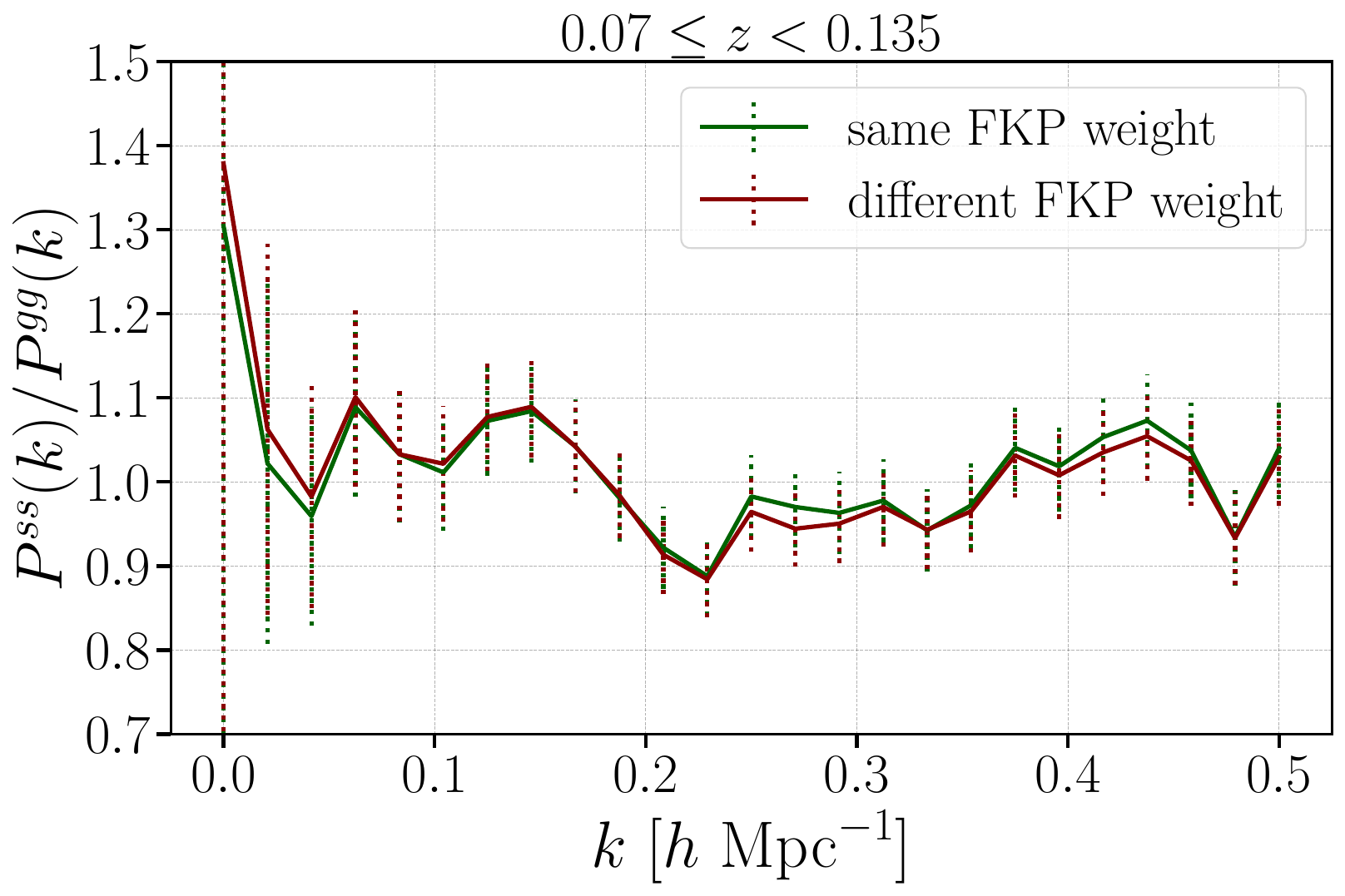}
\includegraphics[width=.47\textwidth]{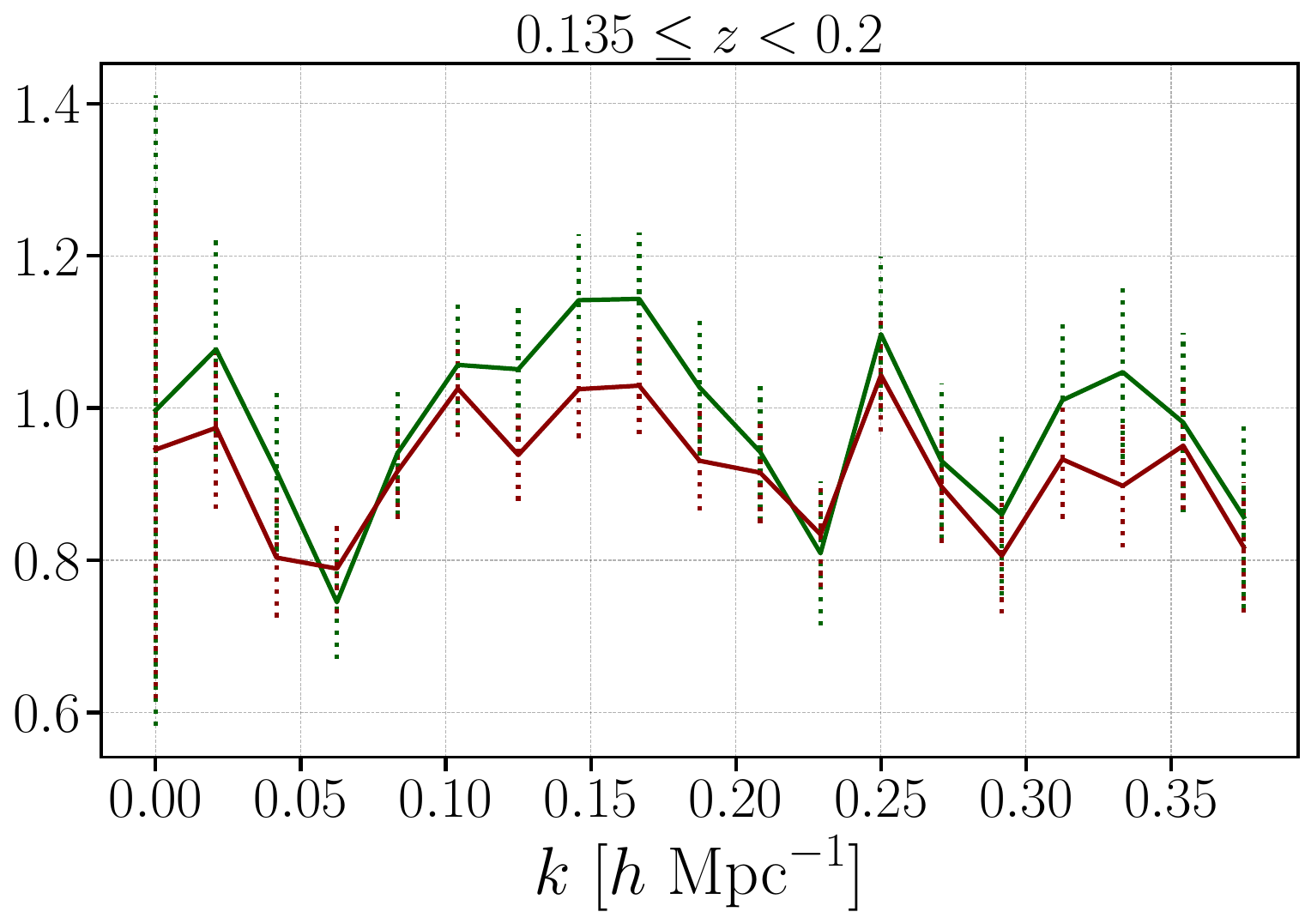}
\caption{ Average ratio of the power spectrum monopole using amplitudes of 900 and 140 for the first and second redshift respectively. $P^{ss}$, and $P^{gg}$ respectively show the siren and galaxy power spectrum monopole, while their ratio represents the ratio of their bias squared. Red lines display cases where FKP weights for sirens and galaxies are their corresponding weights (different FKP weight) and green lines are where both sirens and galaxies weights are set to galaxy weights (same FKP weight).}
\label{fig:diff_same_fkpweight}
\end{figure}

Figure \ref{fig:diff_same_fkpweight} shows the ratio of bias squared obtained through the ratios of the power spectra for different FKP weight strategies. As we see in the plots, the cases of using different FKP weights (red lines) have a small deviation from those of using the same FKP weights (green lines). In particular, there is a noticeable effect in the redshift range of $0.135 \leq z <0.2$. So applying the galaxy FKP weights for both sirens and galaxies avoids any additional offset and improves the fit.   
The values of the average power spectrum monopole ratios, the $\chi^2$'s, and the standard deviations (std) of the power spectrum ratios using the same and different FKP weights are shown in Table \ref{table:sirnumtable}. As we see, $\chi^2$'s are lower when the same FKP weights are applied.
Therefore, to provide the cleanest comparison in the rest of our analysis, we use the galaxy FKP weights for both galaxies and sirens.

\begin{table}
\centering
\begin{tabular}{| c | c | c | c | c |} 
 \hline 
 Redshift & FKP weight & $\textrm{Avg}[P^{ss}/P^{gg}]$ & $\textrm{Std}[P^{ss}/P^{gg}]$ & $\chi^2$ \\[0.5ex]
 \hline
$0.07< z \leq 0.135$ & Same & 1.01 & 0.09 & 21 \\
$0.07< z \leq 0.135$ & Different  & 1.01 & 0.09 & 25 \\
$0.135< z \leq 0.2$ & Same & 0.98 & 0.1 & 28 \\
$0.135< z \leq 0.2$ & Different & 0.91 & 0.08 & 47 \\
\hline
\end{tabular}
\caption{Calculated average $P^{ss}/P^{gg}$ (the bias squared ratio), the $\chi ^2$ (over 25 data points), and the standard deviation (std) for using similar and different FKP weight corresponding to the samples in Table \ref{table:sirnum} and illustrated in Figure \ref{fig:diff_same_fkpweight}. }
\label{table:sirnumtable}
\end{table}

\section{Relation between BH masses and the GW bias}
\label{sec:BH mass}
In this section we review our attempt in modeling the potential correlation between clustering of BBH merger events and their masses. As we see below, given the current physical channels of connections we have considered between the two, the correlation as expected is negligible. 

\subsection{Modeling BH masses in a binary system}
\label{sec:BH  mass modelling}

\paragraph{Modeling the Mass distribution of stellar BHs at formation time (source frame):}

We assume the initial mass distribution of stellar BHs to be dictated by the initial mass function (IMF) of the star population with additional minimum and maximum mass cutoffs. More specifically, we consider the Kroupa IMF, which has been thoroughly examined both theoretically and observationally \cite{2001MNRAS.322..231K} and assume that the mass distribution of BHs at formation time mostly follows the same broken power-law behavior with a lower bound,
\begin{equation}
\label{eq:kroupaIM}
 P_{\rm Kr}(m)\propto
\begin{cases}
0 & m < m_{\mathrm{min}}, \\
m^{-\alpha_K} & m_{\mathrm{min}}\leq m, \\
\end{cases}
\end{equation}
where $m$ is the BH mass after formation and set $m_{\mathrm{min}}=5 M_\odot$ above the maximum mass scale of a stable neutron star and around the minimum mass expected for stellar black holes \citep{McClintock:2003gx}. $\alpha_K$ is a free parameter that for the higher mass ranges of the stellar population that could lead to the formation of BH is estimated to be $\alpha_K=2.3\pm 0.7$ \cite{2001MNRAS.322..231K}; In principle one could also explore varying this parameter for BH populations. However, we did not test this variation since as mentioned above we expected the connection of BH masses to $b_{\rm GW}$ to be at best very minimal, but this may be another interesting avenue to examine for connecting BH masses to their environment.

Note that so far in our assumptions, there is no direct connection between the distribution of BH masses and their environment, and thus the host galaxies. However, next we take into account that stellar mass BHs not only have a minimum mass threshold, but also an upper limit at the time of their formation, due to the occurrence of pair instability supernova (PISN).\footnote{The prevailing theories of stellar evolution suggest the existence of an upper {\it mass gap} for stellar BHs, resulting from the excessive production of electron-positron pairs in more massive stars, followed by a subsequent decrease in internal pressure that triggers a partial collapse of the star, leading to an intensified thermonuclear explosion \cite{1967ApJ...148..803R}. Ultimately, the star is completely blown apart, leaving no compact remnants in its wake.}
The current theoretical estimate for the maximum stellar BH masses due to the PISN process is expected to be around $45 M_{\odot}$ with a small window allowing for a potential dependence on stellar metallicity \cite{2023MNRAS.523.4539K,2022MNRAS.515.5495M,2019ApJ...887...53F}. One can express this as a logarithmic relation  
\begin{equation}
m_{\mathrm{PISN}}(Z)=m_{\mathrm{PISN}}(\tilde{Z})-\alpha_{Z} \log _{10}(Z / \tilde{Z}) ,
\label{eq:pisnm}
\end{equation}
where $\Tilde{Z}=10^{-4}$, $M_{\operatorname{PISN}}(\Tilde{Z})=45 \mathrm{M}_{\odot}$, and $\alpha_Z$ is a free parameter of order one \cite{2019ApJ...887...53F, 2022MNRAS.515.5495M}.  
This physics-driven model for the BBH mass distribution has been shown to fit the GWTC-3 data \cite{2023MNRAS.523.4539K}, which may be evidence towards a concurrent picture of star formation and BBH mergers in the Universe.
This metallicity dependence connects the BBH masses to their with their environments. 

We can impose this upper bound constraint on the BH mass distribution corresponding to the PISN mass gap using the following window function,   
\begin{equation}
\label{eq:wind_s}
    \mathcal{W}_s (m,Z)=
   \left \{
        \begin{array}{ll}
            1 &\quad m_{min}\leq m \leq m_{PISN}(Z),\\
            0 &  \quad \text{otherwise},
        \end{array}
    \right.  
\end{equation}
so the overall mass distributions of the stellar mass BHs in the source frame is given by 
\begin{equation} 
\mathcal{P}_s (m,Z)=\mathcal{W}_s (m,Z)P_{\rm Kr}(m) .
\label{eq:BHs-Mass-source}
\end{equation}

\paragraph{Modeling mass distribution of the BH at merger time:} To determine the mass distribution of BHs at merger time, one needs to take into account the impact of the delay time between the formation of single BHs to the merging of two BHs in a binary system. 
The BH mass distribution at merger redshift is then
\cite{2022MNRAS.515.5495M,2023MNRAS.523.4539K} 
\begin{align}
\label{eq:p_m_zm}
\mathcal{P}_m(m,z_m)&= \mathcal{N}\int_{z_m}^{\infty} P_D(t_d)\frac{d t_f}{dz_f} \mathcal{P}_s (m,Z) dz_f\nonumber\\
&=\mathcal{W}(m,z_m) P_{\rm Kr}(m),
\end{align}
where, $t_d = t_m - t_f$ and $\mathcal{N}$ is a normalization factor so that $\int \mathcal{P}_m(m,z_m) dm=1$.
The last line uses Eq.~\eqref{eq:BHs-Mass-source} for substitution and assumes that the Kroupa mass function is redshift independent, which is a reasonable assumption \citep{2001MNRAS.322..231K}. The BH mass distribution at merger redshift is then a multiplication of Kroupa IMF and the accumulated window function $\mathcal{W}(m,z_m)$ given by
\begin{equation}
\label{eq:ob win}
\mathcal{W}(m,z_m)\equiv\mathcal{N} \int_{z_m}^{\infty} P_D(t_d) \frac{d t_f}{dz_f} \mathcal{W}_s (m,Z) d z_f.
\end{equation}
In order to evaluate Eq.~\eqref{eq:ob win} and substitute it back into Eq.~\eqref{eq:p_m_zm}, we need their metallicity information at the redshift of the BBH formation. 
However, using an observed survey, we only have access to their observed metallicity which in our framework corresponds to metallicity of the host galaxy at the merger time. If we assume the metallicity of the host galaxy does not change significantly between the formation and merger times, then we could substitute the observed metallicity $Z$, from the galaxy survey in Equation \eqref{eq:ob win} and since $\mathcal{W}_s$ no longer depends on the formation redshift, factor it out of the integration. In this case  and we can absorb the whole integration into the normalization constant i.e.
\begin{equation}
\label{eq:ob win_z_indep2}
\mathcal{W}(m,z_m)\equiv \widetilde{\mathcal{N}} \,\mathcal{W}_s(m, Z)  \, .
\end{equation}
leading to the following mass distribution
\begin{equation}
\mathcal{P}_m(m)=\widetilde{\mathcal{N}} \,\mathcal{W}_s(m, Z) P_{\rm Kr}(m),
\end{equation}
with no explicit redshift dependence. The main drawback of this approach is that it could only hold if the delay times are relatively short compared to the galaxy evolution. For the longer delay-time cases, galaxies can have lower metallicities at the formation, and thus a higher PISN mass scale and cut off. However, note that even then the only dependence to formation metallicity is through a logarithmic cutoff, and it is unlikely that it would make a significant change 

So far we have modeled the probability mass distribution of stellar BHs at $z_m$, however, every BBH merger consists of two companion masses. Therefore, the mass distribution for the system involves sampling both a primary BH, $m_1$, out of the distribution, and a secondary BH, $m_2$, out of it. To do this, we choose the conditional probability $P(m_2|m_1)$ as a power law with a positive index $\beta_{m_2}$ along with a minimum and a maximum cut as $[m_{min},m_1]$. 

Eventually, we characterize the binary system with a single mass scale, i.e. the chirp mass
\begin{align}
\label{eq:m_ch}
 m_{ch}\equiv \frac{(m_1 m_2)^{3/5}}{(m_1+m_2)^{1/5}}\,.
\end{align}

\subsection{Results for GW bias vs.\ BH masses} \label{sec: BH mass result}
Following the prescription described in Section \ref{sec:BH mass modelling}, we now explored the relationship between GW bias and BH masses.  The BBH mass parameters were set as $m_\mathrm{min}=5 M_\odot$, $\alpha_K=2.3$, $\alpha_Z=1.5$, and $\beta_{m_2}=1.0$. Within each redshift bin, we also grouped the sirens according to their chirp masses into two sets, those with $4.3 ~ M_{\odot}\leq m_\mathrm{ch} < 6~ M_{\odot}$ and those with $6 ~ M_{\odot} \leq m_\mathrm{ch} < 34~M_{\odot}$. We chose this BH mass binning in order to roughly have the same number of sources in each range. The lower mass range is smaller since the IMF has a decaying power law form and prefers lower chirp masses. Next, we aim to see if we can find any correlation between the GW bias and chirp mass ranges in the binary system. 

\begin{figure}
\centering 
\includegraphics[width=.49\textwidth]{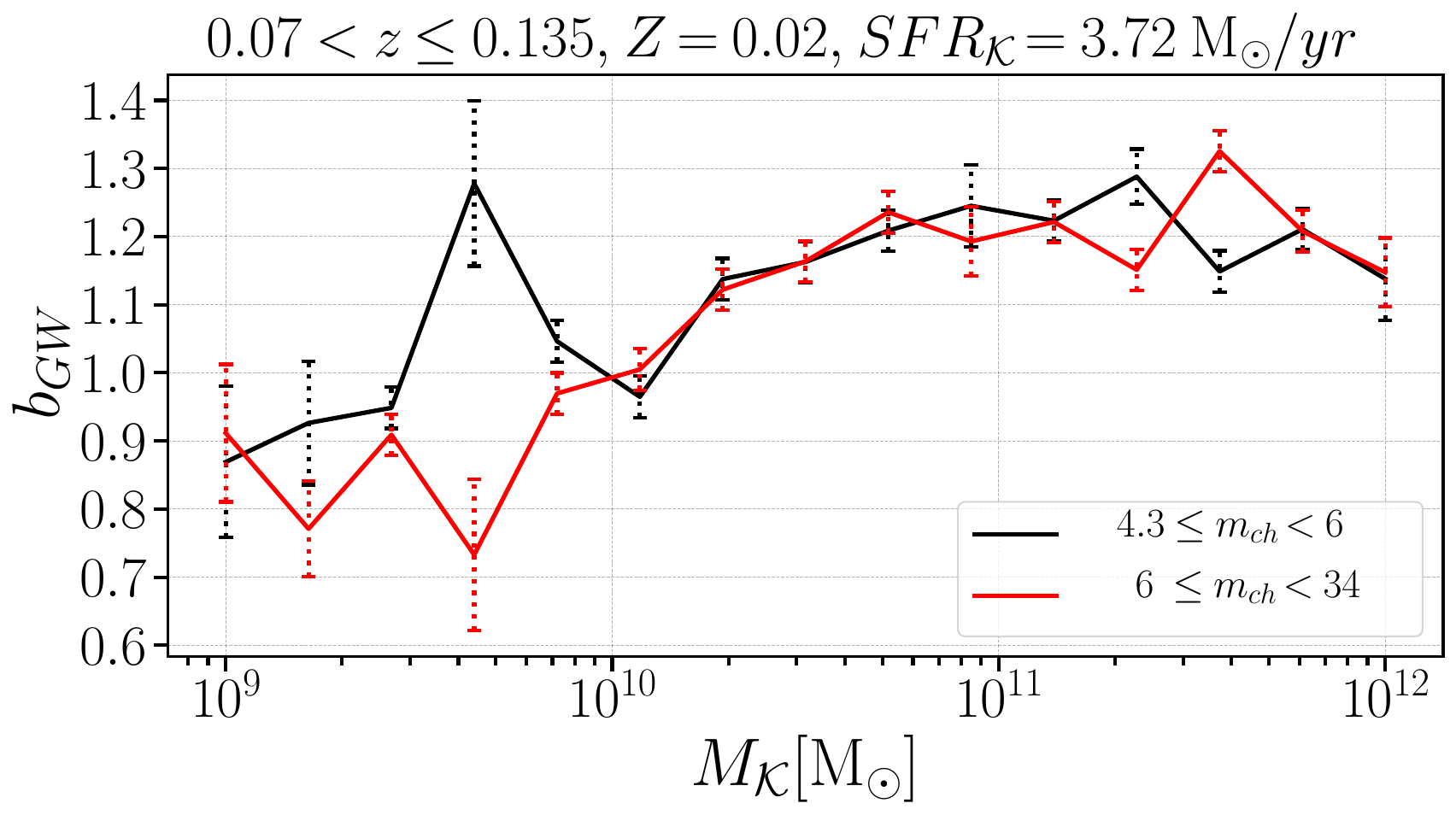}
\includegraphics[width=.49\textwidth]{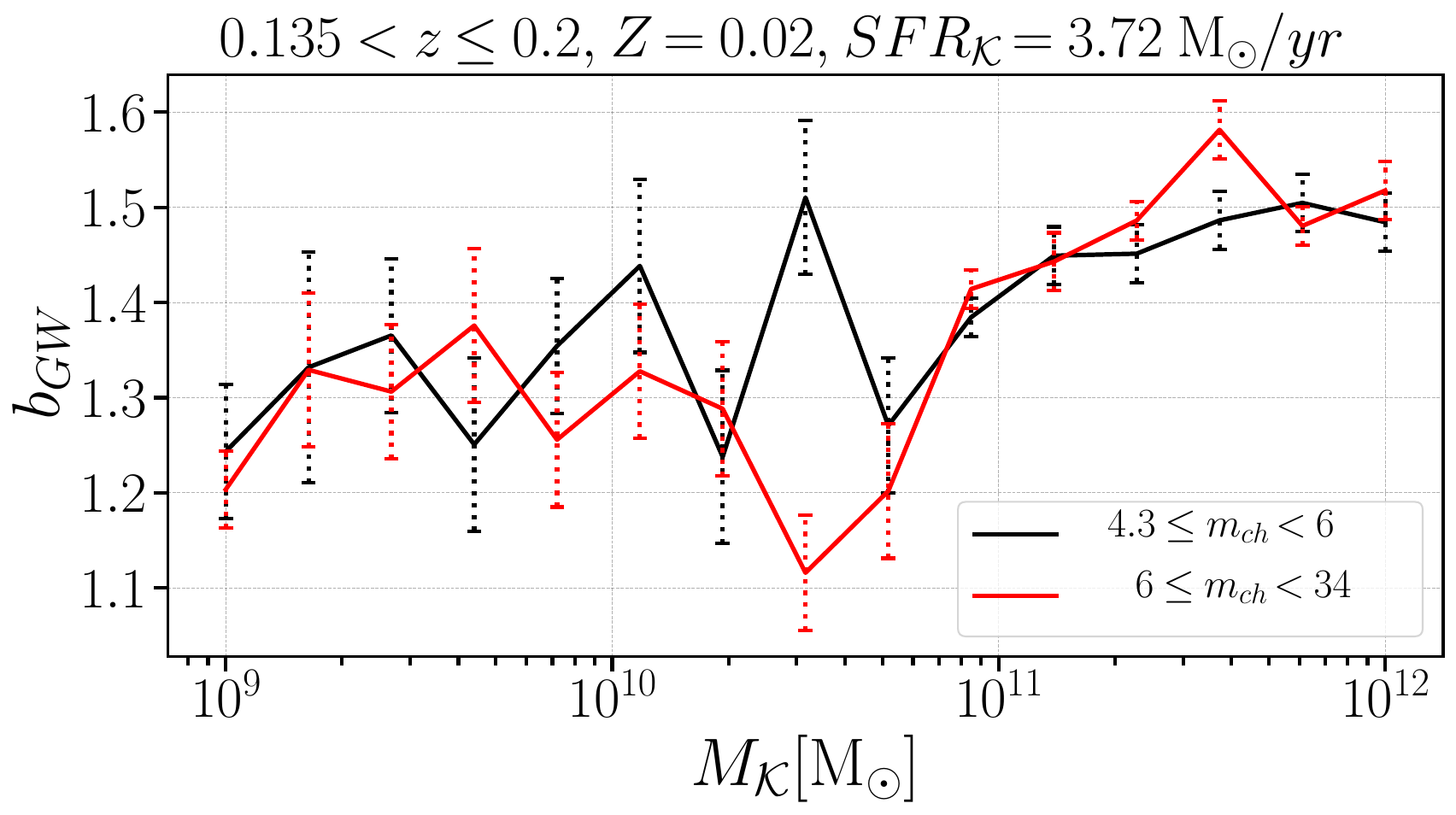}

\caption{GW bias parameter for varying $M_\mathcal{K}$ using $M_*$-$Z$-SFR model in the two different redshift ranges and in each for two BBH chirp mass bins, $4.3 M_\odot \leq m_{ch} < 6 M_\odot$ (red solid line) and $6 M_\odot \leq m_{ch} < 34 M_\odot$ (black solid line). In these cases we have set $\delta_l=\infty$, $\delta_h=0.5, ~\epsilon_l=\infty, ~\epsilon_h=0.5 ,~ \zeta_l=0.5,~\zeta_h=0.25$, $Z_\mathcal{K}= 0.02$, and $\rm{SFR}_\mathcal{K}= 3.72 ~M_\odot/\rm yr$. }
\label{fig:2mbin_MZsfr_z002_sfr372}
\end{figure}

\begin{figure}
\centering 
\includegraphics[width=.49\textwidth]{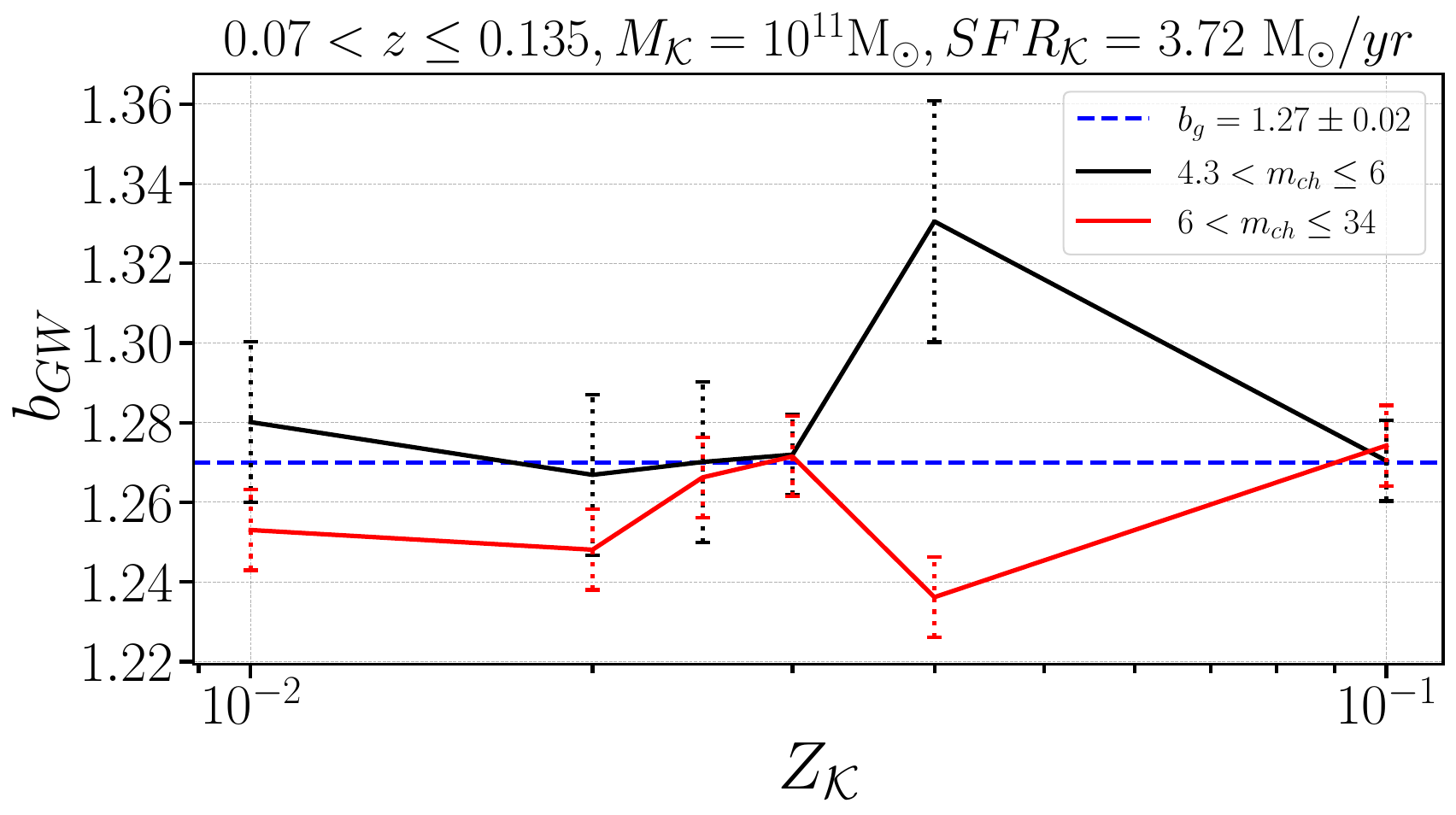}
\includegraphics[width=.49\textwidth]{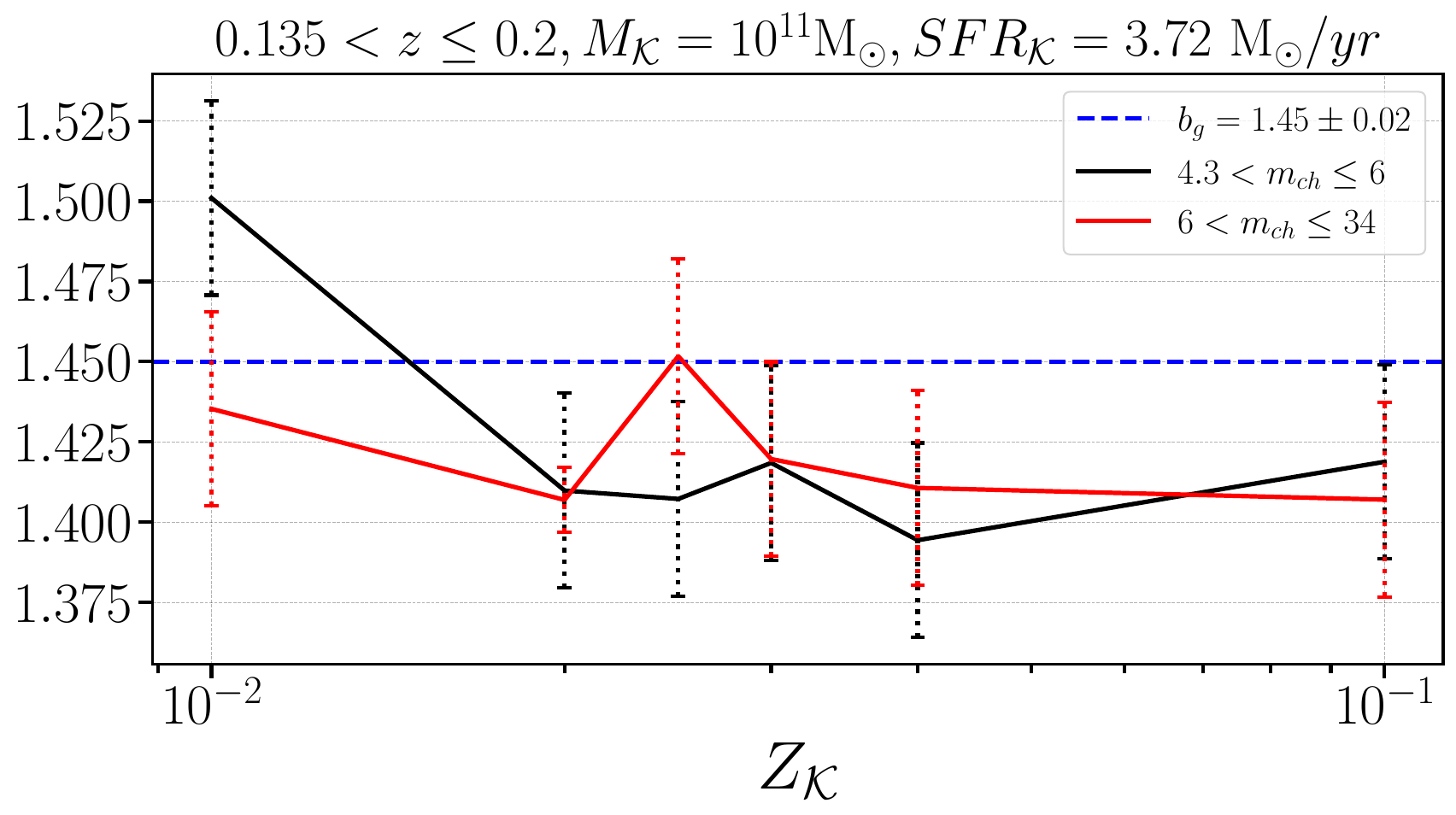}

\caption{GW bias parameter for two different BBH chirp mass ranges while varying $Z_\mathcal{K}$ using $M_*$-$Z$-SFR model. In this cases we have set $\delta_l=1.3$, $\delta_h=2.5, ~\epsilon_l=2, ~\epsilon_h=1 ,~ \zeta_l=1,~\zeta_h=5$, $M_\mathcal{K}= 10^{11}~M_\odot$, and $\rm{SFR}_\mathcal{K}= 3.72 ~M_\odot/\rm yr$.}
\label{fig:2mbin_MZsfr_Mc11_sfr372}
\end{figure}

Figures \ref{fig:2mbin_MZsfr_z002_sfr372} and \ref{fig:2mbin_MZsfr_Mc11_sfr372} present our findings for the GW bias parameter for each mass bin as a function of $M_\mathcal{K}$ and $Z_\mathcal{K}$, within some sample of models where the rest of the shape parameters of the host-galaxy probability are fixed in each case. 

 A key observation from these figures is that there does not seem to be any significant correlation between the GW bias and the masses of BHs. In most plots the GW bias for the two mass bins overlap or follow each other very closely. The GW biases in the right panel of Figure \ref{fig:2mbin_MZsfr_Mc11_sfr372} for both mass bins while overlapping are below the galaxy bias. This can be explained simply as the consequence of $\rm SFR_\mathcal{K}=3.72 \, M_\odot/yr$ around which galaxies have a lower mean stellar mass (see Figure \ref{fig:mean_mass_vs_sfr}). 

 There are some small differences between the GW biases of the two masses in some scenarios that appear to depend on redshift and the slopes of the host-galaxy probability, but we could not find any meaningful pattern for these small variations. For example, the left panel of Figure \ref{fig:2mbin_MZsfr_Mc11_sfr372} shows a slightly lower GW bias for BBH with a higher chirp mass compared to those with a lower chirp mass for $M_\mathcal{K}= 10^{11}\, M_\odot$. We see a similar feature in the left panel of Figure \ref{fig:2mbin_MZsfr_Mc11_sfr372} in the vicinity of $M_\mathcal{K}= 10^{11}\, M_\odot$, however, once we vary $M_\mathcal{K}$ the order can be reversed. Therefore, we do not think this is a meaningful physical effect and probably a consequence of incompleteness edges of survey, i.e. the geometry of the boundary of the observed $M_*$-$SFR$-$Z$ volume in the survey. 
 
From a physical point of view though, it would make sense that higher-mass binary black holes (BBHs) to be more likely to form in low-metallicity environments. That is because less metallicity leads to weaker stellar winds in massive stars and makes them more likely to retain their mass and collapse into more massive black holes. 
Low metallicity galaxies are also typically associated with low-mass galaxies and tend to be less clustered, so we would expect a lower bias. 
However, as we pointed out earlier, our method is limited to having access to the current observed metallicity of the galaxy. 
A model that can include the full evolution of the metallicity of the galaxies might show different results. Second, in our model the PISN mass scale is only a cut-off in the mass distribution of BHs and it also only changes logarithmically with metallicity; at the same time  most of the BHs in the mock catalog have masses below this mass scale since the IMF drops rapidly by the power law function given in Equation \eqref{eq:kroupaIM}. Therefore, the environment of the host galaxy, through our prescription here, has a minor effect on the mass distribution of BHs in the mock catalog.

\begin{figure}
\includegraphics[width=.49\textwidth]{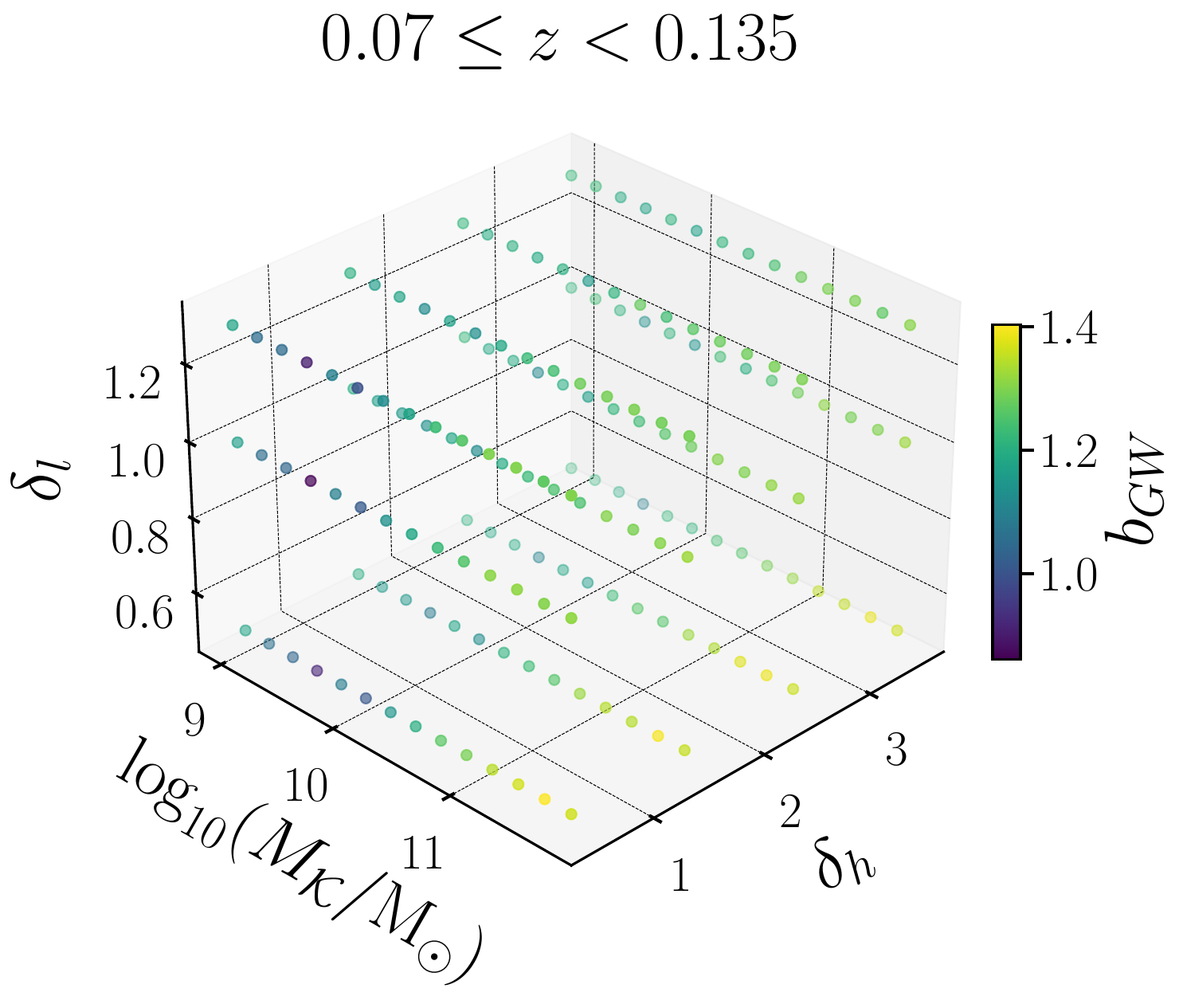}
\includegraphics[width=.49\textwidth]{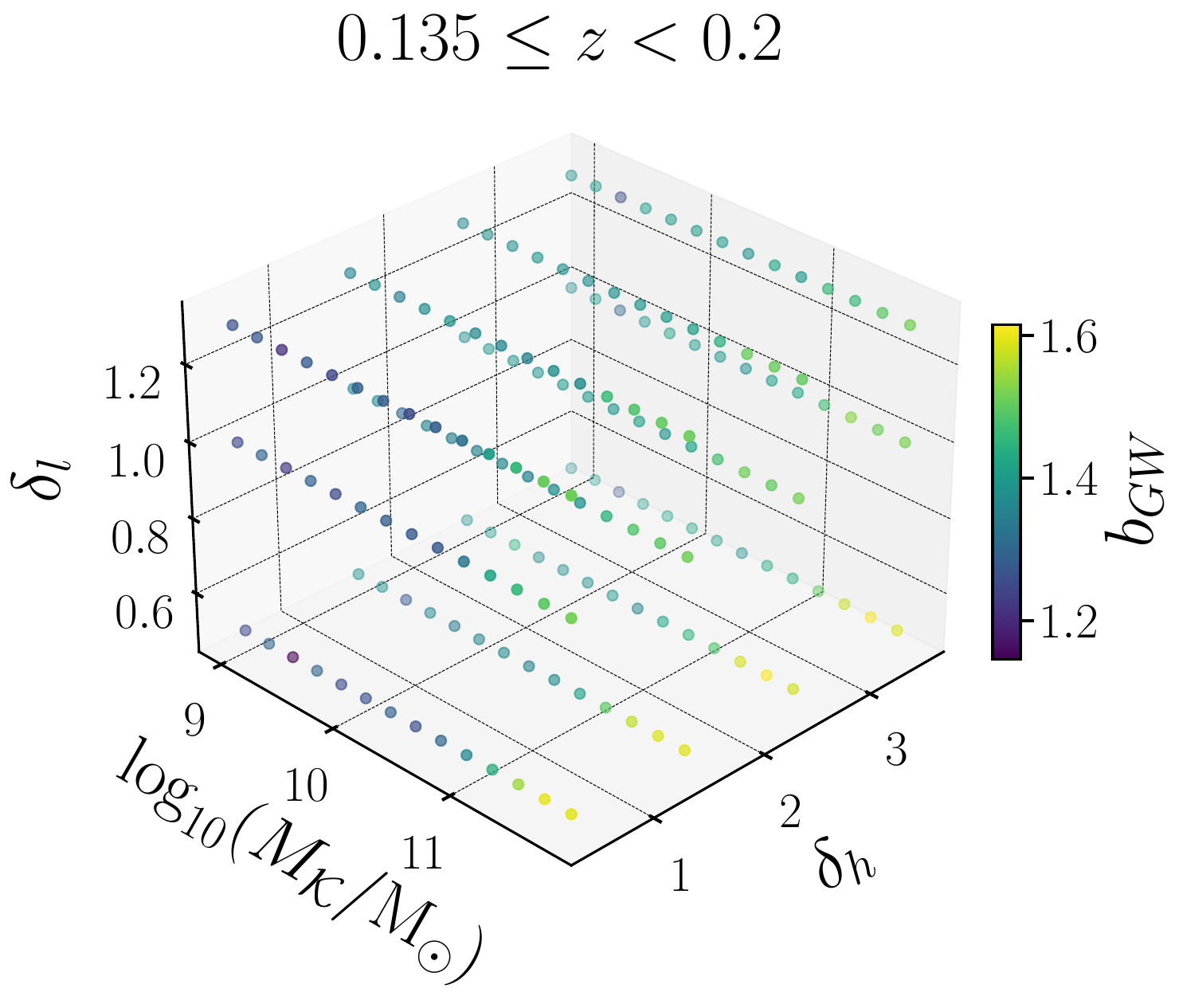}

\caption{ The 3D plot of the GW bias parameter within redshift $0.07 \leq z <0.135$ and $0.135 \leq z <0.2$, respectively. Sirens are populated using $M_*$-$Z$-SFR probability function with parameters: $SFR_\mathcal{K} = 3.72$ $M_\odot$/yr, $Z_\mathcal{K} = 0.025$, $\epsilon_l=2$, $\epsilon_h=1 , \zeta_l=1$, and $\zeta_h=5$. We show the dependence of the GW bias parameter on the parameters of the stellar mass part of the host-galaxy probability function: the slopes $\delta_l$ and $\delta_h$ and the break mass $M_\mathcal{K}$.}
\label{fig:3ddmbias_MZSFRv2_sfr3.72_mc1}
\end{figure}

\bibliographystyle{JHEP}
\bibliography{refs}

@article{dehghani2024,
      title={The Gravitational Wave Bias Parameter from Angular Power Spectra: Bridging Between Galaxies and Binary Black Holes}, 
      author={Amir Dehghani and J. Leo Kim and Dorsa Sadat Hosseini and Alex Krolewski and Suvodip Mukherjee and Ghazal Geshnizjani},
      year={2024},
      eprint={2411.11965},
      archivePrefix={arXiv},
      primaryClass={astro-ph.GA},
      url={https://arxiv.org/abs/2411.11965}, 
}

@ARTICLE{Wadekar19,
       author = {{Wadekar}, Digvijay and {Scoccimarro}, Rom{\'a}n},
        title = "{Galaxy power spectrum multipoles covariance in perturbation theory}",
      journal = {\prd},
         year = 2020,
        month = dec,
       volume = {102},
       number = {12},
          eid = {123517},
        pages = {123517},
          doi = {10.1103/PhysRevD.102.123517},
archivePrefix = {arXiv},
       eprint = {1910.02914},
 primaryClass = {astro-ph.CO},
       adsurl = {https://ui.adsabs.harvard.edu/abs/2020PhRvD.102l3517W}
}

@ARTICLE{Yamamoto06,
       author = {{Yamamoto}, Kazuhiro and {Nakamichi}, Masashi and {Kamino}, Akinari and {Bassett}, Bruce A. and {Nishioka}, Hiroaki},
        title = "{A Measurement of the Quadrupole Power Spectrum in the Clustering of the 2dF QSO Survey}",
      journal = {pasj},
         year = 2006,
        month = feb,
       volume = {58},
        pages = {93-102},
          doi = {10.1093/pasj/58.1.93},
archivePrefix = {arXiv},
       eprint = {astro-ph/0505115},
 primaryClass = {astro-ph},
       adsurl = {https://ui.adsabs.harvard.edu/abs/2006PASJ...58...93Y}
}

@article{LIGOScientific:2020kqk,
    author = "Abbott, R. and others",
    collaboration = "LIGO Scientific, Virgo",
    title = "{Population Properties of Compact Objects from the Second LIGO-Virgo Gravitational-Wave Transient Catalog}",
    eprint = "2010.14533",
    archivePrefix = "arXiv",
    primaryClass = "astro-ph.HE",
    reportNumber = "LIGO-P2000077",
    doi = "10.3847/2041-8213/abe949",
    journal = "Astrophys. J. Lett.",
    volume = "913",
    number = "1",
    pages = "L7",
    year = "2021"
}

@ARTICLE{2005AJ....129.2562B,
       author = {{Blanton}, Michael R. and {Schlegel}, David J. and {Strauss}, Michael A. and {Brinkmann}, J. and {Finkbeiner}, Douglas and {Fukugita}, Masataka and {Gunn}, James E. and {Hogg}, David W. and {Ivezi{\'c}}, {\v{Z}}eljko and {Knapp}, G.~R. and {Lupton}, Robert H. and {Munn}, Jeffrey A. and {Schneider}, Donald P. and {Tegmark}, Max and {Zehavi}, Idit},
        title = "{New York University Value-Added Galaxy Catalog: A Galaxy Catalog Based on New Public Surveys}",
      journal = {\aj},
         year = 2005,
        month = jun,
       volume = {129},
       number = {6},
        pages = {2562-2578},
          doi = {10.1086/429803},
archivePrefix = {arXiv},
       eprint = {astro-ph/0410166},
 primaryClass = {astro-ph},
       adsurl = {https://ui.adsabs.harvard.edu/abs/2005AJ....129.2562B}
}

@ARTICLE{1998AJ....116.3040G,
       author = {{Gunn}, J.~E. and {Carr}, M. and {Rockosi}, C. and {Sekiguchi}, M. and {Berry}, K. and {Elms}, B. and {de Haas}, E. and {Ivezi{\'c}}, {\v{Z}} . and {Knapp}, G. and {Lupton}, R. and {Pauls}, G. and {Simcoe}, R. and {Hirsch}, R. and {Sanford}, D. and {Wang}, S. and {York}, D. and {Harris}, F. and {Annis}, J. and {Bartozek}, L. and {Boroski}, W. and {Bakken}, J. and {Haldeman}, M. and {Kent}, S. and {Holm}, S. and {Holmgren}, D. and {Petravick}, D. and {Prosapio}, A. and {Rechenmacher}, R. and {Doi}, M. and {Fukugita}, M. and {Shimasaku}, K. and {Okada}, N. and {Hull}, C. and {Siegmund}, W. and {Mannery}, E. and {Blouke}, M. and {Heidtman}, D. and {Schneider}, D. and {Lucinio}, R. and {Brinkman}, J.},
        title = "{The Sloan Digital Sky Survey Photometric Camera}",
      journal = {\aj},
         year = 1998,
        month = dec,
       volume = {116},
       number = {6},
        pages = {3040-3081},
          doi = {10.1086/300645},
archivePrefix = {arXiv},
       eprint = {astro-ph/9809085},
 primaryClass = {astro-ph},
       adsurl = {https://ui.adsabs.harvard.edu/abs/1998AJ....116.3040G}
}

@ARTICLE{2006AJ....131.2332G,
       author = {{Gunn}, James E. and {Siegmund}, Walter A. and {Mannery}, Edward J. and {Owen}, Russell E. and {Hull}, Charles L. and {Leger}, R. French and {Carey}, Larry N. and {Knapp}, Gillian R. and {York}, Donald G. and {Boroski}, William N. and {Kent}, Stephen M. and {Lupton}, Robert H. and {Rockosi}, Constance M. and {Evans}, Michael L. and {Waddell}, Patrick and {Anderson}, John E. and {Annis}, James and {Barentine}, John C. and {Bartoszek}, Larry M. and {Bastian}, Steven and {Bracker}, Stephen B. and {Brewington}, Howard J. and {Briegel}, Charles I. and {Brinkmann}, Jon and {Brown}, Yorke J. and {Carr}, Michael A. and {Czarapata}, Paul C. and {Drennan}, Craig C. and {Dombeck}, Thomas and {Federwitz}, Glenn R. and {Gillespie}, Bruce A. and {Gonzales}, Carlos and {Hansen}, Sten U. and {Harvanek}, Michael and {Hayes}, Jeffrey and {Jordan}, Wendell and {Kinney}, Ellyne and {Klaene}, Mark and {Kleinman}, S.~J. and {Kron}, Richard G. and {Kresinski}, Jurek and {Lee}, Glenn and {Limmongkol}, Siriluk and {Lindenmeyer}, Carl W. and {Long}, Daniel C. and {Loomis}, Craig L. and {McGehee}, Peregrine M. and {Mantsch}, Paul M. and {Neilsen}, Eric H., Jr. and {Neswold}, Richard M. and {Newman}, Peter R. and {Nitta}, Atsuko and {Peoples}, John, Jr. and {Pier}, Jeffrey R. and {Prieto}, Peter S. and {Prosapio}, Angela and {Rivetta}, Claudio and {Schneider}, Donald P. and {Snedden}, Stephanie and {Wang}, Shu-i.},
        title = "{The 2.5 m Telescope of the Sloan Digital Sky Survey}",
      journal = {\aj},
         year = 2006,
        month = apr,
       volume = {131},
       number = {4},
        pages = {2332-2359},
          doi = {10.1086/500975},
archivePrefix = {arXiv},
       eprint = {astro-ph/0602326},
 primaryClass = {astro-ph},
       adsurl = {https://ui.adsabs.harvard.edu/abs/2006AJ....131.2332G}
}

@ARTICLE{padmanabhan2008improved,
       author = {{Padmanabhan}, Nikhil and {Schlegel}, David J. and {Finkbeiner}, Douglas P. and {Barentine}, J.~C. and {Blanton}, Michael R. and {Brewington}, Howard J. and {Gunn}, James E. and {Harvanek}, Michael and {Hogg}, David W. and {Ivezi{\'c}}, {\v{Z}}eljko and {Johnston}, David and {Kent}, Stephen M. and {Kleinman}, S.~J. and {Knapp}, Gillian R. and {Krzesinski}, Jurek and {Long}, Dan and {Neilsen}, Jr., Eric H. and {Nitta}, Atsuko and {Loomis}, Craig and {Lupton}, Robert H. and {Roweis}, Sam and {Snedden}, Stephanie A. and {Strauss}, Michael A. and {Tucker}, Douglas L.},
        title = "{An Improved Photometric Calibration of the Sloan Digital Sky Survey Imaging Data}",
      journal = {\apj},
         year = 2008,
        month = feb,
       volume = {674},
       number = {2},
        pages = {1217-1233},
          doi = {10.1086/524677},
archivePrefix = {arXiv},
       eprint = {astro-ph/0703454},
 primaryClass = {astro-ph},
       adsurl = {https://ui.adsabs.harvard.edu/abs/2008ApJ...674.1217P}
}

@ARTICLE{2009ApJS..182..543A,
       author = {{Abazajian}, Kevork N. and {Adelman-McCarthy}, Jennifer K. and {Ag{\"u}eros}, Marcel A. and {Allam}, Sahar S. and {Allende Prieto}, Carlos and {An}, Deokkeun and {Anderson}, Kurt S.~J. and {Anderson}, Scott F. and {Annis}, James and {Bahcall}, Neta A. and {Bailer-Jones}, C.~A.~L. and {Barentine}, J.~C. and {Bassett}, Bruce A. and {Becker}, Andrew C. and {Beers}, Timothy C. and {Bell}, Eric F. and {Belokurov}, Vasily and {Berlind}, Andreas A. and {Berman}, Eileen F. and {Bernardi}, Mariangela and {Bickerton}, Steven J. and {Bizyaev}, Dmitry and {Blakeslee}, John P. and {Blanton}, Michael R. and {Bochanski}, John J. and {Boroski}, William N. and {Brewington}, Howard J. and {Brinchmann}, Jarle and {Brinkmann}, J. and {Brunner}, Robert J. and {Budav{\'a}ri}, Tam{\'a}s and {Carey}, Larry N. and {Carliles}, Samuel and {Carr}, Michael A. and {Castander}, Francisco J. and {Cinabro}, David and {Connolly}, A.~J. and {Csabai}, Istv{\'a}n and {Cunha}, Carlos E. and {Czarapata}, Paul C. and {Davenport}, James R.~A. and {de Haas}, Ernst and {Dilday}, Ben and {Doi}, Mamoru and {Eisenstein}, Daniel J. and {Evans}, Michael L. and {Evans}, N.~W. and {Fan}, Xiaohui and {Friedman}, Scott D. and {Frieman}, Joshua A. and {Fukugita}, Masataka and {G{\"a}nsicke}, Boris T. and {Gates}, Evalyn and {Gillespie}, Bruce and {Gilmore}, G. and {Gonzalez}, Belinda and {Gonzalez}, Carlos F. and {Grebel}, Eva K. and {Gunn}, James E. and {Gy{\"o}ry}, Zsuzsanna and {Hall}, Patrick B. and {Harding}, Paul and {Harris}, Frederick H. and {Harvanek}, Michael and {Hawley}, Suzanne L. and {Hayes}, Jeffrey J.~E. and {Heckman}, Timothy M. and {Hendry}, John S. and {Hennessy}, Gregory S. and {Hindsley}, Robert B. and {Hoblitt}, J. and {Hogan}, Craig J. and {Hogg}, David W. and {Holtzman}, Jon A. and {Hyde}, Joseph B. and {Ichikawa}, Shin-ichi and {Ichikawa}, Takashi and {Im}, Myungshin and {Ivezi{\'c}}, {\v{Z}}eljko and {Jester}, Sebastian and {Jiang}, Linhua and {Johnson}, Jennifer A. and {Jorgensen}, Anders M. and {Juri{\'c}}, Mario and {Kent}, Stephen M. and {Kessler}, R. and {Kleinman}, S.~J. and {Knapp}, G.~R. and {Konishi}, Kohki and {Kron}, Richard G. and {Krzesinski}, Jurek and {Kuropatkin}, Nikolay and {Lampeitl}, Hubert and {Lebedeva}, Svetlana and {Lee}, Myung Gyoon and {Lee}, Young Sun and {French Leger}, R. and {L{\'e}pine}, S{\'e}bastien and {Li}, Nolan and {Lima}, Marcos and {Lin}, Huan and {Long}, Daniel C. and {Loomis}, Craig P. and {Loveday}, Jon and {Lupton}, Robert H. and {Magnier}, Eugene and {Malanushenko}, Olena and {Malanushenko}, Viktor and {Mandelbaum}, Rachel and {Margon}, Bruce and {Marriner}, John P. and {Mart{\'\i}nez-Delgado}, David and {Matsubara}, Takahiko and {McGehee}, Peregrine M. and {McKay}, Timothy A. and {Meiksin}, Avery and {Morrison}, Heather L. and {Mullally}, Fergal and {Munn}, Jeffrey A. and {Murphy}, Tara and {Nash}, Thomas and {Nebot}, Ada and {Neilsen}, Eric H., Jr. and {Newberg}, Heidi Jo and {Newman}, Peter R. and {Nichol}, Robert C. and {Nicinski}, Tom and {Nieto-Santisteban}, Maria and {Nitta}, Atsuko and {Okamura}, Sadanori and {Oravetz}, Daniel J. and {Ostriker}, Jeremiah P. and {Owen}, Russell and {Padmanabhan}, Nikhil and {Pan}, Kaike and {Park}, Changbom and {Pauls}, George and {Peoples}, John, Jr. and {Percival}, Will J. and {Pier}, Jeffrey R. and {Pope}, Adrian C. and {Pourbaix}, Dimitri and {Price}, Paul A. and {Purger}, Norbert and {Quinn}, Thomas and {Raddick}, M. Jordan and {Re Fiorentin}, Paola and {Richards}, Gordon T. and {Richmond}, Michael W. and {Riess}, Adam G. and {Rix}, Hans-Walter and {Rockosi}, Constance M. and {Sako}, Masao and {Schlegel}, David J. and {Schneider}, Donald P. and {Scholz}, Ralf-Dieter and {Schreiber}, Matthias R. and {Schwope}, Axel D. and {Seljak}, Uro{\v{s}} and {Sesar}, Branimir and {Sheldon}, Erin and {Shimasaku}, Kazu and {Sibley}, Valena C. and {Simmons}, A.~E. and {Sivarani}, Thirupathi and {Allyn Smith}, J. and {Smith}, Martin C. and {Smol{\v{c}}i{\'c}}, Vernesa and {Snedden}, Stephanie A. and {Stebbins}, Albert and {Steinmetz}, Matthias and {Stoughton}, Chris and {Strauss}, Michael A. and {SubbaRao}, Mark and {Suto}, Yasushi and {Szalay}, Alexander S. and {Szapudi}, Istv{\'a}n and {Szkody}, Paula and {Tanaka}, Masayuki and {Tegmark}, Max and {Teodoro}, Luis F.~A. and {Thakar}, Aniruddha R. and {Tremonti}, Christy A. and {Tucker}, Douglas L. and {Uomoto}, Alan and {Vanden Berk}, Daniel E. and {Vandenberg}, Jan and {Vidrih}, S. and {Vogeley}, Michael S. and {Voges}, Wolfgang and {Vogt}, Nicole P. and {Wadadekar}, Yogesh and {Watters}, Shannon and {Weinberg}, David H. and {West}, Andrew A. and {White}, Simon D.~M. and {Wilhite}, Brian C. and {Wonders}, Alainna C. and {Yanny}, Brian and {Yocum}, D.~R. and {York}, Donald G. and {Zehavi}, Idit and {Zibetti}, Stefano and {Zucker}, Daniel B.},
        title = "{The Seventh Data Release of the Sloan Digital Sky Survey}",
      journal = {\apjs},
         year = 2009,
        month = jun,
       volume = {182},
       number = {2},
        pages = {543-558},
          doi = {10.1088/0067-0049/182/2/543},
archivePrefix = {arXiv},
       eprint = {0812.0649},
 primaryClass = {astro-ph},
       adsurl = {https://ui.adsabs.harvard.edu/abs/2009ApJS..182..543A}
}

@article{schlegel1998maps,
   title={Maps of Dust Infrared Emission for Use in Estimation of Reddening and Cosmic Microwave Background Radiation Foregrounds},
   volume={500},
   ISSN={1538-4357},
   url={http://dx.doi.org/10.1086/305772},
   DOI={10.1086/305772},
   number={2},
   journal={The Astrophysical Journal},
   publisher={American Astronomical Society},
   author={Schlegel, David J. and Finkbeiner, Douglas P. and Davis, Marc},
   year={1998},
   month=jun, pages={525–553} }

@article{Blanton_2007,
    author = "Blanton, Michael R. and Roweis, Sam",
    title = "{K-corrections and filter transformations in the ultraviolet, optical, and near infrared}",
    eprint = "astro-ph/0606170",
    archivePrefix = "arXiv",
    doi = "10.1086/510127",
    journal = "Astron. J.",
    volume = "133",
    pages = "734--754",
    year = "2007"
}

@article{Thanjavur_2016,
   title={Stellar mass functions of galaxies, discs and spheroids at $z \sim 0.1$},
   volume={459},
   ISSN={1365-2966},
   url={http://dx.doi.org/10.1093/mnras/stw495},
   DOI={10.1093/mnras/stw495},
   number={1},
   journal={Monthly Notices of the Royal Astronomical Society},
   publisher={Oxford University Press (OUP)},
   author={Thanjavur, Karun and Simard, Luc and Bluck, Asa F. L. and Mendel, Trevor},
   year={2016},
   month=mar, pages={44–69} }

@ARTICLE{2016MNRAS.459.2150W,
       author = {{Weigel}, Anna K. and {Schawinski}, Kevin and {Bruderer}, Claudio},
        title = "{Stellar mass functions: methods, systematics and results for the local Universe}",
      journal = {Monthly Notices of the Royal Astronomical Society},
         year = 2016,
        month = jun,
       volume = {459},
       number = {2},
        pages = {2150-2187},
          doi = {10.1093/mnras/stw756},
archivePrefix = {arXiv},
       eprint = {1604.00008},
 primaryClass = {astro-ph.GA},
       adsurl = {https://ui.adsabs.harvard.edu/abs/2016MNRAS.459.2150W}
}

@article{Desjacques_2018,
   title={Large-scale galaxy bias},
   volume={733},
   ISSN={0370-1573},
   url={http://dx.doi.org/10.1016/j.physrep.2017.12.002},
   DOI={10.1016/j.physrep.2017.12.002},
   journal={Physics Reports},
   publisher={Elsevier BV},
   author={Desjacques, Vincent and Jeong, Donghui and Schmidt, Fabian},
   year={2018},
   month=feb, pages={1–193} }

@article{O_Shaughnessy_2010,
   title={BINARY COMPACT OBJECT COALESCENCE RATES: THE ROLE OF ELLIPTICAL GALAXIES},
   volume={716},
   ISSN={1538-4357},
   url={http://dx.doi.org/10.1088/0004-637X/716/1/615},
   DOI={10.1088/0004-637x/716/1/615},
   number={1},
   journal={The Astrophysical Journal},
   publisher={American Astronomical Society},
   author={O’Shaughnessy, R. and Kalogera, V. and Belczynski, Krzysztof},
   year={2010},
   month=may, pages={615–633} }

@article{Dominik_2012,
   title={DOUBLE COMPACT OBJECTS. I. THE SIGNIFICANCE OF THE COMMON ENVELOPE ON MERGER RATES},
   volume={759},
   ISSN={1538-4357},
   url={http://dx.doi.org/10.1088/0004-637X/759/1/52},
   DOI={10.1088/0004-637x/759/1/52},
   number={1},
   journal={The Astrophysical Journal},
   publisher={American Astronomical Society},
   author={Dominik, Michal and Belczynski, Krzysztof and Fryer, Christopher and Holz, Daniel E. and Berti, Emanuele and Bulik, Tomasz and Mandel, Ilya and O’Shaughnessy, Richard},
   year={2012},
   month=oct, pages={52} }

@article{Toffano_2019,
   title={The host galaxies of double compact objects across cosmic time},
   volume={489},
   ISSN={1365-2966},
   url={http://dx.doi.org/10.1093/mnras/stz2415},
   DOI={10.1093/mnras/stz2415},
   number={4},
   journal={Monthly Notices of the Royal Astronomical Society},
   publisher={Oxford University Press (OUP)},
   author={Toffano, Mattia and Mapelli, Michela and Giacobbo, Nicola and Artale, M Celeste and Ghirlanda, Giancarlo},
   year={2019},
   month=sep, pages={4622–4631} }

@article{McCarthy_2020,
   title={Constraining delay time distribution of binary neutron star mergers from host galaxy properties},
   volume={499},
   ISSN={1365-2966},
   url={http://dx.doi.org/10.1093/mnras/staa3206},
   DOI={10.1093/mnras/staa3206},
   number={4},
   journal={Monthly Notices of the Royal Astronomical Society},
   publisher={Oxford University Press (OUP)},
   author={McCarthy, Kevin S and Zheng, Zheng and Ramirez-Ruiz, Enrico},
   year={2020},
   month=oct, pages={5220–5229} }

@article{Mukherjee_2021,
   title={Impact of astrophysical binary coalescence time-scales on the rate of lensed gravitational wave events},
   volume={506},
   ISSN={1365-2966},
   url={http://dx.doi.org/10.1093/mnras/stab1980},
   DOI={10.1093/mnras/stab1980},
   number={3},
   journal={Monthly Notices of the Royal Astronomical Society},
   publisher={Oxford University Press (OUP)},
   author={Mukherjee, Suvodip and Broadhurst, Tom and Diego, Jose M and Silk, Joseph and Smoot, George F},
   year={2021},
   month=jul, pages={3751–3759} }

@article{Belczynski_2002,
   title={A Comprehensive Study of Binary Compact Objects as Gravitational Wave Sources: Evolutionary Channels, Rates, and Physical Properties},
   volume={572},
   ISSN={1538-4357},
   url={http://dx.doi.org/10.1086/340304},
   DOI={10.1086/340304},
   number={1},
   journal={The Astrophysical Journal},
   publisher={American Astronomical Society},
   author={Belczynski, Krzysztof and Kalogera, Vassiliki and Bulik, Tomasz},
   year={2002},
   month=jun, pages={407–431} }

@article{Dominik_2015,
   title={DOUBLE COMPACT OBJECTS. III. GRAVITATIONAL-WAVE DETECTION RATES},
   volume={806},
   ISSN={1538-4357},
   url={http://dx.doi.org/10.1088/0004-637X/806/2/263},
   DOI={10.1088/0004-637x/806/2/263},
   number={2},
   journal={The Astrophysical Journal},
   publisher={American Astronomical Society},
   author={Dominik, Michal and Berti, Emanuele and O’Shaughnessy, Richard and Mandel, Ilya and Belczynski, Krzysztof and Fryer, Christopher and Holz, Daniel E. and Bulik, Tomasz and Pannarale, Francesco},
   year={2015},
   month=jun, pages={263} }

@article{Mapelli_2017,
   title={The cosmic merger rate of stellar black hole binaries from the Illustris simulation},
   volume={472},
   ISSN={1365-2966},
   url={http://dx.doi.org/10.1093/mnras/stx2123},
   DOI={10.1093/mnras/stx2123},
   number={2},
   journal={Monthly Notices of the Royal Astronomical Society},
   publisher={Oxford University Press (OUP)},
   author={Mapelli, Michela and Giacobbo, Nicola and Ripamonti, Emanuele and Spera, Mario},
   year={2017},
   month=aug, pages={2422–2435} }

@ARTICLE{2018MNRAS.474.2959G,
       author = {{Giacobbo}, Nicola and {Mapelli}, Michela and {Spera}, Mario},
        title = "{Merging black hole binaries: the effects of progenitor's metallicity, mass-loss rate and Eddington factor}",
      journal = {Monthly Notices of the Royal Astronomical Society},
         year = 2018,
        month = mar,
       volume = {474},
       number = {3},
        pages = {2959-2974},
          doi = {10.1093/mnras/stx2933},
archivePrefix = {arXiv},
       eprint = {1711.03556},
 primaryClass = {astro-ph.SR},
       adsurl = {https://ui.adsabs.harvard.edu/abs/2018MNRAS.474.2959G}
}

@article{Fishbach_2018,
   title={Does the Black Hole Merger Rate Evolve with Redshift?},
   volume={863},
   ISSN={2041-8213},
   url={http://dx.doi.org/10.3847/2041-8213/aad800},
   DOI={10.3847/2041-8213/aad800},
   number={2},
   journal={The Astrophysical Journal Letters},
   publisher={American Astronomical Society},
   author={Fishbach, Maya and Holz, Daniel E. and Farr, Will M.},
   year={2018},
   month=aug, pages={L41} }

@article{Santoliquido_2022,
   title={Modelling the host galaxies of binary compact object mergers with observational scaling relations},
   volume={516},
   ISSN={1365-2966},
   url={http://dx.doi.org/10.1093/mnras/stac2384},
   DOI={10.1093/mnras/stac2384},
   number={3},
   journal={Monthly Notices of the Royal Astronomical Society},
   publisher={Oxford University Press (OUP)},
   author={Santoliquido, Filippo and Mapelli, Michela and Artale, M Celeste and Boco, Lumen},
   year={2022},
   month=aug, pages={3297–3317} }

@ARTICLE{2020A&A...641A...6P,
       author = {{Planck Collaboration} and {Aghanim}, N. and {Akrami}, Y. and {Ashdown}, M. and {Aumont}, J. and {Baccigalupi}, C. and {Ballardini}, M. and {Banday}, A.~J. and {Barreiro}, R.~B. and {Bartolo}, N. and {Basak}, S. and {Battye}, R. and {Benabed}, K. and {Bernard}, J. -P. and {Bersanelli}, M. and {Bielewicz}, P. and {Bock}, J.~J. and {Bond}, J.~R. and {Borrill}, J. and {Bouchet}, F.~R. and {Boulanger}, F. and {Bucher}, M. and {Burigana}, C. and {Butler}, R.~C. and {Calabrese}, E. and {Cardoso}, J. -F. and {Carron}, J. and {Challinor}, A. and {Chiang}, H.~C. and {Chluba}, J. and {Colombo}, L.~P.~L. and {Combet}, C. and {Contreras}, D. and {Crill}, B.~P. and {Cuttaia}, F. and {de Bernardis}, P. and {de Zotti}, G. and {Delabrouille}, J. and {Delouis}, J. -M. and {Di Valentino}, E. and {Diego}, J.~M. and {Dor{\'e}}, O. and {Douspis}, M. and {Ducout}, A. and {Dupac}, X. and {Dusini}, S. and {Efstathiou}, G. and {Elsner}, F. and {En{\ss}lin}, T.~A. and {Eriksen}, H.~K. and {Fantaye}, Y. and {Farhang}, M. and {Fergusson}, J. and {Fernandez-Cobos}, R. and {Finelli}, F. and {Forastieri}, F. and {Frailis}, M. and {Fraisse}, A.~A. and {Franceschi}, E. and {Frolov}, A. and {Galeotta}, S. and {Galli}, S. and {Ganga}, K. and {G{\'e}nova-Santos}, R.~T. and {Gerbino}, M. and {Ghosh}, T. and {Gonz{\'a}lez-Nuevo}, J. and {G{\'o}rski}, K.~M. and {Gratton}, S. and {Gruppuso}, A. and {Gudmundsson}, J.~E. and {Hamann}, J. and {Handley}, W. and {Hansen}, F.~K. and {Herranz}, D. and {Hildebrandt}, S.~R. and {Hivon}, E. and {Huang}, Z. and {Jaffe}, A.~H. and {Jones}, W.~C. and {Karakci}, A. and {Keih{\"a}nen}, E. and {Keskitalo}, R. and {Kiiveri}, K. and {Kim}, J. and {Kisner}, T.~S. and {Knox}, L. and {Krachmalnicoff}, N. and {Kunz}, M. and {Kurki-Suonio}, H. and {Lagache}, G. and {Lamarre}, J. -M. and {Lasenby}, A. and {Lattanzi}, M. and {Lawrence}, C.~R. and {Le Jeune}, M. and {Lemos}, P. and {Lesgourgues}, J. and {Levrier}, F. and {Lewis}, A. and {Liguori}, M. and {Lilje}, P.~B. and {Lilley}, M. and {Lindholm}, V. and {L{\'o}pez-Caniego}, M. and {Lubin}, P.~M. and {Ma}, Y. -Z. and {Mac{\'\i}as-P{\'e}rez}, J.~F. and {Maggio}, G. and {Maino}, D. and {Mandolesi}, N. and {Mangilli}, A. and {Marcos-Caballero}, A. and {Maris}, M. and {Martin}, P.~G. and {Martinelli}, M. and {Mart{\'\i}nez-Gonz{\'a}lez}, E. and {Matarrese}, S. and {Mauri}, N. and {McEwen}, J.~D. and {Meinhold}, P.~R. and {Melchiorri}, A. and {Mennella}, A. and {Migliaccio}, M. and {Millea}, M. and {Mitra}, S. and {Miville-Desch{\^e}nes}, M. -A. and {Molinari}, D. and {Montier}, L. and {Morgante}, G. and {Moss}, A. and {Natoli}, P. and {N{\o}rgaard-Nielsen}, H.~U. and {Pagano}, L. and {Paoletti}, D. and {Partridge}, B. and {Patanchon}, G. and {Peiris}, H.~V. and {Perrotta}, F. and {Pettorino}, V. and {Piacentini}, F. and {Polastri}, L. and {Polenta}, G. and {Puget}, J. -L. and {Rachen}, J.~P. and {Reinecke}, M. and {Remazeilles}, M. and {Renzi}, A. and {Rocha}, G. and {Rosset}, C. and {Roudier}, G. and {Rubi{\~n}o-Mart{\'\i}n}, J.~A. and {Ruiz-Granados}, B. and {Salvati}, L. and {Sandri}, M. and {Savelainen}, M. and {Scott}, D. and {Shellard}, E.~P.~S. and {Sirignano}, C. and {Sirri}, G. and {Spencer}, L.~D. and {Sunyaev}, R. and {Suur-Uski}, A. -S. and {Tauber}, J.~A. and {Tavagnacco}, D. and {Tenti}, M. and {Toffolatti}, L. and {Tomasi}, M. and {Trombetti}, T. and {Valenziano}, L. and {Valiviita}, J. and {Van Tent}, B. and {Vibert}, L. and {Vielva}, P. and {Villa}, F. and {Vittorio}, N. and {Wandelt}, B.~D. and {Wehus}, I.~K. and {White}, M. and {White}, S.~D.~M. and {Zacchei}, A. and {Zonca}, A.},
        title = "{Planck 2018 results. VI. Cosmological parameters}",
      journal = {\aap},
         year = 2020,
        month = sep,
       volume = {641},
          eid = {A6},
        pages = {A6},
          doi = {10.1051/0004-6361/201833910},
archivePrefix = {arXiv},
       eprint = {1807.06209},
 primaryClass = {astro-ph.CO},
       adsurl = {https://ui.adsabs.harvard.edu/abs/2020A&A...641A...6P}
}

@ARTICLE{1967ApJ...148..803R,
       author = {{Rakavy}, G. and {Shaviv}, G.},
        title = "{Instabilities in Highly Evolved Stellar Models}",
      journal = {\apj},
         year = 1967,
        month = jun,
       volume = {148},
        pages = {803},
          doi = {10.1086/149204},
       adsurl = {https://ui.adsabs.harvard.edu/abs/1967ApJ...148..803R}
}

@article{2020,
   title={Planck2018 results: VI. Cosmological parameters},
   volume={641},
   ISSN={1432-0746},
   url={http://dx.doi.org/10.1051/0004-6361/201833910},
   DOI={10.1051/0004-6361/201833910},
   journal={Astronomy and Astrophysics},
   publisher={EDP Sciences},
   author={Aghanim, N. and Akrami, Y. and Ashdown, M. and Aumont, J. and Baccigalupi, C. and Ballardini, M. and Banday, A. J. and Barreiro, R. B. and Bartolo, N. and Basak, S. and Battye, R. and Benabed, K. and Bernard, J.-P. and Bersanelli, M. and Bielewicz, P. and Bock, J. J. and Bond, J. R. and Borrill, J. and Bouchet, F. R. and Boulanger, F. and Bucher, M. and Burigana, C. and Butler, R. C. and Calabrese, E. and Cardoso, J.-F. and Carron, J. and Challinor, A. and Chiang, H. C. and Chluba, J. and Colombo, L. P. L. and Combet, C. and Contreras, D. and Crill, B. P. and Cuttaia, F. and de Bernardis, P. and de Zotti, G. and Delabrouille, J. and Delouis, J.-M. and Di Valentino, E. and Diego, J. M. and Doré, O. and Douspis, M. and Ducout, A. and Dupac, X. and Dusini, S. and Efstathiou, G. and Elsner, F. and Enßlin, T. A. and Eriksen, H. K. and Fantaye, Y. and Farhang, M. and Fergusson, J. and Fernandez-Cobos, R. and Finelli, F. and Forastieri, F. and Frailis, M. and Fraisse, A. A. and Franceschi, E. and Frolov, A. and Galeotta, S. and Galli, S. and Ganga, K. and Génova-Santos, R. T. and Gerbino, M. and Ghosh, T. and González-Nuevo, J. and Górski, K. M. and Gratton, S. and Gruppuso, A. and Gudmundsson, J. E. and Hamann, J. and Handley, W. and Hansen, F. K. and Herranz, D. and Hildebrandt, S. R. and Hivon, E. and Huang, Z. and Jaffe, A. H. and Jones, W. C. and Karakci, A. and Keihänen, E. and Keskitalo, R. and Kiiveri, K. and Kim, J. and Kisner, T. S. and Knox, L. and Krachmalnicoff, N. and Kunz, M. and Kurki-Suonio, H. and Lagache, G. and Lamarre, J.-M. and Lasenby, A. and Lattanzi, M. and Lawrence, C. R. and Le Jeune, M. and Lemos, P. and Lesgourgues, J. and Levrier, F. and Lewis, A. and Liguori, M. and Lilje, P. B. and Lilley, M. and Lindholm, V. and López-Caniego, M. and Lubin, P. M. and Ma, Y.-Z. and Macías-Pérez, J. F. and Maggio, G. and Maino, D. and Mandolesi, N. and Mangilli, A. and Marcos-Caballero, A. and Maris, M. and Martin, P. G. and Martinelli, M. and Martínez-González, E. and Matarrese, S. and Mauri, N. and McEwen, J. D. and Meinhold, P. R. and Melchiorri, A. and Mennella, A. and Migliaccio, M. and Millea, M. and Mitra, S. and Miville-Deschênes, M.-A. and Molinari, D. and Montier, L. and Morgante, G. and Moss, A. and Natoli, P. and Nørgaard-Nielsen, H. U. and Pagano, L. and Paoletti, D. and Partridge, B. and Patanchon, G. and Peiris, H. V. and Perrotta, F. and Pettorino, V. and Piacentini, F. and Polastri, L. and Polenta, G. and Puget, J.-L. and Rachen, J. P. and Reinecke, M. and Remazeilles, M. and Renzi, A. and Rocha, G. and Rosset, C. and Roudier, G. and Rubiño-Martín, J. A. and Ruiz-Granados, B. and Salvati, L. and Sandri, M. and Savelainen, M. and Scott, D. and Shellard, E. P. S. and Sirignano, C. and Sirri, G. and Spencer, L. D. and Sunyaev, R. and Suur-Uski, A.-S. and Tauber, J. A. and Tavagnacco, D. and Tenti, M. and Toffolatti, L. and Tomasi, M. and Trombetti, T. and Valenziano, L. and Valiviita, J. and Van Tent, B. and Vibert, L. and Vielva, P. and Villa, F. and Vittorio, N. and Wandelt, B. D. and Wehus, I. K. and White, M. and White, S. D. M. and Zacchei, A. and Zonca, A.},
   year={2020},
   month=sep, pages={A6} }

@article{percival2013largescalestructureobservations,
    author = "Percival, Will J.",
    title = "{Large Scale Structure Observations}",
    eprint = "1312.5490",
    archivePrefix = "arXiv",
    primaryClass = "astro-ph.CO",
    year = 2013
}

@article{Artale_2019b,
   title={Host galaxies of merging compact objects: mass, star formation rate, metallicity, and colours},
   volume={487},
   ISSN={1365-2966},
   url={http://dx.doi.org/10.1093/mnras/stz1382},
   DOI={10.1093/mnras/stz1382},
   number={2},
   journal={Monthly Notices of the Royal Astronomical Society},
   publisher={Oxford University Press (OUP)},
   author={Artale, M Celeste and Mapelli, Michela and Giacobbo, Nicola and Sabha, Nadeen B and Spera, Mario and Santoliquido, Filippo and Bressan, Alessandro},
   year={2019},
   month=may, pages={1675–1688} }

@article{Santoliquido_2021,
   title={The cosmic merger rate density of compact objects: impact of star formation, metallicity, initial mass function, and binary evolution},
   volume={502},
   ISSN={1365-2966},
   url={http://dx.doi.org/10.1093/mnras/stab280},
   DOI={10.1093/mnras/stab280},
   number={4},
   journal={Monthly Notices of the Royal Astronomical Society},
   publisher={Oxford University Press (OUP)},
   author={Santoliquido, Filippo and Mapelli, Michela and Giacobbo, Nicola and Bouffanais, Yann and Artale, M Celeste},
   year={2021},
   month=feb, pages={4877–4889} }

@article{Artale_2019,
   title={Mass and star formation rate of the host galaxies of compact binary mergers across cosmic time},
   volume={491},
   ISSN={1365-2966},
   url={http://dx.doi.org/10.1093/mnras/stz3190},
   DOI={10.1093/mnras/stz3190},
   number={3},
   journal={Monthly Notices of the Royal Astronomical Society},
   publisher={Oxford University Press (OUP)},
   author={Artale, M Celeste and Mapelli, Michela and Bouffanais, Yann and Giacobbo, Nicola and Pasquato, Mario and Spera, Mario},
   year={2019},
   month=nov, pages={3419–3434} }

@article{Fishbach_2021,
   title={The Time Delay Distribution and Formation Metallicity of LIGO-Virgo’s Binary Black Holes},
   volume={914},
   ISSN={2041-8213},
   url={http://dx.doi.org/10.3847/2041-8213/ac05c4},
   DOI={10.3847/2041-8213/ac05c4},
   number={2},
   journal={The Astrophysical Journal Letters},
   publisher={American Astronomical Society},
   author={Fishbach, Maya and Kalogera, Vicky},
   year={2021},
   month=jun, pages={L30} }

@article{Chruslinska_2018,
   title={The influence of the distribution of cosmic star formation at different metallicities on the properties of merging double compact objects},
   volume={482},
   ISSN={1365-2966},
   url={http://dx.doi.org/10.1093/mnras/sty3087},
   DOI={10.1093/mnras/sty3087},
   number={4},
   journal={Monthly Notices of the Royal Astronomical Society},
   publisher={Oxford University Press (OUP)},
   author={Chruslinska, Martyna and Nelemans, Gijs and Belczynski, Krzysztof},
   year={2018},
   month=nov, pages={5012–5017} }

@article{Rauf_2023,
   title={Exploring binary black hole mergers and host galaxies with<scp>shark</scp>and COMPAS},
   volume={523},
   ISSN={1365-2966},
   url={http://dx.doi.org/10.1093/mnras/stad1757},
   DOI={10.1093/mnras/stad1757},
   number={4},
   journal={Monthly Notices of the Royal Astronomical Society},
   publisher={Oxford University Press (OUP)},
   author={Rauf, Liana and Howlett, Cullan and Davis, Tamara M and Lagos, Claudia D P},
   year={2023},
   month=jun, pages={5719–5737} }

@article{Naab_2017,
   title={Theoretical Challenges in Galaxy Formation},
   volume={55},
   ISSN={1545-4282},
   url={http://dx.doi.org/10.1146/annurev-astro-081913-040019},
   DOI={10.1146/annurev-astro-081913-040019},
   number={1},
   journal={Annual Review of Astronomy and Astrophysics},
   publisher={Annual Reviews},
   author={Naab, Thorsten and Ostriker, Jeremiah P.},
   year={2017},
   month=aug, pages={59–109} }

@article{Nagamine_2021,
   title={Feedback models in galaxy simulations and probing their impact by cosmological hydrodynamic simulations},
   volume={17},
   ISSN={1743-9221},
   url={http://dx.doi.org/10.1017/S1743921323000133},
   DOI={10.1017/s1743921323000133},
   number={S373},
   journal={Proceedings of the International Astronomical Union},
   publisher={Cambridge University Press (CUP)},
   author={Nagamine, Kentaro},
   year={2021},
   month=aug, pages={283–292} }

@ARTICLE{BeutlerMcDonald20,
       author = {{Beutler}, Florian and {McDonald}, Patrick},
        title = "{Unified galaxy power spectrum measurements from 6dFGS, BOSS, and eBOSS}",
      journal = {JCAP},
         year = 2021,
        month = nov,
       volume = {2021},
       number = {11},
          eid = {031},
        pages = {031},
          doi = {10.1088/1475-7516/2021/11/031},
archivePrefix = {arXiv},
       eprint = {2106.06324},
 primaryClass = {astro-ph.CO},
       adsurl = {https://ui.adsabs.harvard.edu/abs/2021JCAP...11..031B}
}

@article{Jackson1972,
    author = "Jackson, J. C.",
    title = "{Fingers of God: A critique of Rees' theory of primoridal gravitational radiation}",
    eprint = "0810.3908",
    archivePrefix = "arXiv",
    primaryClass = "astro-ph",
    doi = "10.1093/mnras/156.1.1P",
    journal = "Monthly Notices of the Royal Astronomical Society",
    volume = "156",
    pages = "1P--5P",
    year = "1972"
}

@ARTICLE{Sefusatti15,
       author = {{Sefusatti}, E. and {Crocce}, M. and {Scoccimarro}, R. and {Couchman}, H.~M.~P.},
        title = "{Accurate estimators of correlation functions in Fourier space}",
      journal = {Monthly Notices of the Royal Astronomical Society},
         year = 2016,
        month = aug,
       volume = {460},
       number = {4},
        pages = {3624-3636},
          doi = {10.1093/mnras/stw1229},
archivePrefix = {arXiv},
       eprint = {1512.07295},
 primaryClass = {astro-ph.CO},
       adsurl = {https://ui.adsabs.harvard.edu/abs/2016MNRAS.460.3624S}
}

@ARTICLE{Jing05,
       author = {{Jing}, Y.~P.},
        title = "{Correcting for the Alias Effect When Measuring the Power Spectrum Using a Fast Fourier Transform}",
      journal = {\apj},
         year = 2005,
        month = feb,
       volume = {620},
       number = {2},
        pages = {559-563},
          doi = {10.1086/427087},
archivePrefix = {arXiv},
       eprint = {astro-ph/0409240},
 primaryClass = {astro-ph},
       adsurl = {https://ui.adsabs.harvard.edu/abs/2005ApJ...620..559J}
}

@ARTICLE{Hand2018,
       author = {{Hand}, Nick and {Feng}, Yu and {Beutler}, Florian and {Li}, Yin and {Modi}, Chirag and {Seljak}, Uro{\v{s}} and {Slepian}, Zachary},
        title = "{nbodykit: An Open-source, Massively Parallel Toolkit for Large-scale Structure}",
      journal = {\aj},
         year = 2018,
        month = oct,
       volume = {156},
       number = {4},
          eid = {160},
        pages = {160},
          doi = {10.3847/1538-3881/aadae0},
archivePrefix = {arXiv},
       eprint = {1712.05834},
 primaryClass = {astro-ph.IM},
       adsurl = {https://ui.adsabs.harvard.edu/abs/2018AJ....156..160H}
}

@ARTICLE{Hand2017,
       author = {{Hand}, Nick and {Li}, Yin and {Slepian}, Zachary and {Seljak}, Uro{\v{s}}},
        title = "{An optimal FFT-based anisotropic power spectrum estimator}",
      journal = {JCAP},
         year = 2017,
        month = jul,
       volume = {2017},
       number = {7},
          eid = {002},
        pages = {002},
          doi = {10.1088/1475-7516/2017/07/002},
archivePrefix = {arXiv},
       eprint = {1704.02357},
 primaryClass = {astro-ph.CO},
       adsurl = {https://ui.adsabs.harvard.edu/abs/2017JCAP...07..002H}
}

@unpublished{Alves24,
  author = "Alves, Otavio and {DESI Collaboration}",
  title  = "Analytical covariance matrices of DESI galaxy power spectrum multipoles",
  note   = "(in prep.)",
  year   = "2024"
}

@article{Peacock:2000qk,
    author = "Peacock, J. A. and Smith, R. E.",
    title = "{Halo occupation numbers and galaxy bias}",
    eprint = "astro-ph/0005010",
    archivePrefix = "arXiv",
    doi = "10.1046/j.1365-8711.2000.03779.x",
    journal = "Monthly Notices of the Royal Astronomical Society",
    volume = "318",
    pages = "1144",
    year = "2000"
}

@ARTICLE{ValeOstriker04,
       author = {{Vale}, A. and {Ostriker}, J.~P.},
        title = "{Linking halo mass to galaxy luminosity}",
      journal = {Monthly Notices of the Royal Astronomical Society},
         year = 2004,
        month = sep,
       volume = {353},
       number = {1},
        pages = {189-200},
          doi = {10.1111/j.1365-2966.2004.08059.x},
archivePrefix = {arXiv},
       eprint = {astro-ph/0402500},
 primaryClass = {astro-ph},
       adsurl = {https://ui.adsabs.harvard.edu/abs/2004MNRAS.353..189V}
}

@ARTICLE{Cole2000,
       author = {{Cole}, Shaun and {Lacey}, Cedric G. and {Baugh}, Carlton M. and {Frenk}, Carlos S.},
        title = "{Hierarchical galaxy formation}",
      journal = {Monthly Notices of the Royal Astronomical Society},
         year = 2000,
        month = nov,
       volume = {319},
       number = {1},
        pages = {168-204},
          doi = {10.1046/j.1365-8711.2000.03879.x},
archivePrefix = {arXiv},
       eprint = {astro-ph/0007281},
 primaryClass = {astro-ph},
       adsurl = {https://ui.adsabs.harvard.edu/abs/2000MNRAS.319..168C}
}

@ARTICLE{Dalal08,
       author = {{Dalal}, Neal and {White}, Martin and {Bond}, J. Richard and {Shirokov}, Alexander},
        title = "{Halo Assembly Bias in Hierarchical Structure Formation}",
      journal = {\apj},
         year = 2008,
        month = nov,
       volume = {687},
       number = {1},
        pages = {12-21},
          doi = {10.1086/591512},
archivePrefix = {arXiv},
       eprint = {0803.3453},
 primaryClass = {astro-ph},
       adsurl = {https://ui.adsabs.harvard.edu/abs/2008ApJ...687...12D}
}

@ARTICLE{Behroozi_2013,
       author = {{Behroozi}, Peter S. and {Wechsler}, Risa H. and {Conroy}, Charlie},
        title = "{The Average Star Formation Histories of Galaxies in Dark Matter Halos from z = 0-8}",
      journal = {\apj},
         year = 2013,
        month = jun,
       volume = {770},
       number = {1},
          eid = {57},
        pages = {57},
          doi = {10.1088/0004-637X/770/1/57},
archivePrefix = {arXiv},
       eprint = {1207.6105},
 primaryClass = {astro-ph.CO},
       adsurl = {https://ui.adsabs.harvard.edu/abs/2013ApJ...770...57B}
}

@ARTICLE{Behroozi2019,
       author = {{Behroozi}, Peter and {Wechsler}, Risa H. and {Hearin}, Andrew P. and {Conroy}, Charlie},
        title = "{UNIVERSEMACHINE: The correlation between galaxy growth and dark matter halo assembly from z = 0-10}",
      journal = {Monthly Notices of the Royal Astronomical Society},
         year = 2019,
        month = sep,
       volume = {488},
       number = {3},
        pages = {3143-3194},
          doi = {10.1093/mnras/stz1182},
archivePrefix = {arXiv},
       eprint = {1806.07893},
 primaryClass = {astro-ph.GA},
       adsurl = {https://ui.adsabs.harvard.edu/abs/2019MNRAS.488.3143B}
}

@article{Croton:2006ys,
    author = "Croton, Darren J. and Gao, Liang and White, Simon D. M.",
    title = "{Halo assembly bias and its effects on galaxy clustering}",
    eprint = "astro-ph/0605636",
    archivePrefix = "arXiv",
    doi = "10.1111/j.1365-2966.2006.11230.x",
    journal = "Monthly Notices of the Royal Astronomical Society",
    volume = "374",
    pages = "1303--1309",
    year = "2007"
}

@article{Sheth:1999mn,
    author = "Sheth, Ravi K. and Tormen, Giuseppe",
    title = "{Large scale bias and the peak background split}",
    eprint = "astro-ph/9901122",
    archivePrefix = "arXiv",
    doi = "10.1046/j.1365-8711.1999.02692.x",
    journal = "Monthly Notices of the Royal Astronomical Society",
    volume = "308",
    pages = "119",
    year = "1999"
}

@article{Wechsler:2018pic,
    author = "Wechsler, Risa H. and Tinker, Jeremy L.",
    title = "{The Connection between Galaxies and their Dark Matter Halos}",
    eprint = "1804.03097",
    archivePrefix = "arXiv",
    primaryClass = "astro-ph.GA",
    doi = "10.1146/annurev-astro-081817-051756",
    journal = "Ann. Rev. Astron. Astrophys.",
    volume = "56",
    pages = "435--487",
    year = "2018"
}

@article{Gao:2006qz,
    author = "Gao, Liang and White, Simon D. M.",
    title = "{Assembly bias in the clustering of dark matter haloes}",
    eprint = "astro-ph/0611921",
    archivePrefix = "arXiv",
    doi = "10.1111/j.1745-3933.2007.00292.x",
    journal = "Monthly Notices of the Royal Astronomical Society",
    volume = "377",
    pages = "L5--L9",
    year = "2007"
}

@article{Benson:1999mva,
    author = "Benson, A. J. and Cole, S. and Frenk, C. S. and Baugh, C. M. and Lacey, Cedric G.",
    title = "{The Nature of galaxy bias and clustering}",
    eprint = "astro-ph/9903343",
    archivePrefix = "arXiv",
    doi = "10.1046/j.1365-8711.2000.03101.x",
    journal = "Monthly Notices of the Royal Astronomical Society",
    volume = "311",
    pages = "793--808",
    year = "2000"
}

@article{LIGOScientific:2018jsj,
    author = "Abbott, B. P. and others",
    collaboration = "LIGO Scientific, Virgo",
    title = "{Binary Black Hole Population Properties Inferred from the First and Second Observing Runs of Advanced LIGO and Advanced Virgo}",
    eprint = "1811.12940",
    archivePrefix = "arXiv",
    primaryClass = "astro-ph.HE",
    reportNumber = "LIGO-P1800324",
    doi = "10.3847/2041-8213/ab3800",
    journal = "Astrophys. J. Lett.",
    volume = "882",
    number = "2",
    pages = "L24",
    year = "2019"
}

@article{Calore:2020bpd,
    author = "Calore, Francesca and Cuoco, Alessandro and Regimbau, Tania and Sachdev, Surabhi and Serpico, Pasquale Dario",
    title = "{Cross-correlating galaxy catalogs and gravitational waves: a tomographic approach}",
    eprint = "2002.02466",
    archivePrefix = "arXiv",
    primaryClass = "astro-ph.CO",
    reportNumber = "LAPTH-001/20",
    doi = "10.1103/PhysRevResearch.2.023314",
    journal = "Phys. Rev. Res.",
    volume = "2",
    pages = "023314",
    year = "2020"
}

@article{Mukherjee:2019oma,
    author = "Mukherjee, Suvodip and Silk, Joseph",
    title = "{Time-dependence of the astrophysical stochastic gravitational wave background}",
    eprint = "1912.07657",
    archivePrefix = "arXiv",
    primaryClass = "gr-qc",
    doi = "10.1093/mnras/stz3226",
    journal = "Monthly Notices of the Royal Astronomical Society",
    volume = "491",
    number = "4",
    pages = "4690--4701",
    year = "2020"
}

@article{Diaz:2021pem,
    author = "Diaz, Cristina Cigarran and Mukherjee, Suvodip",
    title = "{Mapping the cosmic expansion history from LIGO-Virgo-KAGRA in synergy with DESI and SPHEREx}",
    eprint = "2107.12787",
    archivePrefix = "arXiv",
    primaryClass = "astro-ph.CO",
    doi = "10.1093/mnras/stac208",
    journal = "Monthly Notices of the Royal Astronomical Society",
    volume = "511",
    number = "2",
    pages = "2782--2795",
    year = "2022"
}

@article{Ross_2015,
	doi = {10.1093/mnras/stv154},
  
	url = {https://doi.org/10.1093\%2Fmnras\%2Fstv154},
  
	year = 2015,
	month = {mar},
  
	publisher = {Oxford University Press ({OUP})},
  
	volume = {449},
  
	number = {1},
  
	pages = {835--847},
  
	author = {Ashley J. Ross and Lado Samushia and Cullan Howlett and Will J. Percival and Angela Burden and Marc Manera},
  
	title = {The clustering of the {SDSS} {DR}7 main Galaxy sample {\textendash} I. A 4~per cent distance measure at z~=~0.15},
  
	journal = {Monthly Notices of the Royal Astronomical Society}
}

@article{abbott2020gw190425,
    author = "Abbott, B. P. and others",
    collaboration = "LIGO Scientific, Virgo",
    title = "{GW190425: Observation of a Compact Binary Coalescence with Total Mass $\sim 3.4 M_{\odot}$}",
    eprint = "2001.01761",
    archivePrefix = "arXiv",
    primaryClass = "astro-ph.HE",
    reportNumber = "LIGO-P190425",
    doi = "10.3847/2041-8213/ab75f5",
    journal = "Astrophys. J. Lett.",
    volume = "892",
    number = "1",
    pages = "L3",
    year = "2020"
}

@article{LIGOScientific:2018mvr,
    author = "Abbott, B. P. and others",
    collaboration = "LIGO Scientific, Virgo",
    title = "{GWTC-1: A Gravitational-Wave Transient Catalog of Compact Binary Mergers Observed by LIGO and Virgo during the First and Second Observing Runs}",
    eprint = "1811.12907",
    archivePrefix = "arXiv",
    primaryClass = "astro-ph.HE",
    reportNumber = "LIGO-P1800307",
    doi = "10.1103/PhysRevX.9.031040",
    journal = "Phys. Rev. X",
    volume = "9",
    number = "3",
    pages = "031040",
    year = "2019"
}

@article{LIGOScientific:2020ibl,
    author = "Abbott, R. and others",
    collaboration = "LIGO Scientific, Virgo",
    title = "{GWTC-2: Compact Binary Coalescences Observed by LIGO and Virgo During the First Half of the Third Observing Run}",
    eprint = "2010.14527",
    archivePrefix = "arXiv",
    primaryClass = "gr-qc",
    reportNumber = "P2000061",
    doi = "10.1103/PhysRevX.11.021053",
    journal = "Phys. Rev. X",
    volume = "11",
    pages = "021053",
    year = "2021"
}

@article{LIGOScientific:2021djp,
    author = "Abbott, R. and others",
    collaboration = "LIGO Scientific, VIRGO, KAGRA",
    title = "{GWTC-3: Compact Binary Coalescences Observed by LIGO and Virgo During the Second Part of the Third Observing Run}",
    eprint = "2111.03606",
    archivePrefix = "arXiv",
    primaryClass = "gr-qc",
    reportNumber = "LIGO-P2000318",
    month = "11",
    year = "2021"
}

@article{Artale_2020,
   title={An astrophysically motivated ranking criterion for low-latency electromagnetic follow-up of gravitational wave events},
   volume={495},
   ISSN={1365-2966},
   url={http://dx.doi.org/10.1093/mnras/staa1252},
   DOI={10.1093/mnras/staa1252},
   number={2},
   journal={Monthly Notices of the Royal Astronomical Society},
   publisher={Oxford University Press (OUP)},
   author={Artale, M Celeste and Bouffanais, Yann and Mapelli, Michela and Giacobbo, Nicola and Sabha, Nadeen B and Santoliquido, Filippo and Pasquato, Mario and Spera, Mario},
   year={2020},
   month=may, pages={1841–1852} }

@article{astropy:2013,
Adsurl = {http://adsabs.harvard.edu/abs/2013A%26A...558A..33A},
Archiveprefix = {arXiv},
Author = {{Astropy Collaboration} and {Robitaille}, T.~P. and {Tollerud}, E.~J. and {Greenfield}, P. and {Droettboom}, M. and {Bray}, E. and {Aldcroft}, T. and {Davis}, M. and {Ginsburg}, A. and {Price-Whelan}, A.~M. and {Kerzendorf}, W.~E. and {Conley}, A. and {Crighton}, N. and {Barbary}, K. and {Muna}, D. and {Ferguson}, H. and {Grollier}, F. and {Parikh}, M.~M. and {Nair}, P.~H. and {Unther}, H.~M. and {Deil}, C. and {Woillez}, J. and {Conseil}, S. and {Kramer}, R. and {Turner}, J.~E.~H. and {Singer}, L. and {Fox}, R. and {Weaver}, B.~A. and {Zabalza}, V. and {Edwards}, Z.~I. and {Azalee Bostroem}, K. and {Burke}, D.~J. and {Casey}, A.~R. and {Crawford}, S.~M. and {Dencheva}, N. and {Ely}, J. and {Jenness}, T. and {Labrie}, K. and {Lim}, P.~L. and {Pierfederici}, F. and {Pontzen}, A. and {Ptak}, A. and {Refsdal}, B. and {Servillat}, M. and {Streicher}, O.},
Doi = {10.1051/0004-6361/201322068},
Eid = {A33},
Eprint = {1307.6212},
Journal = {\aap},
Month = oct,
Pages = {A33},
Primaryclass = {astro-ph.IM},
Title = {{Astropy: A community Python package for astronomy}},
Volume = 558,
Year = 2013}

@ARTICLE{astropy:2018,
       author = {{Astropy Collaboration} and {Price-Whelan}, A.~M. and
         {Sip{\H{o}}cz}, B.~M. and {G{\"u}nther}, H.~M. and {Lim}, P.~L. and
         {Crawford}, S.~M. and {Conseil}, S. and {Shupe}, D.~L. and
         {Craig}, M.~W. and {Dencheva}, N. and {Ginsburg}, A. and {Vand
        erPlas}, J.~T. and {Bradley}, L.~D. and {P{\'e}rez-Su{\'a}rez}, D. and
         {de Val-Borro}, M. and {Aldcroft}, T.~L. and {Cruz}, K.~L. and
         {Robitaille}, T.~P. and {Tollerud}, E.~J. and {Ardelean}, C. and
         {Babej}, T. and {Bach}, Y.~P. and {Bachetti}, M. and {Bakanov}, A.~V. and
         {Bamford}, S.~P. and {Barentsen}, G. and {Barmby}, P. and
         {Baumbach}, A. and {Berry}, K.~L. and {Biscani}, F. and {Boquien}, M. and
         {Bostroem}, K.~A. and {Bouma}, L.~G. and {Brammer}, G.~B. and
         {Bray}, E.~M. and {Breytenbach}, H. and {Buddelmeijer}, H. and
         {Burke}, D.~J. and {Calderone}, G. and {Cano Rodr{\'\i}guez}, J.~L. and
         {Cara}, M. and {Cardoso}, J.~V.~M. and {Cheedella}, S. and {Copin}, Y. and
         {Corrales}, L. and {Crichton}, D. and {D'Avella}, D. and {Deil}, C. and
         {Depagne}, {\'E}. and {Dietrich}, J.~P. and {Donath}, A. and
         {Droettboom}, M. and {Earl}, N. and {Erben}, T. and {Fabbro}, S. and
         {Ferreira}, L.~A. and {Finethy}, T. and {Fox}, R.~T. and
         {Garrison}, L.~H. and {Gibbons}, S.~L.~J. and {Goldstein}, D.~A. and
         {Gommers}, R. and {Greco}, J.~P. and {Greenfield}, P. and
         {Groener}, A.~M. and {Grollier}, F. and {Hagen}, A. and {Hirst}, P. and
         {Homeier}, D. and {Horton}, A.~J. and {Hosseinzadeh}, G. and {Hu}, L. and
         {Hunkeler}, J.~S. and {Ivezi{\'c}}, {\v{Z}}. and {Jain}, A. and
         {Jenness}, T. and {Kanarek}, G. and {Kendrew}, S. and {Kern}, N.~S. and
         {Kerzendorf}, W.~E. and {Khvalko}, A. and {King}, J. and {Kirkby}, D. and
         {Kulkarni}, A.~M. and {Kumar}, A. and {Lee}, A. and {Lenz}, D. and
         {Littlefair}, S.~P. and {Ma}, Z. and {Macleod}, D.~M. and
         {Mastropietro}, M. and {McCully}, C. and {Montagnac}, S. and
         {Morris}, B.~M. and {Mueller}, M. and {Mumford}, S.~J. and {Muna}, D. and
         {Murphy}, N.~A. and {Nelson}, S. and {Nguyen}, G.~H. and
         {Ninan}, J.~P. and {N{\"o}the}, M. and {Ogaz}, S. and {Oh}, S. and
         {Parejko}, J.~K. and {Parley}, N. and {Pascual}, S. and {Patil}, R. and
         {Patil}, A.~A. and {Plunkett}, A.~L. and {Prochaska}, J.~X. and
         {Rastogi}, T. and {Reddy Janga}, V. and {Sabater}, J. and
         {Sakurikar}, P. and {Seifert}, M. and {Sherbert}, L.~E. and
         {Sherwood-Taylor}, H. and {Shih}, A.~Y. and {Sick}, J. and
         {Silbiger}, M.~T. and {Singanamalla}, S. and {Singer}, L.~P. and
         {Sladen}, P.~H. and {Sooley}, K.~A. and {Sornarajah}, S. and
         {Streicher}, O. and {Teuben}, P. and {Thomas}, S.~W. and
         {Tremblay}, G.~R. and {Turner}, J.~E.~H. and {Terr{\'o}n}, V. and
         {van Kerkwijk}, M.~H. and {de la Vega}, A. and {Watkins}, L.~L. and
         {Weaver}, B.~A. and {Whitmore}, J.~B. and {Woillez}, J. and
         {Zabalza}, V. and {Astropy Contributors}},
        title = "{The Astropy Project: Building an Open-science Project and Status of the v2.0 Core Package}",
      journal = {\aj},
         year = 2018,
        month = sep,
       volume = {156},
       number = {3},
          eid = {123},
        pages = {123},
          doi = {10.3847/1538-3881/aabc4f},
archivePrefix = {arXiv},
       eprint = {1801.02634},
 primaryClass = {astro-ph.IM},
       adsurl = {https://ui.adsabs.harvard.edu/abs/2018AJ....156..123A}
}

@ARTICLE{astropy:2022,
       author = {{Astropy Collaboration} and {Price-Whelan}, Adrian M. and {Lim}, Pey Lian and {Earl}, Nicholas and {Starkman}, Nathaniel and {Bradley}, Larry and {Shupe}, David L. and {Patil}, Aarya A. and {Corrales}, Lia and {Brasseur}, C.~E. and {N{"o}the}, Maximilian and {Donath}, Axel and {Tollerud}, Erik and {Morris}, Brett M. and {Ginsburg}, Adam and {Vaher}, Eero and {Weaver}, Benjamin A. and {Tocknell}, James and {Jamieson}, William and {van Kerkwijk}, Marten H. and {Robitaille}, Thomas P. and {Merry}, Bruce and {Bachetti}, Matteo and {G{"u}nther}, H. Moritz and {Aldcroft}, Thomas L. and {Alvarado-Montes}, Jaime A. and {Archibald}, Anne M. and {B{'o}di}, Attila and {Bapat}, Shreyas and {Barentsen}, Geert and {Baz{'a}n}, Juanjo and {Biswas}, Manish and {Boquien}, M{'e}d{'e}ric and {Burke}, D.~J. and {Cara}, Daria and {Cara}, Mihai and {Conroy}, Kyle E. and {Conseil}, Simon and {Craig}, Matthew W. and {Cross}, Robert M. and {Cruz}, Kelle L. and {D'Eugenio}, Francesco and {Dencheva}, Nadia and {Devillepoix}, Hadrien A.~R. and {Dietrich}, J{"o}rg P. and {Eigenbrot}, Arthur Davis and {Erben}, Thomas and {Ferreira}, Leonardo and {Foreman-Mackey}, Daniel and {Fox}, Ryan and {Freij}, Nabil and {Garg}, Suyog and {Geda}, Robel and {Glattly}, Lauren and {Gondhalekar}, Yash and {Gordon}, Karl D. and {Grant}, David and {Greenfield}, Perry and {Groener}, Austen M. and {Guest}, Steve and {Gurovich}, Sebastian and {Handberg}, Rasmus and {Hart}, Akeem and {Hatfield-Dodds}, Zac and {Homeier}, Derek and {Hosseinzadeh}, Griffin and {Jenness}, Tim and {Jones}, Craig K. and {Joseph}, Prajwel and {Kalmbach}, J. Bryce and {Karamehmetoglu}, Emir and {Ka{l}uszy{'n}ski}, Miko{l}aj and {Kelley}, Michael S.~P. and {Kern}, Nicholas and {Kerzendorf}, Wolfgang E. and {Koch}, Eric W. and {Kulumani}, Shankar and {Lee}, Antony and {Ly}, Chun and {Ma}, Zhiyuan and {MacBride}, Conor and {Maljaars}, Jakob M. and {Muna}, Demitri and {Murphy}, N.~A. and {Norman}, Henrik and {O'Steen}, Richard and {Oman}, Kyle A. and {Pacifici}, Camilla and {Pascual}, Sergio and {Pascual-Granado}, J. and {Patil}, Rohit R. and {Perren}, Gabriel I. and {Pickering}, Timothy E. and {Rastogi}, Tanuj and {Roulston}, Benjamin R. and {Ryan}, Daniel F. and {Rykoff}, Eli S. and {Sabater}, Jose and {Sakurikar}, Parikshit and {Salgado}, Jes{'u}s and {Sanghi}, Aniket and {Saunders}, Nicholas and {Savchenko}, Volodymyr and {Schwardt}, Ludwig and {Seifert-Eckert}, Michael and {Shih}, Albert Y. and {Jain}, Anany Shrey and {Shukla}, Gyanendra and {Sick}, Jonathan and {Simpson}, Chris and {Singanamalla}, Sudheesh and {Singer}, Leo P. and {Singhal}, Jaladh and {Sinha}, Manodeep and {Sip{H{o}}cz}, Brigitta M. and {Spitler}, Lee R. and {Stansby}, David and {Streicher}, Ole and {{{S}}umak}, Jani and {Swinbank}, John D. and {Taranu}, Dan S. and {Tewary}, Nikita and {Tremblay}, Grant R. and {Val-Borro}, Miguel de and {Van Kooten}, Samuel J. and {Vasovi{'c}}, Zlatan and {Verma}, Shresth and {de Miranda Cardoso}, Jos{'e} Vin{'i}cius and {Williams}, Peter K.~G. and {Wilson}, Tom J. and {Winkel}, Benjamin and {Wood-Vasey}, W.~M. and {Xue}, Rui and {Yoachim}, Peter and {Zhang}, Chen and {Zonca}, Andrea and {Astropy Project Contributors}},
        title = "{The Astropy Project: Sustaining and Growing a Community-oriented Open-source Project and the Latest Major Release (v5.0) of the Core Package}",
      journal = {apj},
         year = 2022,
        month = aug,
       volume = {935},
       number = {2},
          eid = {167},
        pages = {167},
          doi = {10.3847/1538-4357/ac7c74},
archivePrefix = {arXiv},
       eprint = {2206.14220},
 primaryClass = {astro-ph.IM},
       adsurl = {https://ui.adsabs.harvard.edu/abs/2022ApJ...935..167A}
}

@ARTICLE{Banerjee2010,
       author = {{Banerjee}, Sambaran and {Baumgardt}, Holger and {Kroupa}, Pavel},
        title = "{Stellar-mass black holes in star clusters: implications for gravitational wave radiation}",
      journal = {Monthly Notices of the Royal Astronomical Society},
         year = 2010,
        month = feb,
       volume = {402},
       number = {1},
        pages = {371-380},
          doi = {10.1111/j.1365-2966.2009.15880.x},
archivePrefix = {arXiv},
       eprint = {0910.3954},
 primaryClass = {astro-ph.SR},
       adsurl = {https://ui.adsabs.harvard.edu/abs/2010MNRAS.402..371B}
}

@article{Bera:2020jhx,
    author = "Bera, Sayantani and Rana, Divya and More, Surhud and Bose, Sukanta",
    title = "{Incompleteness Matters Not: Inference of $H_0$ from Binary Black Hole\textendash{}Galaxy Cross-correlations}",
    eprint = "2007.04271",
    archivePrefix = "arXiv",
    primaryClass = "astro-ph.CO",
    reportNumber = "LIGO-P2000239-v2",
    doi = "10.3847/1538-4357/abb4e0",
    journal = "Astrophys. J.",
    volume = "902",
    number = "1",
    pages = "79",
    year = "2020"
}

@ARTICLE{Cao2018,
       author = {{Cao}, Liang and {Lu}, Youjun and {Zhao}, Yuetong},
        title = "{Host galaxy properties of mergers of stellar binary black holes and their implications for advanced LIGO gravitational wave sources}",
      journal = {Monthly Notices of the Royal Astronomical Society},
         year = 2018,
        month = mar,
       volume = {474},
       number = {4},
        pages = {4997-5007},
          doi = {10.1093/mnras/stx3087},
archivePrefix = {arXiv},
       eprint = {1711.09190},
 primaryClass = {astro-ph.GA},
       adsurl = {https://ui.adsabs.harvard.edu/abs/2018MNRAS.474.4997C}
}

@article{Chen:2017rfc,
    author = "Chen, Hsin-Yu and Fishbach, Maya and Holz, Daniel E.",
    title = "{A two per cent Hubble constant measurement from standard sirens within five years}",
    eprint = "1712.06531",
    archivePrefix = "arXiv",
    primaryClass = "astro-ph.CO",
    doi = "10.1038/s41586-018-0606-0",
    journal = "Nature",
    volume = "562",
    number = "7728",
    pages = "545--547",
    year = "2018"
}

@article{DES:2019ccw,
    author = "Soares-Santos, M. and others",
    collaboration = "DES, LIGO Scientific, Virgo",
    title = "{First Measurement of the Hubble Constant from a Dark Standard Siren using the Dark Energy Survey Galaxies and the LIGO/Virgo Binary\textendash{}Black-hole Merger GW170814}",
    eprint = "1901.01540",
    archivePrefix = "arXiv",
    primaryClass = "astro-ph.CO",
    reportNumber = "FERMILAB-PUB-18-629-AE",
    doi = "10.3847/2041-8213/ab14f1",
    journal = "Astrophys. J. Lett.",
    volume = "876",
    number = "1",
    pages = "L7",
    year = "2019"
}

@article{Dominik:2012kk,
    author = "Dominik, Michal and Belczynski, Krzysztof and Fryer, Christopher and Holz, Daniel and Berti, Emanuele and Bulik, Tomasz and Mandel, Ilya and O'Shaughnessy, Richard",
    title = "{Double Compact Objects I: The Significance of the Common Envelope on Merger Rates}",
    eprint = "1202.4901",
    archivePrefix = "arXiv",
    primaryClass = "astro-ph.HE",
    doi = "10.1088/0004-637X/759/1/52",
    journal = "Astrophys. J.",
    volume = "759",
    pages = "52",
    year = "2012"
}

@article{Dominik:2014yma,
    author = "Dominik, Michal and Berti, Emanuele and O'Shaughnessy, Richard and Mandel, Ilya and Belczynski, Krzysztof and Fryer, Christopher and Holz, Daniel E. and Bulik, Tomasz and Pannarale, Francesco",
    title = "{Double Compact Objects III: Gravitational Wave Detection Rates}",
    eprint = "1405.7016",
    archivePrefix = "arXiv",
    primaryClass = "astro-ph.HE",
    doi = "10.1088/0004-637X/806/2/263",
    journal = "Astrophys. J.",
    volume = "806",
    number = "2",
    pages = "263",
    year = "2015"
}

@article{Mukherjee:2019wcg,
    author = "Mukherjee, Suvodip and Wandelt, Benjamin D. and Silk, Joseph",
    title = "{Probing the theory of gravity with gravitational lensing of gravitational waves and galaxy surveys}",
    eprint = "1908.08951",
    archivePrefix = "arXiv",
    primaryClass = "astro-ph.CO",
    doi = "10.1093/mnras/staa827",
    journal = "Monthly Notices of the Royal Astronomical Society",
    volume = "494",
    number = "2",
    pages = "1956--1970",
    year = "2020"
}

@article{Afroz:2024joi,
    author = "Afroz, Samsuzzaman and Mukherjee, Suvodip",
    title = "{Prospect of Precision Cosmology and Testing General Relativity using Binary Black Holes- Galaxies Cross-correlation}",
    eprint = "2407.09262",
    archivePrefix = "arXiv",
    primaryClass = "astro-ph.CO",
    month = "7",
    year = "2024"
}

@article{Afroz:2024fzp,
    author = "Afroz, Samsuzzaman and Mukherjee, Suvodip",
    title = "{Phase Space of Binary Black Holes from Gravitational Wave Observations to Unveil its Formation History}",
    eprint = "2411.07304",
    archivePrefix = "arXiv",
    primaryClass = "astro-ph.HE",
    month = "11",
    year = "2024"
}

@article{Mukherjee:2020mha,
    author = "Mukherjee, Suvodip and Wandelt, Benjamin D. and Silk, Joseph",
    title = "{Testing the general theory of relativity using gravitational wave propagation from dark standard sirens}",
    eprint = "2012.15316",
    archivePrefix = "arXiv",
    primaryClass = "astro-ph.CO",
    doi = "10.1093/mnras/stab001",
    journal = "Monthly Notices of the Royal Astronomical Society",
    volume = "502",
    number = "1",
    pages = "1136--1144",
    year = "2021"
}

@article{LIGOScientific:2016aoc,
    author = "Abbott, B. P. and others",
    collaboration = "LIGO Scientific, Virgo",
    title = "{Observation of Gravitational Waves from a Binary Black Hole Merger}",
    eprint = "1602.03837",
    archivePrefix = "arXiv",
    primaryClass = "gr-qc",
    reportNumber = "LIGO-P150914",
    doi = "10.1103/PhysRevLett.116.061102",
    journal = "Phys. Rev. Lett.",
    volume = "116",
    number = "6",
    pages = "061102",
    year = "2016"
}

@ARTICLE{Mannucci2010,
       author = {{Mannucci}, F. and {Cresci}, G. and {Maiolino}, R. and {Marconi}, A. and {Gnerucci}, A.},
        title = "{A fundamental relation between mass, star formation rate and metallicity in local and high-redshift galaxies}",
      journal = {Monthly Notices of the Royal Astronomical Society},
         year = 2010,
        month = nov,
       volume = {408},
       number = {4},
        pages = {2115-2127},
          doi = {10.1111/j.1365-2966.2010.17291.x},
archivePrefix = {arXiv},
       eprint = {1005.0006},
 primaryClass = {astro-ph.CO},
       adsurl = {https://ui.adsabs.harvard.edu/abs/2010MNRAS.408.2115M}
}

@article{Mukherjee:2018ebj,
    author = "Mukherjee, Suvodip and Wandelt, Benjamin D.",
    title = "{Beyond the classical distance-redshift test: cross-correlating redshift-free standard candles and sirens with redshift surveys}",
    eprint = "1808.06615",
    archivePrefix = "arXiv",
    primaryClass = "astro-ph.CO",
    month = "8",
    year = "2018"
}

@article{Libanore22,
   title={Clustering of Gravitational Wave and Supernovae events:  a multitracer analysis  in Luminosity Distance Space},
   volume={2022},
   ISSN={1475-7516},
   url={http://dx.doi.org/10.1088/1475-7516/2022/02/003},
   DOI={10.1088/1475-7516/2022/02/003},
   number={02},
   journal={Journal of Cosmology and Astroparticle Physics},
   publisher={IOP Publishing},
   author={Libanore, S. and Artale, M.C. and Karagiannis, D. and Liguori, M. and Bartolo, N. and Bouffanais, Y. and Mapelli, M. and Matarrese, S.},
   year={2022},
   month=feb, pages={003} }

@article{Mukherjee:2021bmw,
    author = "Mukherjee, Suvodip and Moradinezhad Dizgah, Azadeh",
    title = "{Toward a Precision Measurement of Binary Black Holes Formation Channels Using Gravitational Waves and Emission Lines}",
    eprint = "2111.13166",
    archivePrefix = "arXiv",
    primaryClass = "astro-ph.GA",
    doi = "10.3847/2041-8213/ac903b",
    journal = "Astrophys. J. Lett.",
    volume = "937",
    number = "2",
    pages = "L27",
    year = "2022"
}

@article{Mukherjee:2020hyn,
    author = "Mukherjee, Suvodip and Wandelt, Benjamin D. and Nissanke, Samaya M. and Silvestri, Alessandra",
    title = "{Accurate precision Cosmology with redshift unknown gravitational wave sources}",
    eprint = "2007.02943",
    archivePrefix = "arXiv",
    primaryClass = "astro-ph.CO",
    doi = "10.1103/PhysRevD.103.043520",
    journal = "Phys. Rev. D",
    volume = "103",
    number = "4",
    pages = "043520",
    year = "2021"
}

@article{Mukherjee:2022afz,
    author = "Mukherjee, Suvodip and Krolewski, Alex and Wandelt, Benjamin D. and Silk, Joseph",
    title = "{Cross-correlating dark sirens and galaxies: measurement of $H_0$ from GWTC-3 of LIGO-Virgo-KAGRA}",
    eprint = "2203.03643",
    archivePrefix = "arXiv",
    primaryClass = "astro-ph.CO",
    month = "3",
    year = "2022"
}

@ARTICLE{Neijssel2019,
       author = {{Neijssel}, Coenraad J. and {Vigna-G{\'o}mez}, Alejandro and {Stevenson}, Simon and {Barrett}, Jim W. and {Gaebel}, Sebastian M. and {Broekgaarden}, Floor S. and {de Mink}, Selma E. and {Sz{\'e}csi}, Dorottya and {Vinciguerra}, Serena and {Mandel}, Ilya},
        title = "{The effect of the metallicity-specific star formation history on double compact object mergers}",
      journal = {Monthly Notices of the Royal Astronomical Society},
         year = 2019,
        month = dec,
       volume = {490},
       number = {3},
        pages = {3740-3759},
          doi = {10.1093/mnras/stz2840},
archivePrefix = {arXiv},
       eprint = {1906.08136},
 primaryClass = {astro-ph.SR},
       adsurl = {https://ui.adsabs.harvard.edu/abs/2019MNRAS.490.3740N}
}

@article{Oguri:2016dgk,
    author = "Oguri, Masamune",
    title = "{Measuring the distance-redshift relation with the cross-correlation of gravitational wave standard sirens and galaxies}",
    eprint = "1603.02356",
    archivePrefix = "arXiv",
    primaryClass = "astro-ph.CO",
    doi = "10.1103/PhysRevD.93.083511",
    journal = "Phys. Rev. D",
    volume = "93",
    number = "8",
    pages = "083511",
    year = "2016"
}

@article{LIGOScientific:2017adf,
    author = "Abbott, B. P. and others",
    collaboration = "LIGO Scientific, Virgo, 1M2H, Dark Energy Camera GW-E, DES, DLT40, Las Cumbres Observatory, VINROUGE, MASTER",
    title = "{A gravitational-wave standard siren measurement of the Hubble constant}",
    eprint = "1710.05835",
    archivePrefix = "arXiv",
    primaryClass = "astro-ph.CO",
    reportNumber = "LIGO-P1700296, FERMILAB-PUB-17-472-A-AE",
    doi = "10.1038/nature24471",
    journal = "Nature",
    volume = "551",
    number = "7678",
    pages = "85--88",
    year = "2017"
}

@article{LIGOScientific:2018gmd,
    author = "Fishbach, M. and others",
    collaboration = "LIGO Scientific, Virgo",
    title = "{A Standard Siren Measurement of the Hubble Constant from GW170817 without the Electromagnetic Counterpart}",
    eprint = "1807.05667",
    archivePrefix = "arXiv",
    primaryClass = "astro-ph.CO",
    reportNumber = "LIGO-P1800192",
    doi = "10.3847/2041-8213/aaf96e",
    journal = "Astrophys. J. Lett.",
    volume = "871",
    number = "1",
    pages = "L13",
    year = "2019"
}

@article{LIGOScientific:2021aug,
    author = "Abbott, R. and others",
    collaboration = "LIGO Scientific, Virgo,, KAGRA, VIRGO",
    title = "{Constraints on the Cosmic Expansion History from GWTC\textendash{}3}",
    eprint = "2111.03604",
    archivePrefix = "arXiv",
    primaryClass = "astro-ph.CO",
    reportNumber = "LIGO-P2100185-v6, LIGO-P2100185-v5",
    doi = "10.3847/1538-4357/ac74bb",
    journal = "Astrophys. J.",
    volume = "949",
    number = "2",
    pages = "76",
    year = "2023"
}

@article{Schutz:1986gp,
    author = "Schutz, Bernard F.",
    title = "{Determining the Hubble Constant from Gravitational Wave Observations}",
    doi = "10.1038/323310a0",
    journal = "Nature",
    volume = "323",
    pages = "310--311",
    year = "1986"
}

@article{KAGRA:2021duu,
    author = "Abbott, R. and others",
    collaboration = "KAGRA, VIRGO, LIGO Scientific",
    title = "{Population of Merging Compact Binaries Inferred Using Gravitational Waves through GWTC-3}",
    eprint = "2111.03634",
    archivePrefix = "arXiv",
    primaryClass = "astro-ph.HE",
    reportNumber = "LIGO-P2100239 ; Data release: https://zenodo.org/record/5655785, LIGO-P2100239",
    doi = "10.1103/PhysRevX.13.011048",
    journal = "Phys. Rev. X",
    volume = "13",
    number = "1",
    pages = "011048",
    year = "2023"
}

@ARTICLE{2019ApJ...887...53F,
       author = {{Farmer}, R. and {Renzo}, M. and {de Mink}, S.~E. and {Marchant}, P. and {Justham}, S.},
        title = "{Mind the Gap: The Location of the Lower Edge of the Pair-instability Supernova Black Hole Mass Gap}",
      journal = {\apj},
         year = 2019,
        month = dec,
       volume = {887},
       number = {1},
          eid = {53},
        pages = {53},
          doi = {10.3847/1538-4357/ab518b},
archivePrefix = {arXiv},
       eprint = {1910.12874},
 primaryClass = {astro-ph.SR},
       adsurl = {https://ui.adsabs.harvard.edu/abs/2019ApJ...887...53F}
}

@article{KAGRA:2013rdx,
    author = "Abbott, B. P. and others",
    collaboration = "KAGRA, LIGO Scientific, Virgo",
    title = "{Prospects for observing and localizing gravitational-wave transients with Advanced LIGO, Advanced Virgo and KAGRA}",
    eprint = "1304.0670",
    archivePrefix = "arXiv",
    primaryClass = "gr-qc",
    reportNumber = "LIGO-P1200087, VIR-0288A-12, JGW-P1808427",
    doi = "10.1007/s41114-020-00026-9",
    journal = "Living Rev. Rel.",
    volume = "19",
    pages = "1",
    year = "2016"
}

@article{LIGOScientific:2014pky,
    author = "Aasi, J. and others",
    collaboration = "LIGO Scientific",
    title = "{Advanced LIGO}",
    eprint = "1411.4547",
    archivePrefix = "arXiv",
    primaryClass = "gr-qc",
    doi = "10.1088/0264-9381/32/7/074001",
    journal = "Class. Quant. Grav.",
    volume = "32",
    pages = "074001",
    year = "2015"
}

@article{VIRGO:2014yos,
    author = "Acernese, F. and others",
    collaboration = "VIRGO",
    title = "{Advanced Virgo: a second-generation interferometric gravitational wave detector}",
    eprint = "1408.3978",
    archivePrefix = "arXiv",
    primaryClass = "gr-qc",
    doi = "10.1088/0264-9381/32/2/024001",
    journal = "Class. Quant. Grav.",
    volume = "32",
    number = "2",
    pages = "024001",
    year = "2015"
}

@article{PhysRevD.88.043007,
  title = {Interferometer design of the KAGRA gravitational wave detector},
  author = {Aso, Yoichi and Michimura, Yuta and Somiya, Kentaro and Ando, Masaki and Miyakawa, Osamu and Sekiguchi, Takanori and Tatsumi, Daisuke and Yamamoto, Hiroaki},
  collaboration = {The KAGRA Collaboration},
  journal = {Phys. Rev. D},
  volume = {88},
  issue = {4},
  pages = {043007},
  numpages = {15},
  year = {2013},
  month = {Aug},
  publisher = {American Physical Society},
  doi = {10.1103/PhysRevD.88.043007},
  url = {https://link.aps.org/doi/10.1103/PhysRevD.88.043007}
}

@ARTICLE{2022MNRAS.515.5495M,
       author = {{Mukherjee}, Suvodip},
        title = "{The redshift dependence of black hole mass distribution: is it reliable for standard sirens cosmology?}",
      journal = {Monthly Notices of the Royal Astronomical Society},
         year = 2022,
        month = oct,
       volume = {515},
       number = {4},
        pages = {5495-5505},
          doi = {10.1093/mnras/stac2152},
archivePrefix = {arXiv},
       eprint = {2112.10256},
 primaryClass = {astro-ph.CO},
       adsurl = {https://ui.adsabs.harvard.edu/abs/2022MNRAS.515.5495M}
}

@ARTICLE{2023MNRAS.523.4539K,
       author = {{Karathanasis}, Christos and {Mukherjee}, Suvodip and {Mastrogiovanni}, Simone},
        title = "{Binary black holes population and cosmology in new lights: signature of PISN mass and formation channel in GWTC-3}",
      journal = {Monthly Notices of the Royal Astronomical Society},
         year = 2023,
        month = aug,
       volume = {523},
       number = {3},
        pages = {4539-4555},
          doi = {10.1093/mnras/stad1373},
archivePrefix = {arXiv},
       eprint = {2204.13495},
 primaryClass = {astro-ph.CO},
       adsurl = {https://ui.adsabs.harvard.edu/abs/2023MNRAS.523.4539K}
}

@ARTICLE{2001MNRAS.322..231K,
       author = {{Kroupa}, Pavel},
        title = "{On the variation of the initial mass function}",
      journal = {Monthly Notices of the Royal Astronomical Society},
         year = 2001,
        month = apr,
       volume = {322},
       number = {2},
        pages = {231-246},
          doi = {10.1046/j.1365-8711.2001.04022.x},
archivePrefix = {arXiv},
       eprint = {astro-ph/0009005},
 primaryClass = {astro-ph},
       adsurl = {https://ui.adsabs.harvard.edu/abs/2001MNRAS.322..231K}
}

@ARTICLE{Moster_2010,
       author = {{Moster}, Benjamin P. and {Somerville}, Rachel S. and {Maulbetsch}, Christian and {van den Bosch}, Frank C. and {Macci{\`o}}, Andrea V. and {Naab}, Thorsten and {Oser}, Ludwig},
        title = "{Constraints on the Relationship between Stellar Mass and Halo Mass at Low and High Redshift}",
      journal = {\apj},
         year = 2010,
        month = feb,
       volume = {710},
       number = {2},
        pages = {903-923},
          doi = {10.1088/0004-637X/710/2/903},
archivePrefix = {arXiv},
       eprint = {0903.4682},
 primaryClass = {astro-ph.CO},
       adsurl = {https://ui.adsabs.harvard.edu/abs/2010ApJ...710..903M}
}

@ARTICLE{2023arXiv231203316V,
       author = {{Vijaykumar}, Aditya and {Fishbach}, Maya and {Adhikari}, Susmita and {Holz}, Daniel E.},
        title = "{Inferring host galaxy properties of LIGO-Virgo-KAGRA's black holes}",
      journal = {arXiv e-prints},
         year = 2023,
        month = dec,
          eid = {arXiv:2312.03316},
        pages = {arXiv:2312.03316},
          doi = {10.48550/arXiv.2312.03316},
archivePrefix = {arXiv},
       eprint = {2312.03316},
 primaryClass = {astro-ph.HE},
       adsurl = {https://ui.adsabs.harvard.edu/abs/2023arXiv231203316V}
}

@ARTICLE{2023MNRAS.523.5719R,
       author = {{Rauf}, Liana and {Howlett}, Cullan and {Davis}, Tamara M. and {Lagos}, Claudia D.~P.},
        title = "{Exploring binary black hole mergers and host galaxies with SHARK and COMPAS}",
      journal = {Monthly Notices of the Royal Astronomical Society},
         year = 2023,
        month = aug,
       volume = {523},
       number = {4},
        pages = {5719-5737},
          doi = {10.1093/mnras/stad1757},
archivePrefix = {arXiv},
       eprint = {2302.08172},
 primaryClass = {astro-ph.GA},
       adsurl = {https://ui.adsabs.harvard.edu/abs/2023MNRAS.523.5719R}
}

@ARTICLE{Gagnon23,
       author = {{Gagnon}, E.~L. and {Anbajagane}, D. and {Prat}, J. and {Chang}, C. and {Frieman}, J.},
        title = "{Cosmological Constraints from Combining Galaxy Surveys and Gravitational Wave Observatories}",
      journal = {arXiv e-prints},
         year = 2023,
        month = dec,
          eid = {arXiv:2312.16289},
        pages = {arXiv:2312.16289},
          doi = {10.48550/arXiv.2312.16289},
archivePrefix = {arXiv},
       eprint = {2312.16289},
 primaryClass = {astro-ph.CO},
       adsurl = {https://ui.adsabs.harvard.edu/abs/2023arXiv231216289G}
}

@ARTICLE{Namikawa16,
       author = {{Namikawa}, Toshiya and {Nishizawa}, Atsushi and {Taruya}, Atsushi},
        title = "{Detecting black-hole binary clustering via the second-generation gravitational-wave detectors}",
      journal = {\prd},
         year = 2016,
        month = jul,
       volume = {94},
       number = {2},
          eid = {024013},
        pages = {024013},
          doi = {10.1103/PhysRevD.94.024013},
archivePrefix = {arXiv},
       eprint = {1603.08072},
 primaryClass = {astro-ph.CO},
       adsurl = {https://ui.adsabs.harvard.edu/abs/2016PhRvD..94b4013N}
}

@article{Scelfo20,
    author = "Scelfo, Giulio and Boco, Lumen and Lapi, Andrea and Viel, Matteo",
    title = "{Exploring galaxies-gravitational waves cross-correlations as an astrophysical probe}",
    eprint = "2007.08534",
    archivePrefix = "arXiv",
    primaryClass = "astro-ph.CO",
    doi = "10.1088/1475-7516/2020/10/045",
    journal = "JCAP",
    volume = "10",
    pages = "045",
    year = "2020"
}

@article{Libanore21,
   title={Gravitational Wave mergers as tracers of Large Scale Structures},
   volume={2021},
   ISSN={1475-7516},
   url={http://dx.doi.org/10.1088/1475-7516/2021/02/035},
   DOI={10.1088/1475-7516/2021/02/035},
   number={02},
   journal={Journal of Cosmology and Astroparticle Physics},
   publisher={IOP Publishing},
   author={Libanore, S. and Artale, M. C. and Karagiannis, D. and Liguori, M. and Bartolo, N. and Bouffanais, Y. and Giacobbo, N. and Mapelli, M. and Matarrese, S.},
   year={2021},
   month=feb, pages={035–035} }

@ARTICLE{2024MNRAS.530.1129P,
       author = {{Peron}, Matteo and {Ravenni}, Andrea and {Libanore}, Sarah and {Liguori}, Michele and {Artale}, Maria Celeste},
        title = "{Clustering of binary black hole mergers: a detailed analysis of the EAGLE + MOBSE simulation}",
      journal = {Monthly Notices of the Royal Astronomical Society},
         year = 2024,
        month = may,
       volume = {530},
       number = {1},
        pages = {1129-1143},
          doi = {10.1093/mnras/stae893},
archivePrefix = {arXiv},
       eprint = {2305.18003},
 primaryClass = {astro-ph.CO},
       adsurl = {https://ui.adsabs.harvard.edu/abs/2024MNRAS.530.1129P}
}

@ARTICLE{2020ApJ...893...35D,
       author = {{Doctor}, Z. and {Wysocki}, D. and {O'Shaughnessy}, R. and {Holz}, D.~E. and {Farr}, B.},
        title = "{Black Hole Coagulation: Modeling Hierarchical Mergers in Black Hole Populations}",
      journal = {apj},
         year = 2020,
        month = apr,
       volume = {893},
       number = {1},
          eid = {35},
        pages = {35},
          doi = {10.3847/1538-4357/ab7fac},
archivePrefix = {arXiv},
       eprint = {1911.04424},
 primaryClass = {astro-ph.HE},
       adsurl = {https://ui.adsabs.harvard.edu/abs/2020ApJ...893...35D}
}

@ARTICLE{2017PhRvD..95l4046G,
       author = {{Gerosa}, Davide and {Berti}, Emanuele},
        title = "{Are merging black holes born from stellar collapse or previous mergers?}",
      journal = {prd},
         year = 2017,
        month = jun,
       volume = {95},
       number = {12},
          eid = {124046},
        pages = {124046},
          doi = {10.1103/PhysRevD.95.124046},
archivePrefix = {arXiv},
       eprint = {1703.06223},
 primaryClass = {gr-qc},
       adsurl = {https://ui.adsabs.harvard.edu/abs/2017PhRvD..95l4046G}
}

@ARTICLE{2019PhRvL.123r1101Y,
       author = {{Yang}, Y. and {Bartos}, I. and {Gayathri}, V. and {Ford}, K.~E.~S. and {Haiman}, Z. and {Klimenko}, S. and {Kocsis}, B. and {M{\'a}rka}, S. and {M{\'a}rka}, Z. and {McKernan}, B. and {O'Shaughnessy}, R.},
        title = "{Hierarchical Black Hole Mergers in Active Galactic Nuclei}",
      journal = {prl},
         year = 2019,
        month = nov,
       volume = {123},
       number = {18},
          eid = {181101},
        pages = {181101},
          doi = {10.1103/PhysRevLett.123.181101},
archivePrefix = {arXiv},
       eprint = {1906.09281},
 primaryClass = {astro-ph.HE},
       adsurl = {https://ui.adsabs.harvard.edu/abs/2019PhRvL.123r1101Y}
}

@ARTICLE{2002AJ....124.1810S,
       author = {{Strauss}, Michael A. and {Weinberg}, David H. and {Lupton}, Robert H. and {Narayanan}, Vijay K. and {Annis}, James and {Bernardi}, Mariangela and {Blanton}, Michael and {Burles}, Scott and {Connolly}, A.~J. and {Dalcanton}, Julianne and {Doi}, Mamoru and {Eisenstein}, Daniel and {Frieman}, Joshua A. and {Fukugita}, Masataka and {Gunn}, James E. and {Ivezi{\'c}}, {\v{Z}}eljko and {Kent}, Stephen and {Kim}, Rita S.~J. and {Knapp}, G.~R. and {Kron}, Richard G. and {Munn}, Jeffrey A. and {Newberg}, Heidi Jo and {Nichol}, R.~C. and {Okamura}, Sadanori and {Quinn}, Thomas R. and {Richmond}, Michael W. and {Schlegel}, David J. and {Shimasaku}, Kazuhiro and {SubbaRao}, Mark and {Szalay}, Alexander S. and {Vanden Berk}, Dan and {Vogeley}, Michael S. and {Yanny}, Brian and {Yasuda}, Naoki and {York}, Donald G. and {Zehavi}, Idit},
        title = "{Spectroscopic Target Selection in the Sloan Digital Sky Survey: The Main Galaxy Sample}",
      journal = {\aj},
         year = 2002,
        month = sep,
       volume = {124},
       number = {3},
        pages = {1810-1824},
          doi = {10.1086/342343},
archivePrefix = {arXiv},
       eprint = {astro-ph/0206225},
 primaryClass = {astro-ph},
       adsurl = {https://ui.adsabs.harvard.edu/abs/2002AJ....124.1810S}
}

@ARTICLE{1972ApJ...176....1G,
       author = {{Gunn}, James E. and {Gott}, III, J. Richard},
        title = "{On the Infall of Matter Into Clusters of Galaxies and Some Effects on Their Evolution}",
      journal = {\apj},
         year = 1972,
        month = aug,
       volume = {176},
        pages = {1},
          doi = {10.1086/151605},
       adsurl = {https://ui.adsabs.harvard.edu/abs/1972ApJ...176....1G}
}

@ARTICLE{2001MNRAS.323....1S,
       author = {{Sheth}, Ravi K. and {Mo}, H.~J. and {Tormen}, Giuseppe},
        title = "{Ellipsoidal collapse and an improved model for the number and spatial distribution of dark matter haloes}",
      journal = {Monthly Notices of the Royal Astronomical Society},
         year = 2001,
        month = may,
       volume = {323},
       number = {1},
        pages = {1-12},
          doi = {10.1046/j.1365-8711.2001.04006.x},
archivePrefix = {arXiv},
       eprint = {astro-ph/9907024},
 primaryClass = {astro-ph},
       adsurl = {https://ui.adsabs.harvard.edu/abs/2001MNRAS.323....1S}
}

@ARTICLE{1986ApJ...304...15B,
       author = {{Bardeen}, J.~M. and {Bond}, J.~R. and {Kaiser}, N. and {Szalay}, A.~S.},
        title = "{The Statistics of Peaks of Gaussian Random Fields}",
      journal = {\apj},
         year = 1986,
        month = may,
       volume = {304},
        pages = {15},
          doi = {10.1086/164143},
       adsurl = {https://ui.adsabs.harvard.edu/abs/1986ApJ...304...15B}
}

@ARTICLE{1996MNRAS.282..347M,
       author = {{Mo}, H.~J. and {White}, S.~D.~M.},
        title = "{An analytic model for the spatial clustering of dark matter haloes}",
      journal = {Monthly Notices of the Royal Astronomical Society},
         year = 1996,
        month = sep,
       volume = {282},
       number = {2},
        pages = {347-361},
          doi = {10.1093/mnras/282.2.347},
archivePrefix = {arXiv},
       eprint = {astro-ph/9512127},
 primaryClass = {astro-ph},
       adsurl = {https://ui.adsabs.harvard.edu/abs/1996MNRAS.282..347M}
}

@article{Tinker_2010,
   title={THE LARGE-SCALE BIAS OF DARK MATTER HALOS: NUMERICAL CALIBRATION AND MODEL TESTS},
   volume={724},
   ISSN={1538-4357},
   url={http://dx.doi.org/10.1088/0004-637X/724/2/878},
   DOI={10.1088/0004-637x/724/2/878},
   number={2},
   journal={The Astrophysical Journal},
   publisher={American Astronomical Society},
   author={Tinker, Jeremy L. and Robertson, Brant E. and Kravtsov, Andrey V. and Klypin, Anatoly and Warren, Michael S. and Yepes, Gustavo and Gottlöber, Stefan},
   year={2010},
   month=nov, pages={878–886} }

@article{Benson_2010,
   title={Galaxy formation spanning cosmic history: Galaxy formation spanning cosmic history},
   ISSN={1365-2966},
   url={http://dx.doi.org/10.1111/j.1365-2966.2010.16592.x},
   DOI={10.1111/j.1365-2966.2010.16592.x},
   journal={Monthly Notices of the Royal Astronomical Society},
   publisher={Oxford University Press (OUP)},
   author={Benson, Andrew J. and Bower, Richard},
   year={2010},
   month=apr, pages={no-no} }

@ARTICLE{1977MNRAS.179..541R,
       author = {{Rees}, M.~J. and {Ostriker}, J.~P.},
        title = "{Cooling, dynamics and fragmentation of massive gas clouds: clues to the masses and radii of galaxies and clusters.}",
      journal = {Monthly Notices of the Royal Astronomical Society},
         year = 1977,
        month = jun,
       volume = {179},
        pages = {541-559},
          doi = {10.1093/mnras/179.4.541},
       adsurl = {https://ui.adsabs.harvard.edu/abs/1977MNRAS.179..541R}
}

@ARTICLE{1993MNRAS.264..201K,
       author = {{Kauffmann}, G. and {White}, S.~D.~M. and {Guiderdoni}, B.},
        title = "{The formation and evolution of galaxies within merging dark matter haloes.}",
      journal = {Monthly Notices of the Royal Astronomical Society},
         year = 1993,
        month = sep,
       volume = {264},
        pages = {201-218},
          doi = {10.1093/mnras/264.1.201},
       adsurl = {https://ui.adsabs.harvard.edu/abs/1993MNRAS.264..201K}
}

@ARTICLE{1994MNRAS.271..781C,
       author = {{Cole}, S. and {Aragon-Salamanca}, A. and {Frenk}, C.~S. and {Navarro}, J.~F. and {Zepf}, S.~E.},
        title = "{A recipe for galaxy formation.}",
      journal = {Monthly Notices of the Royal Astronomical Society},
         year = 1994,
        month = dec,
       volume = {271},
        pages = {781-806},
          doi = {10.1093/mnras/271.4.781},
archivePrefix = {arXiv},
       eprint = {astro-ph/9402001},
 primaryClass = {astro-ph},
       adsurl = {https://ui.adsabs.harvard.edu/abs/1994MNRAS.271..781C}
}

@article{Feldman_1994,
   title={Power-spectrum analysis of three-dimensional redshift surveys},
   volume={426},
   ISSN={1538-4357},
   url={http://dx.doi.org/10.1086/174036},
   DOI={10.1086/174036},
   journal={The Astrophysical Journal},
   publisher={American Astronomical Society},
   author={Feldman, Hume A. and Kaiser, Nick and Peacock, John A.},
   year={1994},
   month=may, pages={23} }

@article{Howlett_2015,
   title={The clustering of the SDSS main galaxy sample – II. Mock galaxy catalogues and a measurement of the growth of structure from redshift space distortions at z = 0.15},
   volume={449},
   ISSN={0035-8711},
   url={http://dx.doi.org/10.1093/mnras/stu2693},
   DOI={10.1093/mnras/stu2693},
   number={1},
   journal={Monthly Notices of the Royal Astronomical Society},
   publisher={Oxford University Press (OUP)},
   author={Howlett, Cullan and Ross, Ashley J. and Samushia, Lado and Percival, Will J. and Manera, Marc},
   year={2015},
   month=mar, pages={848–866} }

@article{Ross_2016,
   title={The clustering of galaxies in the completed SDSS-III Baryon Oscillation Spectroscopic Survey: observational systematics and baryon acoustic oscillations in the correlation function},
   volume={464},
   ISSN={1365-2966},
   url={http://dx.doi.org/10.1093/mnras/stw2372},
   DOI={10.1093/mnras/stw2372},
   number={1},
   journal={Monthly Notices of the Royal Astronomical Society},
   publisher={Oxford University Press (OUP)},
   author={Ross, Ashley J. and Beutler, Florian and Chuang, Chia-Hsun and Pellejero-Ibanez, Marcos and Seo, Hee-Jong and Vargas-Magaña, Mariana and Cuesta, Antonio J. and Percival, Will J. and Burden, Angela and Sánchez, Ariel G. and Grieb, Jan Niklas and Reid, Beth and Brownstein, Joel R. and Dawson, Kyle S. and Eisenstein, Daniel J. and Ho, Shirley and Kitaura, Francisco-Shu and Nichol, Robert C. and Olmstead, Matthew D. and Prada, Francisco and Rodríguez-Torres, Sergio A. and Saito, Shun and Salazar-Albornoz, Salvador and Schneider, Donald P. and Thomas, Daniel and Tinker, Jeremy and Tojeiro, Rita and Wang, Yuting and White, Martin and Zhao, Gong-bo},
   year={2016},
   month=sep, pages={1168–1191} }

@ARTICLE{2015ApJS..219....8C,
       author = {{Chang}, Yu-Yen and {van der Wel}, Arjen and {da Cunha}, Elisabete and {Rix}, Hans-Walter},
        title = "{Stellar Masses and Star Formation Rates for 1M Galaxies from SDSS+WISE}",
      journal = {\apjs},
         year = 2015,
        month = jul,
       volume = {219},
       number = {1},
          eid = {8},
        pages = {8},
          doi = {10.1088/0067-0049/219/1/8},
archivePrefix = {arXiv},
       eprint = {1506.00648},
 primaryClass = {astro-ph.GA},
       adsurl = {https://ui.adsabs.harvard.edu/abs/2015ApJS..219....8C}
}

@ARTICLE{1987MNRAS.227....1K,
       author = {{Kaiser}, Nick},
        title = "{Clustering in real space and in redshift space}",
      journal = {Monthly Notices of the Royal Astronomical Society},
         year = 1987,
        month = jul,
       volume = {227},
        pages = {1-21},
          doi = {10.1093/mnras/227.1.1},
       adsurl = {https://ui.adsabs.harvard.edu/abs/1987MNRAS.227....1K}
}

@article{Iacovelli_2022,
   title={Forecasting the Detection Capabilities of Third-generation Gravitational-wave Detectors Using GWFAST},
   volume={941},
   ISSN={1538-4357},
   url={http://dx.doi.org/10.3847/1538-4357/ac9cd4},
   DOI={10.3847/1538-4357/ac9cd4},
   number={2},
   journal={The Astrophysical Journal},
   publisher={American Astronomical Society},
   author={Iacovelli, Francesco and Mancarella, Michele and Foffa, Stefano and Maggiore, Michele},
   year={2022},
   month=dec, pages={208} }

@article{Hild_2011,
   title={Sensitivity studies for third-generation gravitational wave observatories},
   volume={28},
   ISSN={1361-6382},
   url={http://dx.doi.org/10.1088/0264-9381/28/9/094013},
   DOI={10.1088/0264-9381/28/9/094013},
   number={9},
   journal={Classical and Quantum Gravity},
   publisher={IOP Publishing},
   author={Hild, S and Abernathy, M and Acernese, F and Amaro-Seoane, P and Andersson, N and Arun, K and Barone, F and Barr, B and Barsuglia, M and Beker, M and Beveridge, N and Birindelli, S and Bose, S and Bosi, L and Braccini, S and Bradaschia, C and Bulik, T and Calloni, E and Cella, G and Mottin, E Chassande and Chelkowski, S and Chincarini, A and Clark, J and Coccia, E and Colacino, C and Colas, J and Cumming, A and Cunningham, L and Cuoco, E and Danilishin, S and Danzmann, K and De Salvo, R and Dent, T and De Rosa, R and Di Fiore, L and Di Virgilio, A and Doets, M and Fafone, V and Falferi, P and Flaminio, R and Franc, J and Frasconi, F and Freise, A and Friedrich, D and Fulda, P and Gair, J and Gemme, G and Genin, E and Gennai, A and Giazotto, A and Glampedakis, K and Gräf, C and Granata, M and Grote, H and Guidi, G and Gurkovsky, A and Hammond, G and Hannam, M and Harms, J and Heinert, D and Hendry, M and Heng, I and Hennes, E and Hough, J and Husa, S and Huttner, S and Jones, G and Khalili, F and Kokeyama, K and Kokkotas, K and Krishnan, B and Li, T G F and Lorenzini, M and Lück, H and Majorana, E and Mandel, I and Mandic, V and Mantovani, M and Martin, I and Michel, C and Minenkov, Y and Morgado, N and Mosca, S and Mours, B and Müller–Ebhardt, H and Murray, P and Nawrodt, R and Nelson, J and Oshaughnessy, R and Ott, C D and Palomba, C and Paoli, A and Parguez, G and Pasqualetti, A and Passaquieti, R and Passuello, D and Pinard, L and Plastino, W and Poggiani, R and Popolizio, P and Prato, M and Punturo, M and Puppo, P and Rabeling, D and Rapagnani, P and Read, J and Regimbau, T and Rehbein, H and Reid, S and Ricci, F and Richard, F and Rocchi, A and Rowan, S and Rüdiger, A and Santamaría, L and Sassolas, B and Sathyaprakash, B and Schnabel, R and Schwarz, C and Seidel, P and Sintes, A and Somiya, K and Speirits, F and Strain, K and Strigin, S and Sutton, P and Tarabrin, S and Thüring, A and van den Brand, J and van Veggel, M and van den Broeck, C and Vecchio, A and Veitch, J and Vetrano, F and Vicere, A and Vyatchanin, S and Willke, B and Woan, G and Yamamoto, K},
   year={2011},
   month=apr, pages={094013} }

@article{Di_Carlo_2020,
   title={Binary black holes in young star clusters: the impact of metallicity},
   volume={498},
   ISSN={1365-2966},
   url={http://dx.doi.org/10.1093/mnras/staa2286},
   DOI={10.1093/mnras/staa2286},
   number={1},
   journal={Monthly Notices of the Royal Astronomical Society},
   publisher={Oxford University Press (OUP)},
   author={DiCarlo, Ugo N and Mapelli, Michela and Giacobbo, Nicola and Spera, Mario and Bouffanais, Yann and Rastello, Sara and Santoliquido, Filippo and Pasquato, Mario and Ballone, Alessandro and Trani, Alessandro A and Torniamenti, Stefano and Haardt, Francesco},
   year={2020},
   month=aug, pages={495–506} }

@article{Vitale_2019,
   title={Measuring the Star Formation Rate with Gravitational Waves from Binary Black Holes},
   volume={886},
   ISSN={2041-8213},
   url={http://dx.doi.org/10.3847/2041-8213/ab50c0},
   DOI={10.3847/2041-8213/ab50c0},
   number={1},
   journal={The Astrophysical Journal Letters},
   publisher={American Astronomical Society},
   author={Vitale, Salvatore and Farr, Will M. and Ng, Ken K. Y. and Rodriguez, Carl L.},
   year={2019},
   month=nov, pages={L1} }

@article{Salim_2007,
   title={UV Star Formation Rates in the Local Universe},
   volume={173},
   ISSN={1538-4365},
   url={http://dx.doi.org/10.1086/519218},
   DOI={10.1086/519218},
   number={2},
   journal={The Astrophysical Journal Supplement Series},
   publisher={American Astronomical Society},
   author={Salim, Samir and Rich, R. Michael and Charlot, Stephane and Brinchmann, Jarle and Johnson, Benjamin D. and Schiminovich, David and Seibert, Mark and Mallery, Ryan and Heckman, Timothy M. and Forster, Karl and Friedman, Peter G. and Martin, D. Christopher and Morrissey, Patrick and Neff, Susan G. and Small, Todd and Wyder, Ted K. and Bianchi, Luciana and Donas, Jose and Lee, Young‐Wook and Madore, Barry F. and Milliard, Bruno and Szalay, Alex S. and Welsh, Barry Y. and Yi, Sukyoung K.},
   year={2007},
   month=dec, pages={267–292} }

@article{Peng_2010,
   title={MASS AND ENVIRONMENT AS DRIVERS OF GALAXY EVOLUTION IN SDSS AND zCOSMOS AND THE ORIGIN OF THE SCHECHTER FUNCTION},
   volume={721},
   ISSN={1538-4357},
   url={http://dx.doi.org/10.1088/0004-637X/721/1/193},
   DOI={10.1088/0004-637x/721/1/193},
   number={1},
   journal={The Astrophysical Journal},
   publisher={American Astronomical Society},
   author={Peng, Ying-jie and Lilly, Simon J. and Kovač, Katarina and Bolzonella, Micol and Pozzetti, Lucia and Renzini, Alvio and Zamorani, Gianni and Ilbert, Olivier and Knobel, Christian and Iovino, Angela and Maier, Christian and Cucciati, Olga and Tasca, Lidia and Carollo, C. Marcella and Silverman, John and Kampczyk, Pawel and de Ravel, Loic and Sanders, David and Scoville, Nicholas and Contini, Thierry and Mainieri, Vincenzo and Scodeggio, Marco and Kneib, Jean-Paul and Le Fèvre, Olivier and Bardelli, Sandro and Bongiorno, Angela and Caputi, Karina and Coppa, Graziano and de la Torre, Sylvain and Franzetti, Paolo and Garilli, Bianca and Lamareille, Fabrice and Le Borgne, Jean-Francois and Le Brun, Vincent and Mignoli, Marco and Montero, Enrique Perez and Pello, Roser and Ricciardelli, Elena and Tanaka, Masayuki and Tresse, Laurence and Vergani, Daniela and Welikala, Niraj and Zucca, Elena and Oesch, Pascal and Abbas, Ummi and Barnes, Luke and Bordoloi, Rongmon and Bottini, Dario and Cappi, Alberto and Cassata, Paolo and Cimatti, Andrea and Fumana, Marco and Hasinger, Gunther and Koekemoer, Anton and Leauthaud, Alexei and Maccagni, Dario and Marinoni, Christian and McCracken, Henry and Memeo, Pierdomenico and Meneux, Baptiste and Nair, Preethi and Porciani, Cristiano and Presotto, Valentina and Scaramella, Roberto},
   year={2010},
   month=aug, pages={193–221} }

@article{Abbott_2017,
   title={Exploring the sensitivity of next generation gravitational wave
					detectors},
   volume={34},
   ISSN={1361-6382},
   url={http://dx.doi.org/10.1088/1361-6382/aa51f4},
   DOI={10.1088/1361-6382/aa51f4},
   number={4},
   journal={Classical and Quantum Gravity},
   publisher={IOP Publishing},
   author={Abbott, B P and Abbott, R and Abbott, T D and Abernathy, M R and Ackley, K and Adams, C and Addesso, P and Adhikari, R X and Adya, V B and Affeldt, C and Aggarwal, N and Aguiar, O D and Ain, A and Ajith, P and Allen, B and Altin, P A and Anderson, S B and Anderson, W G and Arai, K and Araya, M C and Arceneaux, C C and Areeda, J S and Arun, K G and Ashton, G and Ast, M and Aston, S M and Aufmuth, P and Aulbert, C and Babak, S and Baker, P T and Ballmer, S W and Barayoga, J C and Barclay, S E and Barish, B C and Barker, D and Barr, B and Barsotti, L and Bartlett, J and Bartos, I and Bassiri, R and Batch, J C and Baune, C and Bell, A S and Berger, B K and Bergmann, G and Berry, C P L and Betzwieser, J and Bhagwat, S and Bhandare, R and Bilenko, I A and Billingsley, G and Birch, J and Birney, R and Biscans, S and Bisht, A and Biwer, C and Blackburn, J K and Blair, C D and Blair, D G and Blair, R M and Bock, O and Bogan, C and Bohe, A and Bond, C and Bork, R and Bose, S and Brady, P R and Braginsky, V B and Brau, J E and Brinkmann, M and Brockill, P and Broida, J E and Brooks, A F and Brown, D A and Brown, D D and Brown, N M and Brunett, S and Buchanan, C C and Buikema, A and Buonanno, A and Byer, R L and Cabero, M and Cadonati, L and Cahillane, C and Calderón Bustillo, J and Callister, T and Camp, J B and Cannon, K C and Cao, J and Capano, C D and Caride, S and Caudill, S and Cavaglià, M and Cepeda, C B and Chamberlin, S J and Chan, M and Chao, S and Charlton, P and Cheeseboro, B D and Chen, H Y and Chen, Y and Cheng, C and Cho, H S and Cho, M and Chow, J H and Christensen, N and Chu, Q and Chung, S and Ciani, G and Clara, F and Clark, J A and Collette, C G and Cominsky, L and Constancio, M and Cook, D and Corbitt, T R and Cornish, N and Corsi, A and Costa, C A and Coughlin, M W and Coughlin, S B and Countryman, S T and Couvares, P and Cowan, E E and Coward, D M and Cowart, M J and Coyne, D C and Coyne, R and Craig, K and Creighton, J D E and Cripe, J and Crowder, S G and Cumming, A and Cunningham, L and Dal Canton, T and Danilishin, S L and Danzmann, K and Darman, N S and Dasgupta, A and Da Silva Costa, C F and Dave, I and Davies, G S and Daw, E J and De, S and DeBra, D and Del Pozzo, W and Denker, T and Dent, T and Dergachev, V and DeRosa, R T and DeSalvo, R and Devine, R C and Dhurandhar, S and Díaz, M C and Di Palma, I and Donovan, F and Dooley, K L and Doravari, S and Douglas, R and Downes, T P and Drago, M and Drever, R W P and Driggers, J C and Dwyer, S E and Edo, T B and Edwards, M C and Effler, A and Eggenstein, H-B and Ehrens, P and Eichholz, J and Eikenberry, S S and Engels, W and Essick, R C and Etzel, T and Evans, M and Evans, T M and Everett, R and Factourovich, M and Fair, H and Fairhurst, S and Fan, X and Fang, Q and Farr, B and Farr, W M and Favata, M and Fays, M and Fehrmann, H and Fejer, M M and Fenyvesi, E and Ferreira, E C and Fisher, R P and Fletcher, M and Frei, Z and Freise, A and Frey, R and Fritschel, P and Frolov, V V and Fulda, P and Fyffe, M and Gabbard, H A G and Gair, J R and Gaonkar, S G and Gaur, G and Gehrels, N and Geng, P and George, J and Gergely, L and Ghosh, Abhirup and Ghosh, Archisman and Giaime, J A and Giardina, K D and Gill, K and Glaefke, A and Goetz, E and Goetz, R and Gondan, L and González, G and Gopakumar, A and Gordon, N A and Gorodetsky, M L and Gossan, S E and Graef, C and Graff, P B and Grant, A and Gras, S and Gray, C and Green, A C and Grote, H and Grunewald, S and Guo, X and Gupta, A and Gupta, M K and Gushwa, K E and Gustafson, E K and Gustafson, R and Hacker, J J and Hall, B R and Hall, E D and Hammond, G and Haney, M and Hanke, M M and Hanks, J and Hanna, C and Hannam, M D and Hanson, J and Hardwick, T and Harry, G M and Harry, I W and Hart, M J and Hartman, M T and Haster, C-J and Haughian, K and Heintze, M C and Hendry, M and Heng, I S and Hennig, J and Henry, J and Heptonstall, A W and Heurs, M and Hild, S and Hoak, D and Holt, K and Holz, D E and Hopkins, P and Hough, J and Houston, E A and Howell, E J and Hu, Y M and Huang, S and Huerta, E A and Hughey, B and Husa, S and Huttner, S H and Huynh-Dinh, T and Indik, N and Ingram, D R and Inta, R and Isa, H N and Isi, M and Isogai, T and Iyer, B R and Izumi, K and Jang, H and Jani, K and Jawahar, S and Jian, L and Jiménez-Forteza, F and Johnson, W W and Jones, D I and Jones, R and Ju, L and Haris, K and Kalaghatgi, C V and Kalogera, V and Kandhasamy, S and Kang, G and Kanner, J B and Kapadia, S J and Karki, S and Karvinen, K S and Kasprzack, M and Katsavounidis, E and Katzman, W and Kaufer, S and Kaur, T and Kawabe, K and Kehl, M S and Keitel, D and Kelley, D B and Kells, W and Kennedy, R and Key, J S and Khalili, F Y and Khan, S and Khan, Z and Khazanov, E A and Kijbunchoo, N and Kim, Chi-Woong and Kim, Chunglee and Kim, J and Kim, K and Kim, N and Kim, W and Kim, Y-M and Kimbrell, S J and King, E J and King, P J and Kissel, J S and Klein, B and Kleybolte, L and Klimenko, S and Koehlenbeck, S M and Kondrashov, V and Kontos, A and Korobko, M and Korth, W Z and Kozak, D B and Kringel, V and Krueger, C and Kuehn, G and Kumar, P and Kumar, R and Kuo, L and Lackey, B D and Landry, M and Lange, J and Lantz, B and Lasky, P D and Laxen, M and Lazzarini, A and Leavey, S and Lebigot, E O and Lee, C H and Lee, H K and Lee, H M and Lee, K and Lenon, A and Leong, J R and Levin, Y and Lewis, J B and Li, T G F and Libson, A and Littenberg, T B and Lockerbie, N A and Lombardi, A L and London, L T and Lord, J E and Lormand, M and Lough, J D and Lück, H and Lundgren, A P and Lynch, R and Ma, Y and Machenschalk, B and MacInnis, M and Macleod, D M and Magaña-Sandoval, F and Magaña Zertuche, L and Magee, R M and Mandic, V and Mangano, V and Mansell, G L and Manske, M and Márka, S and Márka, Z and Markosyan, A S and Maros, E and Martin, I W and Martynov, D V and Mason, K and Massinger, T J and Masso-Reid, M and Matichard, F and Matone, L and Mavalvala, N and Mazumder, N and McCarthy, R and McClelland, D E and McCormick, S and McGuire, S C and McIntyre, G and McIver, J and McManus, D J and McRae, T and McWilliams, S T and Meacher, D and Meadors, G D and Melatos, A and Mendell, G and Mercer, R A and Merilh, E L and Meshkov, S and Messenger, C and Messick, C and Meyers, P M and Miao, H and Middleton, H and Mikhailov, E E and Miller, A L and Miller, A and Miller, B B and Miller, J and Millhouse, M and Ming, J and Mirshekari, S and Mishra, C and Mitra, S and Mitrofanov, V P and Mitselmakher, G and Mittleman, R and Mohapatra, S R P and Moore, B C and Moore, C J and Moraru, D and Moreno, G and Morriss, S R and Mossavi, K and Mow-Lowry, C M and Mueller, G and Muir, A W and Mukherjee, Arunava and Mukherjee, D and Mukherjee, S and Mukund, N and Mullavey, A and Munch, J and Murphy, D J and Murray, P G and Mytidis, A and Nayak, R K and Nedkova, K and Nelson, T J N and Neunzert, A and Newton, G and Nguyen, T T and Nielsen, A B and Nitz, A and Nolting, D and Normandin, M E N and Nuttall, L K and Oberling, J and Ochsner, E and O’Dell, J and Oelker, E and Ogin, G H and Oh, J J and Oh, S H and Ohme, F and Oliver, M and Oppermann, P and Oram, Richard J and O’Reilly, B and O’Shaughnessy, R and Ottaway, D J and Overmier, H and Owen, B J and Pai, A and Pai, S A and Palamos, J R and Palashov, O and Pal-Singh, A and Pan, H and Pankow, C and Pannarale, F and Pant, B C and Papa, M A and Paris, H R and Parker, W and Pascucci, D and Patrick, Z and Pearlstone, B L and Pedraza, M and Pekowsky, L and Pele, A and Penn, S and Perreca, A and Perri, L M and Phelps, M and Pierro, V and Pinto, I M and Pitkin, M and Poe, M and Post, A and Powell, J and Prasad, J and Predoi, V and Prestegard, T and Price, L R and Prijatelj, M and Principe, M and Privitera, S and Prokhorov, L and Puncken, O and Pürrer, M and Qi, H and Qin, J and Qiu, S and Quetschke, V and Quintero, E A and Quitzow-James, R and Raab, F J and Rabeling, D S and Radkins, H and Raffai, P and Raja, S and Rajan, C and Rakhmanov, M and Raymond, V and Read, J and Reed, C M and Reid, S and Reitze, D H and Rew, H and Reyes, S D and Riles, K and Rizzo, M and Robertson, N A and Robie, R and Rollins, J G and Roma, V J and Romanov, G and Romie, J H and Rowan, S and Rüdiger, A and Ryan, K and Sachdev, S and Sadecki, T and Sadeghian, L and Sakellariadou, M and Saleem, M and Salemi, F and Samajdar, A and Sammut, L and Sanchez, E J and Sandberg, V and Sandeen, B and Sanders, J R and Sathyaprakash, B S and Saulson, P R and Sauter, O E S and Savage, R L and Sawadsky, A and Schale, P and Schilling, R and Schmidt, J and Schmidt, P and Schnabel, R and Schofield, R M S and Schönbeck, A and Schreiber, E and Schuette, D and Schutz, B F and Scott, J and Scott, S M and Sellers, D and Sengupta, A S and Sergeev, A and Shaddock, D A and Shaffer, T and Shahriar, M S and Shaltev, M and Shapiro, B and Shawhan, P and Sheperd, A and Shoemaker, D H and Shoemaker, D M and Siellez, K and Siemens, X and Sigg, D and Silva, A D and Singer, A and Singer, L P and Singh, A and Singh, R and Sintes, A M and Slagmolen, B J J and Smith, J R and Smith, N D and Smith, R J E and Son, E J and Sorazu, B and Souradeep, T and Srivastava, A K and Staley, A and Steinke, M and Steinlechner, J and Steinlechner, S and Steinmeyer, D and Stephens, B C and Stone, R and Strain, K A and Strauss, N A and Strigin, S and Sturani, R and Stuver, A L and Summerscales, T Z and Sun, L and Sunil, S and Sutton, P J and Szczepańczyk, M J and Talukder, D and Tanner, D B and Tápai, M and Tarabrin, S P and Taracchini, A and Taylor, R and Theeg, T and Thirugnanasambandam, M P and Thomas, E G and Thomas, M and Thomas, P and Thorne, K A and Thrane, E and Tiwari, V and Tokmakov, K V and Toland, K and Tomlinson, C and Tornasi, Z and Torres, C V and Torrie, C I and Töyrä, D and Traylor, G and Trifirò, D and Tse, M and Tuyenbayev, D and Ugolini, D and Unnikrishnan, C S and Urban, A L and Usman, S A and Vahlbruch, H and Vajente, G and Valdes, G and Vander-Hyde, D C and van Veggel, A A and Vass, S and Vaulin, R and Vecchio, A and Veitch, J and Veitch, P J and Venkateswara, K and Vinciguerra, S and Vine, D J and Vitale, S and Vo, T and Vorvick, C and Voss, D V and Vousden, W D and Vyatchanin, S P and Wade, A R and Wade, L E and Wade, M and Walker, M and Wallace, L and Walsh, S and Wang, H and Wang, M and Wang, X and Wang, Y and Ward, R L and Warner, J and Weaver, B and Weinert, M and Weinstein, A J and Weiss, R and Wen, L and Weßels, P and Westphal, T and Wette, K and Whelan, J T and Whiting, B F and Williams, R D and Williamson, A R and Willis, J L and Willke, B and Wimmer, M H and Winkler, W and Wipf, C C and Wittel, H and Woan, G and Woehler, J and Worden, J and Wright, J L and Wu, D S and Wu, G and Yablon, J and Yam, W and Yamamoto, H and Yancey, C C and Yu, H and Zanolin, M and Zevin, M and Zhang, L and Zhang, M and Zhang, Y and Zhao, C and Zhou, M and Zhou, Z and Zhu, X J and Zucker, M E and Zuraw, S E and Zweizig, J and Harms, J},
   year={2017},
   month=jan, pages={044001} }

@article{Maggiore_2020,
   title={Science case for the Einstein telescope},
   volume={2020},
   ISSN={1475-7516},
   url={http://dx.doi.org/10.1088/1475-7516/2020/03/050},
   DOI={10.1088/1475-7516/2020/03/050},
   number={03},
   journal={Journal of Cosmology and Astroparticle Physics},
   publisher={IOP Publishing},
   author={Maggiore, Michele and Broeck, Chris Van Den and Bartolo, Nicola and Belgacem, Enis and Bertacca, Daniele and Bizouard, Marie Anne and Branchesi, Marica and Clesse, Sebastien and Foffa, Stefano and García-Bellido, Juan and Grimm, Stefan and Harms, Jan and Hinderer, Tanja and Matarrese, Sabino and Palomba, Cristiano and Peloso, Marco and Ricciardone, Angelo and Sakellariadou, Mairi},
   year={2020},
   month=mar, pages={050–050} }

@article{McClintock:2003gx,
    author = "McClintock, Jeffrey. E. and Remillard, Ronald. A.",
    title = "{Black hole binaries}",
    eprint = "astro-ph/0306213",
    archivePrefix = "arXiv",
    month = "6",
    year = "2003"
}

@article{Zazzera:2025ord,
    author = "Zazzera, Stefano and Fonseca, Jos\'e and Baker, Tessa and Clarkson, Chris",
    title = "{Exploring future synergies for large-scale structure between gravitational waves and radio sources}",
    eprint = "2505.15645",
    archivePrefix = "arXiv",
    primaryClass = "astro-ph.CO",
    month = "5",
    year = "2025"
}

@ARTICLE{2022NatAs...6..897S,
       author = {{Sales}, Laura V. and {Wetzel}, Andrew and {Fattahi}, Azadeh},
        title = "{Baryonic solutions and challenges for cosmological models of dwarf galaxies}",
      journal = {Nature Astronomy},
         year = 2022,
        month = jun,
       volume = {6},
        pages = {897-910},
          doi = {10.1038/s41550-022-01689-w},
archivePrefix = {arXiv},
       eprint = {2206.05295},
 primaryClass = {astro-ph.GA},
       adsurl = {https://ui.adsabs.harvard.edu/abs/2022NatAs...6..897S}
}

@article{Madau:1996hu,
    author = "Madau, Piero",
    editor = "Holt, Stephen S. and Mundy, Lee G.",
    title = "{Cosmic star formation history}",
    eprint = "astro-ph/9612157",
    archivePrefix = "arXiv",
    doi = "10.1063/1.52821",
    journal = "AIP Conf. Proc.",
    volume = "393",
    number = "1",
    pages = "481",
    year = "1997"
}

@ARTICLE{2003MNRAS.341...33K,
       author = {{Kauffmann}, Guinevere and {Heckman}, Timothy M. and {White}, Simon D.~M. and {Charlot}, St{\'e}phane and {Tremonti}, Christy and {Brinchmann}, Jarle and {Bruzual}, Gustavo and {Peng}, Eric W. and {Seibert}, Mark and {Bernardi}, Mariangela and {Blanton}, Michael and {Brinkmann}, Jon and {Castander}, Francisco and {Cs{\'a}bai}, Istvan and {Fukugita}, Masataka and {Ivezic}, Zeljko and {Munn}, Jeffrey A. and {Nichol}, Robert C. and {Padmanabhan}, Nikhil and {Thakar}, Aniruddha R. and {Weinberg}, David H. and {York}, Donald},
        title = "{Stellar masses and star formation histories for {}10$^{5}$ galaxies from the Sloan Digital Sky Survey}",
      journal = {\mnras},
         year = 2003,
        month = may,
       volume = {341},
       number = {1},
        pages = {33-53},
          doi = {10.1046/j.1365-8711.2003.06291.x},
archivePrefix = {arXiv},
       eprint = {astro-ph/0204055},
 primaryClass = {astro-ph},
       adsurl = {https://ui.adsabs.harvard.edu/abs/2003MNRAS.341...33K}
}

\end{document}